%% file: main.tex
\documentclass[journal]{IEEEtran}
\IEEEoverridecommandlockouts
\usepackage[font={footnotesize}]{caption}
\usepackage[font={footnotesize}]{subcaption}
\usepackage{amsmath}
\usepackage{algpseudocode}
\usepackage[ruled,vlined]{algorithm2e}
\SetKwInput{KwInput}{Input}                
\SetKwInput{KwOutput}{Output}    
\SetKwInput{KwInit}{Initialize} 
\SetKwInput{KWstop}{stop} 
\usepackage{svg}
\usepackage{amsfonts} 
\usepackage{booktabs} 
\usepackage{booktabs}
\usepackage{pgfplots}
\usepackage[utf8]{inputenc}
\usetikzlibrary{patterns}
\usepgfplotslibrary{groupplots} 
\usetikzlibrary{pgfplots.groupplots} 
\usepackage{multirow}
\usepackage{enumerate}
\usepackage{enumitem}
\usepackage{xcolor}

\usepackage{graphicx}

\usepackage{url}
\usepackage{pgfplots}
\pgfplotsset{compat=newest} 

\usepackage{graphicx}
\usepackage{tikz}

\usetikzlibrary{external}
\usepackage{scalerel}
\usetikzlibrary{svg.path}

\definecolor{orcidlogocol}{HTML}{A6CE39}
\tikzset{
  orcidlogo/.pic={
    \fill[orcidlogocol] svg{M256,128c0,70.7-57.3,128-128,128C57.3,256,0,198.7,0,128C0,57.3,57.3,0,128,0C198.7,0,256,57.3,256,128z};
    \fill[white] svg{M86.3,186.2H70.9V79.1h15.4v48.4V186.2z}
                 svg{M108.9,79.1h41.6c39.6,0,57,28.3,57,53.6c0,27.5-21.5,53.6-56.8,53.6h-41.8V79.1z M124.3,172.4h24.5c34.9,0,42.9-26.5,42.9-39.7c0-21.5-13.7-39.7-43.7-39.7h-23.7V172.4z}
                 svg{M88.7,56.8c0,5.5-4.5,10.1-10.1,10.1c-5.6,0-10.1-4.6-10.1-10.1c0-5.6,4.5-10.1,10.1-10.1C84.2,46.7,88.7,51.3,88.7,56.8z};
  }
}

\newcommand\orcidicon[1]{\href{https://orcid.org/#1}{\mbox{\scalerel*{
\begin{tikzpicture}[yscale=-1,transform shape]
\pic{orcidlogo};
\end{tikzpicture}
}{|}}}}
\usepackage[hidelinks]{hyperref}

\begin{document}
\bstctlcite{IEEEexample:BSTcontrol}

\title{Assessing Maintenance of Medium Voltage Cable Networks Under Time-Varying Loading}

\author{\IEEEauthorblockN{Jochen~Lorenz~Cremer~\orcidicon{0000-0001-9284-5083} 
}
\thanks{This research was funded by the Dutch Research Council, Veni Talent Program Grant \#19161 and the Delft AI Initiative Program. Jochen Cremer is with the Department of Electrical Sustainable Energy at Delft University of Technology, the Netherlands and with the Centre for Energy, Austrian Institute of Technology, Austria. 
Corresponding author: Jochen L. Cremer (email: j.l.cremer@tudelft.nl; web: www.jochen-cremer.com).
}
} 

\maketitle
\markboth{IEEE Transactions on Power Delivery}{}

\begin{abstract}
\textcolor{black}{The electrification and ongoing energy transition lead to systematic changes in electricity loading and variability in power systems.} 
Distribution systems were designed for regular operating patterns, assuming constant low loading. Now, operators need to assess whether their assets can withstand more, as well as time-varying loading. Operating the system at or near its \textcolor{black}{ampacity} potentially accelerates thermal ageing, so the question arises: 'how much can one operate at the limits while keeping maintenance and failures low?' 
\textcolor{black}{This paper introduces a novel approach that derives a time-varying Weibull approximation of failure rates using thermal models and provides a shortcut method to quantify maintenance implications under time-varying loading for heterogeneous MV cable populations.}

The case studies investigate a dataset from Denmark and the Oberrhein Medium Voltage (MV) system in Germany, studying ageing assets and the interplay with loading, and replacement paradigms of two different cable insulation types. The studies demonstrate that a small fraction of $25\%$ of old, low-quality cables leads to $82\%$ of failures, and $1.4\%$ of the time of highest loading can cause $46\%$ of cable ageing. The case studies also demonstrate that maintenance needs may be between $10-300$ times higher \textcolor{black}{under future loading conditions associated with the energy transition,}
specifically in networks that have older PILC cables. This paper provides a new tool for operators to plan maintenance under more realistic, future operating conditions. 
\end{abstract}


\section{Introduction} \label{sec:int}
A strategy to decarbonise our economy is the electrification of the private and industrial sectors. Electrification requires operating the distribution and transmission systems more smartly. Smarter tooling ultimately enables Distribution System Operators (DSOs) to operate existing assets at higher capacity, possibly surpassing specified power ratings for short periods. These operating ratings typically ensure the reliability of these assets, providing an expectation of their lifespan \cite{Bil92}. DSOs use these estimations of lifespan in their maintenance plans. \textcolor{black}{However, electrification and renewable integration associated with the energy transition alter electricity operating conditions and increase system loading.}
The actual loadings may surpass even the rated loading, reducing the lifespan of the overloaded assets \cite{Zap21}. The lifespans may currently be overestimated at scale. 

Overestimating the lifespan potentially leads to a large maintenance burden in the future, and in the worst case drastic increase in failures of power network assets, as it can already be experienced in other sectors, like railway and road transportation. The loading of the German railway system increased, and an aged infrastructure (among other reasons) led to more frequent technical failures due to asset fatigue (e.g. \cite{Gro21}). Roads designed decades ago experience types of vehicles (e.g. SUVs) that are heavier than vehicles assumed at the road network's design. Heavier vehicles lead to fast degradation of pavements ($\textrm{degradation} \propto \textrm{(vehicle weight)}^4$ \cite{Gup14}) and to more frequent required maintenance of pavements, as one can observe in many countries. 


Past research statistically analysed the reliability of MV cables, and experiments investigated the impact of loading on ageing. \textcolor{black}{Cable failures account for a major share of MV network faults \cite{Han19} and are typically analysed as a function of service time to support maintenance planning \cite{Han13}.}
Failure rates increase with service time as for most physical assets, such as cables. However, most failure models assume loading does not change over time, i.e., no systematic increase or decrease over years. 
\textcolor{black}{MV networks commonly include XLPE and PILC cables, where ageing PILC assets, often exceeding 50 years of service, contribute disproportionately to failure rates and reinvestment needs.}
In MV networks, the need for reinvestment to replace these old cables is well recognised \cite{Han19}\textcolor{black}{, and a drastic increase in age-related failure rates in PILC cables can currently be observed in Denmark \cite{Sun25,Han23, Den23}. Increasing the loading across the network in the upcoming years through electrification may reduce the lifespan and increase failure rates at scale. The realisation of distributed energy resources (DERs) increases the peaks of loadings \cite{Zap21}, which specifically reduces cable lifetime. This research aims to estimate failure rates under varying loading conditions over time in a distribution of aged cables. }

\textcolor{black}{Planning for the maintenance of MV cables \cite{Raf24} often involves either applying monitoring to the cables or assuming static loading conditions, which does not require monitoring devices.} Parameters that impact ageing and failure rates are load estimates \cite{Her16} and average temperature \cite{Kle19}, among others. The temperature increases with time when the loading increases. Prolonged high cable temperatures lead to faster degradation \textcolor{black}{ ('overheating'). Currently, performing preventive maintenance requires the early identification and localization of overheated sections using power line communication, applying local measurement devices \cite{Bin23a, Bin23b}. If overheated sections are repeatedly identified, the operator acts to replace the specific line or implement other mitigation actions. Recent work also explores data-driven condition monitoring frameworks using machine learning to estimate cable health indices from multi-source inspection and operational data, enabling predictive and automated asset management strategies \cite{Abd26}.} 
The smart cable guard system \cite{Ste14} can monitor the condition of a cable, enabling root cause analysis to prevent failures. The temperatures of cables can be accurately estimated, given the loading \cite{Rie21}; however, these systems can not estimate the lifespan of a single cable or the population of cables, which remains a challenging task.
Through experiments, \cite{Alg20} assesses the relationship between ageing and loading above rated temperature, showing that short-term emergency loading and frequent overloads lead to a thermal memory effect. This effect leads to higher failure rates. For example, \cite{Alg20} states that raising the rated operating temperature from $90$$^{\circ}$C to $95-105$$^{\circ}$C significantly reduces the lifespans of XLPE cables from $40-60$ to $7-30$ years. However, experiments show exposing old $110$kV XLPE cables at the \textcolor{black}{ampacity} $900$A leads to exceeding temperatures $90$$^{\circ}$C after $5$h-$8$h \cite{Qin22}. For PILC cables, \cite{Wei10} shows a clear dependency of the p-factor on the cable age and history, where cables of ages $0-60$ were tested \cite{Mla12}. The Montsinger approach \cite{Mon30} and Arrhenius approach can estimate the lifetime as a function of the stressor (temperature) \cite{Zap21}. The accelerated failure time model assumes accelerated or decelerated ageing relative to a reference, and can predict lifetime and failure rates for varying operating conditions \cite{Wei92}. \textcolor{black}{Additionally, probabilistic prognostic approaches based on dynamic state transition models and Bayesian filtering have been proposed to predict insulation degradation trajectories and remaining useful life under varying loading and environmental conditions \cite{Sid25}.} Frequently, for reliability studies, the failure rate is modelled with a Weibull distribution, providing statistical tools to estimate lifetimes and failure rates. The Common Network Asset Indices Methodology (CNAIM) estimates lifespan based on individual asset data \cite{Ofg21}. These estimations of the lifespan and failure rates for a population of cables with varying ages were possible in the past, as the loading levels remained fairly constant over the years. However, currently, the failure rate is not considered to change with time as the loading patterns change at scale, possibly sometimes exceeding rated temperatures and accelerating ageing. Not estimating the lifespan and failure rate, considering projections of future loading conditions in the electrification can potentially lead to high risks for society, underestimating costs for maintenance and cable replacements. 

\begin{figure}
\centering
    \input{failden.tex}
\caption{Failures of medium voltage cables in Denmark. The increase is due to age-related failure (applied from \cite{Sun25,Han23, Den23}).\vspace{-1em}}
    \label{fig:failurestats1}
\end{figure}
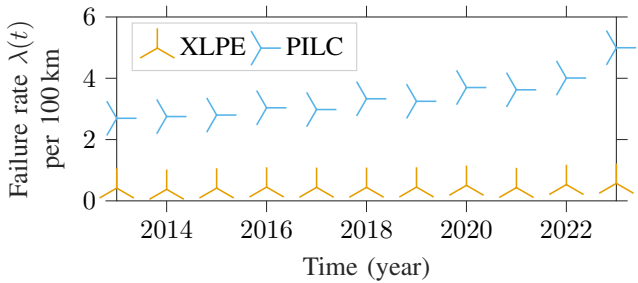

\textcolor{black}{Existing reliability models typically assume stationary loading conditions and therefore do not explicitly capture systematic transitions in loading distributions occurring during the energy transition. Consequently, current approaches cannot fully represent the interaction between thermally driven ageing processes and long-term structural changes in operating conditions. This paper introduces a general approach linking thermal ageing physics with non-stationary reliability theory, allowing analytical estimation of population-level failure evolution under load transition scenarios. More specifically, this paper develops methods to help assess the risk of overestimating expected lifespan and maintenance under loading changes associated with the energy transition.} This paper focuses on a systematic change in the distribution of loading levels toward higher loading levels. Specifically, this paper contributes by

\begin{enumerate}[label=(\roman*)]
  \item method to estimate effective ageing for a distribution of cables with varying age and loading exposure over time
  \item shortcut method to assess maintenance increase at the limits of systemic loading changes for a population of cables. 
        The method is developed for \textit{Run-to-failure}, \textit{Replace-all}, and \textit{Preventive} maintenance strategies.
\end{enumerate}

Case studies focus on power systems with PILC and XLPE MV cables. The studies apply asset data from Danish distribution grids and the Oberrhein MV system in Germany. The studies investigate the impact of thermal ageing on failure rates by applying the Arrhenius and Montsinger models. Then, the studies investigate the impact of time-varying loading \textcolor{black}{associated with energy-transition scenarios}
on a network with diverse cable ages. Studies then focus on the proposed shortcut method to assess three maintenance approaches, comparing the states before and after the energy transition. 

This paper is structured as follows. Sec. \ref{sec:relcable} investigates the reliability of MV cables, developing the methodology for ageing under time-varying loading. Sec. \ref{sec:main} develops the shortcut methods to investigate the three maintenance strategies. Sec. \ref {sec:case} is the case study, and Sec. \ref{sec:cond} concludes. 




\section{Reliability of MV cables} \label{sec:relcable}
\begin{figure}
\centering
        \input{loads.tex}
   \caption{The annual cable utilisation for three energy scenarios (applied from \textcolor{black}{Fig. 10 in} \cite{Zap21}). \vspace{-1em}}
\label{fig:loads}
\end{figure}
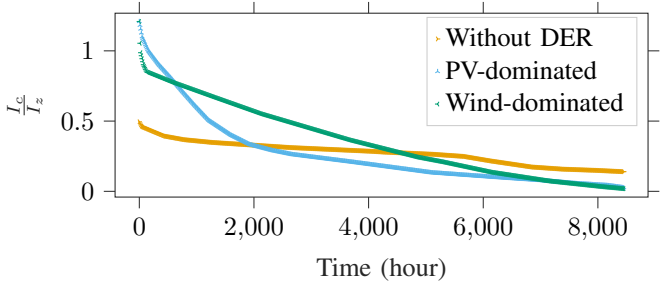

\textcolor{black}{A recent statistic in Fig. \ref{fig:failurestats1} shows a drastic increase in age-related failure rates over the last four years in PILC cables \cite{Sun25, Han23}. These statistics are typically analysed as a function of service time and often do not account for the exposure of cables to different loading levels. However, cables located in parts of the network experiencing higher loading may fail earlier than those operated at lower loading levels. As shown in \cite{Zap21}, even if the average utilisation does not increase or increases only slightly, the expansion of DERs can lead to short periods of high cable utilisation (see Fig. \ref{fig:loads}), which can significantly reduce cable lifetimes. \\
We consider a medium voltage network that} has a set of cables $\Omega^C$ that are considered with a unit length. 
$\rho_A(a)$ is the probability density function of cable service times ('age' $a$). We consider ageing-related failures with the Weibull rate
\begin{equation}\label{eq:Weibull}
\lambda(a) = \frac{\beta}{\hat{\eta}_r}\left(\frac{a}{\hat{\eta}_r}\right)^{\beta-1}
\end{equation}
where $\beta>1$ and the characteristic lifetime $\hat{\eta}_r>1$. {At $a=\hat{\eta}_r$, $63.2\%$ of assets failed.} The expected failures per year are 
\begin{equation} \label{eq:failurerate}
F = |\Omega^C| ~ \int_{0}^{{\infty}} \lambda({a})~ \rho_A({a})\, d{a}
\end{equation}
where $|\Omega^C|$ is the cardinality of the set $\Omega^C$.


Operational conditions impact the service life of medium voltage cables, specifically when exceeding \textcolor{black}{ampacitie}s. As Fig. \ref{fig:loads} shows, \textcolor{black}{ampacitie}s may more frequently be exceeded in grids with DERs than without DERs \cite{Zap21}. However, most maintenance approaches assume constant operation ($\frac{d\lambda(a)}{dt} = 0$) from when the grid was designed without DERs. However, this assumption no longer holds. Failing to consider changes in operational maintenance planning may increase failure rates and lead to higher maintenance needs. 

In response, this paper aims to estimate the number of cable replacements $N(t)$ required to maintain service reliability ($\frac{dF}{dt} = 0$) \textcolor{black}{under evolving loading conditions}. The grid undergoes \textcolor{black}{the transition} from $t_0$ (without DER) to $t_1$ (with wind and PV DERs). We introduce the distribution of loading $\rho_T(t)$ that changes with time $t$, and derive the Weibull failure rate $\lambda(a,t)$ for an age distribution $\rho_A(a)$ of cables. 
As we demonstrate, this extension can be used to estimate
\begin{itemize}
    \item the maintenance increase $\frac{N(t_1)}{N(t_0)}$ to keep reliability constant ($\frac{dF}{dt} = 0$)
    \item the failure rate increase $\frac{F(t_1)}{F(t_0)}$ if the maintenance efforts are not adjusted $N(t_1)=N(t_0)$
\end{itemize}

\subsection{Thermal Ageing of MV cables}

We exemplarily focus on failures caused by thermal ageing, which is the main ageing mechanism for PILC cables, and thermal ageing increases as the load increases. Nonetheless, studies show that failures of cable joints cause $45$ \% MV cable failures \cite{Zap21,Chm15}. These are not investigated in this work. We consider the insulation material as the cause of failure. The conductor temperature $T$ can be estimated with
\begin{equation}\label{eq:conductemp}
    T(\tau) = T_a
    + (T_{max}-T_a
    ) \cdot \left(\frac{|I_c(\tau)|}{I_z}\right)^2 
\end{equation}
that is a function of the conductor current $I_c(\tau)$  at time $\tau$, maximal temperature $T_{max}$, and the rated current $I_z$ for the ambient temperature $T_a$ \cite{Bot22}. \eqref{eq:conductemp} does not consider that the temperature depends on the electrical resistance. \textcolor{black}{Neglecting temperature-dependent electrical resistance introduces a modelling approximation whose impact increases with loading levels. However, for moderate loading ranges typically below ampacity, the approximation provides a tractable analytical representation of temperature dynamics and remains widely applied in thermal rating studies. The sensitivity of this approximation is evaluated in Sec. \ref{sec:csage} and compared with the extended formulation \eqref{eq:conductempepsilon} provided in the Appendix.}
The conductor temperature can also be estimated using real-time temperature rating \cite{Rie21}. The Arrhenius model estimates the thermal ageing acceleration factor as
\begin{equation}\label{eq:XLPEageing}
 r(T) = \frac{\eta_r}{\eta} = \exp{B\left(\frac{1}{T_r}-\frac{1}{T}\right)}
\end{equation}
where the mean lifetime $\eta_r$ and $B>0$ is assumed to be experimentally recorded at the reference temperature $T_r$. The Arrhenius model can be applied to the insulation material in XLPE cables. Montsinger's approach can be applied to estimate the lifetime of insulation material in PILC cables \cite{Mon30}. The Montsinger ageing acceleration factor 
\begin{equation} \label{eq:PILCageing}
 r(T) = \frac{\eta_r}{\eta}= 2^{\frac{T-T_r}{\Delta T}}
\end{equation}
is doubled for increasing increments $\Delta T>0$  of temperature $T$ \cite{Bot22}. \textcolor{black}{The Montsinger model is an empirical approximation derived from accelerated ageing experiments and is primarily valid when thermal degradation is the dominant ageing mechanism. Its applicability may be limited in field conditions where additional factors such as moisture ingress, mechanical stress, or installation defects significantly influence insulation ageing. In this work, the model is applied to represent thermally driven ageing at the system level.} The effective age $A(a,t)$ for a cable that is $a(t)=t-t_0$ years in service at time $t$ is
\begin{equation} \label{eq:effectiveage}
 A(a,t) = \int_{t-a}^{t}r(T(\tau))d\tau
\end{equation}
considering the temperature trajectory $T(t)$. $t_0$ is the installation time of a cable.

The cumulative Weibull hazard with effective age $u$ is
\begin{equation} 
H(u) = \left(\frac{u}{\hat{\eta}_r}\right)^{\beta}
\end{equation}
with the characteristic age $\hat{\eta}_r=\frac{\eta_r}{\Gamma(1+\frac{1}{\beta})}$, i.e. $\eta_r$ is the mean life and $\hat{\eta}_r$ is for the Weibull scale.

We can derive the failure rate as the change of hazard in time, i.e. 
\vspace{-1em}
\begin{equation}\label{eq:genericfailure}
\lambda(a,t) = \frac{d}{dt}H(A(a,t))
           = \left(\frac{\beta}{\hat{\eta}_r}\right)\left(\frac{A(a,t)}{\hat{\eta}_r}\right)^{\beta-1} r(T(t))
\end{equation}
where $u= A(a,t)$. \textcolor{black}{The formulation in \eqref{eq:genericfailure} provides a conceptual extension of classical Weibull reliability modelling by embedding time-dependent stress histories into the hazard formulation through the effective age functional $A(a,t)$. Unlike conventional approaches that consider ageing solely as a function of service time, this approach enables reliability modelling under temporally evolving stress environments while preserving analytical tractability. This formulation therefore establishes a direct connection between physics-of-failure ageing models and population-level survival analysis under non-stationary operating conditions.}
\vspace{-1em}
\subsection{Constant loading}
When assuming a constant loading $I_c(t)=I_c$ and temperature $T(t)=T$, the effective age is $A(a,t)=a \frac{\eta_r}{\eta(T)}$, and the failure rate of the Arrhenius-Weibull model simplifies to
\begin{equation}\label{eq:WeibullArrhenius}
\lambda(a,T) = \frac{\beta}{\hat{\eta}(T)}\left(\frac{a}{\hat{\eta}(T)}\right)^{\beta-1}
\end{equation}
The characteristic lifespan $\hat{\eta}(T)$ can be estimated at a constant temperature $T$ as
\vspace{-1em}
\begin{equation} \label{eq:lifeconstant}
 \hat{\eta}(T) = \frac{\hat{\eta}_r}{r(T)}
\end{equation}
and applying \eqref{eq:XLPEageing} or \eqref{eq:PILCageing} for XLPE or PILC, respectively.

\vspace{-1em}
\subsection{Time-varying loading distribution}






Analogue to the Miner's cumulative damage model \cite{Min45,Kli22}, we consider that different stress levels weighted by their cycle times lead to failure. \textcolor{black}{This formulation assumes cumulative damage accumulation and does not explicitly model potential thermal recovery or relaxation effects during periods of reduced loading. This assumption is consistent with commonly applied accelerated ageing and Miner-type damage accumulation approaches in insulation reliability modelling, where degradation is primarily driven by peak thermal exposure.} As the cumulation, not the sequence, matters, one can also take an estimate of the probability density function of loading conditions $\rho_T(T,\textcolor{black}{t})$ at time $\textcolor{black}{t}$, and
define the effective ageing acceleration as
\begin{equation} \label{eq:Minerprobability}
\overline{r}(\textcolor{black}{t}) = \int_{-\infty}^{\infty}r(T)
\, \rho_T(T,\textcolor{black}{t}) \, dT
\end{equation}
and the effective age
\begin{equation} \label{eq:Minerage}
\overline{A}(a,\textcolor{black}{t}) =  \int_{t-a}^{\textcolor{black}{t}}\overline{r}(s)ds
\end{equation}
Then the cumulative hazard is 
\begin{equation}
H(a,t) = \left(\frac{\overline{A}(a,t) }{\hat{\eta}_r}\right)^{\beta}
\end{equation}
and the failure rate
\vspace{-1em}
\begin{equation}\label{eq:failure}
\lambda(a,t) = \frac{\beta}{\hat{\eta}_r}\left(\frac{\overline{A}(a,t) }{\hat{\eta}_r}\right)^{\beta-1} \overline{r}(t)
\end{equation}

\vspace{-1em}
\subsection{\textcolor{black}{Steady-state} seasonality}\label{sec:stat}
When each year has a similar load distribution $\rho(T)$, we can assume \textcolor{black}{a steady-state} of the load distribution loading $\rho_T(T,\tau) = \rho(T)$ that is constant in time $\tau$ from year to year. Then, one can assume a constant acceleration factor $\overline{r}(\tau) = \overline{r}$ with
\begin{equation} \label{eq:Minerprobability2}
 \overline{r} =  \int_{-\infty}^{\infty}r(T)
\, \rho_T(T) \, dT
\end{equation}
and the characteristic lifetime 
\begin{equation}\label{eq:lifetime}
\hat{\eta} = \frac{\hat{\eta}_r}{\overline{r}}
\end{equation}
\vspace{-1em}
\subsection{\textcolor{black}{Dynamics of} seasonal load}
However, the energy transition changes the loading distribution from year to year, and the above does not hold any longer. We can assume piecewise changes of the distribution of loadings $\rho_{k}(T)$ from season $k$ to season $k+1$ \textcolor{black}{during the transition from low-DER to high-DER loading conditions}. We assume the duration of a season as $\Delta\tau_k = \tau_{k+1} - \tau_k$ and the time interval is $[\tau_k, \tau_{k+1})$. Then, the temperature distribution is 
\begin{equation}
\begin{aligned}\label{eq:loadforesight}
    \rho_T(T,t) &= \rho_k(T), && \text{if $t \in [\tau_k, \tau_{k+1})$}
\end{aligned}
\end{equation}
and we use \eqref{eq:Minerprobability} in season $k$ as the expected ageing acceleration 
\begin{equation}
\overline{r}_{k} = \int_{-\infty}^{\infty}r(T)
\, \rho_{k}(T) \, dT
\end{equation}
We can then obtain the effective age
\begin{equation} 
\begin{aligned}\label{eq:int2}
    \overline{A}(a,t) &= \sum_{k\in\kappa(a,t)} \Delta \tau_k \, \overline{r}_{k},  
    \end{aligned}
\end{equation}
for a cable age $a$ at time $t$ that has been in service during the seasons $\kappa(a,t) = \{k: t-a \leq \tau_k \leq \tau_{k+1} \leq t \}$. Using $\overline{r}_{k}$ and $\overline{A}(a,t)$ in \eqref{eq:failure} gives the failure rate $\lambda(a,t)$.  


\subsection{\textcolor{black}{Time-varying loading in the transition to higher DERs}}


\textcolor{black}{One can model the transition between loading regimes using a logistic function} 
\vspace{-1em}
\begin{equation}\label{eq:energytransition}
{\rho}_{T}(T, \, t) = \rho_0(T) + \frac{\Delta \rho_T(T)}{1 + \exp\{-c_1 ({t} - c_2)\}}
\end{equation}
where $\Delta \rho_T = \rho_1(T) - \rho_0(T)$. 
\textcolor{black}{$c_1>0$ and $c_2$ are stylised descriptors of transition dynamics rather than calibrated physical parameters, where $c_1$ controls the speed of loading changes associated with DER integration and $c_2$ represents the characteristic midpoint of a typically multi-decade transition period.} We then take ${\rho}_{T}(T, \, t)$ in \eqref{eq:Minerprobability} and \eqref{eq:Minerage} and change the order of integration to
\begin{equation}
\overline{A}(a,t)  =  \int_{-\infty}^{\infty}  r(T)\int_{t-a}^{t} \, \rho_T(T,s) \, ds \, dT
\end{equation}
We consider the inner integral 
{\small
\begin{align}
\int_{t-a}^{t} \, \rho_T(T,s) \, ds &= \int_{t-a}^{t} \, \rho_0(T) + \frac{\Delta \rho_T(T)}{1 + e^{-c_1 ({s} - c_2)}} \, ds \nonumber \\  
&= a \,\rho_0(T) + \frac{\Delta \rho_T(T)}{c_1} \ln \left(\frac{1 + e^{c_1 ({t} - c_2)}}{1 + e^{c_1 ({t} -a - c_2)}}\right) 
\end{align}
}
We can now write the effective age over the time window of a cable with age $a$ as
\vspace{-0.25em}
{\small
\begin{align} \label{eq:etaT}
\overline{A}(a,t) &=  a\int_{-\infty}^{\infty}   r(T) \,\rho_0(T) \, dT  \nonumber \\
&+ \frac{1}{c_1 } \ln \left(\frac{1 + e^{c_1 ({t} - c_2)}}{1 + e^{c_1 ({t} -a - c_2)}}\right)\int_{-\infty}^{\infty}  r(T) \Delta \rho_T(T)   \, dT
\end{align}
}
We define the limits of the effective age
\begin{align}\label{eq:limit0}
          \overline{A}_0(a) &:= \lim_{t\rightarrow -\infty } \overline{A}(a,t)  = a\int_{-\infty}^{\infty}  r(T) \,\rho_0(T) \, dT \\ \label{eq:limit1}
     \overline{A}_1(a) &:= \lim_{t\rightarrow +\infty} \overline{A}(a,t) = a \int_{-\infty}^{\infty}  r(T) \,\rho_1(T) \, dT  
\end{align}
before and after the energy transition, respectively. A compact way for the effective age is
\vspace{-0.25em}
{\small
\begin{align} \label{eq:etaT2}
\overline{A}(a,t) &=  \overline{A}_0(a) + \frac{1}{c_1 a} \ln \left(\frac{1 + e^{c_1 ({t} - c_2)}}{1 + e^{c_1 ({t} -a - c_2)}}\right) (\overline{A}_1(a) - \overline{A}_0(a))
\end{align}
}
that in \eqref{eq:failure} can compute the failure rate in time $\lambda(a,t)$.





\vspace{-1em}
\section{Maintenance strategies} \label{sec:main}
DSOs can apply different maintenance and replacement strategies for their assets  \cite{Hig18}, for example:

\subsubsection{Run-to-failure}
This strategy replaces cables after failure. The expected number of cables annually replaced is
\begin{equation} \label{eq:run2failure}
N(t) = |\Omega^C| \int_{0}^{{\infty}} \lambda({a,t})\, \rho_A({a,t})\, d{a}
\end{equation}
where $\rho_A({a,t})$ is the age distribution of the cables that may change with time $t$. $|\Omega^C|$ is the number of cables in the network. 
\textit{Run-to-failure} could be a viable strategy in a grid that runs at very low loading conditions, as some cables operated in low, ideal conditions can last up to $200$ years.

\subsubsection{Replace-all} 
This strategy regularly replaces cables when their failure rate is above a specified threshold $\lambda_R$, or applies a replacement time  $a_R$ that can be converted by $\lambda_R(t) = \lambda(a_R(t),t)$ in \eqref{eq:genericfailure}. At constant loading, one can apply
\begin{equation}
a_R = \hat{\eta}\left(\lambda_R \frac{\hat{\eta}}{\beta} \right)^\frac{1}{\beta-1}
\end{equation}
from \eqref{eq:WeibullArrhenius}. The number of cables annually replaced can be estimated as
\vspace{-0.5em}
{\small
\begin{equation} \label{eq:replace}
N(t) = |\Omega^C| ~ \left(\int_{0}^{a_{R}} \lambda({a,t})~ \rho_A({a,t})\, d{a}  +  \int_{a_{R}}^{{\infty}} \rho_A({a,t})\, d{a}\right)
\end{equation}
}
considering the replacement in response to a failure, and the share of cables that surpassed the replacement time $a_R$. This is typically an expensive strategy. A typical replacement time for PILC is around $a_R= 50$ years \cite{Zap21}.


\subsubsection{Preventive Maintenance}
Applying preventive maintenance reduces failures \cite{Ber05}. 
The strategy regularly assesses the remaining useful life or health index for each cable $c \in \Omega^C$. Subsequently, the cables are assigned to ‘keep’ ${\Omega}^K \gets {\Omega}^K \cup \{c\}$, to perform ‘preventive maintenance’ ${\Omega}^M \gets {\Omega}^M \cup \{c\}$ or to 'replace' ${\Omega}^R \gets {\Omega}^R \cup \{c\}$ \cite{Zho17}. The assignment of cables in these three categories $\Omega^C = \Omega^K \cup \Omega^M \cup \Omega^R$ can be economically optimised \cite{Swa16}. Depending on the level of maintenance, one can expect a reduction in the failure rate
\begin{equation}\label{eq:prev}
\lambda_c(a,t) = \lambda(a,t) ~ (1-x_c),
\end{equation}
where $x_c \geq 0$ 
corresponds to the maintenance carried out on cables $c \in \Omega^M$, e.g. $x_c=0$ if no maintenance was carried out. 
The distribution of maintenance levels carried out for cables at age $a$ is $\rho_{XA}(x,a)$ that is assumed steady in time. The number of cables that are annually expected to fail is

{\small
\begin{equation} \label{eq:prevent}
F(t) = |\Omega^C| \int_{0}^{\infty} \lambda(a,t)  \left( \int_{0}^{\infty} (1 - x ) \rho_{XA}(x, a) \, dx  \right) \rho_A(a,t) \, da
\end{equation}
}

The expected number of cables annually to maintain is $|\Omega^M|$, and the cables to replace is 
\begin{equation} \label{eq:prevman}
N(t) =  F(t) + |\Omega^R| 
\end{equation}
\vspace{-2em}
\subsection{Assessing maintenance strategies with shortcuts}
The maintenance strategies \textit{Replace-all}, \textit{Run-to-failure}, and \textit{Preventive Maintenance} can be analysed for time-varying load with \eqref{eq:run2failure}, \eqref{eq:replace} and \eqref{eq:prevent}, respectively. We start considering a loading profile that is \textcolor{black}{steady-state} in seasonality (Sec. \ref{sec:stat}) and does not change over years, i.e. $\overline{r}_{k} =\overline{r}_{k+1} = \overline{r}  \, \, \,\forall k$, then \eqref{eq:Minerprobability2} simplifies to the constant acceleration factor 
\begin{equation}
    \overline{r}(t) = \overline{r}
\end{equation}
that can be computed with \eqref{eq:Minerprobability2}. The characteristic lifetime $\hat{\eta}$ can be computed by \eqref{eq:lifetime}. The effective age is 
\begin{equation}
    \overline{A}(a) = a \,\overline{r}
\end{equation}
and applying \eqref{eq:failure} leads to the failure rate 
\begin{equation}
\lambda(a) = \frac{\beta \, \overline{r}}{\hat{\eta}_r}\left(\frac{\overline{A}(a) } {\hat{\eta}_r}\right)^{\beta-1} = \frac{\beta}{\hat{\eta}}\left(\frac{a } {\hat{\eta}}\right)^{\beta-1}
\end{equation}  


Then, we consider the limits before and after the energy transition (\eqref{eq:limit0} 
and \eqref{eq:limit1}
): $\hat{\eta}_0$ and $\overline{r}_0$ considers the load distribution $\rho_{0}(T)$ at time $t_0$ and $\hat{\eta}_1$ and $\overline{r}_1$ consider the load distribution $\rho_{1}(T)$ at time $t_1$, respectively.
\subsubsection{Run-to-failure}
To assess the increase in maintenance $\frac{N(t_1)}{N(t_0)}$ in  \eqref{eq:run2failure}, assuming the cable network size $|\Omega^C|$ is constant and the age distribution $\rho_{A}$ also stays constant 
\begin{equation} \label{eq:fail}
\frac{N(t_1)}{N(t_0)} = \frac{\int_{0}^{{\infty}} \frac{\beta}{\hat{\eta}_1 } \left(\frac{a}{\hat{\eta}_1 } \right)^{\beta-1}\, \rho_{A}({a})\, d{a}}{\int_{0}^{{\infty}} \frac{\beta}{\hat{\eta}_0} \left(\frac{a}{\hat{\eta}_0} \right)^{\beta-1}\, \rho_{A}({a})\, d{a}} = \left(\frac{\hat{\eta}_0 }{\hat{\eta}_1 }\right)^{\beta}  = \left(\frac{\overline{r}_1 }{\overline{r}_0 }\right)^{\beta} 
\end{equation}

\subsubsection{Replace-all}
Assessing the increase in maintenance $\frac{N(t_1)}{N(t_0)}$ in \eqref{eq:replace} where we assume the same replacement age $a_R$ leads to
\begin{equation} 
 \frac{N(t_1)}{N(t_0)} =  \frac{\int_{0}^{a_{R}} \textcolor{black}{\frac{\beta}{\hat{\eta}_1 } \left(\frac{a}{\hat{\eta}_1 } \right)^{\beta-1}\,} \rho_A({a})\, d{a}  +  \int_{a_{R}}^{{\infty}} \rho_A({a})\, d{a}}{\int_{0}^{a_{R}} \textcolor{black}{\frac{\beta}{\hat{\eta}_0 } \left(\frac{a}{\hat{\eta}_0 } \right)^{\beta-1}\,} \rho_A({a})\, d{a}  +  \int_{a_{R}}^{{\infty}} \rho_A({a})\, d{a}}
\end{equation}
We can simplify to
\begin{equation}\label{eq:simplifiedreplace}
\frac{N(t_1)}{N(t_0)} =  \frac{ \beta \left(\hat{\eta}_1\right)^{-\beta} \int_{0}^{a_{R}} a^{\beta-1}~ \rho_A({a})\, d{a}  +  \int_{a_{R}}^{{\infty}} \rho_A({a})\, d{a}}{\beta \left(\hat{\eta}_0\right)^{-\beta} \int_{0}^{a_{R}} a^{\beta-1}~ \rho_A({a})\, d{a}  +  \int_{a_{R}}^{{\infty}} \rho_A({a})\, d{a}}
\end{equation}
We introduce 
\vspace{-0.5em}
\begin{equation}\label{eq:rXLPE}
    \mu := \frac{\beta \int_{0}^{a_{R}} a^{\beta-1}~ \rho_A({a})\, d{a}}{\int_{a_{R}}^{{\infty}} \rho_A({a})\, d{a}}
\end{equation}
and obtain 
\vspace{-0.5em}
\begin{equation} \label{eq:rep}
\frac{N(t_1)}{N(t_0)} = \frac{\hat{\eta}_1^{-\beta}\mu+1}{\hat{\eta}_0^{-\beta}\mu+1} = \frac{\overline{r}_1^{\beta}\mu+\hat{\eta}_r}{\overline{r}_0^{\beta}\mu+\hat{\eta}_r}.
\end{equation}
$\frac{N(t_1)}{N(t_0)}$ may be further simplified if one has an analytic function of $\rho_A({a})$ or makes assumptions. 

\subsubsection{Preventive maintenance}
Assessing the increase in maintenance $\frac{N(t_1)}{N(t_0)}$ in the \textit{Preventive Maintenance} strategy in \eqref{eq:prevent} we assume the distribution of maintenance levels $\rho_{XA}(x,a)$ is constant \vspace{-0.5em}
\begin{equation}
\kappa(a) := \int_{0}^{\infty} (1 - x ) \rho_{XA}(x, a) \, dx
\end{equation}
The number of cables that are annually expected to fail in season $t_0$ is
\vspace{-0.5em}
\begin{equation}\label{eq:prevfail}
F(t_0) = |\Omega^C| \int_{0}^{\infty} \frac{\beta}{\hat{\eta}_0} \left(\frac{a}{\hat{\eta}_0 } \right)^{\beta-1}  \kappa(a)\, \rho_A(a) \, da
\end{equation}
and the change of failures is
\begin{equation} 
\frac{F(t_1)}{F(t_0)} = \left(\frac{\hat{\eta}_0}{\hat{\eta}_1}\right)^{\beta} = \left(\frac{\overline{r}_1}{\overline{r}_0}\right)^{\beta}
\end{equation}
Using \eqref{eq:prevman} and assuming maintenance strategy is not changed $(\frac{|\Omega^R|}{dt}=0)$, an estimate of the cables requiring replacement is
\begin{equation} 
N(t_1) = \left(\frac{\hat{\eta}_0}{\hat{\eta}_1}\right)^{\beta} F(t_0) + |\Omega^R|
\end{equation}
or the maintenance increase is
\begin{equation} \label{eq:prev2}
\frac{N(t_1)}{N(t_0)} = \frac{\left(\frac{\hat{\eta}_0}{\hat{\eta}_1}\right)^{\beta} F(t_0) + |\Omega^R|}{F(t_0) + |\Omega^R|}
\end{equation} 
that can also be expressed in terms of $\overline{r}_1$ and $\overline{r}_0$. 



\begin{figure}
\centering
\includegraphics[width=\columnwidth]{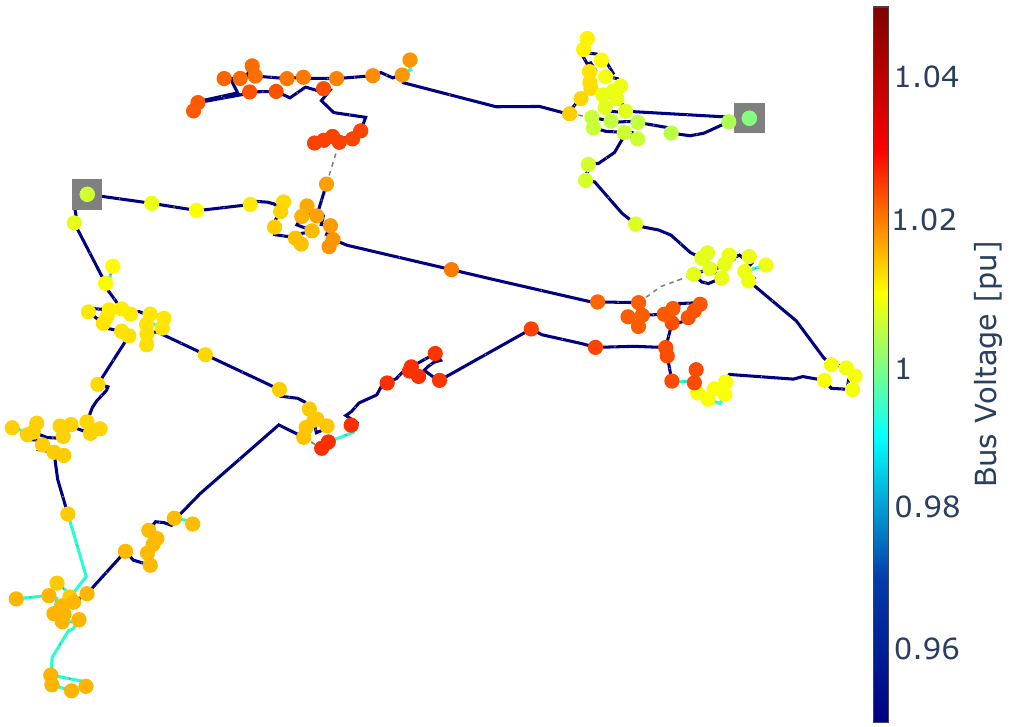}
\caption{Oberrhein MV network with \textcolor[RGB]{103,255,220}{\rule[0.6ex]{1.5em}{1.2pt}}\textcolor{black}{PILC} and
  \textcolor[RGB]{48,48,151}{\rule[0.6ex]{1.5em}{1.2pt}}\textcolor{black}{XLPE} cables. \vspace{-1em}}
    \label{fig:usecaseMV}
\end{figure}

\subsection{Considering individual asset data}
We consider a DSO that applies the CNAIM or a similar collection of asset data. The DSO has access to the age $a_c$, maintenance $x_c$, and health index $h_c$ of cables $c \in \Omega^C$. The cables $\Omega^C = \Omega^{CP} \cup\Omega^{CX}$ can be assigned to the two cable types $\Omega^{CP}$ PILC and $\Omega^{CX}$ XLPE cables, respectively. The DSO also operates a well-monitored MV system. The loading $T_c(t)$ is an actual recording or future estimates at time $t$ and available for each cable $c$. We can compute $F(t)$ the expected number of failures during season $t$ by converting \eqref{eq:prevent} to a summation over all cables, then using \eqref{eq:genericfailure} and \eqref{eq:prev} gives
\begin{equation}
\begin{split}
F(t) &\approx \sum_{c\in \Omega^C} \lambda_c(a_c,t)\,\Delta t \\
     &= \sum_{c\in \Omega^C} \left(\frac{\beta}{\hat{\eta}_r}\right)
        \left(\frac{A(a_c,t)}{\hat{\eta}_r}\right)^{\beta-1}
        r(T_c(t))\, (1-x_c)\, \Delta t 
\end{split}
\end{equation}
where $\Delta t$ is the length of a season $t$, and each cable has an individual effective age
\begin{equation} 
 A(a_c,t) = \int_{t-a_c}^{t}r(T_c(\tau))d\tau
\end{equation}
from \eqref{eq:effectiveage} considering the individual temperature trajectory $T_c(\tau)$ from the records and projections of loads over cable $c$. $r({T_c}(\tau))$ follows \eqref{eq:XLPEageing} $\text{if } c \in \Omega^{CX}$ or \eqref{eq:PILCageing} $\text{if } c \in \Omega^{CP}$, respectively. Also $\hat{\eta}_r$ and $\beta$ differ for these two cable types. Assuming the time unit is in years, the number of annual replacements is then
\begin{equation}\label{eq:failurnumber}
N(t) = F(t) + |\Omega^R|
\end{equation}
where $|\Omega^R|$ are the scheduled replacements during season $t$. The \textit{Run-to-failure} maintenance strategy replaces only post-failure ($\Omega^R = \emptyset$) and does not apply preventive maintenance $x_c = 0 \, \forall c$. The \textit{Replace-all} strategy also does not apply preventive maintenance ($x_c = 0$) and replaces cables ${\Omega}^R \gets {\Omega}^R \cup \{c\}$ if $a_c \geq a_R$. In the \textit{Preventive Maintenance} strategy ${\Omega}^R \gets {\Omega}^R \cup \{c\}$ if $h_c  \leq h_R \, \forall c$ with the health index threshold for replacement $h_R$.





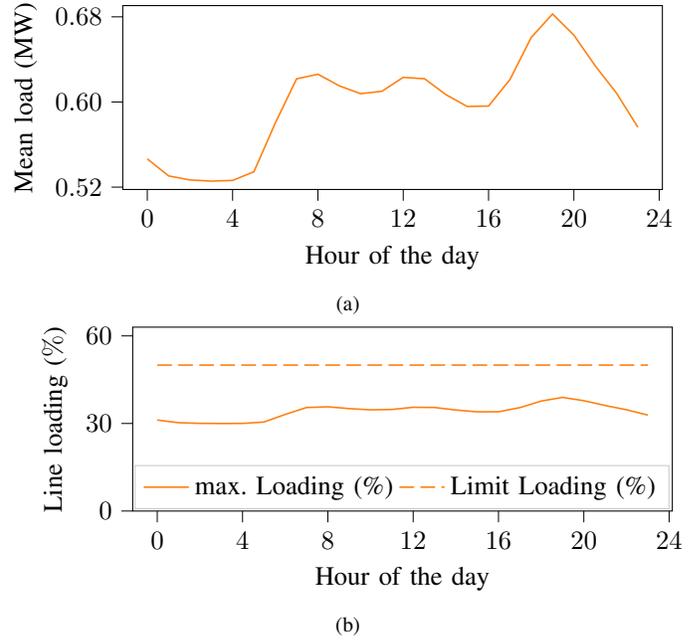
\begin{figure}
\centering
\begin{subfigure}{\columnwidth}
  \centering
  \makebox[\columnwidth][c]{\input{pfoberrhein.tex}}
  \caption{}
  \label{fig:pfoberrhein1}
\end{subfigure}
\begin{subfigure}{\columnwidth}
  \centering
  \makebox[\columnwidth][c]{\hspace*{5mm}\input{pfoberrhein2.tex}} 
  \caption{}
  \label{fig:pfoberrhein2}
\end{subfigure}
\caption{\textcolor{black}{Loading of the Oberrhein MV grid. (a) Mean load at buses over a $24$-hour period. (b) Line loading (in \%) relative to the maximum loading and limit loading.}
\vspace{-1em} }
\label{fig:pfoberrhein}
\end{figure}

\section{Case study} \label{sec:case}
\subsection{Study settings, test networks and data}
The case study examines PILC and XLPE cables using two different thermal models, incorporating partly Danish cable data and thermal data from various sources. \textcolor{black}{The study considers an initial cable population defined at time $t = 0$ based on Danish MV asset statistics, specifying the share, age distribution, and Weibull ageing parameters for PILC and XLPE cables. These cable populations are analysed using thermal ageing models and applied to the Oberrhein MV grid to evaluate failure behaviour under different loading scenarios derived from projected DER integration.} The Oberrhein network is shown in Fig. \ref{fig:usecaseMV} with the two cable types.
For thermal ageing, the Arrhenius model \eqref{eq:XLPEageing} was applied to the XLPE cables, and the Montsinger approach \eqref{eq:PILCageing} to the PILC cables. The cable data is reported in Table \ref{tab:thermal_aging_parameters}. 
The Weibull parameter $\beta \approx 3$ was estimated for XLPE cables from \textcolor{black}{Fig. 8 in} \cite{Han23},  considering only cables older than $10$ years. 
$\beta \approx 3.4$ was assumed for PILC cables 
from \textcolor{black}{Fig. 7 in} \cite{Han19}. These shape parameters $\beta$ need to be estimated separately for each cable population. 
The Danish data features PILC and XLPE cables, where $\Omega^C = \Omega^C_{\mathrm{PILC}} \cup \Omega^C_{\mathrm{XLPE}}$. We assume at $t=0$ that the share of PILC cables in the network is $\tfrac{| \Omega^C_{\mathrm{PILC}}|}{|\Omega^C|} =0.25$  and the share of XLPE cables is $\tfrac{| \Omega^C_{\mathrm{XLPE}}|}{|\Omega^C|} =0.75$ as in the Danish MV grid in 2023 \cite{Han23}. At the simulation time $t=0$ years, the ages for PILC cables were assumed as a skewed normal distribution with $\mu = 58$, $\sigma = 21$ and $\delta = -3$. The ages for XLPE cables were assumed as a skewed normal distribution with $\mu = 10$, $\sigma = 15$ and $\delta = 3$. The cumulative distributions of these two skewed normal distributions roughly match \textcolor{black}{Fig. 1 in }\cite{Han23}. Representative load scenarios for wind, PV, and without DERs are shown in Fig. \ref{fig:loads}, which was obtained from digitising \textcolor{black}{Fig. 10 in}  \cite{Zap21}. 

\begin{table*}[t]
\centering
\caption{Parameters Used in Thermal Aging Models for XLPE (1×240 mm²) and PILC (3×120 mm²) Cables at 12/20 kV}
\begin{tabular}{|l|c|c|c|}
\hline
\textbf{Parameter} & \textbf{Symbol} & \textbf{XLPE Cable} & \textbf{PILC Cable} \\
\hline
Thermal aging model                      & –           & Arrhenius             & Montsinger \\
Aging slope / doubling factor            & $B$, $\Delta T$ & $16237$ K             & 6.5°C \\ 
Reference lifetime                       & $\eta_r$       & {55} years              & {100} years  \\ 
Reference conductor temperature          & $T_r$       & 90°C                 & {15}°C \\ 
Maximum conductor temperature            & $T_{\max}$  & 90°C                 & 65°C \\ 
Ambient temperature (soil)               & $T_a$       & 20°C                 & 20°C \\
Temp. coefficient of resistance          & $\alpha$    & $4.03 \times 10^{-3}$ 1/K (Al) & $3.9 \times 10^{-3}$ 1/K (Cu) \\
Rated current per conductor              & $I_r$       & 313 A                 & 302 A \\
Adjusted ampacity (standard conditions)  & $I_z$       & 313 A                 & 250 A \\ 
\hline
\end{tabular}
\label{tab:thermal_aging_parameters}
\vspace{1mm}
\noindent
\begin{minipage}{2\columnwidth}
\textbf{Notes:} XLPE cable parameters are based on \cite{Hestad2012,Alg20,IEC60502-2,IEC60287-1-1}. PILC cable parameters are based on \cite{Zap21,BS6480}. Adjusted ampacities refer to standard buried installations at a 20°C soil temperature and a 1.0K·m/W thermal resistivity. \vspace{-1em}
\end{minipage} 
\end{table*}

\cite{Zap21} considers three DER scenarios in Germany until $2032$ and used renewable projections from the Erneuerbare-Energien-Gesetz $2014$. The MV system Oberrhein system applied spatial load distribution over a $24$h window. Fig. \ref{fig:pfoberrhein1} shows the mean load across all buses. After solving the AC power flow, Fig. \ref{fig:pfoberrhein2} shows the line loadings. The study assumes a maximal line loading of $I_z=50\%$, and assumes lines have all the unit length. The $25\%$ lines with the lowest average loading are assumed to be PILC and all other lines as XLPE cables (Fig. \ref{fig:usecaseMV}). The ages of the $181$ lines were sampled from the initial skewed normal, age distributions at $t=0$ (also shown later in Fig. \ref{fig:Age_changes}). The \textit{Replace-all} maintenance strategy assumes the replacement age $a_R=50$ years. The \textit{Preventive} strategy assumes $\tfrac{|\Omega^R|}{|\Omega^C|} = 0.002$ and $\kappa(a)=0.98 \text{ if } a\ge 25,\; 0 \text{ otherwise}$. All simulations were implemented \textcolor{black}{using} Python 3.13.5 \textcolor{black}{in Jupyter Notebook scripts with the NumPy 2.2.5 library. The MV Oberrhein grid model and AC power-flow calculations were performed using the Pandapower 3.1.2 package.}

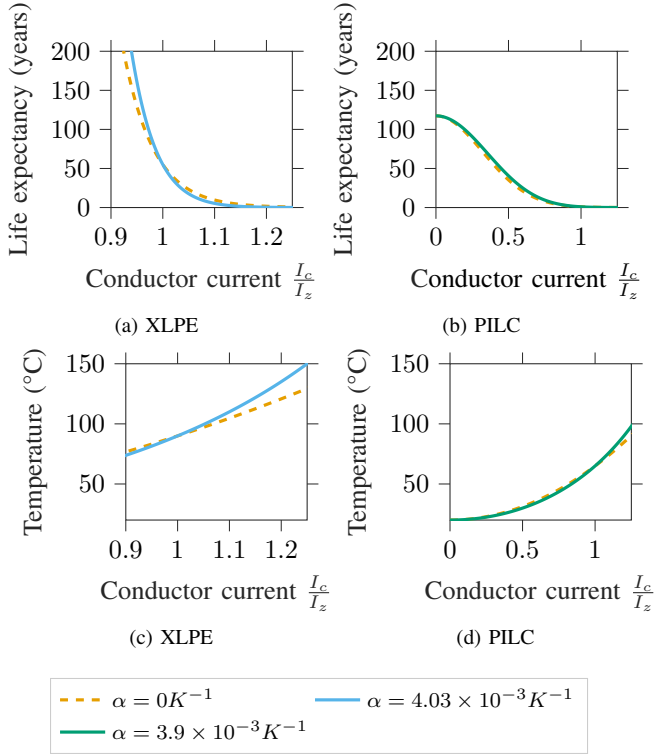
\begin{figure}
    \centering
    \begin{minipage}[b]{0.47\columnwidth}
        \centering
        \input{lifeexpect_XLPE.tex}
        \vspace{-15pt}
        \subcaption{XLPE}\label{fig:1a}
    \end{minipage}
    \hspace{-0.01\columnwidth}
    \begin{minipage}[b]{0.47\columnwidth}
        \centering
        \input{lifeexpect_PILC.tex}
        \vspace{-15pt}
        \subcaption{PILC}\label{fig:1b}
    \end{minipage}
    
    \hspace{0.03\columnwidth}
    \begin{minipage}[b]{0.47\columnwidth}
        \centering
        \input{temperature_XLPE.tex}
        \vspace{-15pt}
        \subcaption{XLPE}\label{fig:1c}
        \vspace{10pt}
    \end{minipage}
    \hspace{-0.01\columnwidth}
    \begin{minipage}[b]{0.47\columnwidth}
        \centering
       \input{temperature_PILC.tex}
       \vspace{-15pt}
        \subcaption{PILC}\label{fig:1d}
        \vspace{10pt}
    \end{minipage}
\definecolor{cornflowerblue86180233}{RGB}{86,180,233}
\definecolor{darkslategray38}{RGB}{38,38,38}
\definecolor{lightgray204}{RGB}{204,204,204}
\definecolor{orange2301590}{RGB}{230,159,0}
\definecolor{mycolor3}{RGB}{0,158,115}
\centering
\ref{mylegendXLPE}
    \caption{\textcolor{black}{Life expectancy and conductor temperature as functions of conductor current for XLPE and PILC cables. (a) Life expectancy of XLPE cables. (b) Life expectancy of PILC cables. (c) Conductor temperature of XLPE cables. (d) Conductor temperature of PILC cables.}    
   \vspace{-1em}}
    \label{fig:2x2}
\end{figure}

\subsection{Thermal ageing of MV cables}\label{sec:csage}
This case study examines the thermal ageing of the insulation materials under different steady loading levels. To compute the conductor temperature $T$, \eqref{eq:conductemp} and \eqref{eq:conductempepsilon} were applied, respectively, when neglecting or considering the temperature coefficient of resistance. To compute one data point ($T,\eta_T$) the current $I_c(\tau)=I_c$, ambient temperature $T_a$ and temperature $T(\tau)=T$ were assumed in steady state, i.e. $\frac{dI(\tau)}{d\tau}=0$, $\frac{dT_a(\tau)}{d\tau}=0$ and $\frac{dT(\tau)}{d\tau}=0$. \eqref{eq:XLPEageing} and \eqref{eq:PILCageing} were used to compute lifespans $\eta$ for PILC and XLPE, respectively. Fig. \ref{fig:2x2} confirms that the temperature increases with the loading, and the lifetime decreases. XLPE has a higher life expectancy than PILC cables. The lifetime drastically reduces when approaching the \textcolor{black}{ampacity} $I_z$ in XLPE and approaching around $50\%$ of $I_z$ in PILC. \textcolor{black}{A quantitative comparison between the simplified temperature model in \eqref{eq:conductemp} and the extended formulation in \eqref{eq:conductempepsilon} shows that neglecting temperature-dependent resistance leads to limited deviations at moderate loading levels but increasingly affects temperature and lifetime estimates close to ampacity.}
This study illustrates the fundamental behaviour of thermal ageing, demonstrating the importance of considering different loading levels in maintenance planning, as one cable at a lower loading level can last significantly longer than a cable at a higher loading level with respect to thermal ageing.

\vspace{-1em}
\subsection{Time-varying loading}
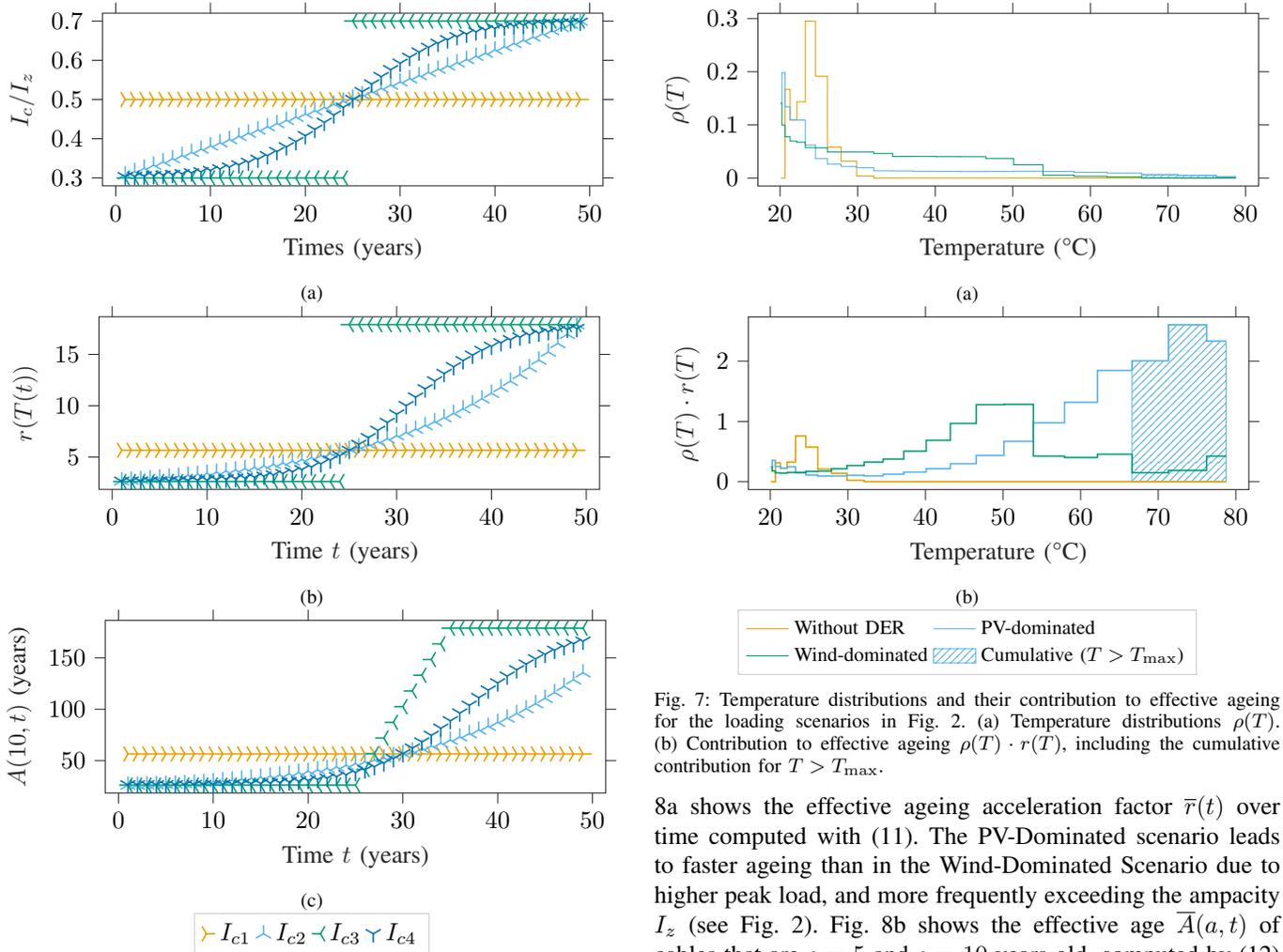
\begin{figure}
\centering
\begin{subfigure}{\columnwidth} 
        \makebox[\columnwidth][c]{\hspace{0cm}\input{current_varying_load.tex}}
   \caption{}
\label{fig:current_varying_load}
\end{subfigure}
\begin{subfigure}{\columnwidth} 
        \makebox[\columnwidth][c]{\hspace{0cm}\input{ageing_varying_load.tex}}
   \caption{}
\label{fig:ageing_varying_load}
\end{subfigure}
\begin{subfigure}{\columnwidth} 
        \makebox[\columnwidth][c]{\hspace{0cm}\input{life_varying_load.tex}}
   \caption{}
\label{fig:life_varying_load}
\end{subfigure}
\centering
\definecolor{cornflowerblue86180233}{RGB}{86,180,233}
\definecolor{darkcyan0114178}{RGB}{0,114,178}
\definecolor{darkcyan0158115}{RGB}{0,158,115}
\definecolor{darkslategray38}{RGB}{38,38,38}
\definecolor{lightgray204}{RGB}{204,204,204}
\definecolor{orange2301590}{RGB}{230,159,0}
\centering
\ref{mylegendIC}
    \caption{\textcolor{black}{Accelerated ageing under time-varying conductor currents. (a) Normalised current profiles $I_c/I_z$. (b) Ageing acceleration factor $r(T(t))$. (c) Effective age $A(10,t)$ of a cable with service age $a=10$ years.}    
   \vspace{-1em}} 
    \label{fig:varying_load}
\end{figure}

This study investigates effective ageing and failure rates under time-varying loading with two simulations: the first is explorative, applying basic loading profiles, and the second applies loading profiles during the energy transition. In the first explorative experiment, Fig \ref{fig:current_varying_load} shows four different current profiles $I_c(t)$ in time, where one of them is constant $\frac{I_{c1}(t)}{I_z}=0.5$. All four currents have the same average current over the $50$ years. $\frac{I_{c2}(t)}{I_z}$ linearly increases in time, and $\frac{I_{c3}(t)}{I_z}$ is a step change at $25$ years. ${I_{c4}(t)}$ applies the logistic function $\frac{I_{c4}(t)}{I_z} = 0.3 + \frac{0.4}{1 + \exp\{-c_1 ({t} - c_2)\}}$ with $c_1 = 0.2$ and $c_2=25$. Fig. \ref{fig:ageing_varying_load} shows the instantaneous ageing acceleration factor $r(T(t))$ from \eqref{eq:PILCageing} for a PILC cable. Fig. \ref{fig:life_varying_load} exemplarily shows the effective age $A(10,t)$ for cables of actual age (service time) $a=10$, i.e. \eqref{eq:effectiveage} integrates $r(T(t))$ over a moving window of the previous $10$ years. One can see that due to the integration of $r(T(t))$ that $A(10,t)$ drastically increases at the step change in $I_{c3}$. When comparing the linear increase $I_{c4}$ with the step change $I_{c3}$, the effective age of $10$ years cables is around $30\%$ lower for the second half of the transition.

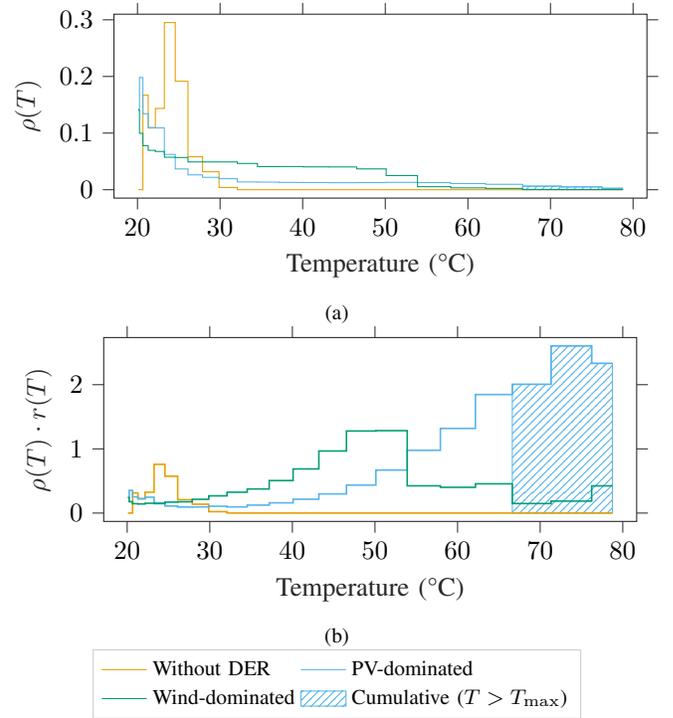
\begin{figure}
\centering
\begin{subfigure}{\columnwidth} 
        \makebox[\columnwidth][c]{\hspace{0cm}\input{rhoT.tex}}
   \caption{}
\label{fig:rhoT}
\end{subfigure}
\begin{subfigure}{\columnwidth} 
        \makebox[\columnwidth][c]{\hspace{0cm}\input{Agecontribution.tex}}
   \caption{}
\label{fig:Agecontribution}
\end{subfigure}
\centering
\definecolor{cornflowerblue86180233}{RGB}{86,180,233}
\definecolor{darkcyan0114178}{RGB}{0,114,178}
\definecolor{darkcyan0158115}{RGB}{0,158,115}
\definecolor{darkslategray38}{RGB}{38,38,38}
\definecolor{lightgray204}{RGB}{204,204,204}
\definecolor{orange2301590}{RGB}{230,159,0}
\centering
\ref{mylegendload}
    \caption{\textcolor{black}{Temperature distributions and their contribution to effective ageing for the loading scenarios in Fig. 2. (a) Temperature distributions $\rho(T)$. (b) Contribution to effective ageing $\rho({T})\cdot r(T)$, including the cumulative contribution for $T>T_{\mathrm{max}}$.} 
    \vspace{-1em}}
    \label{fig:loaddistribution}
\end{figure}

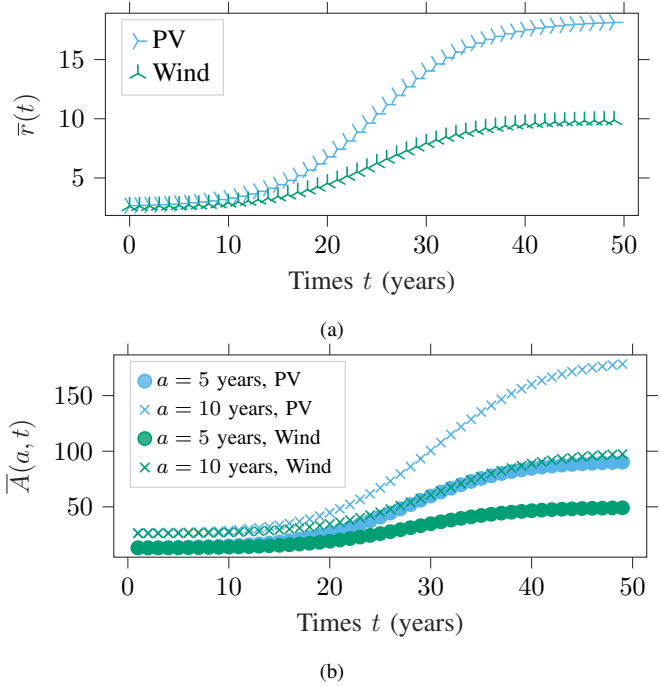
\begin{figure}
\centering
\begin{subfigure}{\columnwidth} 
        \makebox[\columnwidth][c]{\hspace{0cm}\input{rPW.tex}}
   \caption{}
\label{fig:rPW}
\end{subfigure}
\begin{subfigure}{\columnwidth} 
        \makebox[\columnwidth][c]{\hspace{0cm}\input{effAge.tex}}
   \caption{}
\label{fig:effAge}
\end{subfigure}
\centering
    \caption{\textcolor{black}{(a) Time-varying ageing acceleration factor $\overline{r}(t)$ for PV- and wind-dominated scenarios. (b) Effective age $\overline{A}(a,t)$ of cables with service ages $a = 5$ years and $a = 10$ years under the same scenarios.} \vspace{-1em}} 
    \label{fig:Acceleratedage}
\end{figure}

\begin{figure} 
        \makebox[\columnwidth][c]{\hspace{0cm}\input{effAge_limits.tex}}
   \caption{The effective age for different cable ages $a$ before ($\overline{A}_0(a)$) and after ($\overline{A}_1(a)$) the energy transition. \vspace{-1.5em}}
\label{fig:effAge_limits}
\end{figure}
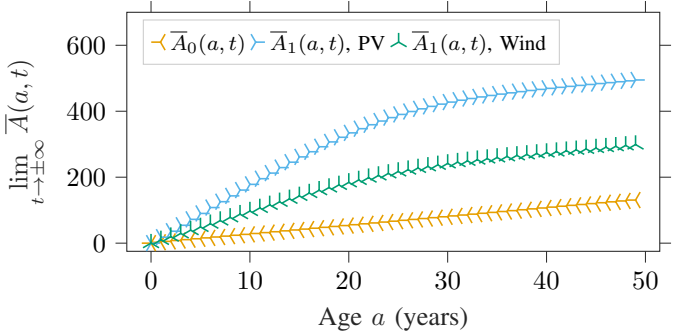

The second experiment applies the annual line loading distributions for the three scenarios (without DER, PV-dominated and Wind-dominated) from Fig. \ref{fig:loads} to PILC cables. We compute the temperatures \eqref{eq:conductemp}. Fig. \ref{fig:rhoT} shows the distribution over temperatures $\rho_T(T)$. We then model the energy transition \eqref{eq:energytransition} with $c_1 = 0.2$ and $c_2=25$ where $\rho_0$ is the initial distribution without DER, and we consider two cases where $\rho_1$ is either Wind-dominated or PV-dominated. As exemplarily shaded in the PV-dominated scenario, the cable is only operated $1.4\%$ of the time above the rated temperature ($\frac{\int_{T_{\max}}^{\infty}\rho_T(T) \, dT}{\int_{-\infty}^{\infty}\rho_T(T) \, dT}=1.4\%$ and see Fig. \ref{fig:rhoT}). Fig. \ref{fig:Agecontribution} shows the contribution of the temperature time-shares to the effective ageing. The 'area under the curve' is the cumulative contribution of the times that the cable is operated above the \textcolor{black}{ampacity}, i.e. applying \eqref{eq:Minerprobability2}. In the PV-dominated case, the $1.4\%$ share of operating above the rated temperature causes $46\%$ of the ageing ($\frac{\int_{T_{\max}}^{\infty}r(T)\, \rho_T(T) \, dT}{\int_{-\infty}^{\infty}r(T)\, \rho_T(T) \, dT} = 46\%$ and see Fig. \ref{fig:Agecontribution}). \textcolor{black}{These results demonstrate that reliability behaviour under non-stationary stress can exhibit nonlinear amplification effects, where relatively small time shares at elevated loading levels disproportionately contribute to long-term degradation. Such behaviour cannot be captured by classical stationary Weibull models that assume constant stress exposure over time.} Fig. \ref{fig:rPW} shows the effective ageing acceleration factor $\overline{r}(t)$ over time computed with \eqref{eq:Minerprobability}. The PV-Dominated scenario leads to faster ageing than in the Wind-Dominated Scenario due to higher peak load, and more frequently exceeding the \textcolor{black}{ampacity} $I_z$ (see Fig. \ref{fig:loads}). Fig. \ref{fig:effAge} shows the effective age $\overline{A}(a,\textcolor{black}{t})$ of cables that are $a=5$ and $a=10$ years old, computed by \eqref{eq:Minerage} for PV and Wind dominated grids, respectively. Although a cable is just $10$ years in service, the effective age surpasses $50$ years already at the intermediate of the transition ($t=25$ years) in the PV-scenario. This shows the fast acceleration of ageing in cables exposed to the energy transition. Fig. \ref{fig:effAge_limits} shows the effective age $\overline{A}(a,t)$ for cables with actual age $a$ 
at the limits (\eqref{eq:limit0}), before and after the energy transition, $A_0(a)$ and $A_1(a)$, respectively. This shows the effective age is around $2-3$ times higher after the energy transition than before. One can also notice the high increase of effective ageing before $30$ years.



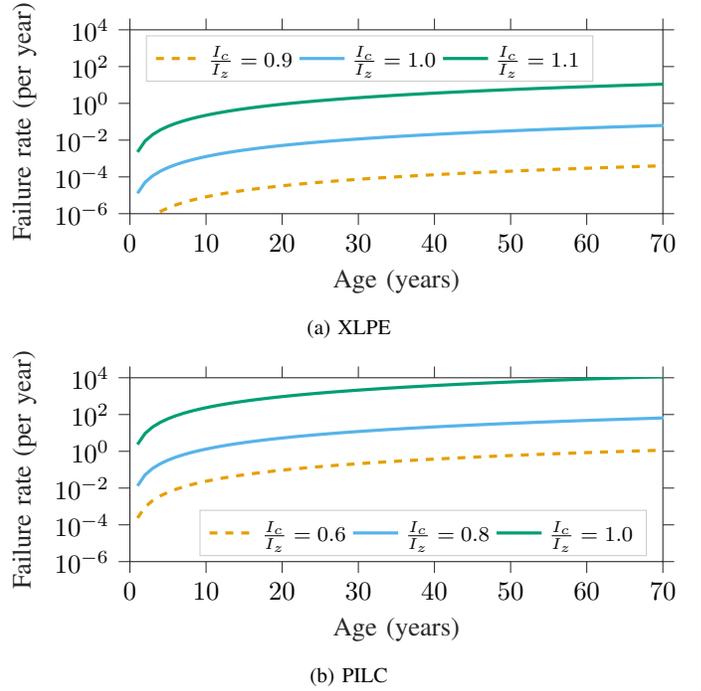
\begin{figure}
\centering
\begin{subfigure}{\columnwidth} 
        \makebox[\columnwidth][c]{\hspace{0cm}\input{failure_XLPE.tex}}
   \caption{XLPE}
\label{fig:failure_XLPE}
\end{subfigure}
\begin{subfigure}{\columnwidth} 
        \makebox[\columnwidth][c]{\hspace{0cm}\input{failure_PILC.tex}}
   \caption{PILC}
\label{fig:failure_PILC}
\end{subfigure}
\centering
    \caption{\textcolor{black}{Failure rate as a function of cable age for constant loading conditions. (a) XLPE cables for different conductor currents $\frac{I_c}{I_z}$. (b) PILC cables for different conductor currents $\frac{I_c}{I_z}$.}
    \vspace{-0.5em}}
    \label{fig:failures} 
\end{figure}

\vspace{-1em}
\subsection{Failure statistics of an aged population of MV cables}
Two studies investigate the failure statistics of a population that has heterogeneous ages and where the age distribution changes with time. The first case study investigates the Weibull failure rate \eqref{eq:WeibullArrhenius} under various operating \textcolor{black}{conditions} for an aged population of cables. Fig. \ref{fig:failures} shows the failure rates for PILC and XLPE cables of varying ages when applying three different conductor currents $I_c$. These currents are steadily (constantly) applied to the cable, assuming $\alpha=0$. These currents lead to different characteristic lives $\hat{\eta}$ using  \eqref{eq:XLPEageing} and \eqref{eq:PILCageing} and $\hat{\eta_r}$. This figure confirms that the failure rates drastically increase when permanently operated above the \textcolor{black}{ampacity}. 

\begin{figure}
\centering
\input{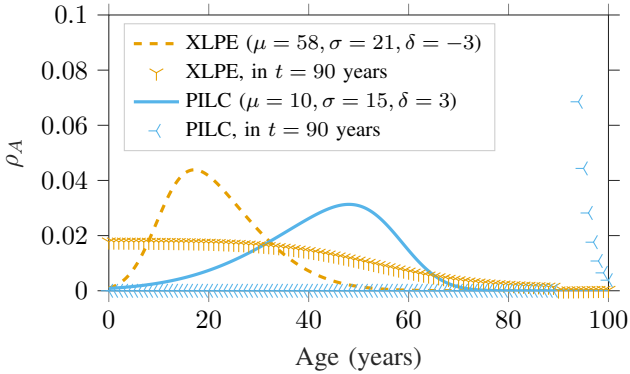}
    \caption{The age distributions of XLPE and PILC.\vspace{-1em}}
    \label{fig:Age_changes}
\end{figure}

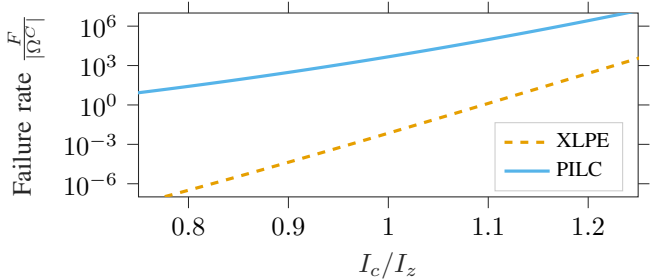
\begin{figure}
\centering
\input{Failures_both.tex}
    \caption{Failures increase with $I_c$ applied to XLPE and PILC cables. \vspace{-1em}}
    \label{fig:failures_both} 
\end{figure}

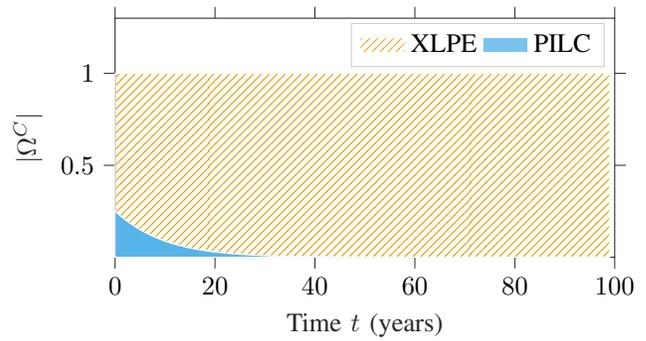
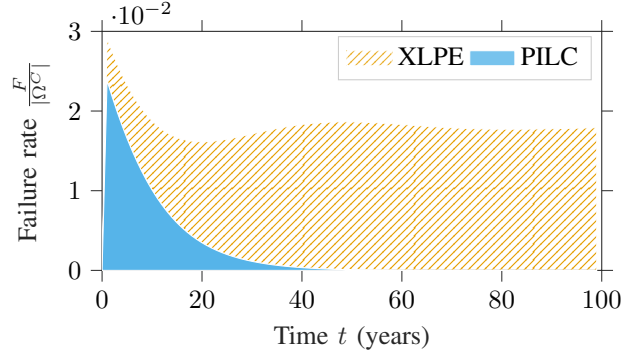
\begin{figure}
\centering
\begin{subfigure}{\columnwidth} 
        \input{Distribution_change.tex}
   \caption{}
\label{fig:Distribution_change}
\end{subfigure}
\begin{subfigure}{\columnwidth} 
        \makebox[\columnwidth][c]{\hspace{0cm}\input{Failures_change.tex}}
   \caption{}
\label{fig:Failures_change}
\end{subfigure}
\centering
    \caption{\textcolor{black}{Evolution of cable populations under \textit{Run-to-failure} replacement. (a) Share of XLPE and PILC cables in the network over time. (b) Normalised failure-rate contribution $\frac{F}{|\Omega^C|}$ over time for XLPE and PILC cables. }
    \vspace{-1em}}
    \label{fig:agechanges}
\end{figure}

The second case study simulates successively replacing PILC with XLPE cables when PILC cables fail, which represents a scenario that the Danish DSOs (and others) currently apply. Fig. \ref{fig:Age_changes} shows the age distributions of PILC and XLPE at the start of the simulation $t=0$ years. The time $t$ and ages $a$ are considered discrete $\mathbb{N}$ with the unit 'years'. Fig. \ref{fig:failures_both} shows the expected failures per year for a conductor current $I_c$ permanently applied to these cables at $t=0$ years. The applied current $I_c$ leads to the operating temperature $T$, the characteristic lifetime $\hat{\eta}(T)$ and the Arrhenius-Weibull failure rate \eqref{eq:WeibullArrhenius} used in \eqref{eq:failurerate}. This figure shows the exponential increase in failure rates for the two cable types. Next, this study investigates the failures of a set of cables of varying ages, and when their age distribution changes with time, applying the \textit{Run-to-failure} maintenance. All XLPE cables are assumed to operate at the constant $I_c=I_z$, and all PILC cables operate at the constant $I_c=0.5 I_z$. With time, all failed PILC cables are replaced with XLPE cables, and all failed XLPE cables are replaced in the following year by new XLPE cables. The replacement changes their age distributions. For example, the PILC distribution change is computed by $\rho_A(0,t) =0$ and $\rho_A(a+1,t) = \rho_A(a,t-1) \left(1- \lambda(a)\right) $ for $t=1,\dots 100$. Fig. \ref{fig:Age_changes} shows the distributions after $t=90$ years, where the XLPE distribution reached some sort of steady-state and only a few very old PILC cables remained. Fig. \ref{fig:Distribution_change} shows the change in the shares of these two cable types. After $20$ years, nearly all PILC cables were replaced by XLPE, due to the high failure rate of PILC cables (Fig. \ref{fig:failures}) and their high age (Fig. \ref{fig:Age_changes}). Fig. \ref{fig:Failures_change} shows the normalised failure rate $\frac{F}{|\Omega^C|}$. Despite only $25\%$ of cables being of type PILC at $t=0$, these cables cause around $82\%$ of total failures in the system. 

\vspace{-1em}
\subsection{Assessing maintenance strategies} \label{sec:assessmain}
\vspace{-0.25em}
Two case studies investigate the shortcut method to assess maintenance strategies during the energy transition: the first study focuses on the energy scenarios from \eqref{fig:loaddistribution}, and the second study considers spatial loading variability, focusing on the MV Oberrhein system. The first study assumes \textcolor{black}{steady-state} $\rho_A(a,t)=\rho_A(a)$ age distribution (from $t=0$, Fig. \ref{fig:Age_changes}). We use the load distributions from Fig. \ref{fig:loaddistribution} for PV, and Wind-dominated at $t_1$ and Without DER at $t_0$, to assess the relative maintenance increase $\tfrac{N(t_1)}{N(t_0)}$ from before to after the energy transition. Table \ref{tab:PILC_XLPE} shows the maintenance increase when applying the three strategies (applying equations \eqref{eq:fail}, \eqref{eq:rep}, and \eqref{eq:prev2}, respectively). For PILC cables, we observe around $250-300$ times increase in maintenance in the PV-Dominated and around times $45-55$ increase in the Wind-Dominated scenarios. The higher increase in the PV-dominated scenario over the Wind-Dominated scenario is due to more frequently operating the assets beyond and near the \textcolor{black}{ampacitie}s. For XLPE cables, the \textit{Run-to-failure} maintenance strategy shows up to a million increases in maintenance as a very low $\overline{r}_0$ appears in \eqref{eq:fail} in the denominator as base with the exponent $\beta =3.4$. Fig. \ref{fig:failures_both} shows that the $\frac{I_c}{I_z}$ needs to be in ranges near $0.9$ to lead to high failure rates in XLPE. However, Fig. \ref{fig:loads} shows that before the transition at $t_0$ (Without DER) $\frac{I_c}{I_z} < 0.5$ meaning $\overline{r}_0$ is extremely low. In other words, a $\Delta\frac{I_c}{I_z}=0.1$ increase can lead to a few magnitudes increase in failure rates (see Fig. \ref{fig:failures_both}) and  $\overline{r}$, where the increase in $\overline{r}$ in \eqref{eq:fail} has a power with the exponent $\beta =3.4$. Further in Table \ref{tab:PILC_XLPE}, the \textit{Preventive} maintenance strategy does not lead to an increase in maintenance of XLPE cables, and the  \textit{Replace-all} strategy has a $10$-fold increase for the PV-Dominated scenario. 

\begin{table}[b]\label{tab:strategies}
\centering
\begin{tabular}{l|cc|cc}
 & \multicolumn{2}{c|}{PILC} & \multicolumn{2}{c}{XLPE} \\
 & PV & Wind & PV & Wind \\
\hline
\textit{Run-to-failure} & 307 & 54 & 1,491,111 & 79,504 \\
\textit{Replace-all}    & 307 & 54 & 10.4 & 1.07 \\
\textit{Preventive}     & 254 & 44 & 1 & 1 \\
\hline
\end{tabular}
\caption{Comparing maintenance increase $\frac{N(t_1)}{N(t_0)}$  \textcolor{black}{(\textendash)} of three strategies applied to PILC and XLPE in PV and Wind-dominated scenarios.\vspace{-1em}}
\label{tab:PILC_XLPE}
\end{table}

The second study on the MV Oberrhein system investigates the \textit{Replace-all} strategy. Each cable $c$ of age $a_c$ in year $t$ is modelled as $Y \sim \mathrm{Bernoulli}\!\big(\lambda(a_c,t)\big)$, 
where $Y=1$ indicates failure (with probability $\lambda(a_c,t)$) 
and $Y=0$ survival (with probability $1-\lambda(a_c,t)$). If a cable fails or $a_c > a_R$, the cable is replaced by the same type (PILC or XLPE) with age $a_c=0$ in the following year. Fig. \ref{fig:Oberrheincableages} shows the histogram of age distribution (aggregated for PILC and XLPE) at three times $t=\{0,25,50\}$ years. The distribution at $t=0$ years and $t=50$ years shows a high similarity. The age replacement $a_R= 50$ years leads to some periodicity in time with period $50$ years. This also shows that scheduled replacement dominates over replacement due to failures. Fig. \ref{fig:Oberrhein_maintenance} shows the maintenance over time when applying \eqref{eq:failurnumber}, where the same periodicity appears. The peak of replacements per year reaches about $19$ as these are at the start of the simulation, where a large fraction of PILC cables are replaced due to their age beyond $50$ years (see Fig. \ref{fig:Age_changes}). However, these simulations show averages of around $5$ replacements per year, which seems high for $181$ cables, demonstrating the inefficiency of the \textit{Replace-all} strategy that does not consider asset health condition $h_c$.

\begin{figure}
\centering
\begin{subfigure}{\columnwidth} 
        \makebox[\columnwidth][c]{\input{Oberrheincableages.tex}}
   \caption{}
\label{fig:Oberrheincableages}
\end{subfigure}
\begin{subfigure}{\columnwidth} 
        \makebox[\columnwidth][c]{\hspace*{3mm}\input{Oberrhein_maintenance.tex}}
   \caption{}
\label{fig:Oberrhein_maintenance}
\end{subfigure}
\centering
    \caption{\textcolor{black}{\textit{Replace-all} strategy applied to the Oberrhein system. (a) Histogram of the cable age distribution at $t=0$, $t=25$, and $t=50$ years. (b) Annual number of replacements $N(t)$.}    
    \vspace{-1em} }
    \label{fig:Oberrhein}
\end{figure}
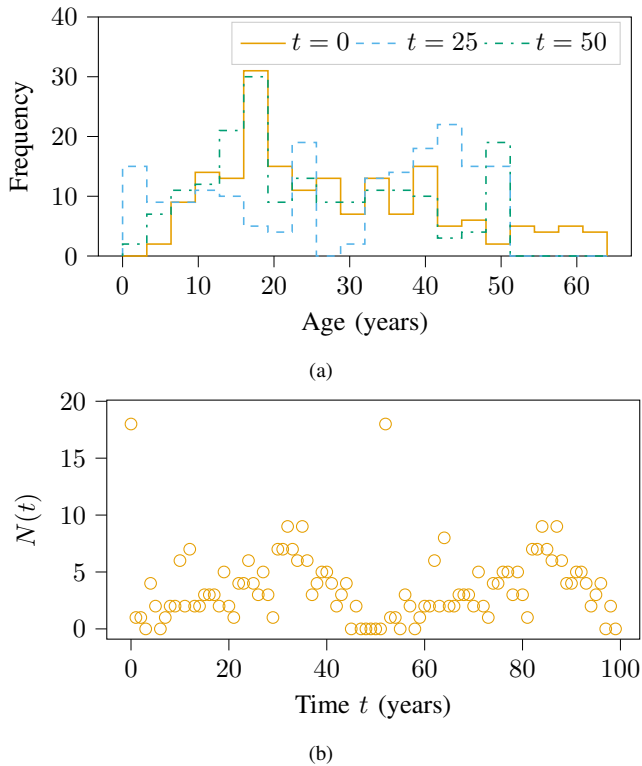

\begin{figure}
\centering
\input{betasenpilc.tex}
    \caption{Sensitivity of PILC results to the selection of $\beta$ in \textit{Run-to-failure}. \vspace{-1em}}
    \label{fig:betasens}
\end{figure}
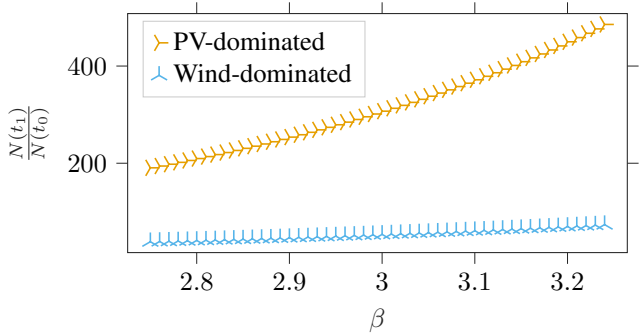

\subsection{Discussion on the sensitivity of results to ageing parameters}
The results and conclusions of this work are highly sensitive to the ageing parameters. For example, when repeating the study in Sec. \ref{sec:assessmain} with varying values for $\beta$, Fig. \ref{fig:betasens} shows the sensitivity of the XLPE results in the \textit{Run-to-failure} maintenance strategy. By varying $\beta$ just a small fraction, one can see that the maintenance increase can change by a factor $2$. This study shows high sensitivity on the shape parameter $\beta$ that appears in the exponent to evaluate maintenance increase \eqref{eq:fail}. Selecting the value $\beta$ is important as it tremendously impacts the results, i.e., some references estimate $\beta$ for all age-related modes of failures, not isolating only thermal ageing of cables. \textcolor{black}{This sensitivity is also structurally supported by the analytical shortcut formulations, where $\beta$ appears as an exponent in maintenance scaling relations (e.g. \eqref{eq:fail}), implying that small parameter variations propagate nonlinearly into maintenance predictions.} At the same time, the literature shows a high variability in reported $\beta$. For example, data from \cite{Tan23} suggest values closer to $\beta=6.3$, based on an average of four manufacturers. Overall, to fit an accurate $\beta$, we recommend conducting ageing (stress-tests) experiments to estimate failure rates under changes in operating conditions in an existing fleet of medium-voltage cables with heterogeneous ages. \textcolor{black}{While $\beta$ represents a dominant statistical uncertainty, other model parameters also influence the robustness of the quantitative predictions. Beyond the Weibull shape parameter $\beta$, the robustness of the quantitative results is mainly governed by parameters that influence the ageing acceleration factor $r(T)$, such as thermal ageing constants, temperature limits, and ampacity assumptions through the temperature model. Small variations in these parameters can lead to disproportionate changes in predicted maintenance because the shortcut formulations amplify changes in $r$, for example, through the relation $\frac{N(t_1)}{N(t_0)} = \left(\frac{r_1}{r_0}\right)^{\beta}$ in the \textit{Run-to-failure} strategy. This implies that uncertainties affecting operation close to ampacity have the strongest impact on results, while uncertainties affecting lower loading regimes have comparatively smaller influence. Overall, the results indicate that while quantitative maintenance estimates are sensitive to ageing and loading parameters, the qualitative findings regarding the dominant influence of high-loading periods and transition-driven ageing trends remain robust across plausible parameter variations. Consequently, the proposed framework is primarily intended to provide structurally reliable system-level insight, while absolute maintenance levels should be interpreted with consideration of parameter uncertainty. The present model assumes monotonic cumulative ageing and does not explicitly capture potential thermal recovery effects during extended low-loading periods. Incorporating recovery dynamics would require physics-based degradation modelling or state-dependent ageing formulations and represents a relevant direction for future research.}

\section{Conclusion}
\label{sec:cond}
Time-varying loading of MV cables impacts the estimations of failure rates, estimated lifespan, and maintenance required. 
\textcolor{black}{Considering time variation and loading variability is essential for reliability assessment in electricity systems undergoing the energy transition.}
The proposed approach combines probability estimates of the loading with shortcut methods to estimate maintenance increase before and after a systematic load change, such as during the energy transition. This method supports estimating on MV grid data from Denmark and Germany, where our case study demonstrates that the top $1.4$\% of times with the highest loading slightly above the \textcolor{black}{ampacity} causes $46$\% of thermal-related cable ageing. These changes specifically impact PILC cables, and operators may expect increasing failure rates. Our studies show up to $10-300$ times higher maintenance needs in the future, meaning that society may face a large financial burden of incorrectly performed maintenance. \textcolor{black}{Overall, the results quantify how systematic increases in loading variability can significantly accelerate population-level ageing and maintenance demand. Qualitatively, the study reveals that reliability degradation under evolving operating conditions is dominated by nonlinear stress accumulation effects that are not captured by stationary reliability modelling assumptions.} These findings also suggest operators need to carefully assess their maintenance strategies, failure models, and accurately estimate model parameters to guide society to a cost-efficient energy transition.

\vspace{-0.5em}

\appendix
\subsection{Temperature-dependent electrical resistivity}
The conductor temperature $T$ can be estimated with \eqref{eq:conductemp} that can be expanded considering the material-specific temperature coefficient $\alpha$ as follows

{\footnotesize
\begin{equation}\label{eq:conductempepsilon}
    T(\tau) = T_a + (T_{max}-T_a) \cdot \left(\frac{|I_c(\tau)|}  {I_z}\right)^2 \cdot \frac{1+\alpha \cdot\left(T (\tau-1)-T_a\right)}{1+\alpha \cdot \left(T_{max} -T_a\right)}
\end{equation}
}
where $T (\tau-1)$ refers to the previous time step. For steady conditions, one can consider $T (\tau) = T (\tau-1)$, and either analytically solve this equation for $T$, or numerically approximate it, for example, using as initial point for $T (\tau-1)$ from \eqref{eq:conductemp}.

\bibliographystyle{IEEEtran}
\bstctlcite{IEEEexample:BSTcontrol}
\bibliography{ref.bib}

\end{document}

%% file: failden.tex
\begin{tikzpicture}

\definecolor{cornflowerblue86180233}{RGB}{86,180,233}
\definecolor{darkslategray38}{RGB}{38,38,38}
\definecolor{lightgray204}{RGB}{204,204,204}
\definecolor{orange2301590}{RGB}{230,159,0}

\begin{axis}[height=0.1\textheight,width=0.75\columnwidth,
scale only axis,
axis line style={darkslategray38},
legend cell align={left},
legend style={
  fill opacity=0.8,
  draw opacity=1,
  text opacity=1,
  at={(0.03,0.97)},
  anchor=north west,
  draw=lightgray204,
  legend columns=2
},
tick align=outside,
x grid style={lightgray204},
xlabel=\textcolor{darkslategray38}{Time (year)},
xmajorticks=true,
xmin=2013, xmax=2023,
xtick style={color=darkslategray38},
y grid style={lightgray204},
ylabel={Failure rate $\lambda(t)$ \\per 100\,km},
ylabel style={align=center, text=darkslategray38},,
ymajorticks=true,
xticklabel style={/pgf/number format/1000 sep=},
ymin=0, ymax=6,
ytick style={color=darkslategray38}
]
\addplot [semithick, orange2301590, mark=Mercedes star, mark size=7.5, mark options={solid}, only marks]
table {%
2013 0.4163
2014 0.3755
2015 0.4163
2016 0.449
2017 0.4408
2018 0.4408
2019 0.449
2020 0.5061
2021 0.4327
2022 0.5306
2023 0.5714
};
\addlegendentry{XLPE}
\addplot [semithick,cornflowerblue86180233 , mark=Mercedes star, mark size=7.5, mark options={solid,rotate=270}, only marks]
table {%
2013 2.694
2014 2.751
2015 2.8
2016 3.037
2017 2.98
2018 3.331
2019 3.249
2020 3.698
2021 3.624
2022 4.008
2023 4.996
};
\addlegendentry{PILC}

\end{axis}

\end{tikzpicture}

%% file: loads.tex
\begin{tikzpicture}

\definecolor{cornflowerblue86180233}{RGB}{86,180,233}
\definecolor{darkcyan0158115}{RGB}{0,158,115}
\definecolor{darkslategray38}{RGB}{38,38,38}
\definecolor{lightgray204}{RGB}{204,204,204}
\definecolor{orange2301590}{RGB}{230,159,0}

\begin{axis}[height=0.1\textheight,width=0.8\columnwidth,
scale only axis,
axis line style={darkslategray38},
legend cell align={left},
legend style={fill opacity=0.8, draw opacity=1, text opacity=1, draw=lightgray204},
tick align=outside,
x grid style={lightgray204},
xlabel=\textcolor{darkslategray38}{Time (hour)},
xmajorticks=true,
xmin=-422.6, xmax=8875,
xtick style={color=darkslategray38},
y grid style={lightgray204},
ylabel=\textcolor{darkslategray38}{$\frac{I_c}{I_z}$},
ymajorticks=true,
ymin=-0.04093, ymax=1.268,
ytick style={color=darkslategray38}
]
\addplot [orange2301590, mark=Mercedes star, mark size=1, mark options={solid,rotate=270}, only marks]
table {%
0 0.4951
12 0.4854
24 0.4757
36 0.4659
48 0.4613
60 0.4593
72 0.4572
84 0.4552
96 0.4531
108 0.451
120 0.449
133 0.4468
145 0.4447
157 0.4427
169 0.4406
181 0.4385
193 0.4365
205 0.4344
217 0.4324
229 0.4303
241 0.4283
253 0.4262
266 0.424
278 0.4219
290 0.4199
302 0.4179
314 0.4159
326 0.4139
338 0.4119
350 0.4099
362 0.4078
374 0.4058
386 0.4038
399 0.4017
411 0.3996
423 0.3976
435 0.3956
447 0.3936
459 0.3927
471 0.3918
483 0.3909
495 0.39
507 0.3891
519 0.3882
532 0.3873
544 0.3864
556 0.3855
568 0.3846
580 0.3837
592 0.3829
604 0.382
616 0.3811
628 0.3802
640 0.3793
652 0.3784
665 0.3775
677 0.3766
689 0.3757
701 0.3748
713 0.3739
725 0.373
737 0.3722
749 0.3713
761 0.3704
773 0.3695
785 0.3686
798 0.3677
810 0.3671
822 0.3667
834 0.3662
846 0.3657
858 0.3652
870 0.3647
882 0.3643
894 0.3638
906 0.3633
918 0.3628
931 0.3623
943 0.3618
955 0.3614
967 0.3609
979 0.3604
991 0.3599
1003 0.3594
1015 0.359
1027 0.3585
1039 0.358
1051 0.3575
1064 0.357
1076 0.3565
1088 0.356
1100 0.3556
1112 0.3551
1124 0.3546
1136 0.3541
1148 0.3536
1160 0.3532
1172 0.3527
1184 0.3522
1197 0.3517
1209 0.3512
1221 0.3507
1233 0.3503
1245 0.3498
1257 0.3493
1269 0.3488
1281 0.3483
1293 0.3479
1305 0.3476
1317 0.3473
1330 0.347
1342 0.3466
1354 0.3463
1366 0.346
1378 0.3457
1390 0.3453
1402 0.345
1414 0.3447
1426 0.3444
1438 0.3441
1450 0.3437
1463 0.3434
1475 0.3431
1487 0.3428
1499 0.3424
1511 0.3421
1523 0.3418
1535 0.3415
1547 0.3412
1559 0.3408
1571 0.3405
1583 0.3402
1596 0.3398
1608 0.3395
1620 0.3392
1632 0.3389
1644 0.3386
1656 0.3382
1668 0.3379
1680 0.3376
1692 0.3373
1704 0.337
1717 0.3366
1729 0.3363
1741 0.336
1753 0.3356
1765 0.3353
1777 0.335
1789 0.3347
1801 0.3344
1813 0.334
1825 0.3337
1837 0.3334
1850 0.3331
1862 0.3327
1874 0.3324
1886 0.3321
1898 0.3318
1910 0.3314
1922 0.3311
1934 0.3308
1946 0.3305
1958 0.3302
1970 0.3298
1983 0.3295
1995 0.3292
2007 0.3289
2019 0.3285
2031 0.3282
2043 0.3279
2055 0.3276
2067 0.3272
2079 0.3269
2091 0.3266
2103 0.3263
2116 0.3259
2128 0.3256
2140 0.3253
2152 0.325
2164 0.3247
2176 0.3243
2188 0.324
2200 0.3237
2212 0.3234
2224 0.3231
2236 0.3227
2249 0.3224
2261 0.3221
2273 0.3217
2285 0.3214
2297 0.3211
2309 0.3208
2321 0.3205
2333 0.3201
2345 0.3198
2357 0.3195
2369 0.3192
2382 0.3188
2394 0.3185
2406 0.3182
2418 0.3179
2430 0.3175
2442 0.3172
2454 0.3169
2466 0.3166
2478 0.3163
2490 0.3159
2502 0.3156
2515 0.3153
2527 0.3149
2539 0.3146
2551 0.3143
2563 0.314
2575 0.3137
2587 0.3133
2599 0.313
2611 0.3127
2623 0.3124
2635 0.3121
2648 0.3117
2660 0.3114
2672 0.3112
2684 0.3109
2696 0.3107
2708 0.3105
2720 0.3103
2732 0.31
2744 0.3098
2756 0.3096
2768 0.3094
2781 0.3091
2793 0.3089
2805 0.3087
2817 0.3084
2829 0.3082
2841 0.308
2853 0.3078
2865 0.3075
2877 0.3073
2889 0.3071
2901 0.3069
2914 0.3066
2926 0.3064
2938 0.3062
2950 0.3059
2962 0.3057
2974 0.3055
2986 0.3053
2998 0.305
3010 0.3048
3022 0.3046
3034 0.3044
3047 0.3041
3059 0.3039
3071 0.3037
3083 0.3034
3095 0.3032
3107 0.303
3119 0.3028
3131 0.3025
3143 0.3023
3155 0.3021
3167 0.3019
3180 0.3016
3192 0.3014
3204 0.3012
3216 0.3009
3228 0.3007
3240 0.3005
3252 0.3003
3264 0.3
3276 0.2998
3288 0.2996
3300 0.2994
3313 0.2991
3325 0.2989
3337 0.2987
3349 0.2985
3361 0.2982
3373 0.298
3385 0.2978
3397 0.2976
3409 0.2973
3421 0.2971
3434 0.2969
3446 0.2966
3458 0.2964
3470 0.2962
3482 0.296
3494 0.2957
3506 0.2955
3518 0.2953
3530 0.2951
3542 0.2948
3554 0.2946
3567 0.2944
3579 0.2941
3591 0.2939
3603 0.2937
3615 0.2935
3627 0.2932
3639 0.293
3651 0.2928
3663 0.2926
3675 0.2923
3687 0.2921
3700 0.2919
3712 0.2916
3724 0.2914
3736 0.2912
3748 0.291
3760 0.2907
3772 0.2905
3784 0.2903
3796 0.2901
3808 0.2898
3820 0.2896
3833 0.2894
3845 0.2891
3857 0.2889
3869 0.2887
3881 0.2885
3893 0.2882
3905 0.288
3917 0.2878
3929 0.2876
3941 0.2873
3953 0.2871
3966 0.2869
3978 0.2866
3990 0.2864
4002 0.2862
4014 0.286
4026 0.2857
4038 0.2855
4050 0.2853
4062 0.285
4074 0.2847
4086 0.2844
4099 0.2841
4111 0.2838
4123 0.2835
4135 0.2832
4147 0.2829
4159 0.2826
4171 0.2823
4183 0.282
4195 0.2817
4207 0.2814
4219 0.2811
4232 0.2808
4244 0.2805
4256 0.2802
4268 0.2799
4280 0.2796
4292 0.2793
4304 0.279
4316 0.2788
4328 0.2786
4340 0.2783
4352 0.2781
4365 0.2779
4377 0.2777
4389 0.2774
4401 0.2772
4413 0.277
4425 0.2768
4437 0.2766
4449 0.2764
4461 0.2761
4473 0.2759
4485 0.2757
4498 0.2755
4510 0.2752
4522 0.275
4534 0.2748
4546 0.2746
4558 0.2744
4570 0.2741
4582 0.2739
4594 0.2737
4606 0.2735
4618 0.2733
4631 0.273
4643 0.2728
4655 0.2726
4667 0.2724
4679 0.2722
4691 0.2719
4703 0.2717
4715 0.2715
4727 0.2713
4739 0.2711
4751 0.2709
4764 0.2706
4776 0.2704
4788 0.2702
4800 0.27
4812 0.2697
4824 0.2695
4836 0.2693
4848 0.2691
4860 0.2689
4872 0.2687
4884 0.2684
4897 0.2682
4909 0.268
4921 0.2678
4933 0.2675
4945 0.2673
4957 0.2671
4969 0.2669
4981 0.2667
4993 0.2664
5005 0.2662
5017 0.266
5030 0.2658
5042 0.2656
5054 0.2653
5066 0.2651
5078 0.2649
5090 0.2647
5102 0.2645
5114 0.2642
5126 0.264
5138 0.2638
5151 0.2635
5163 0.2631
5175 0.2628
5187 0.2624
5199 0.2621
5211 0.2618
5223 0.2614
5235 0.2611
5247 0.2607
5259 0.2604
5271 0.2601
5284 0.2597
5296 0.2594
5308 0.259
5320 0.2587
5332 0.2583
5344 0.258
5356 0.2577
5368 0.2573
5380 0.257
5392 0.2566
5404 0.2563
5417 0.2559
5429 0.2556
5441 0.2553
5453 0.2549
5465 0.2546
5477 0.2542
5489 0.2539
5501 0.2536
5513 0.2532
5525 0.2529
5537 0.2525
5550 0.2522
5562 0.2518
5574 0.2515
5586 0.2512
5598 0.2508
5610 0.2505
5622 0.2501
5634 0.2498
5646 0.2495
5658 0.2491
5670 0.2488
5683 0.2481
5695 0.2473
5707 0.2464
5719 0.2456
5731 0.2448
5743 0.244
5755 0.2431
5767 0.2423
5779 0.2415
5791 0.2407
5803 0.2398
5816 0.2389
5828 0.2381
5840 0.2373
5852 0.2365
5864 0.2356
5876 0.2348
5888 0.234
5900 0.2332
5912 0.2323
5924 0.2315
5936 0.2307
5949 0.2298
5961 0.229
5973 0.2281
5985 0.2273
5997 0.2265
6009 0.2257
6021 0.2248
6033 0.224
6045 0.2232
6057 0.2224
6069 0.2215
6082 0.2206
6094 0.2198
6106 0.219
6118 0.2182
6130 0.2173
6142 0.2165
6154 0.2158
6166 0.215
6178 0.2143
6190 0.2136
6202 0.2129
6215 0.2121
6227 0.2114
6239 0.2106
6251 0.2099
6263 0.2092
6275 0.2085
6287 0.2078
6299 0.207
6311 0.2063
6323 0.2056
6335 0.2049
6348 0.2041
6360 0.2034
6372 0.2027
6384 0.2019
6396 0.2012
6408 0.2005
6420 0.1998
6432 0.199
6444 0.1983
6456 0.1976
6468 0.1969
6481 0.1961
6493 0.1954
6505 0.1947
6517 0.1939
6529 0.1932
6541 0.1925
6553 0.1918
6565 0.1911
6577 0.1903
6589 0.1896
6601 0.1889
6614 0.1881
6626 0.1874
6638 0.1867
6650 0.1859
6662 0.1852
6674 0.1845
6686 0.1838
6698 0.1831
6710 0.1823
6722 0.1816
6734 0.1809
6747 0.1801
6759 0.1794
6771 0.1787
6783 0.1779
6795 0.1772
6807 0.1765
6819 0.1758
6831 0.1751
6843 0.1743
6855 0.1736
6868 0.1729
6880 0.1726
6892 0.1722
6904 0.1719
6916 0.1716
6928 0.1712
6940 0.1709
6952 0.1705
6964 0.1702
6976 0.1699
6988 0.1695
7001 0.1691
7013 0.1688
7025 0.1685
7037 0.1681
7049 0.1678
7061 0.1675
7073 0.1671
7085 0.1668
7097 0.1664
7109 0.1661
7121 0.1658
7134 0.1654
7146 0.1651
7158 0.1647
7170 0.1644
7182 0.164
7194 0.1637
7206 0.1634
7218 0.163
7230 0.1627
7242 0.1623
7254 0.162
7267 0.1616
7279 0.1613
7291 0.161
7303 0.1606
7315 0.1603
7327 0.1599
7339 0.1596
7351 0.1593
7363 0.1589
7375 0.1586
7387 0.1582
7400 0.1579
7412 0.1577
7424 0.1575
7436 0.1573
7448 0.1571
7460 0.157
7472 0.1568
7484 0.1566
7496 0.1564
7508 0.1563
7520 0.1561
7533 0.1559
7545 0.1557
7557 0.1555
7569 0.1554
7581 0.1552
7593 0.155
7605 0.1548
7617 0.1546
7629 0.1545
7641 0.1543
7653 0.1541
7666 0.1539
7678 0.1537
7690 0.1536
7702 0.1534
7714 0.1532
7726 0.153
7738 0.1528
7750 0.1527
7762 0.1525
7774 0.1523
7786 0.1521
7799 0.1519
7811 0.1518
7823 0.1516
7835 0.1514
7847 0.1512
7859 0.1511
7871 0.1509
7883 0.1507
7895 0.1505
7907 0.1503
7919 0.1502
7932 0.15
7944 0.1498
7956 0.1496
7968 0.1494
7980 0.1493
7992 0.149
8004 0.1488
8016 0.1486
8028 0.1483
8040 0.1481
8052 0.1479
8065 0.1476
8077 0.1474
8089 0.1471
8101 0.1469
8113 0.1467
8125 0.1464
8137 0.1462
8149 0.1459
8161 0.1457
8173 0.1455
8185 0.1452
8198 0.145
8210 0.1447
8222 0.1445
8234 0.1443
8246 0.144
8258 0.1438
8270 0.1436
8282 0.1433
8294 0.1431
8306 0.1429
8318 0.1426
8331 0.1424
8343 0.1421
8355 0.1419
8367 0.1416
8379 0.1414
8391 0.1412
8403 0.1409
8415 0.1407
8427 0.1405
8439 0.1402
8452 0.14
};
\addlegendentry{Without DER}
\addplot [cornflowerblue86180233, mark=Mercedes star, mark size=1, mark options={solid}, only marks]
table {%
0 1.209
12 1.182
24 1.155
36 1.127
48 1.1
60 1.089
72 1.078
84 1.067
96 1.056
108 1.044
120 1.033
133 1.021
145 1.01
157 1
169 0.994
181 0.9875
193 0.981
205 0.9745
217 0.968
229 0.9615
241 0.955
253 0.9485
266 0.9415
278 0.935
290 0.9285
302 0.922
314 0.9155
326 0.909
338 0.9027
350 0.8973
362 0.8919
374 0.8865
386 0.8811
399 0.8752
411 0.8698
423 0.8644
435 0.859
447 0.8536
459 0.8482
471 0.8428
483 0.8374
495 0.832
507 0.8265
519 0.8211
532 0.8153
544 0.8099
556 0.8043
568 0.7986
580 0.7928
592 0.7871
604 0.7814
616 0.7756
628 0.7699
640 0.7641
652 0.7584
665 0.7522
677 0.7464
689 0.7407
701 0.7349
713 0.7292
725 0.7234
737 0.7177
749 0.712
761 0.7062
773 0.7007
785 0.6952
798 0.6892
810 0.6837
822 0.6781
834 0.6726
846 0.6671
858 0.6615
870 0.656
882 0.6505
894 0.6449
906 0.6394
918 0.6339
931 0.6279
943 0.6224
955 0.6168
967 0.6113
979 0.6059
991 0.6007
1003 0.5955
1015 0.5903
1027 0.5851
1039 0.5799
1051 0.5747
1064 0.5691
1076 0.5639
1088 0.5587
1100 0.5535
1112 0.5483
1124 0.5431
1136 0.5379
1148 0.5327
1160 0.5275
1172 0.5223
1184 0.5172
1197 0.5115
1209 0.5063
1221 0.5021
1233 0.4989
1245 0.4956
1257 0.4924
1269 0.4891
1281 0.4859
1293 0.4827
1305 0.4794
1317 0.4762
1330 0.4726
1342 0.4694
1354 0.4662
1366 0.4629
1378 0.4597
1390 0.4564
1402 0.4532
1414 0.4499
1426 0.4467
1438 0.4434
1450 0.4402
1463 0.4367
1475 0.4334
1487 0.4302
1499 0.4269
1511 0.4237
1523 0.4204
1535 0.4172
1547 0.4139
1559 0.4107
1571 0.4075
1583 0.4042
1596 0.4007
1608 0.3981
1620 0.3956
1632 0.3932
1644 0.3907
1656 0.3882
1668 0.3858
1680 0.3833
1692 0.3809
1704 0.3784
1717 0.3758
1729 0.3733
1741 0.3709
1753 0.3684
1765 0.366
1777 0.3635
1789 0.361
1801 0.3586
1813 0.3561
1825 0.3537
1837 0.3512
1850 0.3486
1862 0.3461
1874 0.3437
1886 0.3412
1898 0.3387
1910 0.3363
1922 0.3338
1934 0.3321
1946 0.3308
1958 0.3295
1970 0.3282
1983 0.3268
1995 0.3255
2007 0.3242
2019 0.3229
2031 0.3216
2043 0.3203
2055 0.319
2067 0.3177
2079 0.3164
2091 0.3151
2103 0.3138
2116 0.3124
2128 0.3111
2140 0.3098
2152 0.3085
2164 0.3072
2176 0.3059
2188 0.3046
2200 0.3034
2212 0.3021
2224 0.3008
2236 0.2995
2249 0.298
2261 0.2967
2273 0.2955
2285 0.2942
2297 0.2929
2309 0.2916
2321 0.2903
2333 0.2891
2345 0.2882
2357 0.2872
2369 0.2862
2382 0.2851
2394 0.2842
2406 0.2832
2418 0.2822
2430 0.2812
2442 0.2802
2454 0.2793
2466 0.2783
2478 0.2773
2490 0.2763
2502 0.2753
2515 0.2743
2527 0.2733
2539 0.2723
2551 0.2713
2563 0.2703
2575 0.2694
2587 0.2684
2599 0.2674
2611 0.2664
2623 0.2654
2635 0.2644
2648 0.2634
2660 0.2624
2672 0.2615
2684 0.2609
2696 0.2603
2708 0.2596
2720 0.259
2732 0.2584
2744 0.2578
2756 0.2572
2768 0.2565
2781 0.2559
2793 0.2552
2805 0.2546
2817 0.254
2829 0.2534
2841 0.2528
2853 0.2522
2865 0.2515
2877 0.2509
2889 0.2503
2901 0.2497
2914 0.249
2926 0.2484
2938 0.2478
2950 0.2471
2962 0.2465
2974 0.2459
2986 0.2453
2998 0.2447
3010 0.244
3022 0.2434
3034 0.2428
3047 0.2421
3059 0.2415
3071 0.2409
3083 0.2403
3095 0.2397
3107 0.239
3119 0.2384
3131 0.2378
3143 0.2372
3155 0.2366
3167 0.2359
3180 0.2353
3192 0.2347
3204 0.234
3216 0.2334
3228 0.2328
3240 0.2322
3252 0.2316
3264 0.2309
3276 0.2303
3288 0.2297
3300 0.2291
3313 0.2284
3325 0.2278
3337 0.2272
3349 0.2265
3361 0.2259
3373 0.2253
3385 0.2247
3397 0.2241
3409 0.2234
3421 0.2228
3434 0.2222
3446 0.2215
3458 0.2209
3470 0.2203
3482 0.2197
3494 0.2191
3506 0.2184
3518 0.2178
3530 0.2172
3542 0.2166
3554 0.216
3567 0.2153
3579 0.2147
3591 0.2141
3603 0.2134
3615 0.2128
3627 0.2122
3639 0.2116
3651 0.211
3663 0.2103
3675 0.2097
3687 0.2091
3700 0.2084
3712 0.2078
3724 0.2072
3736 0.2065
3748 0.2059
3760 0.2052
3772 0.2046
3784 0.2039
3796 0.2032
3808 0.2026
3820 0.2019
3833 0.2012
3845 0.2006
3857 0.1999
3869 0.1993
3881 0.1986
3893 0.198
3905 0.1973
3917 0.1967
3929 0.196
3941 0.1954
3953 0.1947
3966 0.194
3978 0.1933
3990 0.1927
4002 0.192
4014 0.1914
4026 0.1907
4038 0.1901
4050 0.1894
4062 0.1888
4074 0.1881
4086 0.1875
4099 0.1868
4111 0.1861
4123 0.1855
4135 0.1848
4147 0.1841
4159 0.1835
4171 0.1828
4183 0.1822
4195 0.1815
4207 0.1809
4219 0.1802
4232 0.1795
4244 0.1789
4256 0.1782
4268 0.1776
4280 0.1769
4292 0.1763
4304 0.1756
4316 0.175
4328 0.1743
4340 0.1736
4352 0.173
4365 0.1723
4377 0.1716
4389 0.171
4401 0.1703
4413 0.1697
4425 0.169
4437 0.1684
4449 0.1677
4461 0.1671
4473 0.1664
4485 0.1658
4498 0.165
4510 0.1644
4522 0.1637
4534 0.1631
4546 0.1624
4558 0.1618
4570 0.1611
4582 0.1605
4594 0.1598
4606 0.1592
4618 0.1585
4631 0.1578
4643 0.1572
4655 0.1565
4667 0.1559
4679 0.1552
4691 0.1545
4703 0.1539
4715 0.1532
4727 0.1526
4739 0.1519
4751 0.1513
4764 0.1506
4776 0.1499
4788 0.1493
4800 0.1486
4812 0.148
4824 0.1473
4836 0.1467
4848 0.146
4860 0.1454
4872 0.1447
4884 0.144
4897 0.1433
4909 0.1427
4921 0.142
4933 0.1414
4945 0.1407
4957 0.1401
4969 0.1394
4981 0.1388
4993 0.1381
5005 0.1375
5017 0.1368
5030 0.1361
5042 0.1354
5054 0.1348
5066 0.1341
5078 0.1335
5090 0.1328
5102 0.1322
5114 0.1317
5126 0.1313
5138 0.131
5151 0.1306
5163 0.1303
5175 0.1299
5187 0.1295
5199 0.1292
5211 0.1288
5223 0.1285
5235 0.1281
5247 0.1278
5259 0.1274
5271 0.1271
5284 0.1267
5296 0.1263
5308 0.126
5320 0.1256
5332 0.1253
5344 0.1249
5356 0.1246
5368 0.1242
5380 0.1239
5392 0.1235
5404 0.1231
5417 0.1228
5429 0.1224
5441 0.1221
5453 0.1217
5465 0.1213
5477 0.121
5489 0.1206
5501 0.1203
5513 0.1199
5525 0.1196
5537 0.1192
5550 0.1188
5562 0.1185
5574 0.1181
5586 0.1178
5598 0.1174
5610 0.1171
5622 0.1167
5634 0.1164
5646 0.116
5658 0.1157
5670 0.1153
5683 0.1149
5695 0.1146
5707 0.1142
5719 0.1139
5731 0.1135
5743 0.1132
5755 0.1128
5767 0.1124
5779 0.1121
5791 0.1117
5803 0.1114
5816 0.111
5828 0.1106
5840 0.1103
5852 0.1099
5864 0.1096
5876 0.1092
5888 0.1089
5900 0.1085
5912 0.1082
5924 0.1078
5936 0.1075
5949 0.1071
5961 0.1067
5973 0.1064
5985 0.106
5997 0.1057
6009 0.1053
6021 0.105
6033 0.1046
6045 0.1043
6057 0.1039
6069 0.1035
6082 0.1032
6094 0.1028
6106 0.1025
6118 0.1021
6130 0.1017
6142 0.1014
6154 0.101
6166 0.1007
6178 0.1003
6190 0.09998
6202 0.09962
6215 0.09924
6227 0.09888
6239 0.09853
6251 0.09818
6263 0.09782
6275 0.09747
6287 0.09712
6299 0.09676
6311 0.09641
6323 0.09605
6335 0.0957
6348 0.09532
6360 0.09497
6372 0.09464
6384 0.09431
6396 0.09398
6408 0.09364
6420 0.09331
6432 0.09298
6444 0.09264
6456 0.09231
6468 0.09198
6481 0.09162
6493 0.09128
6505 0.09095
6517 0.09062
6529 0.09029
6541 0.08995
6553 0.08962
6565 0.08929
6577 0.08895
6589 0.08862
6601 0.08829
6614 0.08793
6626 0.08759
6638 0.08726
6650 0.08693
6662 0.0866
6674 0.08626
6686 0.08593
6698 0.0856
6710 0.08526
6722 0.08493
6734 0.0846
6747 0.08424
6759 0.0839
6771 0.08357
6783 0.08324
6795 0.0829
6807 0.08257
6819 0.08224
6831 0.08191
6843 0.08157
6855 0.08124
6868 0.08088
6880 0.08055
6892 0.08021
6904 0.07988
6916 0.07955
6928 0.07921
6940 0.07888
6952 0.07855
6964 0.07822
6976 0.07788
6988 0.07747
7001 0.07701
7013 0.07658
7025 0.07615
7037 0.07572
7049 0.0753
7061 0.07487
7073 0.07444
7085 0.07402
7097 0.07359
7109 0.07316
7121 0.07273
7134 0.07227
7146 0.07184
7158 0.07142
7170 0.07099
7182 0.07056
7194 0.07014
7206 0.06971
7218 0.06928
7230 0.06885
7242 0.06843
7254 0.068
7267 0.06754
7279 0.06711
7291 0.06668
7303 0.06626
7315 0.06583
7327 0.0654
7339 0.06497
7351 0.06455
7363 0.06412
7375 0.06369
7387 0.06327
7400 0.0628
7412 0.06238
7424 0.06195
7436 0.06152
7448 0.0611
7460 0.06067
7472 0.06024
7484 0.05981
7496 0.05939
7508 0.05896
7520 0.05853
7533 0.05807
7545 0.05764
7557 0.05722
7569 0.05679
7581 0.05636
7593 0.05601
7605 0.05571
7617 0.0554
7629 0.05509
7641 0.05479
7653 0.05448
7666 0.05415
7678 0.05384
7690 0.05353
7702 0.05322
7714 0.05292
7726 0.05261
7738 0.0523
7750 0.052
7762 0.05169
7774 0.05138
7786 0.05107
7799 0.05074
7811 0.05043
7823 0.05013
7835 0.04982
7847 0.04951
7859 0.04921
7871 0.0489
7883 0.04859
7895 0.04829
7907 0.04798
7919 0.04767
7932 0.04734
7944 0.04703
7956 0.04672
7968 0.04642
7980 0.04611
7992 0.0458
8004 0.0455
8016 0.04519
8028 0.04488
8040 0.04457
8052 0.04427
8065 0.04394
8077 0.04363
8089 0.04332
8101 0.04301
8113 0.04271
8125 0.0424
8137 0.04209
8149 0.04179
8161 0.04148
8173 0.04117
8185 0.04061
8198 0.03987
8210 0.0392
8222 0.03853
8234 0.03785
8246 0.03718
8258 0.0365
8270 0.03583
8282 0.03515
8294 0.03448
8306 0.0338
8318 0.03313
8331 0.0324
8343 0.03172
8355 0.03105
8367 0.03037
8379 0.0297
8391 0.02903
8403 0.02835
8415 0.02768
8427 0.027
8439 0.02633
8452 0.0256
};
\addlegendentry{PV-dominated}
\addplot [darkcyan0158115, mark=Mercedes star, mark size=1, mark options={solid,rotate=90}, only marks]
table {%
0 1.209
12 1.054
24 0.9868
36 0.9655
48 0.9442
60 0.9228
72 0.9016
84 0.8904
96 0.8793
108 0.8682
120 0.857
133 0.8489
145 0.8469
157 0.8449
169 0.8429
181 0.8409
193 0.839
205 0.837
217 0.835
229 0.833
241 0.831
253 0.829
266 0.8269
278 0.8249
290 0.8229
302 0.8209
314 0.819
326 0.817
338 0.815
350 0.813
362 0.811
374 0.809
386 0.8071
399 0.8049
411 0.8029
423 0.8009
435 0.799
447 0.797
459 0.795
471 0.793
483 0.791
495 0.789
507 0.7871
519 0.7851
532 0.7829
544 0.7809
556 0.779
568 0.7771
580 0.7754
592 0.7736
604 0.7718
616 0.7701
628 0.7683
640 0.7666
652 0.7648
665 0.7629
677 0.7611
689 0.7594
701 0.7576
713 0.7558
725 0.7541
737 0.7523
749 0.7506
761 0.7488
773 0.747
785 0.7453
798 0.7434
810 0.7416
822 0.7398
834 0.7381
846 0.7363
858 0.7346
870 0.7328
882 0.731
894 0.7293
906 0.7275
918 0.7258
931 0.7239
943 0.7221
955 0.7203
967 0.7186
979 0.7168
991 0.715
1003 0.7133
1015 0.7115
1027 0.7098
1039 0.708
1051 0.7062
1064 0.7043
1076 0.7026
1088 0.7008
1100 0.699
1112 0.6973
1124 0.6955
1136 0.6938
1148 0.692
1160 0.6902
1172 0.6885
1184 0.6867
1197 0.6848
1209 0.683
1221 0.6813
1233 0.6795
1245 0.6778
1257 0.676
1269 0.6742
1281 0.6725
1293 0.6707
1305 0.669
1317 0.6672
1330 0.6653
1342 0.6636
1354 0.6619
1366 0.6601
1378 0.6584
1390 0.6567
1402 0.6549
1414 0.6532
1426 0.6514
1438 0.6497
1450 0.648
1463 0.6461
1475 0.6444
1487 0.6426
1499 0.6409
1511 0.6392
1523 0.6374
1535 0.6357
1547 0.6339
1559 0.6322
1571 0.6305
1583 0.6287
1596 0.6269
1608 0.6251
1620 0.6234
1632 0.6217
1644 0.6199
1656 0.6182
1668 0.6164
1680 0.6147
1692 0.613
1704 0.6112
1717 0.6094
1729 0.6076
1741 0.6059
1753 0.6042
1765 0.6024
1777 0.6007
1789 0.5989
1801 0.5972
1813 0.5955
1825 0.5937
1837 0.592
1850 0.5901
1862 0.5884
1874 0.5867
1886 0.5849
1898 0.5832
1910 0.5814
1922 0.5797
1934 0.578
1946 0.5762
1958 0.5745
1970 0.5728
1983 0.5709
1995 0.5692
2007 0.5674
2019 0.5657
2031 0.5639
2043 0.5622
2055 0.5605
2067 0.5587
2079 0.557
2091 0.5553
2103 0.5535
2116 0.5517
2128 0.5502
2140 0.5487
2152 0.5473
2164 0.5458
2176 0.5444
2188 0.5429
2200 0.5415
2212 0.54
2224 0.5386
2236 0.5371
2249 0.5356
2261 0.5341
2273 0.5327
2285 0.5312
2297 0.5298
2309 0.5283
2321 0.5269
2333 0.5254
2345 0.524
2357 0.5225
2369 0.5211
2382 0.5195
2394 0.5181
2406 0.5166
2418 0.5152
2430 0.5138
2442 0.5123
2454 0.5109
2466 0.5094
2478 0.508
2490 0.5065
2502 0.5051
2515 0.5035
2527 0.5021
2539 0.5006
2551 0.4992
2563 0.4977
2575 0.4963
2587 0.4948
2599 0.4934
2611 0.4919
2623 0.4905
2635 0.489
2648 0.4875
2660 0.486
2672 0.4846
2684 0.4831
2696 0.4817
2708 0.4802
2720 0.4788
2732 0.4773
2744 0.4759
2756 0.4745
2768 0.473
2781 0.4714
2793 0.47
2805 0.4685
2817 0.4671
2829 0.4657
2841 0.4642
2853 0.4628
2865 0.4613
2877 0.4599
2889 0.4584
2901 0.457
2914 0.4554
2926 0.454
2938 0.4525
2950 0.4511
2962 0.4496
2974 0.4482
2986 0.4467
2998 0.4453
3010 0.4438
3022 0.4424
3034 0.4409
3047 0.4394
3059 0.4379
3071 0.4365
3083 0.435
3095 0.4336
3107 0.4321
3119 0.4307
3131 0.4293
3143 0.4278
3155 0.4264
3167 0.4249
3180 0.4233
3192 0.4219
3204 0.4205
3216 0.419
3228 0.4176
3240 0.4161
3252 0.4147
3264 0.4132
3276 0.4118
3288 0.4103
3300 0.4089
3313 0.4073
3325 0.4059
3337 0.4044
3349 0.403
3361 0.4015
3373 0.4001
3385 0.3986
3397 0.3972
3409 0.3957
3421 0.3943
3434 0.3927
3446 0.3913
3458 0.3898
3470 0.3884
3482 0.3869
3494 0.3855
3506 0.384
3518 0.3826
3530 0.3812
3542 0.3797
3554 0.3783
3567 0.3767
3579 0.3752
3591 0.3738
3603 0.3724
3615 0.3709
3627 0.3695
3639 0.368
3651 0.3666
3663 0.3652
3675 0.3639
3687 0.3627
3700 0.3614
3712 0.3601
3724 0.3589
3736 0.3576
3748 0.3564
3760 0.3552
3772 0.3539
3784 0.3527
3796 0.3515
3808 0.3502
3820 0.349
3833 0.3476
3845 0.3464
3857 0.3452
3869 0.3439
3881 0.3427
3893 0.3415
3905 0.3402
3917 0.339
3929 0.3378
3941 0.3365
3953 0.3353
3966 0.3339
3978 0.3327
3990 0.3315
4002 0.3302
4014 0.329
4026 0.3278
4038 0.3265
4050 0.3253
4062 0.324
4074 0.3228
4086 0.3216
4099 0.3202
4111 0.319
4123 0.3178
4135 0.3165
4147 0.3153
4159 0.314
4171 0.3128
4183 0.3116
4195 0.3103
4207 0.3091
4219 0.3079
4232 0.3065
4244 0.3053
4256 0.3041
4268 0.3028
4280 0.3016
4292 0.3003
4304 0.2991
4316 0.2979
4328 0.2966
4340 0.2954
4352 0.2942
4365 0.2928
4377 0.2916
4389 0.2903
4401 0.2891
4413 0.2879
4425 0.2866
4437 0.2854
4449 0.2842
4461 0.2829
4473 0.2817
4485 0.2805
4498 0.2791
4510 0.2779
4522 0.2766
4534 0.2754
4546 0.2742
4558 0.2729
4570 0.2717
4582 0.2705
4594 0.2692
4606 0.268
4618 0.2667
4631 0.2654
4643 0.2642
4655 0.2629
4667 0.2617
4679 0.2605
4691 0.2592
4703 0.258
4715 0.2568
4727 0.2555
4739 0.2543
4751 0.253
4764 0.2517
4776 0.2505
4788 0.2492
4800 0.248
4812 0.2468
4824 0.2455
4836 0.2443
4848 0.2434
4860 0.2424
4872 0.2415
4884 0.2406
4897 0.2396
4909 0.2386
4921 0.2377
4933 0.2368
4945 0.2359
4957 0.2349
4969 0.234
4981 0.2331
4993 0.2322
5005 0.2312
5017 0.2303
5030 0.2293
5042 0.2284
5054 0.2274
5066 0.2265
5078 0.2256
5090 0.2247
5102 0.2237
5114 0.2228
5126 0.2219
5138 0.221
5151 0.2199
5163 0.219
5175 0.2181
5187 0.2172
5199 0.2162
5211 0.2153
5223 0.2144
5235 0.2135
5247 0.2125
5259 0.2116
5271 0.2107
5284 0.2097
5296 0.2086
5308 0.2076
5320 0.2065
5332 0.2055
5344 0.2045
5356 0.2034
5368 0.2024
5380 0.2013
5392 0.2003
5404 0.1992
5417 0.1981
5429 0.1971
5441 0.196
5453 0.195
5465 0.194
5477 0.1929
5489 0.1919
5501 0.1908
5513 0.1898
5525 0.1887
5537 0.1877
5550 0.1866
5562 0.1855
5574 0.1845
5586 0.1835
5598 0.1824
5610 0.1814
5622 0.1803
5634 0.1793
5646 0.1782
5658 0.1772
5670 0.1762
5683 0.175
5695 0.174
5707 0.173
5719 0.1719
5731 0.1709
5743 0.1698
5755 0.1688
5767 0.1677
5779 0.1667
5791 0.1657
5803 0.1646
5816 0.1635
5828 0.1625
5840 0.1614
5852 0.1604
5864 0.1593
5876 0.1583
5888 0.1572
5900 0.1562
5912 0.1552
5924 0.1541
5936 0.1531
5949 0.152
5961 0.1509
5973 0.1499
5985 0.1488
5997 0.1478
6009 0.1468
6021 0.1457
6033 0.1447
6045 0.1436
6057 0.1426
6069 0.1415
6082 0.1404
6094 0.1394
6106 0.1383
6118 0.1373
6130 0.1363
6142 0.1355
6154 0.1347
6166 0.134
6178 0.1333
6190 0.1325
6202 0.1318
6215 0.1309
6227 0.1302
6239 0.1295
6251 0.1287
6263 0.128
6275 0.1272
6287 0.1265
6299 0.1257
6311 0.125
6323 0.1242
6335 0.1235
6348 0.1227
6360 0.1219
6372 0.1212
6384 0.1204
6396 0.1197
6408 0.1189
6420 0.1182
6432 0.1174
6444 0.1167
6456 0.1159
6468 0.1152
6481 0.1144
6493 0.1136
6505 0.1129
6517 0.1121
6529 0.1114
6541 0.1107
6553 0.1099
6565 0.1092
6577 0.1084
6589 0.1077
6601 0.1069
6614 0.1061
6626 0.1054
6638 0.1046
6650 0.1039
6662 0.1031
6674 0.1024
6686 0.1016
6698 0.1009
6710 0.1001
6722 0.09938
6734 0.09864
6747 0.09783
6759 0.09708
6771 0.09633
6783 0.09558
6795 0.09484
6807 0.09409
6819 0.09334
6831 0.0926
6843 0.09185
6855 0.0911
6868 0.09029
6880 0.08954
6892 0.0888
6904 0.08805
6916 0.0873
6928 0.08656
6940 0.08581
6952 0.08506
6964 0.08431
6976 0.08357
6988 0.08282
7001 0.08201
7013 0.08126
7025 0.08052
7037 0.07977
7049 0.07902
7061 0.07828
7073 0.07753
7085 0.07678
7097 0.07603
7109 0.07529
7121 0.07454
7134 0.07373
7146 0.07314
7158 0.07262
7170 0.0721
7182 0.07157
7194 0.07105
7206 0.07053
7218 0.07001
7230 0.06948
7242 0.06896
7254 0.06844
7267 0.06787
7279 0.06735
7291 0.06683
7303 0.0663
7315 0.06578
7327 0.06526
7339 0.06474
7351 0.06421
7363 0.06369
7375 0.06317
7387 0.06265
7400 0.06208
7412 0.06156
7424 0.06103
7436 0.06051
7448 0.05999
7460 0.05947
7472 0.05894
7484 0.05842
7496 0.0579
7508 0.05737
7520 0.05685
7533 0.05629
7545 0.05576
7557 0.05524
7569 0.05472
7581 0.05419
7593 0.05367
7605 0.05315
7617 0.05263
7629 0.0521
7641 0.05158
7653 0.05106
7666 0.05049
7678 0.04997
7690 0.04945
7702 0.04892
7714 0.0484
7726 0.04788
7738 0.04736
7750 0.04683
7762 0.04631
7774 0.04579
7786 0.04527
7799 0.0447
7811 0.04418
7823 0.04365
7835 0.04313
7847 0.04261
7859 0.04209
7871 0.04156
7883 0.04104
7895 0.04052
7907 0.03999
7919 0.03947
7932 0.03891
7944 0.03838
7956 0.03786
7968 0.03734
7980 0.03682
7992 0.03629
8004 0.03577
8016 0.03525
8028 0.03472
8040 0.0342
8052 0.03368
8065 0.03311
8077 0.03259
8089 0.03212
8101 0.03167
8113 0.03122
8125 0.03078
8137 0.03033
8149 0.02988
8161 0.02943
8173 0.02899
8185 0.02854
8198 0.02805
8210 0.0276
8222 0.02716
8234 0.02671
8246 0.02626
8258 0.02581
8270 0.02537
8282 0.02492
8294 0.02447
8306 0.02402
8318 0.02358
8331 0.02309
8343 0.02264
8355 0.02219
8367 0.02175
8379 0.0213
8391 0.02085
8403 0.0204
8415 0.01996
8427 0.01951
8439 0.01906
8452 0.01858
};
\addlegendentry{Wind-dominated}
\end{axis}

\end{tikzpicture}

%% file: pfoberrhein.tex
\begin{tikzpicture}

\definecolor{darkgray176}{RGB}{176,176,176}
\definecolor{darkorange25512714}{RGB}{255,127,14}
\definecolor{lightgray204}{RGB}{204,204,204}
\definecolor{steelblue31119180}{RGB}{31,119,180}

\begin{axis}[height=0.1\textheight,width=0.81\columnwidth,
scale only axis,
tick align=outside,
tick pos=left,
x grid style={darkgray176},
xlabel={Hour of the day},
xmin=-1.15, xmax=24.15,
xtick style={color=black},
xtick={0,4,8,12,16,20,24},
xtick={0,4,8,12,16,20,24},
y grid style={darkgray176},
ylabel={ Mean load (MW)},
ymin=0.517776416505442, ymax=0.690563680134013,
ytick style={color=black},
ytick={0.52,0.6,0.68},
yticklabels={
  $\mathrm{0.52}$,
  $\mathrm{0.60}$,
  $\mathrm{0.68}$,
}
]
\addplot [semithick, darkorange25512714]
table {%
0 0.546567341891156
1 0.530521152462584
2 0.526740917319728
3 0.525630383034014
4 0.526291401319727
5 0.534486378462585
6 0.580167042748299
7 0.621750218462584
8 0.6260857207483
9 0.615194236748298
10 0.607818757319727
11 0.610039236748299
12 0.623124983319729
13 0.621829458176872
14 0.606893508462585
15 0.595764010748298
16 0.596187015319727
17 0.620957232176871
18 0.660637476462584
19 0.682709713605442
20 0.662805522176871
21 0.633990545034014
22 0.608295079319728
23 0.576281056462585
};
\end{axis}
\end{tikzpicture}

%% file: pfoberrhein2.tex
\begin{tikzpicture}

\definecolor{darkgray176}{RGB}{176,176,176}
\definecolor{darkorange25512714}{RGB}{255,127,14}
\definecolor{lightgray204}{RGB}{204,204,204}
\definecolor{steelblue31119180}{RGB}{31,119,180}

\begin{axis}[height=0.1\textheight,width=0.81\columnwidth,
scale only axis,
legend cell align={left},
legend columns=2,
legend style={
  fill opacity=0.8,
  draw opacity=1,
  text opacity=1,
  at={(0.499,0.25)},
  anchor=north,
  draw=lightgray204
},
tick align=outside,
tick pos=left,
x grid style={darkgray176},
xlabel={Hour of the day},
xmin=-1.15, xmax=24.15,
xtick style={color=black},
xtick={0,4,8,12,16,20,24},
xtick={0,4,8,12,16,20,24},
y grid style={darkgray176},
ylabel={Line loading ($\%$)},
ymin=0, ymax=62.997978833613,
ytick style={color=black},
ytick={0,30,60},
yticklabels={
  $\mathrm{0}$,
  $\mathrm{30}$,
  $\mathrm{60}$,
}
]
\addplot [semithick, darkorange25512714]
table {%
0 31.1753441490035
1 30.2588780379683
2 30.0429074031252
3 29.9794562303071
4 30.01722423728
5 30.4853909973917
6 33.0928506292574
7 35.462972001539
8 35.709885718093
9 35.0895267378668
10 34.6692982947759
11 34.7958246506741
12 35.541271118166
13 35.4674851481681
14 34.6165732977398
15 33.9822303166479
16 34.0063445628204
17 35.4178060666598
18 37.6762974422942
19 38.931186592459
20 37.7996036889292
21 36.1599803420174
22 34.6964406864928
23 32.8711887619633
};
\addlegendentry{max. Loading (\%)}
\addplot [semithick, darkorange25512714, dash pattern=on 5.55pt off 2.4pt]
table {%
0 50
1 50
2 50
3 50
4 50
5 50
6 50
7 50
8 50
9 50
10 50
11 50
12 50
13 50
14 50
15 50
16 50
17 50
18 50
19 50
20 50
21 50
22 50
23 50
};
\addlegendentry{Limit Loading (\%)}
\end{axis}
\end{tikzpicture}

%% file: lifeexpect_XLPE.tex
\begin{tikzpicture}

\definecolor{cornflowerblue86180233}{RGB}{86,180,233}
\definecolor{darkslategray38}{RGB}{38,38,38}
\definecolor{lightgray204}{RGB}{204,204,204}
\definecolor{orange2301590}{RGB}{230,159,0}

\begin{axis}[height=0.085\textheight,scale only axis,
axis line style={darkslategray38},
tick align=outside,
x grid style={lightgray204},
xlabel=\textcolor{darkslategray38}{Conductor current $\frac{I_c}{I_z}$},
xmajorticks=true,
xmin=0.9, xmax=1.25,
xtick style={color=darkslategray38},
y grid style={lightgray204},
ylabel=\textcolor{darkslategray38}{Life expectancy (years)},
ymajorticks=true,
unbounded coords=discard,
ymin=0, ymax=200,
restrict y to domain*=1e-6:inf,
ytick style={color=darkslategray38}
]
\addplot [restrict y to domain*=0:300, very thick, orange2301590,dashed]
table {%
0 2383902.86810344
0.00319488817891374 2383581.04923903
0.00638977635782748 2382615.86269724
0.00958466453674121 2381008.11824185
0.012779552715655 2378759.164178
0.0159744408945687 2375870.8854034
0.0191693290734824 2372345.70068586
0.0223642172523962 2368186.55917236
0.0255591054313099 2363396.93613756
0.0287539936102236 2357980.82798025
0.0319488817891374 2351942.74647805
0.0351437699680511 2345287.71231281
0.0383386581469649 2338021.24787934
0.0415335463258786 2330149.36939324
0.0447284345047923 2321678.57831323
0.0479233226837061 2312615.85209665
0.0511182108626198 2302968.63430588
0.0543130990415335 2292744.82408716
0.0575079872204473 2281952.76504242
0.060702875399361 2270601.23351712
0.0638977635782748 2258699.42632772
0.0670926517571885 2246256.94795353
0.0702875399361022 2233283.79721857
0.073482428115016 2219790.35349041
0.0766773162939297 2205787.36242275
0.0798722044728434 2191285.9212707
0.0830670926517572 2176297.46380712
0.0862619808306709 2160833.7448695
0.0894568690095847 2144906.82456797
0.0926517571884984 2128529.05218392
0.0958466453674121 2111713.04979117
0.0990415335463259 2094471.69562987
0.10223642172524 2076818.10726518
0.105431309904153 2058765.6245619
0.108626198083067 2040327.79250661
0.111821086261981 2021518.34390911
0.115015974440895 2002351.18201442
0.118210862619808 1982840.36305664
0.121405750798722 1963000.07878613
0.124600638977636 1942844.63900009
0.12779552715655 1922388.45410777
0.130990415335463 1901646.0177595
0.134185303514377 1880631.8895696
0.137380191693291 1859360.6779616
0.140575079872204 1837847.02316426
0.143769968051118 1816105.58038586
0.146964856230032 1794151.00319357
0.150159744408946 1771997.92712375
0.153354632587859 1749660.95354868
0.156549520766773 1727154.63382343
0.159744408945687 1704493.45373672
0.162939297124601 1681691.81828757
0.166134185303514 1658764.03680938
0.169329073482428 1635724.3084616
0.172523961661342 1612586.70810793
0.175718849840256 1589365.17259927
0.178913738019169 1566073.48747842
0.182108626198083 1542725.2741218
0.185303514376997 1519333.97733337
0.188498402555911 1495912.8534036
0.191693290734824 1472474.95864591
0.194888178913738 1449033.13842164
0.198083067092652 1425600.01666301
0.201277955271566 1402187.98590295
0.204472843450479 1378809.19781905
0.207667731629393 1355475.55429761
0.210862619808307 1332198.69902326
0.21405750798722 1308990.00959735
0.217252396166134 1285860.59018827
0.220447284345048 1262821.26471507
0.223642172523962 1239882.57056458
0.226837060702875 1217054.75284182
0.230031948881789 1194347.75915138
0.233226837060703 1171771.23490752
0.236421725239617 1149334.51916891
0.23961661341853 1127046.64099326
0.242811501597444 1104916.31630634
0.246006389776358 1082951.94527853
0.249201277955272 1061161.61020169
0.252396166134185 1039553.07385789
0.255591054313099 1018133.77837098
0.258785942492013 996910.844531168
0.261980830670927 975891.071582226
0.26517571884984 955080.937459818
0.268370607028754 934486.599469581
0.271565495207668 914113.895392145
0.274760383386581 893968.34500258
0.277955271565495 874055.151990648
0.281150159744409 854379.206268082
0.284345047923323 834945.086648758
0.287539936102236 815757.063887131
0.29073482428115 796819.10405997
0.293929712460064 778134.872276488
0.297124600638978 759707.736701205
0.300319488817891 741540.772874278
0.303514376996805 723636.768313372
0.306709265175719 705998.227381538
0.309904153354633 688627.376404997
0.313099041533546 671526.169025232
0.31629392971246 654696.291769291
0.319488817891374 638139.169822746
0.322683706070288 621855.972989394
0.325878594249201 605847.621822306
0.329073482428115 590114.793910578
0.332268370607029 574657.930306765
0.335463258785943 559477.242079762
0.338658146964856 544572.716978428
0.34185303514377 529944.126191435
0.345047923322684 515591.031188953
0.348242811501597 501512.790632445
0.351437699680511 487708.567338734
0.354632587859425 474177.335285286
0.357827476038339 460917.886643698
0.361022364217252 447928.838828892
0.364217252396166 435208.641552014
0.36741214057508 422755.583865229
0.370607028753994 410567.801187201
0.373801916932907 398643.282298478
0.376996805111821 386979.876296275
0.380191693290735 375575.299498858
0.383386581469649 364427.142289946
0.386581469648562 353532.875894153
0.389776357827476 342889.85907495
0.39297124600639 332495.344746968
0.396166134185303 322346.486495124
0.399361022364217 312440.344993312
0.402555910543131 302773.894316054
0.405750798722045 293344.028136788
0.408945686900958 284147.565807079
0.412140575079872 275181.258311392
0.415335463258786 266441.794092529
0.4185303514377 257925.804743315
0.421725239616613 249629.870560477
0.424920127795527 241550.52595715
0.428115015974441 233684.2647308
0.431309904153355 226027.545183788
0.434504792332268 218576.795094198
0.437699680511182 211328.416534875
0.440894568690096 204278.790539098
0.44408945686901 197424.281611522
0.447284345047923 190761.242083523
0.450479233226837 184286.016312292
0.453674121405751 177994.944723404
0.456869009584665 171884.367696877
0.460063897763578 165950.629297044
0.463258785942492 160190.080846815
0.466453674121406 154599.084347187
0.469648562300319 149174.015743114
0.472843450479233 143911.268037086
0.476038338658147 138807.254252
0.479233226837061 133858.410245072
0.482428115015974 129061.19737482
0.485623003194888 124412.105023261
0.488817891373802 119907.652975648
0.492012779552716 115544.393660292
0.495207667731629 111318.914251064
0.498402555910543 107227.838635386
0.501597444089457 103267.829250604
0.504792332268371 99435.5887917679
0.507987220447284 95727.8617938896
0.511182108626198 92141.4360919275
0.514376996805112 88673.1441617233
0.517571884984026 85319.8643452615
0.520766773162939 82078.5219636227
0.523961661341853 78946.0903210828
0.527156549520767 75919.591603823
0.530351437699681 72996.097676758
0.533546325878594 70172.7307819984
0.536741214057508 67446.6641424709
0.539936102236422 64815.1224742352
0.543130990415335 62275.3824110216
0.546325878594249 59824.7728444915
0.549520766773163 57460.6751837217
0.552715654952077 55180.5235373809
0.55591054313099 52981.8048220255
0.559105431309904 50862.0587999156
0.562300319488818 48818.8780497118
0.565495207667732 46849.9078733478
0.568690095846645 44952.8461423504
0.571884984025559 43125.4430867909
0.575079872204473 41365.5010300212
0.578274760383387 39670.8740722675
0.5814696485623 38039.4677260906
0.584664536741214 36469.2385066572
0.587859424920128 34958.1934797011
0.591054313099042 33504.3897699577
0.594249201277955 32105.9340328082
0.597444089456869 30760.9818917765
0.600638977635783 29467.7373444423
0.603833865814696 28224.4521392662
0.60702875399361 27029.4251257316
0.610223642172524 25881.0015801349
0.613418530351438 24777.5725092734
0.616613418530351 23717.5739341924
0.619808306709265 22699.4861560856
0.623003194888179 21721.833006347
0.626198083067093 20783.1810827058
0.629392971246006 19882.1389732843
0.63258785942492 19017.3564703504
0.635782747603834 18187.5237754492
0.638977635782748 17391.3706975275
0.642172523961661 16627.6658455851
0.645367412140575 15895.2158173124
0.648562300319489 15192.8643851042
0.651757188498403 14519.491680761
0.654952076677316 13874.0133801235
0.65814696485623 13255.3798888157
0.661341853035144 12662.575530203
0.664536741214057 12094.6177366076
0.667731629392971 11550.5562447598
0.670926517571885 11029.4722963992
0.674121405750799 10530.4778448827
0.677316293929712 10052.7147685938
0.680511182108626 9595.35409189529
0.68370607028754 9157.59521430789
0.686900958466454 8738.66514854909
0.690095846645367 8337.81776801101
0.693290734824281 7954.33306421023
0.696485623003195 7587.51641469332
0.699680511182109 7236.69786183714
0.702875399361022 6901.23140293938
0.706070287539936 6580.49429195216
0.70926517571885 6273.88635317298
0.712460063897764 5980.82930716836
0.715654952076677 5700.76610916883
0.718849840255591 5433.16030014008
0.722044728434505 5177.49537070213
0.725239616613419 4933.2741380356
0.728434504792332 4700.01813588712
0.731629392971246 4477.26701775627
0.73482428115016 4264.57797331891
0.738019169329074 4061.52515812016
0.741214057507987 3867.69913654329
0.744408945686901 3682.70633804115
0.747603833865815 3506.1685265947
0.750798722044728 3337.72228334473
0.753993610223642 3177.0185023248
0.757188498402556 3023.721899206
0.76038338658147 2877.5105329496
0.763578274760383 2738.07534024838
0.766773162939297 2605.11968262494
0.769968051118211 2478.35890604272
0.773162939297125 2357.51991287426
0.776357827476038 2242.34074606152
0.779552715654952 2132.57018529358
0.782747603833866 2027.96735501872
0.78594249201278 1928.30134410066
0.789137380191693 1833.35083692213
0.792332268370607 1742.90375573296
0.795527156549521 1656.75691403507
0.798722044728435 1574.71568079196
0.801916932907348 1496.59365524743
0.805111821086262 1422.21235213389
0.808306709265176 1351.40089704974
0.811501597444089 1283.99573178194
0.814696485623003 1219.84032934963
0.817891373801917 1158.7849185428
0.821086261980831 1100.68621773071
0.824281150159744 1045.40717771375
0.827476038338658 992.816733393201
0.830670926517572 942.789564034669
0.833865814696486 895.205861901221
0.837060702875399 849.951109034432
0.840255591054313 806.915861962975
0.843450479233227 765.995544120453
0.846645367412141 727.090245756721
0.849840255591054 690.104531129082
0.853035143769968 654.947252762768
0.856230031948882 621.531372572994
0.859424920127795 589.773789643839
0.862619808306709 559.595174462628
0.865814696485623 530.919809411711
0.869009584664537 503.675435323208
0.87220447284345 477.793103905728
0.875399361022364 453.207035855882
0.878594249201278 429.854484471189
0.881789137380192 407.675604584687
0.884984025559105 386.613326645613
0.888178913738019 366.613235774326
0.891373801916933 347.623455623686
0.894568690095847 329.594536882992
0.89776357827476 312.479350264672
0.900958466453674 296.232983817857
0.904153354632588 280.812644416911
0.907348242811502 266.177563277096
0.910543130990415 252.288905353333
0.913738019169329 239.109682482103
0.916932907348243 226.604670130302
0.920127795527157 214.740327618819
0.92332268370607 203.484721692326
0.926517571884984 192.807453310636
0.929712460063898 182.679587540612
0.932907348242812 173.073586431331
0.936102236421725 163.963244758798
0.939297124600639 155.323628530026
0.942492012779553 147.131016139826
0.945686900958466 139.362842077026
0.94888178913738 131.997643080227
0.952076677316294 125.015006646452
0.955271565495208 118.395521799287
0.958466453674121 112.120732026236
0.961661341853035 106.173090298109
0.964856230031949 100.535916086224
0.968051118210863 95.1933542961709
0.971246006389776 90.1303360397217
0.97444089456869 85.3325411692355
0.977635782747604 80.7863625016665
0.980830670926518 76.4788716618545
0.984025559105431 72.3977864774018
0.987220447284345 68.5314398598936
0.990415335463259 64.8687501096539
0.993610223642173 61.3991925835915
0.996805111821086 58.1127726679489
1 55
1.00319488817891 52.051863884882
1.00638977635783 49.2598098558221
1.00958466453674 46.6157173280353
1.01277955271565 44.1118782985242
1.01597444089457 41.7409770458839
1.01916932907348 39.4960707860538
1.0223642172524 37.3705712417112
1.02555910543131 35.3582270847051
1.02875399361022 33.4531072125821
1.03194888178914 31.6495848218311
1.03514376996805 29.9423222420208
1.03833865814696 28.3262564964698
1.04153354632588 26.7965855565272
1.04472843450479 25.3487552579004
1.04792332268371 23.9784468488128
1.05111821086262 22.6815651410267
1.05431309904153 21.4542272360111
1.05750798722045 20.2927517997119
1.06070287539936 19.1936488605188
1.06389776357827 18.1536101061168
1.06709265175719 17.1694996559657
1.0702875399361 16.2383452871629
1.07348242811502 15.3573300924115
1.07667731629393 14.5237845497578
1.07987220447284 13.7351789846541
1.08306709265176 12.9891164057641
1.08626198083067 12.2833256967591
1.08945686900958 11.6156551471415
1.0926517571885 10.9840663058973
1.09584664536741 10.386628142506
1.09904153354633 9.82151150053952
1.10223642172524 9.28698382974914
1.10543130990415 8.78140418318444
1.10862619808307 8.30321846650477
1.11182108626198 7.85095492723065
1.11501597444089 7.42321987225082
1.11821086261981 7.01869360243895
1.12140575079872 6.6361265537534
1.12460063897764 6.27433563468652
1.12779552715655 5.93220075040616
1.13099041533546 5.60866150438409
1.13418530351438 5.3027140687379
1.13738019169329 5.01340821493047
1.1405750798722 4.73984449686384
1.14376996805112 4.48117157878487
1.14696485623003 4.23658370077981
1.15015974440895 4.00531827498207
1.15335463258786 3.78665360594567
1.15654952076677 3.57990672895206
1.15974440894569 3.38443136031983
1.1629392971246 3.19961595407145
1.16613418530351 3.02488185958703
1.16932907348243 2.85968157513519
1.17252396166134 2.7034970924204
1.17571884984026 2.55583832752419
1.17891373801917 2.41624163384352
1.18210862619808 2.28426839284675
1.185303514377 2.15950367867185
1.18849840255591 2.04155499278996
1.19169329073482 1.93005106514224
1.19488817891374 1.82464071833788
1.19808306709265 1.72499179166965
1.20127795527157 1.63079012186583
1.20447284345048 1.54173857765091
1.20766773162939 1.45755614533412
1.21086261980831 1.3779770627844
1.21405750798722 1.30274999928347
1.21725239616613 1.23163727887443
1.22044728434505 1.16441414494452
1.22364217252396 1.10086806389429
1.22683706070288 1.04079806585495
1.23003194888179 0.984014120518578
1.2332268370607 0.930336546245086
1.23642172523962 0.879595450702839
1.23961661341853 0.831630201389284
1.24281150159744 0.786288924462362
1.24600638977636 0.743428030394116
1.24920127795527 0.702911765034139
1.25239616613419 0.664611784743483
1.2555910543131 0.628406754328375
1.25878594249201 0.594181966568931
1.26198083067093 0.561828982200208
1.26517571884984 0.531245289262261
1.26837060702875 0.502333980791893
1.27156549520767 0.475003449882394
1.27476038338658 0.449167101188072
1.2779552715655 0.424743077998481
1.28115015974441 0.401654004053168
1.28434504792332 0.379826739310754
1.28753993610224 0.359192148927626
1.29073482428115 0.339684884740389
1.29392971246006 0.321243178583337
1.29712460063898 0.303808646807361
1.30031948881789 0.287326105400013
1.30351437699681 0.271743395138194
1.30670926517572 0.257011216234764
1.30990415335463 0.243082971969045
1.31309904153355 0.229914620817976
1.31629392971246 0.21746453663043
1.31948881789137 0.205693376411368
1.32268370607029 0.194563955305545
1.3258785942492 0.184041128392321
1.32907348242811 0.174091678923741
1.33226837060703 0.1646842126577
1.33546325878594 0.15578905795655
1.33865814696486 0.147378171339108
1.34185303514377 0.139425048190713
1.34504792332268 0.131904638351768
1.3482428115016 0.124793266320177
1.35143769968051 0.118068555817279
1.35463258785942 0.111709358480315
1.35782747603834 0.105695686457165
1.36102236421725 0.100008648691203
1.36421725239617 0.0946303906954589
1.36741214057508 0.0895440376261712
1.37060702875399 0.0847336404759739
1.37380191693291 0.0801841252167231
1.37699680511182 0.0758812447310932
1.38019169329073 0.0718115333808063
1.38338658146965 0.0679622640675595
1.38658146964856 0.0643214076505205
1.38977635782748 0.0608775945916121
1.39297124600639 0.0576200787068167
1.3961661341853 0.0545387029083115
1.39936102236422 0.0516238668285053
1.40255591054313 0.048866496222967
1.40575079872204 0.046258014054832
1.40894568690096 0.0437903131685649
1.41214057507987 0.0414557304659842
1.41533546325879 0.0392470225021781
1.4185303514377 0.0371573424234477
1.42172523961661 0.0351802181736466
1.42492012779553 0.0333095318993021
1.42811501597444 0.0315395004877146
1.43130990415335 0.0298646571758099
1.43450479233227 0.0282798341709293
1.43769968051118 0.0267801462279575
1.4408945686901 0.0253609751302185
1.44408945686901 0.0240179550244648
1.44728434504792 0.0227469585629846
1.45047923322684 0.021544083808439
1.45367412140575 0.0204056418594692
1.45686900958466 0.0193281451574102
1.46006389776358 0.0183082964366277
1.46325878594249 0.017342978283049
1.46645367412141 0.0164292432674032
1.46964856230032 0.0155643046215251
1.47284345047923 0.0147455274278154
1.47603833865815 0.0139704202935918
1.47923322683706 0.0132366274836197
1.48242811501597 0.0125419214855822
1.48562300319489 0.0118841959846315
1.4888178913738 0.0112614592244826
1.49201277955272 0.0106718277337466
1.49520766773163 0.0101135203973732
1.49840255591054 0.0095848528541844
1.50159744408946 0.00908423220252271
1.50479233226837 0.00861015199703238
};
\addplot [restrict y to domain*=0:300, very thick, cornflowerblue86180233]
table {%
0 2383902.86810344
0.00319488817891374 2383651.85450112
0.00638977635782748 2382898.97122789
0.00958466453674121 2381644.69072128
0.012779552715655 2379889.79983186
0.0159744408945687 2377635.39900595
0.0191693290734824 2374882.9011432
0.0223642172523962 2371634.03013073
0.0255591054313099 2367890.81905551
0.0287539936102236 2363655.60809794
0.0319488817891374 2358931.04210992
0.0351437699680511 2353720.0678799
0.0383386581469649 2348025.93109022
0.0415335463258786 2341852.17296977
0.0447284345047923 2335202.6266479
0.0479233226837061 2328081.41321375
0.0511182108626198 2320492.93748771
0.0543130990415335 2312441.88351023
0.0575079872204473 2303933.20975455
0.060702875399361 2294972.14407059
0.0638977635782748 2285564.1783667
0.0670926517571885 2275715.06303695
0.0702875399361022 2265430.80114193
0.073482428115016 2254717.64235101
0.0766773162939297 2243582.07665474
0.0798722044728434 2232030.82785638
0.0830670926517572 2220070.84685106
0.0862619808306709 2207709.30470245
0.0894568690095847 2194953.58552629
0.0926517571884984 2181811.27919071
0.0958466453674121 2168290.17384319
0.0990415335463259 2154398.24827455
0.10223642172524 2140143.66413032
0.105431309904153 2125534.75798015
0.108626198083067 2110580.03325565
0.111821086261981 2095288.15206798
0.115015974440895 2079667.92691559
0.118210862619808 2063728.31229362
0.121405750798722 2047478.39621564
0.124600638977636 2030927.39165909
0.12779552715655 2014084.62794572
0.130990415335463 1996959.54206753
0.134185303514377 1979561.66997044
0.137380191693291 1961900.63780585
0.140575079872204 1943986.15316178
0.143769968051118 1925827.99628423
0.146964856230032 1907436.01130026
0.150159744408946 1888820.09745296
0.153354632587859 1869990.20035962
0.156549520766773 1850956.30330347
0.159744408945687 1831728.41856945
0.162939297124601 1812316.57883436
0.166134185303514 1792730.82862168
0.169329073482428 1772981.21583058
0.172523961661342 1753077.7833494
0.175718849840256 1733030.56076265
0.178913738019169 1712849.55616108
0.182108626198083 1692544.74806366
0.185303514376997 1672126.07746058
0.188498402555911 1651603.43998518
0.191693290734824 1630986.67822379
0.194888178913738 1610285.57417059
0.198083067092652 1589509.84183597
0.201277955271566 1568669.1200149
0.204472843450479 1547772.9652229
0.207667731629393 1526830.84480594
0.210862619808307 1505852.13023092
0.21405750798722 1484846.09056221
0.217252396166134 1463821.88613059
0.220447284345048 1442788.56239932
0.223642172523962 1421755.04403254
0.226837060702875 1400730.12917057
0.230031948881789 1379722.48391644
0.233226837060703 1358740.63703716
0.236421725239617 1337792.97488371
0.23961661341853 1316887.73653259
0.242811501597444 1296033.00915175
0.246006389776358 1275236.72359338
0.249201277955272 1254506.65021555
0.252396166134185 1233850.39493448
0.255591054313099 1213275.39550856
0.258785942492013 1192788.91805547
0.261980830670927 1172398.0538026
0.26517571884984 1152109.71607137
0.268370607028754 1131930.63749524
0.271565495207668 1111867.36747107
0.274760383386581 1091926.26984309
0.277955271565495 1072113.5208184
0.281150159744409 1052435.10711273
0.284345047923323 1032896.82432464
0.287539936102236 1013504.27553643
0.29073482428115 994262.870139106
0.293929712460064 975177.822879367
0.297124600638978 956254.153125291
0.300319488817891 937496.68434805
0.303514376996805 918910.043815941
0.306709265175719 900498.662497539
0.309904153354633 882266.775169764
0.313099041533546 864218.420727153
0.31629392971246 846357.442687888
0.319488817891374 828687.489892125
0.322683706070288 811212.017388114
0.325878594249201 793934.287501173
0.329073482428115 776857.371080503
0.332268370607029 759984.148918811
0.335463258785943 743317.313339337
0.338658146964856 726859.369944891
0.34185303514377 710612.639523407
0.345047923322684 694579.260104241
0.348242811501597 678761.189159589
0.351437699680511 663160.205945077
0.354632587859425 647777.913973637
0.357827476038339 632615.743616743
0.361022364217252 617674.954826864
0.364217252396166 602956.639975147
0.36741214057508 588461.726798157
0.370607028753994 574190.981447657
0.373801916932907 560145.011637119
0.376996805111821 546324.269878955
0.380191693290735 532729.056806332
0.383386581469649 519359.52457341
0.386581469648562 506215.680327998
0.389776357827476 493297.389750477
0.39297124600639 480604.380653173
0.396166134185303 468136.246634055
0.399361022364217 455892.450779101
0.402555910543131 443872.32940731
0.405750798722045 432075.095852872
0.408945686900958 420499.844278688
0.412140575079872 409145.553515842
0.415335463258786 398011.090923563
0.4185303514377 387095.216264294
0.421725239616613 376396.585588782
0.424920127795527 365913.755126031
0.428115015974441 355645.185173153
0.431309904153355 345589.243980379
0.434504792332268 335744.211626415
0.437699680511182 326108.283879711
0.440894568690096 316679.576041138
0.44408945686901 307456.126763895
0.447284345047923 298435.901846446
0.450479233226837 289616.797994618
0.453674121405751 280996.646548963
0.456869009584665 272573.217173806
0.460063897763578 264344.221504448
0.463258785942492 256307.316749196
0.466453674121406 248460.109243065
0.469648562300319 240800.157950105
0.472843450479233 233324.977911543
0.476038338658147 226032.043637033
0.479233226837061 218918.792436466
0.482428115015974 211982.627690004
0.485623003194888 205220.922054125
0.488817891373802 198631.020601605
0.492012779552716 192210.243893576
0.495207667731629 185955.8909819
0.498402555910543 179865.242340295
0.501597444089457 173935.562722763
0.504792332268371 168164.103948068
0.507987220447284 162548.107609085
0.511182108626198 157084.80770608
0.514376996805112 151771.433203026
0.517571884984026 146605.210506291
0.520766773162939 141583.365865071
0.523961661341853 136703.127693173
0.527156549520767 131961.728811797
0.530351437699681 127356.40861315
0.533546325878594 122884.415144799
0.536741214057508 118543.007114839
0.539936102236422 114329.455818008
0.543130990415335 110241.046983037
0.546325878594249 106275.082541598
0.549520766773163 102428.882319341
0.552715654952077 98699.7856495678
0.55591054313099 95085.1529102305
0.559105431309904 91582.3669849939
0.562300319488818 88188.8346492072
0.565495207667732 84901.9878816902
0.568690095846645 81719.2851033383
0.571884984025559 78638.2123435975
0.575079872204473 75656.2843359536
0.578274760383387 72771.0455436099
0.5814696485623 69980.0711166224
0.584664536741214 67280.9677817849
0.587859424920128 64671.3746666235
0.591054313099042 62148.9640589065
0.594249201277955 59711.4421031062
0.597444089456869 57356.5494352976
0.600638977635783 55082.0617580094
0.603833865814696 52885.7903565793
0.60702875399361 50765.5825585776
0.610223642172524 48719.3221379029
0.613418530351438 46744.9296651657
0.616613418530351 44840.3628059905
0.619808306709265 43003.6165688829
0.623003194888179 41232.7235043219
0.626198083067093 39525.7538567309
0.629392971246006 37880.8156710024
0.63258785942492 36296.0548552369
0.635782747603834 34769.6552013639
0.638977635782748 33299.8383653039
0.642172523961661 31884.8638083238
0.645367412140575 30523.0287012289
0.648562300319489 29212.6677930233
0.651757188498403 27952.1532456533
0.654952076677316 26739.8944364348
0.65814696485623 25574.3377297472
0.661341853035144 24453.9662195548
0.664536741214057 23377.2994442936
0.667731629392971 22342.893075642
0.670926517571885 21349.3385826642
0.674121405750799 20395.2628727901
0.677316293929712 19479.3279110673
0.680511182108626 18600.2303190958
0.68370607028754 17756.7009550142
0.686900958466454 16947.5044758921
0.690095846645367 16171.4388838326
0.693290734824281 15427.3350570709
0.696485623003195 14714.0562673153
0.699680511182109 14030.4976845373
0.702875399361022 13375.5858703944
0.706070287539936 12748.2782614239
0.70926517571885 12147.5626431139
0.712460063897764 11572.4566159243
0.715654952076677 11022.0070542891
0.718849840255591 10495.2895596003
0.722044728434505 9991.40790813474
0.725239616613419 9509.49349484839
0.728434504792332 9048.70477392916
0.731629392971246 8608.22669696094
0.73482428115016 8187.270149516
0.738019169329074 7785.0713869604
0.741214057507987 7400.89147021583
0.744408945686901 7034.01570219413
0.747603833865815 6683.75306558002
0.750798722044728 6349.43566260631
0.753993610223642 6030.4181574341
0.757188498402556 5726.07722171363
0.76038338658147 5435.81098387326
0.763578274760383 5159.0384826498
0.766773162939297 4895.19912534354
0.769968051118211 4643.752151251
0.773162939297125 4404.17610069834
0.776357827476038 4175.96829007024
0.779552715654952 3958.64429320019
0.782747603833866 3751.73742946147
0.78594249201278 3554.79825887069
0.789137380191693 3367.39408449086
0.792332268370607 3189.10846239563
0.795527156549521 3019.5407194311
0.798722044728435 2858.30547899048
0.801916932907348 2705.03219499188
0.805111821086262 2559.36469422921
0.808306709265176 2420.96072724496
0.811501597444089 2289.49152785302
0.814696485623003 2164.64138142099
0.817891373801917 2046.10720200232
0.821086261980831 1933.59811839154
0.824281150159744 1826.83506915821
0.827476038338658 1725.55040669969
0.830670926517572 1629.48751033684
0.833865814696486 1538.400408463
0.837060702875399 1452.05340974168
0.840255591054313 1370.22074333602
0.843450479233227 1292.68620814063
0.846645367412141 1219.24283097411
0.849840255591054 1149.6925336805
0.853035143769968 1083.84580907653
0.856230031948882 1021.52140567288
0.859424920127795 962.546021087757
0.862619808306709 906.754004063416
0.865814696485623 853.987064987929
0.869009584664537 804.093994817589
0.87220447284345 756.930392288727
0.875399361022364 712.358399301345
0.878594249201278 670.246444351855
0.881789137380192 630.468993886818
0.884984025559105 592.906311445221
0.888178913738019 557.444224452765
0.891373801916933 523.973898528012
0.894568690095847 492.391619157027
0.89776357827476 462.598580590573
0.900958466453674 434.500681815373
0.904153354632588 408.008329449182
0.907348242811502 383.036247407709
0.910543130990415 359.50329319024
0.913738019169329 337.332280629941
0.916932907348243 316.449808954078
0.920127795527157 296.786097999341
0.92332268370607 278.274829427156
0.926517571884984 260.852993784358
0.929712460063898 244.460743254912
0.932907348242812 229.041249949213
0.936102236421725 214.54056957838
0.939297124600639 200.907510362085
0.942492012779553 188.093507019867
0.945686900958466 176.052499697302
0.94888178913738 164.740817680108
0.952076677316294 154.117067751108
0.955271565495208 144.142027046863
0.958466453674121 134.778540272905
0.961661341853035 125.991421138717
0.964856230031949 117.747357875804
0.968051118210863 110.014822704699
0.971246006389776 102.763985119119
0.97444089456869 95.9666288580739
0.977635782747604 89.5960724393061
0.980830670926518 83.6270931300887
0.984025559105431 78.0358542341225
0.987220447284345 72.7998355759885
0.990415335463259 67.8977670673687
0.993610223642173 63.3095652420644
0.996805111821086 59.0162726495919
1 55
1.00319488817891 51.2438709553142
1.00638977635783 47.7319694658629
1.00958466453674 44.4492895525176
1.01277955271565 41.3816874386824
1.01597444089457 38.5158359386397
1.01916932907348 35.8391810116039
1.0223642172524 33.3399003935744
1.02555910543131 31.006864221764
1.02875399361022 28.8295975690579
1.03194888178914 26.7982448085787
1.03514376996805 24.9035357310361
1.03833865814696 23.1367533400897
1.04153354632588 21.4897032534594
1.04472843450479 19.9546846399956
1.04792332268371 18.5244626253315
1.05111821086262 17.1922421011192
1.05431309904153 15.9516428751692
1.05750798722045 14.7966761020934
1.06070287539936 13.7217219362656
1.06389776357827 12.7215083510883
1.06709265175719 11.791091070673
1.0702875399361 10.9258345620972
1.07348242811502 10.1213940384238
1.07667731629393 9.37369842461316
1.07987220447284 8.67893424037089
1.08306709265176 8.03353035581523
1.08626198083067 7.43414357765184
1.08945686900958 6.87764502527787
1.0926517571885 6.36110725793366
1.09584664536741 5.88179211565224
1.09904153354633 5.43713923834437
1.10223642172524 5.02475522888987
1.10543130990415 4.64240342758794
1.10862619808307 4.28799426675373
1.11182108626198 3.95957617563099
1.11501597444089 3.65532700712565
1.11821086261981 3.37354595915508
1.12140575079872 3.11264596464749
1.12460063897764 2.87114652542207
1.12779552715655 2.6476669663331
1.13099041533546 2.44092008716731
1.13418530351438 2.24970619085007
1.13738019169329 2.07290746753971
1.1405750798722 1.90948271517178
1.14376996805112 1.75846237795998
1.14696485623003 1.61894388526534
1.15015974440895 1.49008727411417
1.15335463258786 1.37111107947739
1.15654952076677 1.26128847722077
1.15974440894569 1.15994366539978
1.1629392971246 1.06644847030181
1.16613418530351 0.980219164338206
1.16932907348243 0.900713483555399
1.17252396166134 0.827427833172708
1.17571884984026 0.759894670163299
1.17891373801917 0.697680052476376
1.18210862619808 0.640381345053157
1.185303514377 0.587625073317935
1.18849840255591 0.539064915329783
1.19169329073482 0.494379824260244
1.19488817891374 0.453272273319725
1.19808306709265 0.415466615690053
1.20127795527157 0.380707552434728
1.20447284345048 0.348758701751562
1.20766773162939 0.319401263306435
1.21086261980831 0.29243277174197
1.21405750798722 0.267665933792159
1.21725239616613 0.244927543753864
1.22044728434505 0.224057472369822
1.22364217252396 0.204907724465414
1.22683706070288 0.187341560954364
1.23003194888179 0.171232681086812
1.2332268370607 0.156464461057934
1.23642172523962 0.14292924532683
1.23961661341853 0.13052768721446
1.24281150159744 0.119168135556581
1.24600638977636 0.108766064383463
1.24920127795527 0.0992435427831532
1.25239616613419 0.0905287422797829
1.2555910543131 0.0825554792233752
1.25878594249201 0.0752627898432049
1.26198083067093 0.068594535763604
1.26517571884984 0.0624990379194646
1.26837060702875 0.0569287369391442
1.27156549520767 0.0518398781853245
1.27476038338658 0.0471922197600664
1.2779552715655 0.0429487618891854
1.28115015974441 0.039075496203526
1.28434504792332 0.0355411735310313
1.28753993610224 0.0323170889040859
1.29073482428115 0.0293768825717019
1.29392971246006 0.0266963558860388
1.29712460063898 0.024253301007808
1.30031948881789 0.0220273434455238
1.30351437699681 0.0199997965096504
1.30670926517572 0.0181535268246403
1.30990415335463 0.0164728300999331
1.31309904153355 0.0149433164154122
1.31629392971246 0.0135518043277686
1.31948881789137 0.0122862231519422
1.32268370607029 0.0111355228164656
1.3258785942492 0.0100895907333081
1.32907348242811 0.00913917516187426
1.33226837060703 0.00827581458333285
1.33546325878594 0.00749177263555794
1.33865814696486 0.00677997819083323
1.34185303514377 0.00613397018821653
1.34504792332268 0.00554784686022429
1.3482428115016 0.00501621901939406
1.35143769968051 0.00453416709443578
1.35463258785942 0.00409720162819324
1.35782747603834 0.00370122697061514
1.36102236421725 0.00334250791947195
1.36421725239617 0.00301763907974747
1.36741214057508 0.00272351672956185
1.37060702875399 0.00245731299623545
1.37380191693291 0.0022164521607488
1.37699680511182 0.00199858892247128
1.38019169329073 0.0018015884686832
1.38338658146965 0.00162350820516978
1.38658146964856 0.00146258101507771
1.38977635782748 0.00131719992335453
1.39297124600639 0.00118590405348835
1.3961661341853 0.00106736577198195
1.39936102236422 0.000960378924075098
1.40255591054313 0.000863848071717806
1.40575079872204 0.000776778651733983
1.40894568690096 0.000698267978539238
1.41214057507987 0.000627497021722501
1.41533546325879 0.00056372289430372
1.4185303514377 0.000506271992569092
1.42172523961661 0.000454533733090776
1.42492012779553 0.000407954836887371
1.42811501597444 0.00036603411469941
1.43130990415335 0.000328317711065088
1.43450479233227 0.000294394768307341
1.43769968051118 0.000263893474704814
1.4408945686901 0.000236477464035791
1.44408945686901 0.00021184253637341
1.44728434504792 0.000189713672489514
1.45047923322684 0.000169842316508647
1.45367412140575 0.000152003903557669
1.45686900958466 0.000135995611093528
1.46006389776358 0.000121634314374709
1.46325878594249 0.000108754728182074
1.46645367412141 9.72077184034549e-05
1.46964856230032 8.68587684831553e-05
1.47284345047923 7.75865870119701e-05
1.47603833865815 6.92818439041659e-05
1.47923322683706 6.18460236828955e-05
1.48242811501597 5.51903853823956e-05
1.48562300319489 4.92350194808343e-05
1.4888178913738 4.39079931082261e-05
1.49201277955272 3.91445755353846e-05
1.49520766773163 3.48865366478226e-05
1.49840255591054 3.10815117479995e-05
1.50159744408946 2.76824266149478e-05
1.50479233226837 2.46469772864624e-05
};
\end{axis}

\end{tikzpicture}

%% file: lifeexpect_PILC.tex
\begin{tikzpicture}

\definecolor{cornflowerblue86180233}{RGB}{86,180,233}
\definecolor{darkslategray38}{RGB}{38,38,38}
\definecolor{lightgray204}{RGB}{204,204,204}
\definecolor{orange2301590}{RGB}{230,159,0}
\definecolor{mycolor3}{RGB}{0,158,115}

\begin{axis}[height=0.085\textheight,scale only axis,
axis line style={darkslategray38},
tick align=outside,
x grid style={lightgray204},
xlabel=\textcolor{darkslategray38}Conductor current {$\frac{I_c}{I_z}$},
xmajorticks=true,
xmin=0, xmax=1.25,
xtick style={color=darkslategray38},
y grid style={lightgray204},
ylabel=\textcolor{darkslategray38}{Life expectancy (years)},
ymajorticks=true,
ymin=0, ymax=200,
ytick style={color=darkslategray38}
]
\addplot [very thick, orange2301590,dashed]
table {%
0 117.346046000463
0.004 117.337036589678
0.008 117.310012507375
0.012 117.264986200517
0.016 117.201978403428
0.02 117.121018121881
0.024 117.02214261084
0.028 116.905397345894
0.032 116.770835988395
0.036 116.618520344357
0.04 116.448520317141
0.044 116.260913853987
0.048 116.055786886458
0.052 115.833233264838
0.056 115.593354686575
0.06 115.336260618832
0.064 115.062068215242
0.068 114.770902226934
0.072 114.462894907954
0.076 114.138185915149
0.08 113.79692220265
0.084 113.439257911042
0.088 113.065354251356
0.092 112.675379383989
0.096 112.269508292697
0.1 111.847922653773
0.104 111.410810700566
0.108 110.958367083471
0.112 110.490792725536
0.116 110.008294673843
0.12 109.511085946807
0.124 108.999385377564
0.128 108.473417453591
0.132 107.933412152738
0.136 107.379604775836
0.14 106.812235776042
0.144 106.231550585106
0.148 105.637799436732
0.152 105.031237187206
0.156 104.412123133472
0.16 103.780720828847
0.164 103.137297896542
0.168 102.482125841183
0.172 101.815479858524
0.176 101.137638643514
0.18 100.448884196934
0.184 99.7495016307707
0.188 99.0397789725212
0.192 98.3200069686142
0.196 97.5904788871362
0.2 96.8514903200465
0.204 96.1033389850677
0.208 95.3463245274348
0.212 94.5807483216867
0.216 93.8069132736814
0.22 93.0251236230153
0.224 92.2356847460231
0.228 91.4389029595367
0.232 90.6350853255741
0.236 89.8245394571322
0.24 89.0075733252496
0.244 88.1844950675074
0.248 87.355612798129
0.252 86.5212344198399
0.256 85.6816674376426
0.26 84.8372187746599
0.264 83.9881945901959
0.268 83.1349001001595
0.272 82.2776393999917
0.276 81.4167152902349
0.28 80.552429104875
0.284 79.6850805425886
0.288 78.8149675010154
0.292 77.942385914179
0.296 77.0676295931692
0.3 76.1909900701967
0.304 75.3127564461257
0.308 74.4332152415857
0.312 73.5526502517562
0.316 72.6713424049181
0.32 71.7895696248538
0.324 70.9076066971793
0.328 70.0257251396822
0.332 69.1441930767364
0.336 68.263275117857
0.34 67.3832322404574
0.344 66.504321676861
0.348 65.6267968056176
0.352 64.7509070471683
0.356 63.876897763898
0.36 63.0050101646085
0.364 62.135481213441
0.368 61.2685435432711
0.372 60.4044253735948
0.376 59.5433504329182
0.38 58.6855378856596
0.384 57.8312022635664
0.388 56.9805534016458
0.392 56.1337963786037
0.396 55.2911314617782
0.4 54.4527540565542
0.404 53.6188546602385
0.408 52.7896188203691
0.412 51.9652270974314
0.416 51.1458550319466
0.42 50.3316731158956
0.424 49.5228467684365
0.428 48.7195363158701
0.432 47.9218969758038
0.436 47.1300788454621
0.44 46.3442268940857
0.444 45.5644809593601
0.448 44.7909757478098
0.452 44.0238408390924
0.456 43.2632006941224
0.46 42.5091746669521
0.464 41.7618770203354
0.468 41.0214169448958
0.472 40.2878985818183
0.476 39.561421048983
0.48 38.8420784704553
0.484 38.1299600092457
0.488 37.4251499032501
0.492 36.7277275042808
0.496 36.037767320095
0.5 35.3553390593274
0.504 34.6805076792319
0.508 34.0133334361362
0.512 33.3538719385107
0.516 32.7021742025557
0.52 32.0582867102055
0.524 31.4222514694504
0.528 30.7941060768772
0.532 30.1738837823256
0.536 29.5616135555619
0.54 28.9573201548661
0.544 28.3610241974344
0.548 27.7727422314934
0.552 27.1924868100266
0.556 26.6202665660129
0.56 26.0560862890765
0.564 25.499947003449
0.568 24.9518460471456
0.572 24.4117771522566
0.576 23.8797305262584
0.58 23.3556929342463
0.584 22.8396477819961
0.588 22.3315751997596
0.592 21.8314521267016
0.596 21.3392523958875
0.6 20.8549468197327
0.604 20.3785032758238
0.608 19.9098867930261
0.612 19.4490596377936
0.616 18.9959814005955
0.62 18.5506090823807
0.624 18.1128971809989
0.628 17.6827977775009
0.632 17.2602606222434
0.636 16.8452332207231
0.64 16.4376609190692
0.644 16.0374869891249
0.648 15.6446527130495
0.652 15.259097467376
0.656 14.8807588064623
0.66 14.5095725452722
0.664 14.14547284143
0.668 13.7883922764909
0.672 13.4382619363725
0.676 13.0950114908964
0.68 12.7585692723888
0.684 12.4288623532932
0.688 12.1058166227503
0.692 11.7893568621002
0.696 11.4794068192689
0.7 11.1758892819976
0.704 10.8787261498803
0.708 10.5878385051748
0.712 10.303146682356
0.716 10.0245703363801
0.72 9.75202850963379
0.724 9.48543969754236
0.728 9.22472191281265
0.732 8.96979274829089
0.736 8.72056943841511
0.74 8.4769689192455
0.744 8.23890788705718
0.748 8.00630285548232
0.752 7.77907021119007
0.756 7.55712626809485
0.76 7.34038732008533
0.764 7.1287696922682
0.768 6.92218979072253
0.772 6.72056415076226
0.776 6.52380948370615
0.78 6.33184272215575
0.784 6.14458106378389
0.788 5.96194201363759
0.792 5.78384342496056
0.796 5.61020353854215
0.8 5.44094102060077
0.804 5.27597499921121
0.808 5.11522509928652
0.812 4.9586114761263
0.816 4.80605484754445
0.82 4.65747652459052
0.824 4.5127984408798
0.828 4.37194318054843
0.832 4.23483400485064
0.836 4.10139487741612
0.84 3.97155048818654
0.844 3.84522627605095
0.848 3.72234845020032
0.852 3.6028440102227
0.856 3.48664076496067
0.86 3.37366735015352
0.864 3.26385324488728
0.868 3.15712878687607
0.872 3.05342518659883
0.876 2.95267454031583
0.88 2.85480984198982
0.884 2.75976499413697
0.888 2.66747481763323
0.892 2.57787506050148
0.896 2.49090240570594
0.9 2.40649447797955
0.904 2.32458984971081
0.908 2.24512804591651
0.912 2.16804954832674
0.916 2.09329579860863
0.92 2.02080920075559
0.924 1.95053312266821
0.928 1.88241189695352
0.932 1.81639082096879
0.936 1.75241615613621
0.94 1.69043512655445
0.944 1.63039591693303
0.948 1.57224766987532
0.952 1.51594048253543
0.956 1.46142540267456
0.96 1.40865442414156
0.964 1.35758048180245
0.968 1.3081574459434
0.972 1.26034011617114
0.976 1.2140842148346
0.98 1.16934637999109
0.984 1.12608415794013
0.988 1.08425599534742
0.992 1.04382123098129
0.996 1.00474008708337
1 0.966973660394866
1.004 0.930483912859457
1.008 0.89523366202323
1.012 0.861186571151791
1.016 0.828307139084122
1.02 0.796560689842336
1.024 0.765913362015993
1.028 0.73633209793918
1.032 0.707784632678069
1.036 0.680239482846208
1.04 0.653665935264278
1.044 0.628034035480606
1.048 0.603314576168215
1.052 0.57947908541371
1.056 0.556499814912816
1.06 0.53434972808691
1.064 0.513002488134386
1.068 0.492432446030225
1.072 0.472614628486676
1.076 0.453524725887458
1.08 0.435139080207438
1.084 0.417434672929275
1.088 0.400389112968057
1.092 0.383980624614499
1.096 0.368188035506831
1.1 0.352990764641059
1.104 0.338368810428834
1.108 0.324302738811758
1.112 0.310773671440507
1.116 0.297763273926751
1.12 0.285253744175453
1.124 0.273227800804698
1.128 0.261668671659842
1.132 0.250560082428373
1.136 0.239886245361508
1.14 0.229631848108171
1.144 0.219782042666649
1.148 0.210322434458884
1.152 0.20123907153199
1.156 0.192518433891284
1.16 0.184147422968781
1.164 0.17611335123079
1.168 0.168403931927951
1.172 0.161007268990753
1.176 0.153911847073299
1.18 0.147106521747783
1.184 0.140580509851906
1.188 0.134323379991184
1.192 0.128325043197842
1.196 0.122575743747782
1.2 0.117066050136842
1.204 0.111786846217363
1.208 0.106729322495873
1.212 0.101884967592462
1.216 0.0972455598622665
1.22 0.0928031591792609
1.224 0.0885500988823823
1.228 0.0844789778838569
1.232 0.0805826529394118
1.236 0.0768542310799151
1.24 0.0732870622038308
1.244 0.0698747318297406
1.248 0.0666110540080428
1.252 0.0634900643908232
1.256 0.0605060134587639
1.26 0.0576533599038508
1.264 0.0549267641665309
1.268 0.0523210821258743
1.272 0.0498313589412012
1.276 0.047452823043551
1.28 0.0451808802752857
1.284 0.0430111081760511
1.288 0.0409392504132445
1.292 0.0389612113550806
1.296 0.0370730507842841
1.3 0.0352709787503869
1.304 0.0335513505585632
1.308 0.0319106618928841
1.312 0.030345544071846
1.316 0.0288527594339844
1.32 0.0274291968513585
1.324 0.0260718673686676
1.328 0.0247778999657359
1.332 0.0235445374410858
1.336 0.0223691324143068
1.34 0.021249143444913
1.344 0.0201821312653794
1.348 0.0191657551260367
1.352 0.0181977692495101
1.356 0.0172760193923826
1.36 0.0163984395117716
1.364 0.0155630485345128
1.368 0.0147679472266542
1.372 0.0140113151609763
1.376 0.0132914077802648
1.38 0.0126065535540822
1.384 0.0119551512267979
1.388 0.0113356671546584
1.392 0.0107466327297002
1.396 0.0101866418883277
1.4 0.00965434870240619
1.404 0.00914846505074309
1.408 0.00866775836885655
1.412 0.00821104947496026
1.416 0.00777721047012041
1.42 0.00736516271057069
1.424 0.006973874850202
1.428 0.00660236095127442
1.432 0.00624967866143145
1.436 0.00591492745512819
1.44 0.00559724693761898
1.444 0.00529581520968324
1.448 0.00500984729130204
1.452 0.00473859360253227
1.456 0.00448133849985983
1.46 0.00423739886634767
1.464 0.00400612275392959
1.468 0.00378688807623562
1.472 0.00357910135036938
1.476 0.00338219648609305
1.48 0.00319563362091059
1.484 0.00301889799957403
1.488 0.00285149889657283
1.492 0.00269296858020039
1.496 0.00254286131682622
1.5 0.00240075241403597
1.504 0.0022662373013354
};
\addplot [very thick, mycolor3]
table {%
0 117.346046000463
0.004 117.338381615994
0.008 117.315391246373
0.012 117.2770832414
0.016 117.223471512257
0.02 117.154575523778
0.024 117.070420283632
0.028 116.971036328438
0.032 116.856459706806
0.036 116.726731959328
0.04 116.581900095528
0.044 116.422016567784
0.048 116.247139242249
0.052 116.057331366782
0.056 115.85266153592
0.06 115.633203652911
0.064 115.399036888832
0.068 115.150245638831
0.072 114.886919475507
0.076 114.609153099481
0.08 114.317046287175
0.084 114.01070383585
0.088 113.690235505924
0.092 113.355755960633
0.096 113.007384703056
0.1 112.64524601056
0.104 112.269468866712
0.108 111.880186890695
0.112 111.477538264288
0.116 111.061665656454
0.12 110.632716145597
0.124 110.190841139526
0.128 109.736196293202
0.132 109.268941424308
0.136 108.789240426714
0.14 108.297261181883
0.144 107.793175468291
0.148 107.277158868923
0.152 106.749390676894
0.156 106.21005379928
0.16 105.659334659219
0.164 105.097423096334
0.168 104.52451226557
0.172 103.940798534499
0.176 103.346481379167
0.18 102.741763278553
0.184 102.126849607716
0.188 101.501948529698
0.192 100.867270886257
0.196 100.223030087505
0.2 99.569442000531
0.204 98.9067248370688
0.208 98.2350990403027
0.212 97.5547871708764
0.216 96.8660137921835
0.22 96.1690053550187
0.224 95.4639900816631
0.228 94.751197849483
0.232 94.030860074118
0.236 93.3032095923361
0.24 92.5684805446324
0.244 91.8269082576495
0.248 91.0787291264947
0.252 90.3241804970322
0.256 89.5635005482268
0.26 88.7969281746135
0.264 88.0247028689712
0.268 87.2470646052737
0.272 86.4642537219942
0.276 85.676510805837
0.28 84.8840765759702
0.284 84.0871917688329
0.288 83.2860970235882
0.292 82.4810327682945
0.296 81.672239106866
0.3 80.8599557068911
0.304 80.044421688379
0.308 79.2258755135015
0.312 78.4045548773975
0.316 77.5806966001063
0.32 76.754536519693
0.324 75.9263093866322
0.328 75.0962487595095
0.332 74.2645869021037
0.336 73.4315546819098
0.34 72.5973814701591
0.344 71.762295043396
0.348 70.9265214866654
0.352 70.0902850983661
0.356 69.2538082968214
0.36 68.4173115286197
0.364 67.5810131787733
0.368 66.745129482744
0.372 65.9098744403817
0.376 65.0754597318208
0.38 64.2420946353767
0.384 63.409985947484
0.388 62.5793379047169
0.392 61.750352107927
0.396 60.9232274485371
0.4 60.0981600370229
0.404 59.2753431336165
0.408 58.4549670812608
0.412 57.6372192408444
0.416 56.8222839287422
0.42 56.0103423566883
0.424 55.2015725740026
0.428 54.396149412192
0.432 53.594244431946
0.436 52.796025872544
0.44 52.0016586036874
0.444 51.2113040797724
0.448 50.4251202966126
0.452 49.643261750622
0.456 48.865879400465
0.46 48.0931206311799
0.464 47.3251292207777
0.468 46.5620453093202
0.472 45.8040053704747
0.476 45.0511421855448
0.48 44.3035848199732
0.484 43.561458602309
0.488 42.8248851056345
0.492 42.0939821314388
0.496 41.3688636959294
0.5 40.6496400187673
0.504 39.9364175142111
0.508 39.2292987846531
0.512 38.5283826165293
0.516 37.8337639785834
0.52 37.1455340224624
0.524 36.463780085621
0.528 35.7885856965095
0.532 35.1200305820194
0.536 34.4581906771568
0.54 33.8031381369163
0.544 33.154941350323
0.548 32.5136649566096
0.552 31.8793698634966
0.556 31.2521132675387
0.56 30.6319486765009
0.564 30.0189259337282
0.568 29.4130912444676
0.572 28.8144872041034
0.576 28.2231528282647
0.58 27.6391235847618
0.584 27.0624314273083
0.588 26.4931048309847
0.592 25.9311688293973
0.596 25.3766450534869
0.6 24.8295517719392
0.604 24.2899039331491
0.608 23.7577132086907
0.612 23.2329880382423
0.616 22.7157336759166
0.62 22.2059522379461
0.624 21.7036427516707
0.628 21.2088012057769
0.632 20.7214206017359
0.636 20.2414910063867
0.64 19.7689996056133
0.644 19.3039307590595
0.648 18.8462660558305
0.652 18.3959843711252
0.656 17.9530619237465
0.66 17.5174723344337
0.664 17.089186684965
0.668 16.6681735779731
0.672 16.2543991974225
0.676 15.8478273696916
0.68 15.4484196252072
0.684 15.0561352605778
0.688 14.67093140117
0.692 14.2927630640771
0.696 13.9215832214254
0.7 13.5573428639665
0.704 13.1999910649014
0.708 12.8494750438876
0.712 12.5057402311755
0.716 12.1687303318243
0.72 11.8383873899471
0.724 11.5146518529361
0.728 11.1974626356181
0.732 10.8867571842936
0.736 10.5824715406101
0.74 10.2845404052254
0.744 9.99289720121281
0.748 9.70747413716499
0.752 9.42820226995149
0.756 9.15501156708704
0.76 8.88783096866858
0.764 8.62658844883948
0.768 8.37121107674104
0.772 8.12162507691188
0.776 7.87775588909714
0.78 7.63952822743056
0.784 7.40686613895318
0.788 7.17969306143431
0.792 6.95793188046058
0.796 6.74150498576096
0.8 6.53033432673623
0.804 6.32434146716275
0.808 6.12344763904172
0.812 5.92757379556611
0.816 5.73664066317891
0.82 5.55056879269733
0.824 5.3692786094792
0.828 5.19269046260862
0.832 5.02072467307968
0.836 4.85330158095775
0.84 4.69034159149991
0.844 4.53176522021644
0.848 4.37749313685728
0.852 4.22744620830843
0.856 4.08154554038427
0.86 3.93971251850351
0.864 3.80186884723738
0.868 3.66793658872001
0.872 3.53783819991233
0.876 3.4114965687119
0.88 3.28883504890239
0.884 3.16977749393752
0.888 3.05424828955578
0.892 2.94217238522292
0.896 2.83347532440102
0.9 2.72808327364341
0.904 2.62592305051643
0.908 2.52692215034983
0.912 2.43100877181872
0.916 2.33811184136118
0.92 2.24816103643654
0.924 2.1610868076306
0.928 2.07682039961466
0.932 1.99529387096681
0.936 1.91644011286418
0.94 1.84019286665646
0.944 1.7664867403314
0.948 1.69525722388405
0.952 1.62644070360258
0.956 1.55997447528378
0.96 1.49579675639278
0.964 1.43384669718176
0.968 1.37406439078346
0.972 1.3163908822959
0.976 1.2607681768753
0.98 1.20713924685503
0.984 1.15544803790876
0.988 1.10563947427683
0.992 1.05765946307513
0.996 1.01145489770647
1 0.966973660394866
1.004 0.92416462386344
1.008 0.882977652177348
1.012 0.84336360077321
1.016 0.805274315697038
1.02 0.768662632072942
1.024 0.733482371825109
1.028 0.699688340675879
1.032 0.667236324442863
1.036 0.636083084658318
1.04 0.606186353534066
1.044 0.577504828295432
1.048 0.549998164907677
1.052 0.523626971218583
1.056 0.49835279954072
1.06 0.474138138697078
1.064 0.450946405553598
1.068 0.428741936062157
1.072 0.407489975837413
1.076 0.38715667029088
1.08 0.367709054345366
1.084 0.349115041752863
1.088 0.331343414038662
1.092 0.314363809094369
1.096 0.29814670944218
1.1 0.282663430192568
1.104 0.267886106717205
1.108 0.253787682058718
1.112 0.240341894098467
1.116 0.227523262503267
1.12 0.21530707547158
1.124 0.203669376299349
1.128 0.19258694978524
1.132 0.182037308494665
1.136 0.171998678901527
1.14 0.162449987426211
1.144 0.153370846387874
1.148 0.144741539888656
1.152 0.13654300964694
1.156 0.128756840796328
1.16 0.121365247666512
1.164 0.114351059561694
1.168 0.107697706551762
1.172 0.10138920529085
1.176 0.0954101448774696
1.18 0.0897456727698143
1.184 0.0843814807693608
1.188 0.0793037910853516
1.192 0.0744993424922115
1.196 0.0699553765914336
1.2 0.0656596241889331
1.204 0.06160029179835
1.208 0.0577660482802452
1.212 0.0541460116266252
1.216 0.0507297358996957
1.22 0.0475071983332404
1.224 0.0444687866045037
1.228 0.0416052862839484
1.232 0.0389078684697647
1.236 0.0363680776135044
1.24 0.0339778195427334
1.244 0.0317293496861144
1.248 0.0296152615058571
1.252 0.0276284751420152
1.256 0.0257622262726518
1.26 0.0240100551934554
1.264 0.0223657961199491
1.268 0.0208235667150185
1.272 0.0193777578440668
1.276 0.0180230235597081
1.28 0.016754271317517
1.284 0.015566652423977
1.288 0.0144555527174057
1.292 0.0134165834822785
1.296 0.0124455725970344
1.3 0.0115385559151185
1.304 0.0106917688787017
1.308 0.009901638364215
1.312 0.00916477475854667
1.316 0.00847796426447655
1.32 0.00783816143365705
1.324 0.00724248192520174
1.328 0.00668819548770611
1.332 0.00617271916230207
1.336 0.00569361070413663
1.34 0.00524856221946909
1.344 0.0048353940153961
1.348 0.00445204865904173
1.352 0.00409658524289179
1.356 0.00376717385280349
1.36 0.00346209023508758
1.364 0.0031797106589374
1.368 0.00291850697036799
1.372 0.00267704183372887
1.376 0.00245396415676615
1.38 0.00224800469513168
1.384 0.0020579718321706
1.388 0.00188274752976186
1.392 0.00172128344594004
1.396 0.00157259721498983
1.4 0.00143576888567731
1.404 0.00130993751326359
1.408 0.00119429790093657
1.412 0.00108809748629526
1.416 0.000990633368527583
1.42 0.00090124947193629
1.424 0.000819333841489498
1.428 0.000744316066099833
1.432 0.000675664825371206
1.436 0.000612885555592768
1.44 0.000555518230805962
1.444 0.000503135254822582
1.448 0.000455339460128413
1.452 0.000411762209668573
1.456 0.000372061597576456
1.46 0.000335920744977862
1.464 0.000303046187075252
1.468 0.000273166347793647
1.472 0.000246030098349218
1.476 0.000221405396183765
1.48 0.000199078000792701
1.484 0.00017885026306062
1.488 0.000160539984806637
1.492 0.000143979345331239
1.496 0.000129013891847061
1.5 0.00011550159076755
1.504 0.000103311936919628
};
\end{axis}

\end{tikzpicture}

%% file: temperature_XLPE.tex
\begin{tikzpicture}

\definecolor{cornflowerblue86180233}{RGB}{86,180,233}
\definecolor{darkslategray38}{RGB}{38,38,38}
\definecolor{lightgray204}{RGB}{204,204,204}
\definecolor{orange2301590}{RGB}{230,159,0}

\begin{axis}[height=0.085\textheight, scale only axis,
axis line style={darkslategray38},
legend cell align={left},
legend style={fill opacity=0.8, draw opacity=1, text opacity=1, draw=lightgray204},
tick align=outside,
x grid style={lightgray204},
xlabel=\textcolor{darkslategray38}{Conductor current $\frac{I_c}{I_z}$},
xmajorticks=true,
legend style={
  fill opacity=0.8,
  draw opacity=1,
  text opacity=1,
  at={(0.03,0.97)},
  anchor=north west,
  draw=lightgray204,
  font=\footnotesize
},
legend columns=2, 
legend to name={mylegendXLPE},
xmin=0.9, xmax=1.25,
xtick style={color=darkslategray38},
y grid style={lightgray204},
ylabel=\textcolor{darkslategray38}{Temperature (°C)},
ymajorticks=true,
ymin=20, ymax=150,
ytick style={color=darkslategray38}
]
\addplot [very thick, orange2301590, dashed]
table {%
0 20
0.00319488817891374 20.0007145117333
0.00638977635782748 20.0028580469332
0.00958466453674121 20.0064306055997
0.012779552715655 20.0114321877329
0.0159744408945687 20.0178627933326
0.0191693290734824 20.0257224223989
0.0223642172523962 20.0350110749319
0.0255591054313099 20.0457287509314
0.0287539936102236 20.0578754503976
0.0319488817891374 20.0714511733303
0.0351437699680511 20.0864559197297
0.0383386581469649 20.1028896895957
0.0415335463258786 20.1207524829283
0.0447284345047923 20.1400442997275
0.0479233226837061 20.1607651399933
0.0511182108626198 20.1829150037257
0.0543130990415335 20.2064938909247
0.0575079872204473 20.2315018015903
0.060702875399361 20.2579387357225
0.0638977635782748 20.2858046933214
0.0670926517571885 20.3150996743868
0.0702875399361022 20.3458236789188
0.073482428115016 20.3779767069175
0.0766773162939297 20.4115587583828
0.0798722044728434 20.4465698333146
0.0830670926517572 20.4830099317131
0.0862619808306709 20.5208790535782
0.0894568690095847 20.5601771989099
0.0926517571884984 20.6009043677082
0.0958466453674121 20.6430605599731
0.0990415335463259 20.6866457757046
0.10223642172524 20.7316600149027
0.105431309904153 20.7781032775674
0.108626198083067 20.8259755636987
0.111821086261981 20.8752768732967
0.115015974440895 20.9260072063612
0.118210862619808 20.9781665628923
0.121405750798722 21.0317549428901
0.124600638977636 21.0867723463545
0.12779552715655 21.1432187732854
0.130990415335463 21.201094223683
0.134185303514377 21.2603986975472
0.137380191693291 21.321132194878
0.140575079872204 21.3832947156754
0.143769968051118 21.4468862599394
0.146964856230032 21.51190682767
0.150159744408946 21.5783564188672
0.153354632587859 21.646235033531
0.156549520766773 21.7155426716614
0.159744408945687 21.7862793332585
0.162939297124601 21.8584450183221
0.166134185303514 21.9320397268524
0.169329073482428 22.0070634588492
0.172523961661342 22.0835162143127
0.175718849840256 22.1613979932428
0.178913738019169 22.2407087956394
0.182108626198083 22.3214486215027
0.185303514376997 22.4036174708326
0.188498402555911 22.4872153436291
0.191693290734824 22.5722422398922
0.194888178913738 22.6586981596219
0.198083067092652 22.7465831028182
0.201277955271566 22.8358970694812
0.204472843450479 22.9266400596107
0.207667731629393 23.0188120732068
0.210862619808307 23.1124131102696
0.21405750798722 23.2074431707989
0.217252396166134 23.3039022547949
0.220447284345048 23.4017903622575
0.223642172523962 23.5011074931866
0.226837060702875 23.6018536475824
0.230031948881789 23.7040288254448
0.233226837060703 23.8076330267738
0.236421725239617 23.9126662515694
0.23961661341853 24.0191284998316
0.242811501597444 24.1270197715604
0.246006389776358 24.2363400667558
0.249201277955272 24.3470893854178
0.252396166134185 24.4592677275465
0.255591054313099 24.5728750931417
0.258785942492013 24.6879114822036
0.261980830670927 24.804376894732
0.26517571884984 24.9222713307271
0.268370607028754 25.0415947901887
0.271565495207668 25.162347273117
0.274760383386581 25.2845287795119
0.277955271565495 25.4081393093734
0.281150159744409 25.5331788627015
0.284345047923323 25.6596474394962
0.287539936102236 25.7875450397575
0.29073482428115 25.9168716634854
0.293929712460064 26.0476273106799
0.297124600638978 26.179811981341
0.300319488817891 26.3134256754688
0.303514376996805 26.4484683930631
0.306709265175719 26.5849401341241
0.309904153354633 26.7228408986516
0.313099041533546 26.8621706866458
0.31629392971246 27.0029294981065
0.319488817891374 27.1451173330339
0.322683706070288 27.2887341914279
0.325878594249201 27.4337800732885
0.329073482428115 27.5802549786157
0.332268370607029 27.7281589074095
0.335463258785943 27.8774918596699
0.338658146964856 28.0282538353969
0.34185303514377 28.1804448345905
0.345047923322684 28.3340648572508
0.348242811501597 28.4891139033776
0.351437699680511 28.645591972971
0.354632587859425 28.8034990660311
0.357827476038339 28.9628351825577
0.361022364217252 29.123600322551
0.364217252396166 29.2857944860109
0.36741214057508 29.4494176729374
0.370607028753994 29.6144698833304
0.373801916932907 29.7809511171901
0.376996805111821 29.9488613745164
0.380191693290735 30.1182006553093
0.383386581469649 30.2889689595688
0.386581469648562 30.461166287295
0.389776357827476 30.6347926384877
0.39297124600639 30.809848013147
0.396166134185303 30.986332411273
0.399361022364217 31.1642458328655
0.402555910543131 31.3435882779246
0.405750798722045 31.5243597464504
0.408945686900958 31.7065602384428
0.412140575079872 31.8901897539017
0.415335463258786 32.0752482928273
0.4185303514377 32.2617358552195
0.421725239616613 32.4496524410783
0.424920127795527 32.6389980504037
0.428115015974441 32.8297726831957
0.431309904153355 33.0219763394543
0.434504792332268 33.2156090191795
0.437699680511182 33.4106707223714
0.440894568690096 33.6071614490298
0.44408945686901 33.8050811991548
0.447284345047923 34.0044299727465
0.450479233226837 34.2052077698047
0.453674121405751 34.4074145903296
0.456869009584665 34.6110504343211
0.460063897763578 34.8161153017791
0.463258785942492 35.0226091927038
0.466453674121406 35.2305321070951
0.469648562300319 35.439884044953
0.472843450479233 35.6506650062775
0.476038338658147 35.8628749910686
0.479233226837061 36.0765139993263
0.482428115015974 36.2915820310506
0.485623003194888 36.5080790862416
0.488817891373802 36.7260051648991
0.492012779552716 36.9453602670232
0.495207667731629 37.166144392614
0.498402555910543 37.3883575416713
0.501597444089457 37.6119997141953
0.504792332268371 37.8370709101859
0.507987220447284 38.0635711296431
0.511182108626198 38.2915003725668
0.514376996805112 38.5208586389572
0.517571884984026 38.7516459288142
0.520766773162939 38.9838622421378
0.523961661341853 39.217507578928
0.527156549520767 39.4525819391848
0.530351437699681 39.6890853229083
0.533546325878594 39.9270177300983
0.536741214057508 40.1663791607549
0.539936102236422 40.4071696148782
0.543130990415335 40.649389092468
0.546325878594249 40.8930375935245
0.549520766773163 41.1381151180476
0.552715654952077 41.3846216660372
0.55591054313099 41.6325572374935
0.559105431309904 41.8819218324164
0.562300319488818 42.1327154508059
0.565495207667732 42.384938092662
0.568690095846645 42.6385897579847
0.571884984025559 42.893670446774
0.575079872204473 43.1501801590299
0.578274760383387 43.4081188947524
0.5814696485623 43.6674866539416
0.584664536741214 43.9282834365973
0.587859424920128 44.1905092427196
0.591054313099042 44.4541640723086
0.594249201277955 44.7192479253641
0.597444089456869 44.9857608018863
0.600638977635783 45.2537027018751
0.603833865814696 45.5230736253305
0.60702875399361 45.7938735722525
0.610223642172524 46.066102542641
0.613418530351438 46.3397605364962
0.616613418530351 46.614847553818
0.619808306709265 46.8913635946065
0.623003194888179 47.1693086588615
0.626198083067093 47.4486827465831
0.629392971246006 47.7294858577713
0.63258785942492 48.0117179924262
0.635782747603834 48.2953791505476
0.638977635782748 48.5804693321357
0.642172523961661 48.8669885371903
0.645367412140575 49.1549367657116
0.648562300319489 49.4443140176995
0.651757188498403 49.735120293154
0.654952076677316 50.027355592075
0.65814696485623 50.3210199144627
0.661341853035144 50.616113260317
0.664536741214057 50.9126356296379
0.667731629392971 51.2105870224255
0.670926517571885 51.5099674386796
0.674121405750799 51.8107768784003
0.677316293929712 52.1130153415876
0.680511182108626 52.4166828282416
0.68370607028754 52.7217793383621
0.686900958466454 53.0283048719493
0.690095846645367 53.3362594290031
0.693290734824281 53.6456430095234
0.696485623003195 53.9564556135104
0.699680511182109 54.268697240964
0.702875399361022 54.5823678918842
0.706070287539936 54.897467566271
0.70926517571885 55.2139962641244
0.712460063897764 55.5319539854444
0.715654952076677 55.851340730231
0.718849840255591 56.1721564984842
0.722044728434505 56.494401290204
0.725239616613419 56.8180751053905
0.728434504792332 57.1431779440435
0.731629392971246 57.4697098061632
0.73482428115016 57.7976706917494
0.738019169329074 58.1270606008023
0.741214057507987 58.4578795333218
0.744408945686901 58.7901274893078
0.747603833865815 59.1238044687605
0.750798722044728 59.4589104716798
0.753993610223642 59.7954454980657
0.757188498402556 60.1334095479182
0.76038338658147 60.4728026212373
0.763578274760383 60.8136247180231
0.766773162939297 61.1558758382754
0.769968051118211 61.4995559819943
0.773162939297125 61.8446651491798
0.776357827476038 62.191203339832
0.779552715654952 62.5391705539507
0.782747603833866 62.8885667915361
0.78594249201278 63.2393920525881
0.789137380191693 63.5916463371066
0.792332268370607 63.9453296450918
0.795527156549521 64.3004419765436
0.798722044728435 64.656983331462
0.801916932907348 65.014953709847
0.805111821086262 65.3743531116986
0.808306709265176 65.7351815370168
0.811501597444089 66.0974389858016
0.814696485623003 66.4611254580531
0.817891373801917 66.8262409537711
0.821086261980831 67.1927854729557
0.824281150159744 67.560759015607
0.827476038338658 67.9301615817248
0.830670926517572 68.3009931713093
0.833865814696486 68.6732537843604
0.837060702875399 69.046943420878
0.840255591054313 69.4220620808623
0.843450479233227 69.7986097643132
0.846645367412141 70.1765864712307
0.849840255591054 70.5559922016148
0.853035143769968 70.9368269554655
0.856230031948882 71.3190907327828
0.859424920127795 71.7027835335667
0.862619808306709 72.0879053578173
0.865814696485623 72.4744562055344
0.869009584664537 72.8624360767181
0.87220447284345 73.2518449713685
0.875399361022364 73.6426828894855
0.878594249201278 74.034949831069
0.881789137380192 74.4286457961192
0.884984025559105 74.8237707846359
0.888178913738019 75.2203247966193
0.891373801916933 75.6183078320693
0.894568690095847 76.0177198909859
0.89776357827476 76.4185609733691
0.900958466453674 76.8208310792189
0.904153354632588 77.2245302085354
0.907348242811502 77.6296583613184
0.910543130990415 78.036215537568
0.913738019169329 78.4442017372842
0.916932907348243 78.8536169604671
0.920127795527157 79.2644612071165
0.92332268370607 79.6767344772326
0.926517571884984 80.0904367708153
0.929712460063898 80.5055680878645
0.932907348242812 80.9221284283804
0.936102236421725 81.3401177923629
0.939297124600639 81.759536179812
0.942492012779553 82.1803835907277
0.945686900958466 82.60266002511
0.94888178913738 83.0263654829589
0.952076677316294 83.4514999642744
0.955271565495208 83.8780634690565
0.958466453674121 84.3060559973053
0.961661341853035 84.7354775490206
0.964856230031949 85.1663281242026
0.968051118210863 85.5986077228511
0.971246006389776 86.0323163449663
0.97444089456869 86.467453990548
0.977635782747604 86.9040206595964
0.980830670926518 87.3420163521114
0.984025559105431 87.781441068093
0.987220447284345 88.2222948075411
0.990415335463259 88.664577570456
0.993610223642173 89.1082893568374
0.996805111821086 89.5534301666854
1 90
1.00319488817891 90.4479988567812
1.00638977635783 90.8974267370291
1.00958466453674 91.3482836407435
1.01277955271565 91.8005695679245
1.01597444089457 92.2542845185722
1.01916932907348 92.7094284926865
1.0223642172524 93.1660014902673
1.02555910543131 93.6240035113148
1.02875399361022 94.0834345558289
1.03194888178914 94.5442946238096
1.03514376996805 95.0065837152569
1.03833865814696 95.4703018301708
1.04153354632588 95.9354489685513
1.04472843450479 96.4020251303984
1.04792332268371 96.8700303157121
1.05111821086262 97.3394645244924
1.05431309904153 97.8103277567394
1.05750798722045 98.2826200124529
1.06070287539936 98.7563412916331
1.06389776357827 99.2314915942798
1.06709265175719 99.7080709203932
1.0702875399361 100.186079269973
1.07348242811502 100.66551664302
1.07667731629393 101.146383039533
1.07987220447284 101.628678459513
1.08306709265176 102.112402902959
1.08626198083067 102.597556369872
1.08945686900958 103.084138860252
1.0926517571885 103.572150374098
1.09584664536741 104.061590911411
1.09904153354633 104.55246047219
1.10223642172524 105.044759056436
1.10543130990415 105.538486664149
1.10862619808307 106.033643295328
1.11182108626198 106.530228949974
1.11501597444089 107.028243628086
1.11821086261981 107.527687329666
1.12140575079872 108.028560054711
1.12460063897764 108.530861803223
1.12779552715655 109.034592575202
1.13099041533546 109.539752370648
1.13418530351438 110.04634118956
1.13738019169329 110.554359031939
1.1405750798722 111.063805897784
1.14376996805112 111.574681787096
1.14696485623003 112.086986699874
1.15015974440895 112.60072063612
1.15335463258786 113.115883595831
1.15654952076677 113.63247557901
1.15974440894569 114.150496585655
1.1629392971246 114.669946615766
1.16613418530351 115.190825669344
1.16932907348243 115.713133746389
1.17252396166134 116.236870846901
1.17571884984026 116.762036970879
1.17891373801917 117.288632118323
1.18210862619808 117.816656289234
1.185303514377 118.346109483612
1.18849840255591 118.876991701457
1.19169329073482 119.409302942768
1.19488817891374 119.943043207545
1.19808306709265 120.478212495789
1.20127795527157 121.0148108075
1.20447284345048 121.552838142678
1.20766773162939 122.092294501322
1.21086261980831 122.633179883433
1.21405750798722 123.17549428901
1.21725239616613 123.719237718054
1.22044728434505 124.264410170564
1.22364217252396 124.811011646541
1.22683706070288 125.359042145985
1.23003194888179 125.908501668895
1.2332268370607 126.459390215272
1.23642172523962 127.011707785116
1.23961661341853 127.565454378426
1.24281150159744 128.120629995203
1.24600638977636 128.677234635446
1.24920127795527 129.235268299156
1.25239616613419 129.794730986332
1.2555910543131 130.355622696976
1.25878594249201 130.917943431085
1.26198083067093 131.481693188662
1.26517571884984 132.046871969705
1.26837060702875 132.613479774214
1.27156549520767 133.18151660219
1.27476038338658 133.750982453633
1.2779552715655 134.321877328543
1.28115015974441 134.894201226919
1.28434504792332 135.467954148761
1.28753993610224 136.043136094071
1.29073482428115 136.619747062846
1.29392971246006 137.197787055089
1.29712460063898 137.777256070798
1.30031948881789 138.358154109974
1.30351437699681 138.940481172616
1.30670926517572 139.524237258725
1.30990415335463 140.1094223683
1.31309904153355 140.696036501342
1.31629392971246 141.284079657851
1.31948881789137 141.873551837826
1.32268370607029 142.464453041268
1.3258785942492 143.056783268177
1.32907348242811 143.650542518552
1.33226837060703 144.245730792394
1.33546325878594 144.842348089702
1.33865814696486 145.440394410477
1.34185303514377 146.039869754718
1.34504792332268 146.640774122426
1.3482428115016 147.243107513601
1.35143769968051 147.846869928243
1.35463258785942 148.452061366351
1.35782747603834 149.058681827925
1.36102236421725 149.666731312966
1.36421725239617 150.276209821474
1.36741214057508 150.887117353449
1.37060702875399 151.49945390889
1.37380191693291 152.113219487797
1.37699680511182 152.728414090171
1.38019169329073 153.345037716012
1.38338658146965 153.96309036532
1.38658146964856 154.582572038094
1.38977635782748 155.203482734334
1.39297124600639 155.825822454042
1.3961661341853 156.449591197215
1.39936102236422 157.074788963856
1.40255591054313 157.701415753963
1.40575079872204 158.329471567537
1.40894568690096 158.958956404577
1.41214057507987 159.589870265084
1.41533546325879 160.222213149057
1.4185303514377 160.855985056497
1.42172523961661 161.491185987404
1.42492012779553 162.127815941777
1.42811501597444 162.765874919617
1.43130990415335 163.405362920924
1.43450479233227 164.046279945697
1.43769968051118 164.688625993937
1.4408945686901 165.332401065643
1.44408945686901 165.977605160816
1.44728434504792 166.624238279456
1.45047923322684 167.272300421562
1.45367412140575 167.921791587135
1.45686900958466 168.572711776174
1.46006389776358 169.22506098868
1.46325878594249 169.878839224653
1.46645367412141 170.534046484092
1.46964856230032 171.190682766998
1.47284345047923 171.84874807337
1.47603833865815 172.508242403209
1.47923322683706 173.169165756515
1.48242811501597 173.831518133287
1.48562300319489 174.495299533526
1.4888178913738 175.160509957231
1.49201277955272 175.827149404403
1.49520766773163 176.495217875042
1.49840255591054 177.164715369147
1.50159744408946 177.835641886719
1.50479233226837 178.507997427758
};
\addlegendentry{$\alpha = 0 {K}^{-1}$}
\addplot [very thick, cornflowerblue86180233]
table {%
0 20
0.00319488817891374 20.0005572992263
0.00638977635782748 20.0022292119251
0.00958466453674121 20.0050157831569
0.012779552715655 20.0089170880262
0.0159744408945687 20.0139332316862
0.0191693290734824 20.0200643493462
0.0223642172523962 20.0273106062802
0.0255591054313099 20.0356721978387
0.0287539936102236 20.0451493494614
0.0319488817891374 20.0557423166927
0.0351437699680511 20.0674513851985
0.0383386581469649 20.0802768707861
0.0415335463258786 20.0942191194248
0.0447284345047923 20.1092785072697
0.0479233226837061 20.1254554406872
0.0511182108626198 20.1427503562818
0.0543130990415335 20.1611637209263
0.0575079872204473 20.1806960317928
0.060702875399361 20.2013478163867
0.0638977635782748 20.2231196325818
0.0670926517571885 20.2460120686588
0.0702875399361022 20.2700257433442
0.073482428115016 20.2951613058529
0.0766773162939297 20.3214194359318
0.0798722044728434 20.3488008439057
0.0830670926517572 20.3773062707258
0.0862619808306709 20.4069364880196
0.0894568690095847 20.4376922981432
0.0926517571884984 20.4695745342361
0.0958466453674121 20.5025840602774
0.0990415335463259 20.5367217711444
0.10223642172524 20.5719885926738
0.105431309904153 20.6083854817245
0.108626198083067 20.6459134262423
0.111821086261981 20.6845734453279
0.115015974440895 20.7243665893055
0.118210862619808 20.7652939397952
0.121405750798722 20.807356609786
0.124600638977636 20.8505557437123
0.12779552715655 20.8948925175319
0.130990415335463 20.9403681388065
0.134185303514377 20.986983846784
0.137380191693291 21.0347409124837
0.140575079872204 21.083640638783
0.143769968051118 21.1336843605073
0.146964856230032 21.1848734445209
0.150159744408946 21.2372092898215
0.153354632587859 21.2906933276361
0.156549520766773 21.3453270215193
0.159744408945687 21.4011118674544
0.162939297124601 21.4580493939566
0.166134185303514 21.516141162178
0.169329073482428 21.5753887660159
0.172523961661342 21.6357938322227
0.175718849840256 21.6973580205185
0.178913738019169 21.7600830237063
0.182108626198083 21.8239705677892
0.185303514376997 21.8890224120901
0.188498402555911 21.9552403493743
0.191693290734824 22.0226262059739
0.194888178913738 22.0911818419152
0.198083067092652 22.1609091510483
0.201277955271566 22.2318100611794
0.204472843450479 22.3038865342053
0.207667731629393 22.3771405662511
0.210862619808307 22.4515741878095
0.21405750798722 22.5271894638839
0.217252396166134 22.6039884941327
0.220447284345048 22.6819734130174
0.223642172523962 22.761146389953
0.226837060702875 22.8415096294604
0.230031948881789 22.9230653713226
0.233226837060703 23.0058158907429
0.236421725239617 23.0897634985056
0.23961661341853 23.1749105411403
0.242811501597444 23.2612594010882
0.246006389776358 23.3488124968713
0.249201277955272 23.437572283265
0.252396166134185 23.5275412514729
0.255591054313099 23.6187219293044
0.258785942492013 23.7111168813559
0.261980830670927 23.8047287091941
0.26517571884984 23.8995600515428
0.268370607028754 23.9956135844723
0.271565495207668 24.0928920215922
0.274760383386581 24.1913981142463
0.277955271565495 24.2911346517121
0.281150159744409 24.3921044614016
0.284345047923323 24.4943104090664
0.287539936102236 24.5977553990057
0.29073482428115 24.702442374277
0.293929712460064 24.8083743169103
0.297124600638978 24.9155542481259
0.300319488817891 25.0239852285542
0.303514376996805 25.1336703584605
0.306709265175719 25.2446127779717
0.309904153354633 25.3568156673068
0.313099041533546 25.4702822470109
0.31629392971246 25.5850157781926
0.319488817891374 25.7010195627644
0.322683706070288 25.8182969436872
0.325878594249201 25.9368513052177
0.329073482428115 26.05668607316
0.332268370607029 26.1778047151199
0.335463258785943 26.3002107407635
0.338658146964856 26.4239077020795
0.34185303514377 26.5488991936445
0.345047923322684 26.6751888528924
0.348242811501597 26.8027803603882
0.351437699680511 26.9316774401043
0.354632587859425 27.0618838597014
0.357827476038339 27.1934034308138
0.361022364217252 27.3262400093374
0.364217252396166 27.4603974957224
0.36741214057508 27.59587983527
0.370607028753994 27.7326910184332
0.373801916932907 27.8708350811209
0.376996805111821 28.0103161050077
0.380191693290735 28.151138217846
0.383386581469649 28.2933055937839
0.386581469648562 28.4368224536864
0.389776357827476 28.5816930654615
0.39297124600639 28.7279217443903
0.396166134185303 28.8755128534619
0.399361022364217 29.0244708037119
0.402555910543131 29.174800054567
0.405750798722045 29.3265051141924
0.408945686900958 29.4795905398451
0.412140575079872 29.6340609382313
0.415335463258786 29.7899209658687
0.4185303514377 29.9471753294537
0.421725239616613 30.1058287862331
0.424920127795527 30.2658861443808
0.428115015974441 30.4273522633801
0.431309904153355 30.59023205441
0.434504792332268 30.7545304807372
0.437699680511182 30.9202525581134
0.440894568690096 31.0874033551771
0.44408945686901 31.2559879938614
0.447284345047923 31.4260116498066
0.450479233226837 31.5974795527786
0.453674121405751 31.7703969870926
0.456869009584665 31.944769292042
0.460063897763578 32.1206018623336
0.463258785942492 32.2979001485277
0.466453674121406 32.4766696574847
0.469648562300319 32.6569159528166
0.472843450479233 32.8386446553454
0.476038338658147 33.0218614435666
0.479233226837061 33.2065720541193
0.482428115015974 33.3927822822617
0.485623003194888 33.5804979823541
0.488817891373802 33.7697250683465
0.492012779552716 33.9604695142738
0.495207667731629 34.1527373547566
0.498402555910543 34.3465346855095
0.501597444089457 34.5418676638548
0.504792332268371 34.7387425092436
0.507987220447284 34.9371655037837
0.511182108626198 35.1371429927738
0.514376996805112 35.3386813852454
0.517571884984026 35.5417871545109
0.520766773162939 35.7464668387196
0.523961661341853 35.9527270414203
0.527156549520767 36.1605744321316
0.530351437699681 36.3700157469194
0.533546325878594 36.5810577889821
0.536741214057508 36.7937074292431
0.539936102236422 37.0079716069512
0.543130990415335 37.2238573302892
0.546325878594249 37.4413716769895
0.549520766773163 37.6605217949586
0.552715654952077 37.8813149029092
0.55591054313099 38.1037582910008
0.559105431309904 38.3278593214886
0.562300319488818 38.5536254293811
0.565495207667732 38.7810641231054
0.568690095846645 39.0101829851825
0.571884984025559 39.2409896729107
0.575079872204473 39.4734919190577
0.578274760383387 39.7076975325623
0.5814696485623 39.9436143992452
0.584664536741214 40.1812504825291
0.587859424920128 40.4206138241677
0.591054313099042 40.6617125449857
0.594249201277955 40.9045548456265
0.597444089456869 41.149149007312
0.600638977635783 41.3955033926105
0.603833865814696 41.6436264462158
0.60702875399361 41.8935266957366
0.610223642172524 42.1452127524957
0.613418530351438 42.3986933123402
0.616613418530351 42.6539771564628
0.619808306709265 42.9110731522332
0.623003194888179 43.169990254041
0.626198083067093 43.43073750415
0.629392971246006 43.6933240335636
0.63258785942492 43.9577590629016
0.635782747603834 44.2240519032892
0.638977635782748 44.4922119572572
0.642172523961661 44.7622487196553
0.645367412140575 45.0341717785762
0.648562300319489 45.3079908162936
0.651757188498403 45.5837156102117
0.654952076677316 45.8613560338279
0.65814696485623 46.1409220577087
0.661341853035144 46.4224237504787
0.664536741214057 46.7058712798221
0.667731629392971 46.9912749134993
0.670926517571885 47.2786450203758
0.674121405750799 47.5679920714665
0.677316293929712 47.8593266409925
0.680511182108626 48.1526594074541
0.68370607028754 48.448001154717
0.686900958466454 48.7453627731143
0.690095846645367 49.0447552605629
0.693290734824281 49.3461897236957
0.696485623003195 49.6496773790091
0.699680511182109 49.9552295540263
0.702875399361022 50.2628576884767
0.706070287539936 50.5725733354916
0.70926517571885 50.8843881628162
0.712460063897764 51.1983139540386
0.715654952076677 51.514362609836
0.718849840255591 51.8325461492377
0.722044728434505 52.1528767109061
0.725239616613419 52.4753665544357
0.728434504792332 52.8000280616693
0.731629392971246 53.1268737380344
0.73482428115016 53.4559162138961
0.738019169329074 53.7871682459309
0.741214057507987 54.1206427185189
0.744408945686901 54.4563526451556
0.747603833865815 54.7943111698843
0.750798722044728 55.1345315687477
0.753993610223642 55.4770272512611
0.757188498402556 55.8218117619062
0.76038338658147 56.1688987816458
0.763578274760383 56.5183021294605
0.766773162939297 56.8700357639072
0.769968051118211 57.2241137847002
0.773162939297125 57.5805504343142
0.776357827476038 57.9393600996112
0.779552715654952 58.3005573134909
0.782747603833866 58.6641567565635
0.78594249201278 59.0301732588486
0.789137380191693 59.3986218014976
0.792332268370607 59.7695175185409
0.795527156549521 60.1428756986616
0.798722044728435 60.5187117869939
0.801916932907348 60.8970413869486
0.805111821086262 61.2778802620649
0.808306709265176 61.6612443378893
0.811501597444089 62.0471497038828
0.814696485623003 62.4356126153551
0.817891373801917 62.8266494954287
0.821086261980831 63.2202769370306
0.824281150159744 63.6165117049149
0.827476038338658 64.0153707377144
0.830670926517572 64.4168711500237
0.833865814696486 64.8210302345125
0.837060702875399 65.227865464071
0.840255591054313 65.6373944939878
0.843450479233227 66.0496351641596
0.846645367412141 66.4646055013352
0.849840255591054 66.8823237213927
0.853035143769968 67.3028082316515
0.856230031948882 67.7260776332191
0.859424920127795 68.1521507233739
0.862619808306709 68.5810464979839
0.865814696485623 69.0127841539632
0.869009584664537 69.4473830917644
0.87220447284345 69.8848629179117
0.875399361022364 70.3252434475704
0.878594249201278 70.7685447071578
0.881789137380192 71.2147869369941
0.884984025559105 71.6639905939935
0.888178913738019 72.1161763543981
0.891373801916933 72.571365116554
0.894568690095847 73.0295780037307
0.89776357827476 73.4908363669843
0.900958466453674 73.9551617880664
0.904153354632588 74.4225760823775
0.907348242811502 74.8931013019686
0.910543130990415 75.3667597385882
0.913738019169329 75.8435739267796
0.916932907348243 76.3235666470254
0.920127795527157 76.8067609289441
0.92332268370607 77.2931800545354
0.926517571884984 77.7828475614798
0.929712460063898 78.2757872464889
0.932907348242812 78.7720231687113
0.936102236421725 79.2715796531923
0.939297124600639 79.77448129439
0.942492012779553 80.2807529597481
0.945686900958466 80.7904197933271
0.94888178913738 81.3035072194939
0.952076677316294 81.8200409466726
0.955271565495208 82.3400469711559
0.958466453674121 82.8635515809795
0.961661341853035 83.3905813598596
0.964856230031949 83.9211631911968
0.968051118210863 84.4553242621441
0.971246006389776 84.9930920677445
0.97444089456869 85.5344944151357
0.977635782747604 86.079559427826
0.980830670926518 86.6283155500413
0.984025559105431 87.180791551145
0.987220447284345 87.7370165301318
0.990415335463259 88.297019920198
0.993610223642173 88.8608314933885
0.996805111821086 89.4284813653225
1 90
1.00319488817891 90.5754182146897
1.00638977635783 91.1547671849005
1.00958466453674 91.7380784494383
1.01277955271565 92.3253839155497
1.01597444089457 92.9167158641545
1.01916932907348 93.5121069551677
1.0223642172524 94.111590232915
1.02555910543131 94.7151991316415
1.02875399361022 95.3229674811166
1.03194888178914 95.9349295123369
1.03514376996805 96.5511198633293
1.03833865814696 97.1715735850553
1.04153354632588 97.7963261474201
1.04472843450479 98.4254134453876
1.04792332268371 99.058871805204
1.05111821086262 99.6967379907314
1.05431309904153 100.339049209896
1.05750798722045 100.985843121247
1.06070287539936 101.637157840643
1.06389776357827 102.293031948045
1.06709265175719 102.953504494444
1.0702875399361 103.618615008905
1.07348242811502 104.288403505748
1.07667731629393 104.962910491851
1.07987220447284 105.642176974088
1.08306709265176 106.326244466912
1.08626198083067 107.015155000062
1.08945686900958 107.708951126427
1.0926517571885 108.407675930045
1.09584664536741 109.111373034253
1.09904153354633 109.820086609993
1.10223642172524 110.533861384266
1.10543130990415 111.252742648749
1.10862619808307 111.97677626857
1.11182108626198 112.706008691254
1.11501597444089 113.440486955828
1.11821086261981 114.180258702111
1.12140575079872 114.925372180167
1.12460063897764 115.675876259952
1.12779552715655 116.431820441132
1.13099041533546 117.193254863101
1.13418530351438 117.960230315181
1.13738019169329 118.73279824703
1.1405750798722 119.511010779241
1.14376996805112 120.294920714154
1.14696485623003 121.084581546873
1.15015974440895 121.880047476506
1.15335463258786 122.68137341762
1.15654952076677 123.488615011918
1.15974440894569 124.301828640155
1.1629392971246 125.121071434284
1.16613418530351 125.946401289848
1.16932907348243 126.77787687861
1.17252396166134 127.615557661447
1.17571884984026 128.459503901496
1.17891373801917 129.309776677564
1.18210862619808 130.166437897814
1.185303514377 131.029550313721
1.18849840255591 131.899177534322
1.19169329073482 132.775384040745
1.19488817891374 133.658235201044
1.19808306709265 134.547797285335
1.20127795527157 135.444137481249
1.20447284345048 136.347323909692
1.20766773162939 137.257425640951
1.21086261980831 138.174512711116
1.21405750798722 139.098656138861
1.21725239616613 140.029927942566
1.22044728434505 140.968401157806
1.22364217252396 141.914149855207
1.22683706070288 142.867249158677
1.23003194888179 143.827775264034
1.2332268370607 144.795805458017
1.23642172523962 145.771418137723
1.23961661341853 146.754692830442
1.24281150159744 147.745710213938
1.24600638977636 148.744552137154
1.24920127795527 149.751301641374
1.25239616613419 150.76604298185
1.2555910543131 151.788861649896
1.25878594249201 152.819844395467
1.26198083067093 153.859079250242
1.26517571884984 154.906655551211
1.26837060702875 155.962663964789
1.27156549520767 157.027196511471
1.27476038338658 158.100346591029
1.2779552715655 159.182209008285
1.28115015974441 160.272879999461
1.28434504792332 161.372457259119
1.28753993610224 162.481039967727
1.29073482428115 163.598728819839
1.29392971246006 164.725626052926
1.29712460063898 165.861835476875
1.30031948881789 167.007462504155
1.30351437699681 168.162614180693
1.30670926517572 169.327399217463
1.30990415335463 170.50192802281
1.31309904153355 171.686312735537
1.31629392971246 172.880667258761
1.31948881789137 174.085107294575
1.32268370607029 175.299750379532
1.3258785942492 176.524715920965
1.32907348242811 177.760125234185
1.33226837060703 179.006101580563
1.33546325878594 180.262770206533
1.33865814696486 181.530258383535
1.34185303514377 182.80869544893
1.34504792332268 184.09821284791
1.3482428115016 185.398944176437
1.35143769968051 186.711025225228
1.35463258785942 188.034594024837
1.35782747603834 189.369790891847
1.36102236421725 190.716758476206
1.36421725239617 192.075641809761
1.36741214057508 193.446588355987
1.37060702875399 194.829748060992
1.37380191693291 196.225273405792
1.37699680511182 197.633319459926
1.38019169329073 199.054043936435
1.38338658146965 200.487607248247
1.38658146964856 201.934172566014
1.38977635782748 203.393905877447
1.39297124600639 204.866976048184
1.3961661341853 206.353554884245
1.39936102236422 207.853817196132
1.40255591054313 209.367940864594
1.40575079872204 210.896106908153
1.40894568690096 212.438499552396
1.41214057507987 213.995306301134
1.41533546325879 215.56671800945
1.4185303514377 217.152928958716
1.42172523961661 218.754136933637
1.42492012779553 220.370543301384
1.42811501597444 222.002353092874
1.43130990415335 223.649775086287
1.43450479233227 225.313021892871
1.43769968051118 226.992310045117
1.4408945686901 228.687860087388
1.44408945686901 230.399896669059
1.44728434504792 232.128648640283
1.45047923322684 233.874349150437
1.45367412140575 235.637235749359
1.45686900958466 237.417550491458
1.46006389776358 239.215540042794
1.46325878594249 241.031455791229
1.46645367412141 242.865553959751
1.46964856230032 244.718095723079
1.47284345047923 246.58934732766
1.47603833865815 248.479580215175
1.47923322683706 250.389071149668
1.48242811501597 252.318102348432
1.48562300319489 254.26696161678
1.4888178913738 256.235942486824
1.49201277955272 258.225344360424
1.49520766773163 260.235472656431
1.49840255591054 262.266638962393
1.50159744408946 264.319161190872
1.50479233226837 266.39336374054
};
\addlegendentry{$\alpha = 4.03 \times 10^{-3} {K}^{-1}$}
 \addlegendimage{very thick, color=mycolor3}
  \addlegendentry{$\alpha = 3.9 \times 10^{-3} {K}^{-1}$}
\end{axis}

\end{tikzpicture}

%% file: temperature_PILC.tex
\begin{tikzpicture}

\definecolor{cornflowerblue86180233}{RGB}{86,180,233}
\definecolor{darkslategray38}{RGB}{38,38,38}
\definecolor{lightgray204}{RGB}{204,204,204}
\definecolor{orange2301590}{RGB}{230,159,0}
\definecolor{mycolor3}{RGB}{0,158,115}

\begin{axis}[height=0.085\textheight, scale only axis,
axis line style={darkslategray38},
legend cell align={left},
legend style={fill opacity=0.8, draw opacity=1, text opacity=1, draw=lightgray204},
tick align=outside,
x grid style={lightgray204},
xlabel=\textcolor{darkslategray38}{Conductor current $\frac{I_c}{I_z}$},
xmajorticks=true,
xmin=0, xmax=1.25,
xtick style={color=darkslategray38},
y grid style={lightgray204},
ylabel=\textcolor{darkslategray38}{Temperature (°C)},
ymajorticks=true,
ymin=20, ymax=150,
ytick style={color=darkslategray38}
]
\addplot [very thick, orange2301590, dashed]
table {%
0 20
0.004 20.00072
0.008 20.00288
0.012 20.00648
0.016 20.01152
0.02 20.018
0.024 20.02592
0.028 20.03528
0.032 20.04608
0.036 20.05832
0.04 20.072
0.044 20.08712
0.048 20.10368
0.052 20.12168
0.056 20.14112
0.06 20.162
0.064 20.18432
0.068 20.20808
0.072 20.23328
0.076 20.25992
0.08 20.288
0.084 20.31752
0.088 20.34848
0.092 20.38088
0.096 20.41472
0.1 20.45
0.104 20.48672
0.108 20.52488
0.112 20.56448
0.116 20.60552
0.12 20.648
0.124 20.69192
0.128 20.73728
0.132 20.78408
0.136 20.83232
0.14 20.882
0.144 20.93312
0.148 20.98568
0.152 21.03968
0.156 21.09512
0.16 21.152
0.164 21.21032
0.168 21.27008
0.172 21.33128
0.176 21.39392
0.18 21.458
0.184 21.52352
0.188 21.59048
0.192 21.65888
0.196 21.72872
0.2 21.8
0.204 21.87272
0.208 21.94688
0.212 22.02248
0.216 22.09952
0.22 22.178
0.224 22.25792
0.228 22.33928
0.232 22.42208
0.236 22.50632
0.24 22.592
0.244 22.67912
0.248 22.76768
0.252 22.85768
0.256 22.94912
0.26 23.042
0.264 23.13632
0.268 23.23208
0.272 23.32928
0.276 23.42792
0.28 23.528
0.284 23.62952
0.288 23.73248
0.292 23.83688
0.296 23.94272
0.3 24.05
0.304 24.15872
0.308 24.26888
0.312 24.38048
0.316 24.49352
0.32 24.608
0.324 24.72392
0.328 24.84128
0.332 24.96008
0.336 25.08032
0.34 25.202
0.344 25.32512
0.348 25.44968
0.352 25.57568
0.356 25.70312
0.36 25.832
0.364 25.96232
0.368 26.09408
0.372 26.22728
0.376 26.36192
0.38 26.498
0.384 26.63552
0.388 26.77448
0.392 26.91488
0.396 27.05672
0.4 27.2
0.404 27.34472
0.408 27.49088
0.412 27.63848
0.416 27.78752
0.42 27.938
0.424 28.08992
0.428 28.24328
0.432 28.39808
0.436 28.55432
0.44 28.712
0.444 28.87112
0.448 29.03168
0.452 29.19368
0.456 29.35712
0.46 29.522
0.464 29.68832
0.468 29.85608
0.472 30.02528
0.476 30.19592
0.48 30.368
0.484 30.54152
0.488 30.71648
0.492 30.89288
0.496 31.07072
0.5 31.25
0.504 31.43072
0.508 31.61288
0.512 31.79648
0.516 31.98152
0.52 32.168
0.524 32.35592
0.528 32.54528
0.532 32.73608
0.536 32.92832
0.54 33.122
0.544 33.31712
0.548 33.51368
0.552 33.71168
0.556 33.91112
0.56 34.112
0.564 34.31432
0.568 34.51808
0.572 34.72328
0.576 34.92992
0.58 35.138
0.584 35.34752
0.588 35.55848
0.592 35.77088
0.596 35.98472
0.6 36.2
0.604 36.41672
0.608 36.63488
0.612 36.85448
0.616 37.07552
0.62 37.298
0.624 37.52192
0.628 37.74728
0.632 37.97408
0.636 38.20232
0.64 38.432
0.644 38.66312
0.648 38.89568
0.652 39.12968
0.656 39.36512
0.66 39.602
0.664 39.84032
0.668 40.08008
0.672 40.32128
0.676 40.56392
0.68 40.808
0.684 41.05352
0.688 41.30048
0.692 41.54888
0.696 41.79872
0.7 42.05
0.704 42.30272
0.708 42.55688
0.712 42.81248
0.716 43.06952
0.72 43.328
0.724 43.58792
0.728 43.84928
0.732 44.11208
0.736 44.37632
0.74 44.642
0.744 44.90912
0.748 45.17768
0.752 45.44768
0.756 45.71912
0.76 45.992
0.764 46.26632
0.768 46.54208
0.772 46.81928
0.776 47.09792
0.78 47.378
0.784 47.65952
0.788 47.94248
0.792 48.22688
0.796 48.51272
0.8 48.8
0.804 49.08872
0.808 49.37888
0.812 49.67048
0.816 49.96352
0.82 50.258
0.824 50.55392
0.828 50.85128
0.832 51.15008
0.836 51.45032
0.84 51.752
0.844 52.05512
0.848 52.35968
0.852 52.66568
0.856 52.97312
0.86 53.282
0.864 53.59232
0.868 53.90408
0.872 54.21728
0.876 54.53192
0.88 54.848
0.884 55.16552
0.888 55.48448
0.892 55.80488
0.896 56.12672
0.9 56.45
0.904 56.77472
0.908 57.10088
0.912 57.42848
0.916 57.75752
0.92 58.088
0.924 58.41992
0.928 58.75328
0.932 59.08808
0.936 59.42432
0.94 59.762
0.944 60.10112
0.948 60.44168
0.952 60.78368
0.956 61.12712
0.96 61.472
0.964 61.81832
0.968 62.16608
0.972 62.51528
0.976 62.86592
0.98 63.218
0.984 63.57152
0.988 63.92648
0.992 64.28288
0.996 64.64072
1 65
1.004 65.36072
1.008 65.72288
1.012 66.08648
1.016 66.45152
1.02 66.818
1.024 67.18592
1.028 67.55528
1.032 67.92608
1.036 68.29832
1.04 68.672
1.044 69.04712
1.048 69.42368
1.052 69.80168
1.056 70.18112
1.06 70.562
1.064 70.94432
1.068 71.32808
1.072 71.71328
1.076 72.09992
1.08 72.488
1.084 72.87752
1.088 73.26848
1.092 73.66088
1.096 74.05472
1.1 74.45
1.104 74.84672
1.108 75.24488
1.112 75.64448
1.116 76.04552
1.12 76.448
1.124 76.85192
1.128 77.25728
1.132 77.66408
1.136 78.07232
1.14 78.482
1.144 78.89312
1.148 79.30568
1.152 79.71968
1.156 80.13512
1.16 80.552
1.164 80.97032
1.168 81.39008
1.172 81.81128
1.176 82.23392
1.18 82.658
1.184 83.08352
1.188 83.51048
1.192 83.93888
1.196 84.36872
1.2 84.8
1.204 85.23272
1.208 85.66688
1.212 86.10248
1.216 86.53952
1.22 86.978
1.224 87.41792
1.228 87.85928
1.232 88.30208
1.236 88.74632
1.24 89.192
1.244 89.63912
1.248 90.08768
1.252 90.53768
1.256 90.98912
1.26 91.442
1.264 91.89632
1.268 92.35208
1.272 92.80928
1.276 93.26792
1.28 93.728
1.284 94.18952
1.288 94.65248
1.292 95.11688
1.296 95.58272
1.3 96.05
1.304 96.51872
1.308 96.98888
1.312 97.46048
1.316 97.93352
1.32 98.408
1.324 98.88392
1.328 99.36128
1.332 99.84008
1.336 100.32032
1.34 100.802
1.344 101.28512
1.348 101.76968
1.352 102.25568
1.356 102.74312
1.36 103.232
1.364 103.72232
1.368 104.21408
1.372 104.70728
1.376 105.20192
1.38 105.698
1.384 106.19552
1.388 106.69448
1.392 107.19488
1.396 107.69672
1.4 108.2
1.404 108.70472
1.408 109.21088
1.412 109.71848
1.416 110.22752
1.42 110.738
1.424 111.24992
1.428 111.76328
1.432 112.27808
1.436 112.79432
1.44 113.312
1.444 113.83112
1.448 114.35168
1.452 114.87368
1.456 115.39712
1.46 115.922
1.464 116.44832
1.468 116.97608
1.472 117.50528
1.476 118.03592
1.48 118.568
1.484 119.10152
1.488 119.63648
1.492 120.17288
1.496 120.71072
1.5 121.25
1.504 121.79072
};
\addplot [very thick, mycolor3]
table {%
0 20
0.004 20.00061250678
0.008 20.0024500446779
0.012 20.0055126663684
0.016 20.009800459647
0.02 20.015313547436
0.024 20.0220520877937
0.028 20.0300162739253
0.032 20.0392063341971
0.036 20.0496225321529
0.04 20.0612651665323
0.044 20.0741345712931
0.048 20.0882311156343
0.052 20.1035552040231
0.056 20.1201072762241
0.06 20.1378878073302
0.064 20.1568973077975
0.068 20.1771363234812
0.072 20.1986054356755
0.076 20.2213052611549
0.08 20.2452364522187
0.084 20.2703996967381
0.088 20.2967957182053
0.092 20.3244252757858
0.096 20.3532891643733
0.1 20.3833882146463
0.104 20.4147232931287
0.108 20.4472953022519
0.112 20.48110518042
0.116 20.5161539020775
0.12 20.5524424777795
0.124 20.5899719542654
0.128 20.6287434145336
0.132 20.6687579779207
0.136 20.7100168001822
0.14 20.7525210735763
0.144 20.79627202695
0.148 20.8412709258291
0.152 20.887519072509
0.156 20.9350178061501
0.16 20.9837685028751
0.164 21.0337725758687
0.168 21.085031475481
0.172 21.1375466893327
0.176 21.1913197424244
0.18 21.2463521972471
0.184 21.3026456538969
0.188 21.3602017501919
0.192 21.4190221617921
0.196 21.4791086023222
0.2 21.5404628234972
0.204 21.6030866152512
0.208 21.666981805869
0.212 21.7321502621202
0.216 21.7985938893972
0.22 21.8663146318553
0.224 21.9353144725565
0.228 22.0055954336157
0.232 22.0771595763506
0.236 22.1500090014342
0.24 22.2241458490505
0.244 22.2995722990537
0.248 22.3762905711296
0.252 22.4543029249614
0.256 22.5336116603978
0.26 22.6142191176246
0.264 22.6961276773392
0.268 22.7793397609295
0.272 22.8638578306541
0.276 22.9496843898281
0.28 23.0368219830103
0.284 23.1252731961949
0.288 23.2150406570061
0.292 23.3061270348965
0.296 23.3985350413484
0.3 23.4922674300792
0.304 23.5873269972497
0.308 23.6837165816764
0.312 23.7814390650473
0.316 23.8804973721408
0.32 23.980894471049
0.324 24.0826333734042
0.328 24.1857171346089
0.332 24.2901488540702
0.336 24.3959316754373
0.34 24.5030687868428
0.344 24.6115634211488
0.348 24.7214188561954
0.352 24.8326384150544
0.356 24.9452254662858
0.36 25.0591834241997
0.364 25.1745157491204
0.368 25.2912259476564
0.372 25.409317572973
0.376 25.52879422507
0.38 25.6496595510633
0.384 25.7719172454705
0.388 25.8955710505009
0.392 26.0206247563498
0.396 26.1470822014977
0.4 26.2749472730125
0.404 26.4042239068578
0.408 26.5349160882046
0.412 26.6670278517478
0.416 26.8005632820274
0.42 26.9355265137542
0.424 27.0719217321405
0.428 27.2097531732349
0.432 27.3490251242626
0.436 27.48974192397
0.44 27.6319079629744
0.444 27.7755276841188
0.448 27.9206055828314
0.452 28.0671462074905
0.456 28.2151541597938
0.46 28.3646340951341
0.464 28.5155907229789
0.468 28.6680288072563
0.472 28.8219531667458
0.476 28.9773686754743
0.48 29.134280263118
0.484 29.29269291541
0.488 29.4526116745526
0.492 29.6140416396365
0.496 29.7769879670645
0.5 29.941455870982
0.504 30.107450623713
0.508 30.2749775562018
0.512 30.4440420584611
0.516 30.6146495800261
0.52 30.7868056304147
0.524 30.9605157795937
0.528 31.1357856584522
0.532 31.3126209592799
0.536 31.4910274362537
0.54 31.6710109059291
0.544 31.8525772477396
0.548 32.0357324045022
0.552 32.2204823829293
0.556 32.4068332541489
0.56 32.5947911542298
0.564 32.7843622847158
0.568 32.9755529131659
0.572 33.1683693737018
0.576 33.3628180675634
0.58 33.558905463671
0.584 33.7566380991958
0.588 33.9560225801371
0.592 34.1570655819083
0.596 34.3597738499297
0.6 34.5641542002301
0.604 34.7702135200561
0.608 34.9779587684893
0.612 35.1873969770721
0.616 35.398535250442
0.62 35.6113807669738
0.624 35.8259407794313
0.628 36.0422226156268
0.632 36.2602336790897
0.636 36.4799814497448
0.64 36.7014734845986
0.644 36.9247174184354
0.648 37.1497209645224
0.652 37.3764919153252
0.656 37.6050381432311
0.66 37.8353676012841
0.664 38.0674883239283
0.668 38.3014084277621
0.672 38.5371361123022
0.676 38.7746796607579
0.68 39.014047440816
0.684 39.255247905436
0.688 39.4982895936565
0.692 39.7431811314116
0.696 39.9899312323594
0.7 40.238548698721
0.704 40.4890424221308
0.708 40.7414213844988
0.712 40.9956946588842
0.716 41.2518714103811
0.72 41.5099608970155
0.724 41.769972470656
0.728 42.0319155779351
0.732 42.2957997611842
0.736 42.5616346593812
0.74 42.8294300091107
0.744 43.0991956455371
0.748 43.3709415033921
0.752 43.6446776179745
0.756 43.9204141261642
0.76 44.1981612674504
0.764 44.4779293849733
0.768 44.7597289265808
0.772 45.0435704458991
0.776 45.3294646034185
0.78 45.6174221675943
0.784 45.9074540159626
0.788 46.1995711362714
0.792 46.4937846276279
0.796 46.7901057016617
0.8 47.0885456837036
0.804 47.3891160139813
0.808 47.6918282488319
0.812 47.9966940619306
0.816 48.3037252455379
0.82 48.6129337117627
0.824 48.9243314938443
0.828 49.2379307474522
0.832 49.5537437520036
0.836 49.8717829120004
0.84 50.1920607583844
0.844 50.5145899499118
0.848 50.8393832745476
0.852 51.1664536508791
0.856 51.4958141295502
0.86 51.8274778947159
0.864 52.1614582655171
0.868 52.4977686975774
0.872 52.8364227845203
0.876 53.1774342595092
0.88 53.5208169968088
0.884 53.8665850133691
0.888 54.2147524704329
0.892 54.5653336751656
0.896 54.918343082309
0.9 55.2737952958596
0.904 55.6317050707704
0.908 55.9920873146784
0.912 56.3549570896565
0.916 56.7203296139913
0.92 57.0882202639877
0.924 57.4586445757987
0.928 57.8316182472831
0.932 58.20715713989
0.936 58.5852772805718
0.94 58.9659948637249
0.944 59.3493262531592
0.948 59.7352879840977
0.952 60.1238967652046
0.956 60.515169480645
0.96 60.9091231921742
0.964 61.3057751412594
0.968 61.7051427512323
0.972 62.107243629475
0.976 62.5120955696383
0.98 62.9197165538937
0.984 63.3301247552197
0.988 63.7433385397223
0.992 64.1593764689916
0.996 64.5782573024938
1 65
1.004 65.424623724052
1.008 65.8521478424662
1.012 66.2825919308761
1.016 66.7159757753138
1.02 67.1523193748314
1.024 67.5916429441636
1.028 68.0339669164306
1.032 68.4793119458846
1.036 68.9276989106974
1.04 69.3791489157933
1.044 69.8336832957252
1.048 70.2913236175967
1.052 70.7520916840298
1.056 71.2160095361796
1.06 71.683099456797
1.064 72.1533839733396
1.068 72.6268858611325
1.072 73.1036281465791
1.076 73.5836341104238
1.08 74.0669272910665
1.084 74.5535314879305
1.088 75.0434707648851
1.092 75.5367694537228
1.096 76.0334521576931
1.1 76.533543755094
1.104 77.0370694029214
1.108 77.5440545405786
1.112 78.0545248936457
1.116 78.5685064777118
1.12 79.0860256022697
1.124 79.6071088746743
1.128 80.1317832041678
1.132 80.6600758059702
1.136 81.1920142054383
1.14 81.7276262422946
1.144 82.2669400749256
1.148 82.8099841847533
1.152 83.3567873806794
1.156 83.9073788036042
1.16 84.4617879310227
1.164 85.0200445816975
1.168 85.5821789204114
1.172 86.1482214628013
1.176 86.7182030802738
1.18 87.2921550050057
1.184 87.8701088350297
1.188 88.452096539408
1.192 89.0381504634953
1.196 89.6283033342922
1.2 90.2225882658922
1.204 90.8210387650232
1.208 91.4236887366859
1.212 92.0305724898902
1.216 92.6417247434932
1.22 93.2571806321393
1.224 93.8769757123056
1.228 94.5011459684536
1.232 95.129727819291
1.236 95.7627581241443
1.24 96.4002741894451
1.244 97.0423137753336
1.248 97.6889151023791
1.252 98.3401168584227
1.256 98.9959582055435
1.26 99.6564787871503
1.264 100.321718735203
1.268 100.991718677564
1.272 101.666519745488
1.276 102.346163581244
1.28 103.030692345878
1.284 103.720148727126
1.288 104.414575947464
1.292 105.114017772309
1.296 105.818518518384
1.3 106.528123062219
1.304 107.242876848832
1.308 107.962825900556
1.312 108.688016826044
1.316 109.418496829439
1.32 110.154313719719
1.324 110.895515920218
1.328 111.642152478332
1.332 112.394273075407
1.336 113.151928036822
1.34 113.915168342259
1.344 114.684045636177
1.348 115.458612238488
1.352 116.238921155438
1.356 117.025026090704
1.36 117.816981456706
1.364 118.61484238614
1.368 119.418664743742
1.372 120.228505138277
1.376 121.044420934779
1.38 121.866470267015
1.384 122.69471205022
1.388 123.529205994069
1.392 124.370012615915
1.396 125.217193254303
1.4 126.07081008274
1.404 126.930926123761
1.408 127.797605263269
1.412 128.670912265177
1.416 129.550912786344
1.42 130.437673391821
1.424 131.331261570409
1.428 132.231745750544
1.432 133.139195316507
1.436 134.053680624975
1.44 134.975273021916
1.444 135.904044859842
1.448 136.840069515421
1.452 137.783421407466
1.456 138.734176015303
1.46 139.692409897527
1.464 140.658200711168
1.468 141.631627231258
1.472 142.612769370823
1.476 143.601708201309
1.48 144.598525973448
1.484 145.603306138577
1.488 146.616133370425
1.492 147.637093587376
1.496 148.666273975219
1.5 149.703763010408
1.504 150.749650483827
};
\end{axis}

\end{tikzpicture}

%% file: current_varying_load.tex
\begin{tikzpicture}

\definecolor{cornflowerblue86180233}{RGB}{86,180,233}
\definecolor{darkcyan0114178}{RGB}{0,114,178}
\definecolor{darkcyan0158115}{RGB}{0,158,115}
\definecolor{darkslategray38}{RGB}{38,38,38}
\definecolor{lightgray204}{RGB}{204,204,204}
\definecolor{orange2301590}{RGB}{230,159,0}

\begin{axis}[height=0.1\textheight,width=0.8\columnwidth,
scale only axis,
axis line style={darkslategray38},
legend cell align={left},
legend style={
  fill opacity=0.8,
  draw opacity=1,
  text opacity=1,
  at={(0.03,0.97)},
  anchor=north west,
  draw=lightgray204,
},
legend columns=4, 
legend to name={mylegendIC},
tick align=outside,
x grid style={lightgray204},
xlabel=\textcolor{darkslategray38}{Times (years)},
xmajorticks=true,
xmin=-1.4, xmax=51.4,
xtick style={color=darkslategray38},
y grid style={lightgray204},
ylabel=\textcolor{darkslategray38}{$I_c/I_z$},
ymajorticks=true,
ymin=0.28, ymax=0.72,
ytick style={color=darkslategray38}
]
\addplot [semithick, orange2301590, mark=Mercedes star, mark size=3.5, mark options={solid,rotate=270}, only marks]
table {%
1 0.5
2 0.5
3 0.5
4 0.5
5 0.5
6 0.5
7 0.5
8 0.5
9 0.5
10 0.5
11 0.5
12 0.5
13 0.5
14 0.5
15 0.5
16 0.5
17 0.5
18 0.5
19 0.5
20 0.5
21 0.5
22 0.5
23 0.5
24 0.5
25 0.5
26 0.5
27 0.5
28 0.5
29 0.5
30 0.5
31 0.5
32 0.5
33 0.5
34 0.5
35 0.5
36 0.5
37 0.5
38 0.5
39 0.5
40 0.5
41 0.5
42 0.5
43 0.5
44 0.5
45 0.5
46 0.5
47 0.5
48 0.5
49 0.5
};
\addlegendentry{$I_{c1}$}
\addplot [semithick, cornflowerblue86180233, mark=Mercedes star, mark size=3.5, mark options={solid}, only marks]
table {%
1 0.308163265306122
2 0.316326530612245
3 0.324489795918367
4 0.33265306122449
5 0.340816326530612
6 0.348979591836735
7 0.357142857142857
8 0.36530612244898
9 0.373469387755102
10 0.381632653061224
11 0.389795918367347
12 0.397959183673469
13 0.406122448979592
14 0.414285714285714
15 0.422448979591837
16 0.430612244897959
17 0.438775510204082
18 0.446938775510204
19 0.455102040816327
20 0.463265306122449
21 0.471428571428571
22 0.479591836734694
23 0.487755102040816
24 0.495918367346939
25 0.504081632653061
26 0.512244897959184
27 0.520408163265306
28 0.528571428571429
29 0.536734693877551
30 0.544897959183673
31 0.553061224489796
32 0.561224489795918
33 0.569387755102041
34 0.577551020408163
35 0.585714285714286
36 0.593877551020408
37 0.602040816326531
38 0.610204081632653
39 0.618367346938775
40 0.626530612244898
41 0.63469387755102
42 0.642857142857143
43 0.651020408163265
44 0.659183673469388
45 0.66734693877551
46 0.675510204081633
47 0.683673469387755
48 0.691836734693878
49 0.7
};
\addlegendentry{$I_{c2}$}
\addplot [semithick, darkcyan0158115, mark=Mercedes star, mark size=3.5, mark options={solid,rotate=90}, only marks]
table {%
1 0.3
2 0.3
3 0.3
4 0.3
5 0.3
6 0.3
7 0.3
8 0.3
9 0.3
10 0.3
11 0.3
12 0.3
13 0.3
14 0.3
15 0.3
16 0.3
17 0.3
18 0.3
19 0.3
20 0.3
21 0.3
22 0.3
23 0.3
24 0.3
25 0.7
26 0.7
27 0.7
28 0.7
29 0.7
30 0.7
31 0.7
32 0.7
33 0.7
34 0.7
35 0.7
36 0.7
37 0.7
38 0.7
39 0.7
40 0.7
41 0.7
42 0.7
43 0.7
44 0.7
45 0.7
46 0.7
47 0.7
48 0.7
49 0.7
};
\addlegendentry{$I_{c3}$}
\addplot [semithick, darkcyan0114178, mark=Mercedes star flipped, mark size=3.5, mark options={solid}, only marks]
table {%
1 0.303265028461264
2 0.303980720746762
3 0.30485137399371
4 0.305909612677309
5 0.307194483984837
6 0.308752508374452
7 0.310638797430746
8 0.31291818587938
9 0.315666289118706
10 0.318970349271027
11 0.322929670359547
12 0.327655368137339
13 0.333269078597569
14 0.339900195647874
15 0.347681168808847
16 0.356740425960195
17 0.36719264594643
18 0.379126444576567
19 0.392590086600393
20 0.407576568547998
21 0.424010207548955
22 0.441737477509682
23 0.460524935955019
24 0.480066401075009
25 0.5
26 0.519933598924991
27 0.539475064044981
28 0.558262522490318
29 0.575989792451045
30 0.592423431452002
31 0.607409913399607
32 0.620873555423433
33 0.63280735405357
34 0.643259574039805
35 0.652318831191153
36 0.660099804352126
37 0.666730921402431
38 0.672344631862661
39 0.677070329640453
40 0.681029650728973
41 0.684333710881294
42 0.68708181412062
43 0.689361202569254
44 0.691247491625548
45 0.692805516015163
46 0.694090387322691
47 0.69514862600629
48 0.696019279253238
49 0.696734971538736
};
\addlegendentry{$I_{c4}$}
\end{axis}

\end{tikzpicture}

%% file: ageing_varying_load.tex
\begin{tikzpicture}

\definecolor{cornflowerblue86180233}{RGB}{86,180,233}
\definecolor{darkcyan0114178}{RGB}{0,114,178}
\definecolor{darkcyan0158115}{RGB}{0,158,115}
\definecolor{darkslategray38}{RGB}{38,38,38}
\definecolor{lightgray204}{RGB}{204,204,204}
\definecolor{orange2301590}{RGB}{230,159,0}

\begin{axis}[height=0.1\textheight,width=0.8\columnwidth,
scale only axis,
axis line style={darkslategray38},
legend cell align={left},
legend style={
  fill opacity=0.8,
  draw opacity=1,
  text opacity=1,
  at={(0.03,0.97)},
  anchor=north west,
  draw=lightgray204
},
tick align=outside,
x grid style={lightgray204},
xlabel=\textcolor{darkslategray38}{Time $t$ (years) },
xmajorticks=true,
xmin=-1.4, xmax=51.4,
xtick style={color=darkslategray38},
y grid style={lightgray204},
ylabel=\textcolor{darkslategray38}{$r(T(t))$},
ymajorticks=true,
ymin=1.86144802889484, ymax=18.659201891066,
ytick style={color=darkslategray38}
]
\addplot [semithick, orange2301590, mark=Mercedes star, mark size=3.5, mark options={solid,rotate=270}, only marks]
table {%
1 5.65685424949238
2 5.65685424949238
3 5.65685424949238
4 5.65685424949238
5 5.65685424949238
6 5.65685424949238
7 5.65685424949238
8 5.65685424949238
9 5.65685424949238
10 5.65685424949238
11 5.65685424949238
12 5.65685424949238
13 5.65685424949238
14 5.65685424949238
15 5.65685424949238
16 5.65685424949238
17 5.65685424949238
18 5.65685424949238
19 5.65685424949238
20 5.65685424949238
21 5.65685424949238
22 5.65685424949238
23 5.65685424949238
24 5.65685424949238
25 5.65685424949238
26 5.65685424949238
27 5.65685424949238
28 5.65685424949238
29 5.65685424949238
30 5.65685424949238
31 5.65685424949238
32 5.65685424949238
33 5.65685424949238
34 5.65685424949238
35 5.65685424949238
36 5.65685424949238
37 5.65685424949238
38 5.65685424949238
39 5.65685424949238
40 5.65685424949238
41 5.65685424949238
42 5.65685424949238
43 5.65685424949238
44 5.65685424949238
45 5.65685424949238
46 5.65685424949238
47 5.65685424949238
48 5.65685424949238
49 5.65685424949238
};
\addplot [semithick, cornflowerblue86180233, mark=Mercedes star, mark size=3.5, mark options={solid}, only marks]
table {%
1 2.68826989254483
2 2.75484466598857
3 2.82487426170682
4 2.89853724714104
5 2.97602385166876
6 3.05753676398188
7 3.14329199091518
8 3.23351978271026
9 3.32846563013672
10 3.42839133936877
11 3.5335761910369
12 3.64431819044251
13 3.76093541654613
14 3.88376747801967
15 4.0131770853978
16 4.14955174917793
17 4.29330561461037
18 4.44488144489717
19 4.60475276558864
20 4.77342618414043
21 4.95144389988042
22 5.13938642104736
23 5.33787550711196
24 5.54757735629309
25 5.76920606004997
26 6.00352734838459
27 6.25136265204577
28 6.51359351020776
29 6.79116635492717
30 7.0850977066864
31 7.3964798186401
32 7.72648681082508
33 8.07638133960866
34 8.44752185207624
35 8.84137047993862
36 9.25950163292407
37 9.70361135756071
38 10.1755275338159
39 10.6772209893043
40 11.2108176187821
41 11.778611605497
42 12.3830798507495
43 13.0268977288481
44 13.7129562966301
45 14.4443810999837
46 15.224552734511
47 16.0571293337591
48 16.9460711765038
49 17.8956676246037
};
\addplot [semithick, darkcyan0158115, mark=Mercedes star, mark size=3.5, mark options={solid,rotate=90}, only marks]
table {%
1 2.62498229535717
2 2.62498229535717
3 2.62498229535717
4 2.62498229535717
5 2.62498229535717
6 2.62498229535717
7 2.62498229535717
8 2.62498229535717
9 2.62498229535717
10 2.62498229535717
11 2.62498229535717
12 2.62498229535717
13 2.62498229535717
14 2.62498229535717
15 2.62498229535717
16 2.62498229535717
17 2.62498229535717
18 2.62498229535717
19 2.62498229535717
20 2.62498229535717
21 2.62498229535717
22 2.62498229535717
23 2.62498229535717
24 2.62498229535717
25 17.8956676246037
26 17.8956676246037
27 17.8956676246037
28 17.8956676246037
29 17.8956676246037
30 17.8956676246037
31 17.8956676246037
32 17.8956676246037
33 17.8956676246037
34 17.8956676246037
35 17.8956676246037
36 17.8956676246037
37 17.8956676246037
38 17.8956676246037
39 17.8956676246037
40 17.8956676246037
41 17.8956676246037
42 17.8956676246037
43 17.8956676246037
44 17.8956676246037
45 17.8956676246037
46 17.8956676246037
47 17.8956676246037
48 17.8956676246037
49 17.8956676246037
};
\addplot [semithick, darkcyan0114178, mark=Mercedes star flipped, mark size=3.5, mark options={solid}, only marks]
table {%
1 2.64991102751998
2 2.65544325047936
3 2.66220652290823
4 2.67047632184423
5 2.68059044702091
6 2.69296339110641
7 2.70810410803203
8 2.72663798897778
9 2.74933402699593
10 2.77713833216878
11 2.8112153227242
12 2.85299800941124
13 2.90424871120004
14 2.9671311203975
15 3.04429360387991
16 3.13896158255018
17 3.25503322996185
18 3.39716693787745
19 3.57084046702937
20 3.78235040922793
21 4.03870780567462
22 4.34737518795293
23 4.71578907611217
24 5.15063045721369
25 5.65685424949238
26 6.23656965776453
27 6.88796113277243
28 7.60451615835373
29 8.37483130292708
30 9.1831688741224
31 10.0107479679912
32 10.8375440987927
33 11.6442299315637
34 12.4138737537116
35 13.1331171431189
36 13.7927211180772
37 14.3875296400569
38 14.9160023835798
39 15.3795030916379
40 15.7815109801819
41 16.1268762851305
42 16.4211897143071
43 16.6702928668915
44 16.8799275796052
45 17.0555059948788
46 17.2019767498384
47 17.3237624819445
48 17.4247468966848
49 17.5082939264169
};
\end{axis}

\end{tikzpicture}

%% file: life_varying_load.tex
\begin{tikzpicture}

\definecolor{cornflowerblue86180233}{RGB}{86,180,233}
\definecolor{darkcyan0114178}{RGB}{0,114,178}
\definecolor{darkcyan0158115}{RGB}{0,158,115}
\definecolor{darkslategray38}{RGB}{38,38,38}
\definecolor{lightgray204}{RGB}{204,204,204}
\definecolor{orange2301590}{RGB}{230,159,0}

\begin{axis}[height=0.1\textheight,width=0.8\columnwidth,
scale only axis,
axis line style={darkslategray38},
legend cell align={left},
legend style={
  fill opacity=0.8,
  draw opacity=1,
  text opacity=1,
  at={(0.03,0.97)},
  anchor=north west,
  draw=lightgray204
},
tick align=outside,
x grid style={lightgray204},
xlabel=\textcolor{darkslategray38}{Time $t$ (years)},
xmajorticks=true,
xmin=-1.4, xmax=51.4,
xtick style={color=darkslategray38},
y grid style={lightgray204},
ylabel=\textcolor{darkslategray38}{$A(10,t)$ (years)},
ymajorticks=true,
ymin=18.6144802889484, ymax=186.59201891066,
ytick style={color=darkslategray38}
]
\addplot [semithick, orange2301590, mark=Mercedes star, mark size=3.5, mark options={solid,rotate=270}, only marks]
table {%
1 56.5685424949238
2 56.5685424949238
3 56.5685424949238
4 56.5685424949238
5 56.5685424949238
6 56.5685424949238
7 56.5685424949238
8 56.5685424949238
9 56.5685424949238
10 56.5685424949238
11 56.5685424949238
12 56.5685424949238
13 56.5685424949238
14 56.5685424949238
15 56.5685424949238
16 56.5685424949238
17 56.5685424949238
18 56.5685424949238
19 56.5685424949238
20 56.5685424949238
21 56.5685424949238
22 56.5685424949238
23 56.5685424949238
24 56.5685424949238
25 56.5685424949238
26 56.5685424949238
27 56.5685424949238
28 56.5685424949238
29 56.5685424949238
30 56.5685424949238
31 56.5685424949238
32 56.5685424949238
33 56.5685424949238
34 56.5685424949238
35 56.5685424949238
36 56.5685424949238
37 56.5685424949238
38 56.5685424949238
39 56.5685424949238
40 56.5685424949238
41 56.5685424949238
42 56.5685424949238
43 56.5685424949238
44 56.5685424949238
45 56.5685424949238
46 56.5685424949238
47 56.5685424949238
48 56.5685424949238
49 56.5685424949238
};
\addplot [semithick, cornflowerblue86180233, mark=Mercedes star, mark size=3.5, mark options={solid}, only marks]
table {%
1 26.2498229535717
2 26.3131105507593
3 26.4429729213907
4 26.6428648877404
5 26.9164198395243
6 27.2674613958359
7 27.7000158644606
8 28.2183255600186
9 28.8268630473717
10 29.5303463821512
11 30.3337554261628
12 31.1790617246549
13 32.0685352491088
14 33.0045964039481
15 33.9898266348268
16 35.0269798685558
17 36.1189948537519
18 37.2690084774471
19 38.480370139634
20 39.7566572750859
21 41.1016921198576
22 42.5195598287011
23 44.0146280593059
24 45.5915681498718
25 47.2553780281452
26 49.0114070027973
27 50.865382602004
28 52.8234396394394
29 54.89215170475
30 57.0785652940885
31 59.3902368166345
32 61.8352727353942
33 64.4223731251719
34 67.1608789576686
35 70.0608234534517
36 73.1329878733404
37 76.3889621578799
38 79.8412108633948
39 83.503144887003
40 87.3891995213801
41 91.5149194334758
42 95.8970512203327
43 100.553644260257
44 105.504160649497
45 110.76959509405
46 116.372605714096
47 122.337656815682
48 128.691174791881
49 135.461718434569
};
\addplot [semithick, darkcyan0158115, mark=Mercedes star, mark size=3.5, mark options={solid,rotate=90}, only marks]
table {%
1 26.2498229535717
2 26.2498229535717
3 26.2498229535717
4 26.2498229535717
5 26.2498229535717
6 26.2498229535717
7 26.2498229535717
8 26.2498229535717
9 26.2498229535717
10 26.2498229535717
11 26.2498229535717
12 26.2498229535717
13 26.2498229535717
14 26.2498229535717
15 26.2498229535717
16 26.2498229535717
17 26.2498229535717
18 26.2498229535717
19 26.2498229535717
20 26.2498229535717
21 26.2498229535717
22 26.2498229535717
23 26.2498229535717
24 26.2498229535717
25 26.2498229535717
26 41.5205082828182
27 56.7911936120647
28 72.0618789413112
29 87.3325642705577
30 102.603249599804
31 117.873934929051
32 133.144620258297
33 148.415305587544
34 163.68599091679
35 178.956676246037
36 178.956676246037
37 178.956676246037
38 178.956676246037
39 178.956676246037
40 178.956676246037
41 178.956676246037
42 178.956676246037
43 178.956676246037
44 178.956676246037
45 178.956676246037
46 178.956676246037
47 178.956676246037
48 178.956676246037
49 178.956676246037
};
\addplot [semithick, darkcyan0114178, mark=Mercedes star flipped, mark size=3.5, mark options={solid}, only marks]
table {%
1 26.4538506861445
2 26.45837664505
3 26.4684348269149
4 26.4852562812087
5 26.5103475344385
6 26.545552912845
7 26.5931312353369
8 26.6558502747545
9 26.7371031951178
10 26.8410521534993
11 26.9728054170536
12 27.1341097122579
13 27.3316644711898
14 27.5737066594816
15 27.8703614580348
16 28.2340646148938
17 28.6800628063376
18 29.2269919282674
19 29.8975208771671
20 30.7190273172005
21 31.7242393942597
22 32.9517318772101
23 34.4461090557518
24 36.2576494206639
25 38.4411487574801
26 41.0537094030926
27 44.1513174783069
28 47.7842453811175
29 51.9915946015938
30 56.7955854374915
31 62.196403902386
32 68.1684440647025
33 74.6586129755423
34 81.5870538309939
35 88.8502971274918
36 96.3265600211183
37 103.882711481431
38 111.382279988715
39 118.693766213941
40 125.698438002652
41 132.296780108712
42 138.412908425851
43 143.996554041365
44 149.022616976693
45 153.488670802587
46 157.411059654347
47 160.820315286108
48 163.756548127996
49 166.265292641101
};
\end{axis}

\end{tikzpicture}

%% file: rhoT.tex
\begin{tikzpicture}

\definecolor{cornflowerblue86180233}{RGB}{86,180,233}
\definecolor{darkcyan0158115}{RGB}{0,158,115}
\definecolor{darkslategray38}{RGB}{38,38,38}
\definecolor{lightgray204}{RGB}{204,204,204}
\definecolor{orange2301590}{RGB}{230,159,0}

\begin{axis}[height=0.1\textheight,width=0.8\columnwidth,
scale only axis,
axis line style={darkslategray38},
legend cell align={left},
legend style={fill opacity=0.8, draw opacity=1, text opacity=1,  font=\footnotesize, draw=lightgray204},
legend columns=2, 
legend to name={mylegendload},
tick align=outside,
x grid style={lightgray204},
xlabel=\textcolor{darkslategray38}{Temperature (°C)},
xmajorticks=true,
xmin=17.1577775, xmax=81.6810275,
xtick style={color=darkslategray38},
y grid style={lightgray204},
ylabel=\textcolor{darkslategray38}{$\rho(T)$},
ymajorticks=true,
ymin=-0.014758074056548, ymax=0.309919555187507,
ytick style={color=darkslategray38}
]
\path [draw=cornflowerblue86180233,pattern=north east lines, pattern color=cornflowerblue86180233, opacity=1, line width=0pt]
(axis cs:66.6262775,0)
--(axis cs:66.6262775,0)
--(axis cs:71.3175275,0)
--(axis cs:71.3175275,0)
--(axis cs:76.2337775,0)
--(axis cs:76.2337775,0)
--(axis cs:78.7481525,0)
--(axis cs:78.7481525,0.00260262628652549)
--(axis cs:78.7481525,0.00260262628652549)
--(axis cs:76.2337775,0.00260262628652549)
--(axis cs:76.2337775,0.00496865018336685)
--(axis cs:71.3175275,0.00496865018336685)
--(axis cs:71.3175275,0.00638826452147166)
--(axis cs:66.6262775,0.00638826452147166)
--(axis cs:66.6262775,0)
--cycle;
\addplot [thin, orange2301590, const plot mark mid]
table {%
20.0906525 0
20.3944025 0
20.9231525 0.166922985922158
21.6769025 0.108955400449545
22.6556525 0.143499349343428
23.8594025 0.295161481130959
25.2881525 0.191529634449308
26.9419025 0.0582041878622974
28.8206525 0.0317047202176742
30.9244025 0.00402224062463031
33.2531525 0
35.8069025 0
38.5856525 0
41.5894025 0
44.8181525 0
48.2719025 0
51.9506525 0
55.8544025 0
59.9831525 0
64.3369025 0
68.9156525 0
73.7194025 0
78.7481525 0
};
\addlegendentry{Without DER}
\addplot [thin, cornflowerblue86180233, const plot mark mid]
table {%
20.0906525 0.141724831420797
20.3944025 0.198272802555306
20.9231525 0.134271856145747
21.6769025 0.109901810008281
22.6556525 0.109192002839229
23.8594025 0.0623447296817698
25.2881525 0.036673370401041
26.9419025 0.0262628652549391
28.8206525 0.0218857210457826
30.9244025 0.0195196971489412
33.2531525 0.0137229386016799
35.8069025 0.0134863362119957
38.5856525 0.0128948302377854
41.5894025 0.0126582278481013
44.8181525 0.0124216254584171
48.2719025 0.0125399266532592
51.9506525 0.0130131314326275
55.8544025 0.0125399266532592
59.9831525 0.0108837099254702
64.3369025 0.0095823967822075
68.9156525 0.00638826452147166
73.7194025 0.00496865018336685
78.7481525 0.00260262628652549
};
\addlegendentry{PV-dominated}
\addplot [thin, darkcyan0158115, const plot mark mid]
table {%
20.0906525 0.140541819472377
20.3944025 0.0996096060570211
20.9231525 0.0778421862060807
21.6769025 0.0695611025671359
22.6556525 0.0675499822548208
23.8594025 0.0573760794984029
25.2881525 0.0569028747190347
26.9419025 0.0490949958594582
28.8206525 0.0490949958594582
30.9244025 0.0490949958594582
33.2531525 0.0463740683780906
35.8069025 0.0408139122205134
38.5856525 0.0409322134153555
41.5894025 0.0403407074411452
44.8181525 0.0402224062463031
48.2719025 0.0367916715958831
51.9506525 0.0249615521116763
55.8544025 0.00544185496273512
59.9831525 0.0033124334555779
64.3369025 0.00236602389684136
68.9156525 0.000473204779368272
73.7194025 0.000354903584526204
78.7481525 0.000473204779368272
};
\addlegendentry{Wind-dominated}
\addlegendimage{area legend, draw=cornflowerblue86180233, pattern=north east lines, pattern color=cornflowerblue86180233, opacity=1, line width=0pt}
\addlegendentry{Cumulative ($T>T_{\mathrm{max}}$)}
\end{axis}

\end{tikzpicture}

%% file: Agecontribution.tex
\begin{tikzpicture}

\definecolor{cornflowerblue86180233}{RGB}{86,180,233}
\definecolor{darkcyan0158115}{RGB}{0,158,115}
\definecolor{darkslategray38}{RGB}{38,38,38}
\definecolor{lightgray204}{RGB}{204,204,204}
\definecolor{orange2301590}{RGB}{230,159,0}

\begin{axis}[height=0.1\textheight,width=0.8\columnwidth,
scale only axis,
axis line style={darkslategray38},
legend cell align={left},
legend style={
  fill opacity=0.8,
  draw opacity=1,
  text opacity=1,
  at={(0.03,0.97)},
  anchor=north west,
  draw=lightgray204
},
tick align=outside,
x grid style={lightgray204},
xlabel=\textcolor{darkslategray38}{Temperature (°C)},
xmajorticks=true,
xmin=17.1577775, xmax=81.6810275,
xtick style={color=darkslategray38},
y grid style={lightgray204},
ylabel=\textcolor{darkslategray38}{$\rho({T})\cdot r(T)$},
ymajorticks=true,
ymin=-0.130208526027382, ymax=2.73437904657503,
ytick style={color=darkslategray38}
]
\path [draw=cornflowerblue86180233,pattern=north east lines, pattern color=cornflowerblue86180233, opacity=1, line width=0pt]
(axis cs:66.6262775,0)
--(axis cs:66.6262775,0)
--(axis cs:71.3175275,0)
--(axis cs:71.3175275,0)
--(axis cs:76.2337775,0)
--(axis cs:76.2337775,0)
--(axis cs:78.7481525,0)
--(axis cs:78.7481525,2.33203907313008)
--(axis cs:78.7481525,2.33203907313008)
--(axis cs:76.2337775,2.33203907313008)
--(axis cs:76.2337775,2.60417052054765)
--(axis cs:71.3175275,2.60417052054765)
--(axis cs:71.3175275,2.00604717983737)
--(axis cs:66.6262775,2.00604717983737)
--(axis cs:66.6262775,0)
--cycle;
\addlegendimage{area legend, draw=cornflowerblue86180233, pattern=north east lines, pattern color=cornflowerblue86180233, opacity=1, line width=0pt}

\addplot [semithick, orange2301590, const plot mark mid]
table {%
20.0906525 0
20.3944025 0
20.9231525 0.313928685771744
21.6769025 0.222060605547658
22.6556525 0.324639126354985
23.8594025 0.759204982523421
25.2881525 0.57372537313602
26.9419025 0.207975074641855
28.8206525 0.138417253270839
30.9244025 0.0219767701410244
33.2531525 0
35.8069025 0
38.5856525 0
41.5894025 0
44.8181525 0
48.2719025 0
51.9506525 0
55.8544025 0
59.9831525 0
64.3369025 0
68.9156525 0
73.7194025 0
78.7481525 0
};
\addplot [semithick, cornflowerblue86180233, const plot mark mid]
table {%
20.0906525 0.243896636206147
20.3944025 0.352444201723044
20.9231525 0.252522365946796
21.6769025 0.223989470742426
22.6556525 0.247025485264345
23.8594025 0.160361132581099
25.2881525 0.109854765702388
26.9419025 0.0938424117286419
28.8206525 0.0955492233399446
30.9244025 0.106651972743206
33.2531525 0.096115343640006
35.8069025 0.124025115224135
38.5856525 0.15948564860049
41.5894025 0.215669872163521
44.8181525 0.298624794295796
48.2719025 0.435706019707145
51.9506525 0.669347706369518
55.8544025 0.978040025271025
59.9831525 1.31841011981941
64.3369025 1.84663016756885
68.9156525 2.00604717983737
73.7194025 2.60417052054765
78.7481525 2.33203907313008
};
\addplot [semithick, darkcyan0158115, const plot mark mid]
table {%
20.0906525 0.241860771129301
20.3944025 0.177063256474226
20.9231525 0.146396226249332
21.6769025 0.141771591815443
22.6556525 0.152818582975018
23.8594025 0.147580928466477
25.2881525 0.170452071944673
26.9419025 0.175426130033272
28.8206525 0.21434014965447
30.9244025 0.268245870838974
33.2531525 0.32480357505933
35.8069025 0.375339164494093
38.5856525 0.506257196474951
41.5894025 0.687321742128605
44.8181525 0.966975524386388
48.2719025 1.27834502008417
51.9506525 1.28393060039971
55.8544025 0.42443246379686
59.9831525 0.40125525385808
64.3369025 0.455958066066383
68.9156525 0.148596087395361
73.7194025 0.186012180039118
78.7481525 0.424007104205468
};
\end{axis}

\end{tikzpicture}

%% file: rPW.tex
\begin{tikzpicture}

\definecolor{cornflowerblue86180233}{RGB}{86,180,233}
\definecolor{darkcyan0158115}{RGB}{0,158,115}
\definecolor{darkslategray38}{RGB}{38,38,38}
\definecolor{lightgray204}{RGB}{204,204,204}

\begin{axis}[height=0.11\textheight,width=0.8\columnwidth,
scale only axis,
axis line style={darkslategray38},
legend cell align={left},
legend style={
  fill opacity=0.8,
  draw opacity=1,
  text opacity=1,
  at={(0.03,0.97)},
  anchor=north west,
  draw=lightgray204,
},
tick align=outside,
x grid style={lightgray204},
xlabel=\textcolor{darkslategray38}{Times $t$ (years)},
xmajorticks=true,
xmin=-2.45, xmax=51.45,
xtick style={color=darkslategray38},
y grid style={lightgray204},
ylabel=\textcolor{darkslategray38}{$\overline{r}(t)$},
ymajorticks=true,
ymin=1.83316413932658, ymax=18.9154023142278,
ytick style={color=darkslategray38}
]
\addplot [semithick, cornflowerblue86180233, mark=Mercedes star, mark size=3.5, mark options={solid,rotate=270}, only marks]
table {%
0 2.66536424599881
1 2.688449002607
2 2.71655228017725
3 2.75074045065531
4 2.79229458790591
5 2.8427479681214
6 2.90392732239114
7 2.97799673183001
8 3.06750207869543
9 3.17541257276031
10 3.30515397343074
11 3.46062567795397
12 3.6461908978865
13 3.86662594582498
14 4.12701176643777
15 4.43254928995242
16 4.78828152776408
17 5.19871160427977
18 5.66731921982368
19 6.19599959904416
20 6.78447773484278
21 7.42978176800286
22 8.12588314895133
23 8.86361523050051
24 9.63095509083529
25 10.4136929726242
26 11.1964308544131
27 11.9637707147479
28 12.701502796297
29 13.3976041772455
30 14.0429082104056
31 14.6313863462042
32 15.1600667254247
33 15.6286743409686
34 16.0391044174843
35 16.394836655296
36 16.7003741788106
37 16.9607599994234
38 17.1811950473619
39 17.3667602672944
40 17.5222319718176
41 17.6519733724881
42 17.7598838665529
43 17.8493892134184
44 17.9234586228572
45 17.984637977127
46 18.0350913573425
47 18.0766454945931
48 18.1108336650711
49 18.1389369426414
};
\addlegendentry{PV}
\addplot [semithick, darkcyan0158115, mark=Mercedes star, mark size=3.5, mark options={solid}, only marks]
table {%
0 2.609629510913
1 2.62047516813988
2 2.63367862185837
3 2.64974087386789
4 2.6692637970084
5 2.69296775356476
6 2.72171097650765
7 2.75651019225882
8 2.79856150467038
9 2.84925990490372
10 2.9102148798116
11 2.98325844030295
12 3.0704405060608
13 3.17400507948515
14 3.29633928491078
15 3.4398866181677
16 3.60701638181825
17 3.7998442292608
18 4.02000498981624
19 4.26838907603049
20 4.54486728434403
21 4.84804337883647
22 5.17508503238583
23 5.52168557999215
24 5.88219642457097
25 6.24994155216599
26 6.617686679761
27 6.97819752433983
28 7.32479807194615
29 7.65183972549551
30 7.95501581998794
31 8.23149402830149
32 8.47987811451574
33 8.70003887507118
34 8.89286672251374
35 9.05999648616427
36 9.20354381942119
37 9.32587802484682
38 9.42944259827118
39 9.51662466402903
40 9.58966822452037
41 9.65062319942827
42 9.7013215996616
43 9.74337291207315
44 9.77817212782433
45 9.80691535076722
46 9.83061930732358
47 9.85014223046409
48 9.8662044824736
49 9.87940793619211
};
\addlegendentry{Wind}
\end{axis}

\end{tikzpicture}

%% file: effAge.tex
\begin{tikzpicture}

\definecolor{cornflowerblue86180233}{RGB}{86,180,233}
\definecolor{darkcyan0158115}{RGB}{0,158,115}
\definecolor{darkslategray38}{RGB}{38,38,38}
\definecolor{lightgray204}{RGB}{204,204,204}

\begin{axis}[height=0.11\textheight,width=0.8\columnwidth,
scale only axis,
axis line style={darkslategray38},
legend cell align={left},
legend style={
  fill opacity=0.8,
  draw opacity=1,
  text opacity=1,
  at={(0.03,0.97)},
  anchor=north west,
  draw=lightgray204,
  font=\footnotesize
},
tick align=outside,
x grid style={lightgray204},
xlabel=\textcolor{darkslategray38}{Times $t$ (years)},
xmajorticks=true,
xmin=-1.4, xmax=51.4,
xtick style={color=darkslategray38},
y grid style={lightgray204},
ylabel=\textcolor{darkslategray38}{$\overline{A}(a,t)$},
ymajorticks=true,
ymin=4.78650964186513, ymax=186.542543721262,
ytick style={color=darkslategray38}
]
\addplot [semithick, cornflowerblue86180233, mark=*, mark size=2.5, mark options={solid}, only marks]
table {%
1 13.326821229994
2 13.3499059866022
3 13.4010940207807
4 13.4864702254372
5 13.6134005673443
6 13.7907842894669
7 14.006262609251
8 14.2677070609038
9 14.5844686889439
10 14.9675866737983
11 15.4299926791076
12 15.9866910346705
13 16.654885200727
14 17.4540090678565
15 18.405608261534
16 19.5330035780556
17 20.8606594278657
18 22.413180134259
19 24.2138734082577
20 26.2828612408641
21 28.6347896857545
22 31.2762899259932
23 34.2034614706648
24 37.3997574813416
25 40.8347129731328
26 44.4639282109142
27 48.2305772973244
28 52.068464863121
29 55.9063524289175
30 59.6730015153277
31 63.3022167531091
32 66.7371722449003
33 69.9334682555771
34 72.8606398002486
35 75.5021400404874
36 77.8540684853778
37 79.9230563179842
38 81.7237495919829
39 83.2762702983761
40 84.6039261481862
41 85.7313214647079
42 86.6829206583854
43 87.4820445255149
44 88.1502386915714
45 88.7069370471342
46 89.1693430524436
47 89.552461037298
48 89.8692226653381
49 90.1306671169908
};
\addlegendentry{$a=5$ years, PV}
\addplot [semithick, cornflowerblue86180233, mark=x, mark size=2.5, mark options={solid}, only marks]
table {%
1 26.6536424599881
2 26.6767272165963
3 26.7279152507747
4 26.8132914554312
5 26.9402217973383
6 27.1176055194609
7 27.3561685958532
8 27.6688010816844
9 28.0709389143811
10 28.5809872411426
11 29.2207769685745
12 29.9929536439215
13 30.9225922616307
14 32.0384777568004
15 33.3731949353322
16 34.9629962571633
17 36.8473504625362
18 39.068065334986
19 41.6678824761142
20 44.6884695023981
21 48.1677932638101
22 52.136949353859
23 56.6166416049238
24 61.6136308895993
25 67.1175742139969
26 73.0987178966686
27 79.5068672233177
28 86.2719263337857
29 93.3061099102591
30 100.50771448846
31 107.766144964023
32 114.967749542225
33 122.001933118698
34 128.766992229166
35 135.175141555815
36 141.156285238487
37 146.660228562884
38 151.65721784756
39 156.136910098625
40 160.106066188674
41 163.585389950086
42 166.60597697637
43 169.205794117498
44 171.426508989948
45 173.31086319532
46 174.900664517151
47 176.235381695683
48 177.351267190853
49 178.280905808562
};
\addlegendentry{$a=10$ years, PV}
\addplot [semithick, darkcyan0158115, mark=*, mark size=2.5, mark options={solid}, only marks]
table {%
1 13.048147554565
2 13.0589932117919
3 13.0830423227372
4 13.1231536856921
5 13.1827879717875
6 13.2661262144393
7 13.3673620228071
8 13.4901935932075
9 13.63901422401
10 13.8190103319053
11 14.0362574581522
12 14.2978049219475
13 14.6117352357494
14 14.9871788105642
15 15.4342581905713
16 15.9639299289274
17 16.5876878704427
18 17.3170915936427
19 18.1630915039738
20 19.1351412950935
21 20.2401219612698
22 21.481148958288
23 22.8563897614131
24 24.358070351589
25 25.9718777001295
26 27.6769519679514
27 29.4465952688759
28 31.2497077608299
29 33.0528202527839
30 34.8224635537085
31 36.5275378215304
32 38.1413451700709
33 39.6430257602468
34 41.0182665633719
35 42.2592935603901
36 43.3642742265664
37 44.3363240176861
38 45.1823239280172
39 45.9117276512172
40 46.5354855927325
41 47.0651573310886
42 47.5122367110957
43 47.8876802859105
44 48.2016105997124
45 48.4631580635077
46 48.6804051897546
47 48.8604012976499
48 49.0092219284524
49 49.1320534988528
};
\addlegendentry{$a=5$ years, Wind}
\addplot [semithick, darkcyan0158115, mark=x, mark size=2.5, mark options={solid}, only marks]
table {%
1 26.09629510913
2 26.1071407663569
3 26.1311898773022
4 26.1713012402571
5 26.2309355263525
6 26.3142737690043
7 26.4263552345989
8 26.5732359159448
9 26.7621679097021
10 27.0017983036929
11 27.3023836725915
12 27.6651669447545
13 28.101928828957
14 28.6261930345742
15 29.2532685224766
16 30.0001873870796
17 30.8854927923902
18 31.9288268293921
19 33.150270314538
20 34.5693994856648
21 36.2040518901972
22 38.0688368287307
23 40.1734813550557
24 42.5211618555627
25 45.1070189952229
26 47.9170739292212
27 50.927744227164
28 54.106097522243
29 57.4108906043729
30 60.7943412538379
31 64.2044897894818
32 67.5879404389469
33 70.8927335210768
34 74.0710868161558
35 77.0817571140986
36 79.8918120480969
37 82.477669187757
38 84.825349688264
39 86.9299942145891
40 88.7947791531226
41 90.429431557655
42 91.8485607287818
43 93.0700042139277
44 94.1133382509296
45 94.9986436562402
46 95.7455625208432
47 96.3726380087455
48 96.8969022143628
49 97.3336640985653
};
\addlegendentry{$a=10$ years, Wind}
\end{axis}

\end{tikzpicture}

%% file: effAge_limits.tex
\begin{tikzpicture}

\definecolor{cornflowerblue86180233}{RGB}{86,180,233}
\definecolor{darkcyan0158115}{RGB}{0,158,115}
\definecolor{darkslategray38}{RGB}{38,38,38}
\definecolor{lightgray204}{RGB}{204,204,204}
\definecolor{orange2301590}{RGB}{230,159,0}

\begin{axis}[height=0.13\textheight,width=0.8\columnwidth,
scale only axis,
axis line style={darkslategray38},
legend cell align={left},
legend style={
  fill opacity=0.8,
  draw opacity=1,
  text opacity=1,
  at={(0.03,0.97)},
  anchor=north west,
  draw=lightgray204,
  font=\footnotesize
},
legend columns=3, 
tick align=outside,
x grid style={lightgray204},
xlabel=\textcolor{darkslategray38}{Age $a$ (years)},
xmajorticks=true,
xmin=-2.45, xmax=51.45,
xtick style={color=darkslategray38},
y grid style={lightgray204},
ylabel=\textcolor{darkslategray38}{$\lim\limits_{t \to \pm\infty} \overline{A}(a,t)$},
ymajorticks=true,
ymin=-24.7398691480971, ymax=700,
ytick style={color=darkslategray38}
]
\addplot [semithick, orange2301590, mark=Mercedes star, mark size=3.5, mark options={solid,rotate=90}, only marks]
table {%
0 0
1 2.66536424599881
2 5.33072849199761
3 7.99609273799642
4 10.6614569839952
5 13.326821229994
6 15.9921854759928
7 18.6575497219916
8 21.3229139679904
9 23.9882782139893
10 26.6536424599881
11 29.3190067059869
12 31.9843709519857
13 34.6497351979845
14 37.3150994439833
15 39.9804636899821
16 42.6458279359809
17 45.3111921819797
18 47.9765564279785
19 50.6419206739773
20 53.3072849199761
21 55.9726491659749
22 58.6380134119737
23 61.3033776579725
24 63.9687419039713
25 66.6341061499702
26 69.299470395969
27 71.9648346419678
28 74.6301988879666
29 77.2955631339654
30 79.9609273799642
31 82.626291625963
32 85.2916558719618
33 87.9570201179606
34 90.6223843639594
35 93.2877486099582
36 95.953112855957
37 98.6184771019558
38 101.283841347955
39 103.949205593953
40 106.614569839952
41 109.279934085951
42 111.94529833195
43 114.610662577949
44 117.276026823947
45 119.941391069946
46 122.606755315945
47 125.272119561944
48 127.937483807943
49 130.602848053941
};
\addlegendentry{ $\overline{A}_0(a,t)$}
\addplot[semithick, cornflowerblue86180233, mark=Mercedes star, mark size=3.5, mark options={solid,rotate=270}, only marks]
table {%
0 0
1 18.1108336650711
2 36.1874791596642
3 54.2225705170066
4 72.2072084941336
5 90.1306671169908
6 107.980056330409
7 125.739940196962
8 143.39191356945
9 160.914145541268
10 178.280905808562
11 195.462100855924
12 212.422860855348
13 229.123235034158
14 245.518071689454
15 261.557176106938
16 277.185850447907
17 292.345917173332
18 306.977303519536
19 321.020211729941
20 334.417815907187
21 347.119318703484
22 359.083089418232
23 370.279520272645
24 380.693213245269
25 390.324168336104
26 399.187783566605
27 407.313666715556
28 414.743448483559
29 421.527926218402
30 427.723925817446
31 433.39124503727
32 438.58995664155
33 443.378238169314
34 447.810787459266
35 451.937799225704
36 455.804425171529
37 459.450616069415
38 462.911241747369
39 466.2163957208
40 469.39180829356
41 472.459310372256
42 475.437307104086
43 478.341234426477
44 481.183982394598
45 483.976276982504
46 486.72701743316
47 489.443569713337
48 492.132018715944
49 494.797382961943
};
\addlegendentry{ $\overline{A}_1(a,t)$, PV}
\addplot [semithick, darkcyan0158115, mark=Mercedes star, mark size=3.5, mark options={solid}, only marks]
table {%
0 0
1 9.8662044824736
2 19.7163467129377
3 29.5469660202613
4 39.3538813710285
5 49.1320534988528
6 58.875426410926
7 68.5767480105876
8 78.2273712100158
9 87.8170394345362
10 97.3336640985653
11 106.763106696836
12 116.088984721683
13 125.292528541104
14 134.352525027269
15 143.245391749782
16 151.945430624854
17 160.425308739369
18 168.656802767671
19 176.611818587659
20 184.263658313154
21 191.5884563851
22 198.56665390944
23 205.184340589201
24 211.434282141367
25 217.316478565938
26 222.83816414593
27 228.013249178316
28 232.861292557153
29 237.406159841497
30 241.674548917527
31 245.694553907343
32 249.494398136604
33 253.101414518423
34 256.54130113659
35 259.837640421501
36 263.011645500986
37 266.082086007047
38 269.06534444735
39 271.975559327162
40 274.824819232065
41 277.623380736736
42 280.379890928994
43 283.101601905502
44 285.794569659067
45 288.463833456075
46 291.113574329943
47 293.747252951801
48 296.367728119941
49 298.977357630854
};
\addlegendentry{ $\overline{A}_1(a,t)$, Wind}
\end{axis}

\end{tikzpicture}

%% file: failure_XLPE.tex
\begin{tikzpicture}

\definecolor{cornflowerblue86180233}{RGB}{86,180,233}
\definecolor{darkcyan0158115}{RGB}{0,158,115}
\definecolor{darkslategray38}{RGB}{38,38,38}
\definecolor{lightgray204}{RGB}{204,204,204}
\definecolor{orange2301590}{RGB}{230,159,0}

\begin{axis}[height=0.1\textheight,width=0.8\columnwidth, scale only axis,
axis line style={darkslategray38},
legend cell align={left},
legend pos=north west, 
  legend style={
  legend columns=3,
    fill opacity=0.8, draw opacity=1, text opacity=1,
    font=\footnotesize, draw=lightgray204
  },
log basis y={10},
tick align=outside,
x grid style={lightgray204},
xlabel=\textcolor{darkslategray38}{Age (years)},
xmajorticks=true,
xmin=0, xmax=70,
xtick style={color=darkslategray38},
y grid style={lightgray204},
ylabel=\textcolor{darkslategray38}{Failure rate (per year)},
ymajorticks=true,
ymin=0.000001, ymax=10000,
ymode=log,
ytick style={color=darkslategray38},
ytick={1e-06,0.0001,0.01,1,100,10000},
yticklabels={
  ${10^{-6}}$,
  ${10^{-4}}$,
  ${10^{-2}}$,
  ${10^{0}}$,
  ${10^{2}}$,
 ${10^{4}}$
}
]
\addplot [very thick, orange2301590,dashed]
table {%
0 0
1 8.21760949244733e-08
2 3.28704379697893e-07
3 7.39584854320259e-07
4 1.31481751879157e-06
5 2.05440237311183e-06
6 2.95833941728104e-06
7 4.02662865129919e-06
8 5.25927007516629e-06
9 6.65626368888234e-06
10 8.21760949244733e-06
11 9.94330748586127e-06
12 1.18333576691242e-05
13 1.3887760042236e-05
14 1.61065146051968e-05
15 1.84896213580065e-05
16 2.10370803006652e-05
17 2.37488914331728e-05
18 2.66250547555293e-05
19 2.96655702677349e-05
20 3.28704379697893e-05
21 3.62396578616927e-05
22 3.97732299434451e-05
23 4.34711542150464e-05
24 4.73334306764966e-05
25 5.13600593277958e-05
26 5.55510401689439e-05
27 5.9906373199941e-05
28 6.4426058420787e-05
29 6.9110095831482e-05
30 7.39584854320259e-05
31 7.89712272224188e-05
32 8.41483212026606e-05
33 8.94897673727514e-05
34 9.49955657326911e-05
35 0.00010066571628248
36 0.000106500219022117
37 0.000112499073951604
38 0.000118662281070939
39 0.000124989840380124
40 0.000131481751879157
41 0.00013813801556804
42 0.000144958631446771
43 0.000151943599515351
44 0.00015909291977378
45 0.000166406592222058
46 0.000173884616860185
47 0.000181526993688161
48 0.000189333722705986
49 0.00019730480391366
50 0.000205440237311183
51 0.000213740022898555
52 0.000222204160675776
53 0.000230832650642845
54 0.000239625492799764
55 0.000248582687146532
56 0.000257704233683148
57 0.000266990132409614
58 0.000276440383325928
59 0.000286054986432091
60 0.000295833941728104
61 0.000305777249213965
62 0.000315884908889675
63 0.000326156920755234
64 0.000336593284810643
65 0.0003471940010559
66 0.000357959069491006
67 0.000368888490115961
68 0.000379982262930764
69 0.000391240387935417
70 0.000402662865129919
71 0.00041424969451427
72 0.000426000876088469
73 0.000437916409852518
74 0.000449996295806416
75 0.000462240533950162
76 0.000474649124283758
77 0.000487222066807202
78 0.000499959361520495
79 0.000512861008423638
80 0.000525927007516629
81 0.000539157358799469
82 0.000552552062272158
83 0.000566111117934697
84 0.000579834525787083
85 0.00059372228582932
86 0.000607774398061404
87 0.000621990862483338
88 0.000636371679095121
89 0.000650916847896753
90 0.000665626368888234
91 0.000680500242069563
92 0.000695538467440742
93 0.000710741045001769
94 0.000726107974752646
95 0.000741639256693371
96 0.000757334890823946
97 0.000773194877144369
98 0.000789219215654641
99 0.000805407906354763
100 0.000821760949244733
101 0.000838278344324552
102 0.00085496009159422
103 0.000871806191053737
104 0.000888816642703103
105 0.000905991446542318
106 0.000923330602571382
107 0.000940834110790295
108 0.000958501971199056
109 0.000976334183797667
110 0.000994330748586127
111 0.00101249166556444
112 0.00103081693473259
113 0.0010493065560906
114 0.00106796052963845
115 0.00108677885537616
116 0.00110576153330371
117 0.00112490856342111
118 0.00114421994572837
119 0.00116369568022547
120 0.00118333576691242
121 0.00120314020578921
122 0.00122310899685586
123 0.00124324214011236
124 0.0012635396355587
125 0.0012840014831949
126 0.00130462768302094
127 0.00132541823503683
128 0.00134637313924257
129 0.00136749239563816
130 0.0013887760042236
131 0.00141022396499889
132 0.00143183627796402
133 0.00145361294311901
134 0.00147555396046384
135 0.00149765932999853
136 0.00151992905172306
137 0.00154236312563744
138 0.00156496155174167
139 0.00158772433003575
140 0.00161065146051968
141 0.00163374294319345
142 0.00165699877805708
143 0.00168041896511055
144 0.00170400350435388
145 0.00172775239578705
146 0.00175166563941007
147 0.00177574323522294
148 0.00179998518322566
149 0.00182439148341823
150 0.00184896213580065
151 0.00187369714037292
152 0.00189859649713503
153 0.00192366020608699
154 0.00194888826722881
155 0.00197428068056047
156 0.00199983744608198
157 0.00202555856379334
158 0.00205144403369455
159 0.00207749385578561
160 0.00210370803006652
161 0.00213008655653727
162 0.00215662943519788
163 0.00218333666604833
164 0.00221020824908863
165 0.00223724418431879
166 0.00226444447173879
167 0.00229180911134864
168 0.00231933810314833
169 0.00234703144713788
170 0.00237488914331728
171 0.00240291119168652
172 0.00243109759224562
173 0.00245944834499456
174 0.00248796344993335
175 0.00251664290706199
176 0.00254548671638048
177 0.00257449487788882
178 0.00260366739158701
179 0.00263300425747505
180 0.00266250547555293
181 0.00269217104582067
182 0.00272200096827825
183 0.00275199524292569
184 0.00278215386976297
185 0.0028124768487901
186 0.00284296418000708
187 0.00287361586341391
188 0.00290443189901058
189 0.00293541228679711
190 0.00296655702677349
191 0.00299786611893971
192 0.00302933956329578
193 0.00306097735984171
194 0.00309277950857748
195 0.0031247460095031
196 0.00315687686261857
197 0.00318917206792388
198 0.00322163162541905
199 0.00325425553510407
200 0.00328704379697893
201 0.00331999641104364
202 0.00335311337729821
203 0.00338639469574262
204 0.00341984036637688
205 0.00345345038920099
206 0.00348722476421495
207 0.00352116349141876
208 0.00355526657081241
209 0.00358953400239592
210 0.00362396578616927
211 0.00365856192213248
212 0.00369332241028553
213 0.00372824725062843
214 0.00376333644316118
215 0.00379858998788378
216 0.00383400788479623
217 0.00386959013389852
218 0.00390533673519067
219 0.00394124768867266
220 0.00397732299434451
221 0.0040135626522062
222 0.00404996666225774
223 0.00408653502449913
224 0.00412326773893037
225 0.00416016480555146
226 0.0041972262243624
227 0.00423445199536318
228 0.00427184211855382
229 0.0043093965939343
230 0.00434711542150464
231 0.00438499860126482
232 0.00442304613321485
233 0.00446125801735473
234 0.00449963425368446
235 0.00453817484220404
236 0.00457687978291346
237 0.00461574907581274
238 0.00465478272090186
239 0.00469398071818084
240 0.00473334306764966
241 0.00477286976930833
242 0.00481256082315685
243 0.00485241622919522
244 0.00489243598742344
245 0.00493262009784151
246 0.00497296856044943
247 0.00501348137524719
248 0.00505415854223481
249 0.00509500006141227
250 0.00513600593277958
251 0.00517717615633674
252 0.00521851073208375
253 0.00526000966002061
254 0.00530167294014732
255 0.00534350057246388
256 0.00538549255697028
257 0.00542764889366654
258 0.00546996958255264
259 0.00551245462362859
260 0.00555510401689439
261 0.00559791776235004
262 0.00564089585999554
263 0.00568403830983089
264 0.00572734511185609
265 0.00577081626607114
266 0.00581445177247603
267 0.00585825163107078
268 0.00590221584185537
269 0.00594634440482981
270 0.0059906373199941
271 0.00603509458734824
272 0.00607971620689223
273 0.00612450217862607
274 0.00616945250254976
275 0.00621456717866329
276 0.00625984620696668
277 0.00630528958745991
278 0.00635089732014299
279 0.00639666940501592
280 0.00644260584207871
281 0.00648870663133133
282 0.00653497177277381
283 0.00658140126640614
284 0.00662799511222832
285 0.00667475331024034
286 0.00672167586044222
287 0.00676876276283394
288 0.00681601401741551
289 0.00686342962418693
290 0.0069110095831482
291 0.00695875389429932
292 0.00700666255764029
293 0.00705473557317111
294 0.00710297294089177
295 0.00715137466080229
296 0.00719994073290265
297 0.00724867115719286
298 0.00729756593367293
299 0.00734662506234283
300 0.0073958485432026
301 0.0074452363762522
302 0.00749478856149166
303 0.00754450509892097
304 0.00759438598854012
305 0.00764443123034913
306 0.00769464082434798
307 0.00774501477053668
308 0.00779555306891523
309 0.00784625571948363
310 0.00789712272224188
311 0.00794815407718998
312 0.00799934978432793
313 0.00805070984365572
314 0.00810223425517337
315 0.00815392301888086
316 0.0082057761347782
317 0.0082577936028654
318 0.00830997542314243
319 0.00836232159560933
320 0.00841483212026606
321 0.00846750699711265
322 0.00852034622614909
323 0.00857334980737537
324 0.00862651774079151
325 0.00867985002639749
326 0.00873334666419332
327 0.008787007654179
328 0.00884083299635453
329 0.00889482269071991
330 0.00894897673727514
331 0.00900329513602022
332 0.00905777788695514
333 0.00911242499007992
334 0.00916723644539454
335 0.00922221225289901
336 0.00927735241259334
337 0.0093326569244775
338 0.00938812578855152
339 0.00944375900481539
340 0.00949955657326911
341 0.00955551849391268
342 0.00961164476674609
343 0.00966793539176936
344 0.00972439036898247
345 0.00978100969838543
346 0.00983779337997824
347 0.00989474141376091
348 0.00995185379973341
349 0.0100091305378958
350 0.010066571628248
351 0.01012417707079
352 0.0101819468655219
353 0.0102398810124437
354 0.0102979795115553
355 0.0103562423628567
356 0.010414669566348
357 0.0104732611220292
358 0.0105320170299002
359 0.010590937289961
360 0.0106500219022117
361 0.0107092708666523
362 0.0107686841832827
363 0.0108282618521029
364 0.010888003873113
365 0.010947910246313
366 0.0110079809717027
367 0.0110682160492824
368 0.0111286154790519
369 0.0111891792610112
370 0.0112499073951604
371 0.0113107998814994
372 0.0113718567200283
373 0.011433077910747
374 0.0114944634536556
375 0.0115560133487541
376 0.0116177275960423
377 0.0116796061955205
378 0.0117416491471884
379 0.0118038564510463
380 0.0118662281070939
381 0.0119287641153315
382 0.0119914644757588
383 0.0120543291883761
384 0.0121173582531831
385 0.0121805516701801
386 0.0122439094393668
387 0.0123074315607434
388 0.0123711180343099
389 0.0124349688600662
390 0.0124989840380124
391 0.0125631635681484
392 0.0126275074504743
393 0.01269201568499
394 0.0127566882716955
395 0.0128215252105909
396 0.0128865265016762
397 0.0129516921449513
398 0.0130170221404163
399 0.0130825164880711
400 0.0131481751879157
401 0.0132139982399502
402 0.0132799856441746
403 0.0133461374005888
404 0.0134124535091928
405 0.0134789339699867
406 0.0135455787829705
407 0.0136123879481441
408 0.0136793614655075
409 0.0137464993350608
410 0.013813801556804
411 0.013881268130737
412 0.0139488990568598
413 0.0140166943351725
414 0.014084653965675
415 0.0141527779483674
416 0.0142210662832496
417 0.0142895189703217
418 0.0143581360095837
419 0.0144269174010355
420 0.0144958631446771
421 0.0145649732405086
422 0.0146342476885299
423 0.0147036864887411
424 0.0147732896411421
425 0.014843057145733
426 0.0149129890025137
427 0.0149830852114843
428 0.0150533457726447
429 0.015123770685995
430 0.0151943599515351
431 0.0152651135692651
432 0.0153360315391849
433 0.0154071138612946
434 0.0154783605355941
435 0.0155497715620835
436 0.0156213469407627
437 0.0156930866716317
438 0.0157649907546906
439 0.0158370591899394
440 0.015909291977378
441 0.0159816891170065
442 0.0160542506088248
443 0.016126976452833
444 0.016199866649031
445 0.0162729211974188
446 0.0163461400979965
447 0.0164195233507641
448 0.0164930709557215
449 0.0165667829128687
450 0.0166406592222058
451 0.0167146998837328
452 0.0167889048974496
453 0.0168632742633562
454 0.0169378079814527
455 0.0170125060517391
456 0.0170873684742153
457 0.0171623952488813
458 0.0172375863757372
459 0.017312941854783
460 0.0173884616860185
461 0.017464145869444
462 0.0175399944050593
463 0.0176160072928644
464 0.0176921845328594
465 0.0177685261250442
466 0.0178450320694189
467 0.0179217023659835
468 0.0179985370147378
469 0.0180755360156821
470 0.0181526993688161
471 0.0182300270741401
472 0.0183075191316539
473 0.0183851755413575
474 0.018462996303251
475 0.0185409814173343
476 0.0186191308836075
477 0.0186974447020705
478 0.0187759228727234
479 0.0188545653955661
480 0.0189333722705986
481 0.0190123434978211
482 0.0190914790772333
483 0.0191707790088354
484 0.0192502432926274
485 0.0193298719286092
486 0.0194096649167809
487 0.0194896222571424
488 0.0195697439496938
489 0.019650029994435
490 0.019730480391366
491 0.0198110951404869
492 0.0198918742417977
493 0.0199728176952983
494 0.0200539255009888
495 0.0201351976588691
496 0.0202166341689392
497 0.0202982350311992
498 0.0203800002456491
499 0.0204619298122888
};
\addlegendentry{$\frac{I_c}{I_z} = 0.9$}
\addplot [very thick, cornflowerblue86180233]
table {%
0 0
1 1.2839782587926e-05
2 5.1359130351704e-05
3 0.000115558043291334
4 0.000205436521406816
5 0.00032099456469815
6 0.000462232173165336
7 0.000629149346808375
8 0.000821746085627265
9 0.00104002238962201
10 0.0012839782587926
11 0.00155361369313905
12 0.00184892869266135
13 0.0021699232573595
14 0.0025165973872335
15 0.00288895108228335
16 0.00328698434250906
17 0.00371069716791062
18 0.00416008955848803
19 0.00463516151424129
20 0.00513591303517041
21 0.00566234412127537
22 0.00621445477255619
23 0.00679224498901286
24 0.00739571477064538
25 0.00802486411745376
26 0.00867969302943798
27 0.00936020150659806
28 0.010066389548934
29 0.0107982571564458
30 0.0115558043291334
31 0.0123390310669969
32 0.0131479373700362
33 0.0139825232382514
34 0.0148427886716425
35 0.0157287336702094
36 0.0166403582339521
37 0.0175776623628707
38 0.0185406460569652
39 0.0195293093162355
40 0.0205436521406816
41 0.0215836745303036
42 0.0226493764851015
43 0.0237407580050752
44 0.0248578190902248
45 0.0260005597405502
46 0.0271689799560514
47 0.0283630797367286
48 0.0295828590825815
49 0.0308283179936104
50 0.032099456469815
51 0.0333962745111956
52 0.0347187721177519
53 0.0360669492894842
54 0.0374408060263923
55 0.0388403423284762
56 0.040265558195736
57 0.0417164536281716
58 0.0431930286257831
59 0.0446952831885704
60 0.0462232173165336
61 0.0477768310096727
62 0.0493561242679876
63 0.0509610970914783
64 0.0525917494801449
65 0.0542480814339874
66 0.0559300929530057
67 0.0576377840371999
68 0.0593711546865699
69 0.0611302049011158
70 0.0629149346808374
71 0.064725344025735
72 0.0665614329358084
73 0.0684232014110577
74 0.0703106494514828
75 0.0722237770570838
76 0.0741625842278607
77 0.0761270709638133
78 0.0781172372649419
79 0.0801330831312462
80 0.0821746085627265
81 0.0842418135593826
82 0.0863346981212145
83 0.0884532622482223
84 0.090597505940406
85 0.0927674291977654
86 0.0949630320203008
87 0.097184314408012
88 0.099431276360899
89 0.101703917878962
90 0.104002238962201
91 0.106326239610615
92 0.108675919824206
93 0.111051279602972
94 0.113452318946914
95 0.115879037856032
96 0.118331436330326
97 0.120809514369796
98 0.123313271974441
99 0.125842709144263
100 0.12839782587926
101 0.130978622179433
102 0.133585098044782
103 0.136217253475307
104 0.138875088471008
105 0.141558603031884
106 0.144267797157937
107 0.147002670849165
108 0.149763224105569
109 0.152549456927149
110 0.155361369313905
111 0.158198961265836
112 0.161062232782944
113 0.163951183865227
114 0.166865814512686
115 0.169806124725322
116 0.172772114503132
117 0.175763783846119
118 0.178781132754282
119 0.18182416122762
120 0.184892869266135
121 0.187987256869825
122 0.191107324038691
123 0.194253070772733
124 0.19742449707195
125 0.200621602936344
126 0.203844388365913
127 0.207092853360659
128 0.21036699792058
129 0.213666822045677
130 0.21699232573595
131 0.220343508991398
132 0.223720371812023
133 0.227122914197823
134 0.230551136148799
135 0.234005037664952
136 0.23748461874628
137 0.240989879392783
138 0.244520819604463
139 0.248077439381318
140 0.25165973872335
141 0.255267717630557
142 0.25890137610294
143 0.262560714140499
144 0.266245731743234
145 0.269956428911144
146 0.273692805644231
147 0.277454861942493
148 0.281242597805931
149 0.285056013234545
150 0.288895108228335
151 0.292759882787301
152 0.296650336911443
153 0.30056647060076
154 0.304508283855253
155 0.308475776674922
156 0.312468949059767
157 0.316487801009788
158 0.320532332524985
159 0.324602543605357
160 0.328698434250906
161 0.33282000446163
162 0.33696725423753
163 0.341140183578606
164 0.345338792484858
165 0.349563080956286
166 0.353813048992889
167 0.358088696594669
168 0.362390023761624
169 0.366717030493755
170 0.371069716791062
171 0.375448082653545
172 0.379852128081203
173 0.384281853074038
174 0.388737257632048
175 0.393218341755234
176 0.397725105443596
177 0.402257548697134
178 0.406815671515848
179 0.411399473899737
180 0.416008955848803
181 0.420644117363044
182 0.425304958442461
183 0.429991479087054
184 0.434703679296823
185 0.439441559071768
186 0.444205118411888
187 0.448994357317185
188 0.453809275787657
189 0.458649873823305
190 0.463516151424129
191 0.468408108590129
192 0.473325745321305
193 0.478269061617656
194 0.483238057479183
195 0.488232732905887
196 0.493253087897766
197 0.498299122454821
198 0.503370836577051
199 0.508468230264458
200 0.51359130351704
201 0.518740056334799
202 0.523914488717733
203 0.529114600665843
204 0.534340392179129
205 0.539591863257591
206 0.544869013901228
207 0.550171844110042
208 0.555500353884031
209 0.560854543223196
210 0.566234412127537
211 0.571639960597054
212 0.577071188631747
213 0.582528096231615
214 0.58801068339666
215 0.59351895012688
216 0.599052896422276
217 0.604612522282848
218 0.610197827708596
219 0.615808812699519
220 0.621445477255619
221 0.627107821376894
222 0.632795845063346
223 0.638509548314973
224 0.644248931131776
225 0.650013993513754
226 0.655804735460909
227 0.66162115697324
228 0.667463258050746
229 0.673331038693428
230 0.679224498901286
231 0.68514363867432
232 0.69108845801253
233 0.697058956915915
234 0.703055135384477
235 0.709076993418214
236 0.715124531017127
237 0.721197748181216
238 0.727296644910481
239 0.733421221204922
240 0.739571477064538
241 0.745747412489331
242 0.751949027479299
243 0.758176322034443
244 0.764429296154763
245 0.770707949840259
246 0.777012283090931
247 0.783342295906778
248 0.789697988287801
249 0.796079360234001
250 0.802486411745376
251 0.808919142821927
252 0.815377553463653
253 0.821861643670556
254 0.828371413442635
255 0.834906862779889
256 0.841467991682319
257 0.848054800149925
258 0.854667288182707
259 0.861305455780665
260 0.867969302943798
261 0.874658829672108
262 0.881374035965593
263 0.888114921824254
264 0.894881487248092
265 0.901673732237104
266 0.908491656791293
267 0.915335260910658
268 0.922204544595198
269 0.929099507844914
270 0.936020150659806
271 0.942966473039874
272 0.949938474985118
273 0.956936156495538
274 0.963959517571133
275 0.971008558211905
276 0.978083278417852
277 0.985183678188975
278 0.992309757525274
279 0.999461516426749
280 1.0066389548934
281 1.01384207292523
282 1.02107087052223
283 1.02832534768441
284 1.03560550441176
285 1.04291134070429
286 1.050242856562
287 1.05760005198488
288 1.06498292697294
289 1.07239148152617
290 1.07982571564458
291 1.08728562932816
292 1.09477122257692
293 1.10228249539086
294 1.10981944776997
295 1.11738207971426
296 1.12497039122373
297 1.13258438229837
298 1.14022405293818
299 1.14788940314317
300 1.15558043291334
301 1.16329714224868
302 1.1710395311492
303 1.1788075996149
304 1.18660134764577
305 1.19442077524182
306 1.20226588240304
307 1.21013666912944
308 1.21803313542101
309 1.22595528127776
310 1.23390310669969
311 1.24187661168679
312 1.24987579623907
313 1.25790066035652
314 1.26595120403915
315 1.27402742728696
316 1.28212933009994
317 1.2902569124781
318 1.29841017442143
319 1.30658911592994
320 1.31479373700362
321 1.32302403764248
322 1.33128001784652
323 1.33956167761573
324 1.34786901695012
325 1.35620203584969
326 1.36456073431443
327 1.37294511234434
328 1.38135516993943
329 1.3897909070997
330 1.39825232382514
331 1.40673942011576
332 1.41525219597156
333 1.42379065139253
334 1.43235478637867
335 1.44094460093
336 1.4495600950465
337 1.45820126872817
338 1.46686812197502
339 1.47556065478705
340 1.48427886716425
341 1.49302275910662
342 1.50179233061418
343 1.51058758168691
344 1.51940851232481
345 1.52825512252789
346 1.53712741229615
347 1.54602538162958
348 1.55494903052819
349 1.56389835899198
350 1.57287336702094
351 1.58187405461507
352 1.59090042177438
353 1.59995246849887
354 1.60903019478854
355 1.61813360064338
356 1.62726268606339
357 1.63641745104858
358 1.64559789559895
359 1.65480401971449
360 1.66403582339521
361 1.67329330664111
362 1.68257646945218
363 1.69188531182842
364 1.70121983376984
365 1.71058003527644
366 1.71996591634822
367 1.72937747698517
368 1.73881471718729
369 1.74827763695459
370 1.75776623628707
371 1.76728051518472
372 1.77682047364755
373 1.78638611167556
374 1.79597742926874
375 1.8055944264271
376 1.81523710315063
377 1.82490545943934
378 1.83459949529322
379 1.84431921071228
380 1.85406460569652
381 1.86383568024593
382 1.87363243436052
383 1.88345486804028
384 1.89330298128522
385 1.90317677409533
386 1.91307624647062
387 1.92300139841109
388 1.93295222991673
389 1.94292874098755
390 1.95293093162355
391 1.96295880182472
392 1.97301235159106
393 1.98309158092258
394 1.99319648981928
395 2.00332707828116
396 2.01348334630821
397 2.02366529390043
398 2.03387292105783
399 2.04410622778041
400 2.05436521406816
401 2.06464987992109
402 2.07496022533919
403 2.08529625032248
404 2.09565795487093
405 2.10604533898456
406 2.11645840266337
407 2.12689714590736
408 2.13736156871652
409 2.14785167109085
410 2.15836745303036
411 2.16890891453505
412 2.17947605560491
413 2.19006887623995
414 2.20068737644017
415 2.21133155620556
416 2.22200141553612
417 2.23269695443187
418 2.24341817289278
419 2.25416507091888
420 2.26493764851015
421 2.27573590566659
422 2.28655984238822
423 2.29740945867501
424 2.30828475452699
425 2.31918572994414
426 2.33011238492646
427 2.34106471947396
428 2.35204273358664
429 2.36304642726449
430 2.37407580050752
431 2.38513085331572
432 2.3962115856891
433 2.40731799762766
434 2.41845008913139
435 2.4296078602003
436 2.44079131083438
437 2.45200044103364
438 2.46323525079808
439 2.47449574012769
440 2.48578190902248
441 2.49709375748244
442 2.50843128550758
443 2.51979449309789
444 2.53118338025338
445 2.54259794697405
446 2.55403819325989
447 2.56550411911091
448 2.5769957245271
449 2.58851300950847
450 2.60005597405502
451 2.61162461816674
452 2.62321894184364
453 2.63483894508571
454 2.64648462789296
455 2.65815599026538
456 2.66985303220298
457 2.68157575370576
458 2.69332415477371
459 2.70509823540684
460 2.71689799560514
461 2.72872343536862
462 2.74057455469728
463 2.75245135359111
464 2.76435383205012
465 2.7762819900743
466 2.78823582766366
467 2.8002153448182
468 2.81222054153791
469 2.82425141782279
470 2.83630797367286
471 2.84839020908809
472 2.86049812406851
473 2.8726317186141
474 2.88479099272486
475 2.89697594640081
476 2.90918657964192
477 2.92142289244822
478 2.93368488481969
479 2.94597255675633
480 2.95828590825815
481 2.97062493932515
482 2.98298964995732
483 2.99538004015467
484 3.0077961099172
485 3.0202378592449
486 3.03270528813777
487 3.04519839659582
488 3.05771718461905
489 3.07026165220746
490 3.08283179936104
491 3.09542762607979
492 3.10804913236372
493 3.12069631821283
494 3.13336918362711
495 3.14606772860657
496 3.15879195315121
497 3.17154185726102
498 3.184317440936
499 3.19711870417617
};
\addlegendentry{$\frac{I_c}{I_z} = 1.0$}
\addplot [very thick, darkcyan0158115]
table {%
0 0
1 0.0022548141416153
2 0.00901925656646122
3 0.0202933272745377
4 0.0360770262658449
5 0.0563703535403826
6 0.081173309098151
7 0.11048589293915
8 0.14430810506338
9 0.18263994547084
10 0.22548141416153
11 0.272832511135452
12 0.324693236392604
13 0.381063589932987
14 0.4419435717566
15 0.507333181863444
16 0.577232420253518
17 0.651641286926823
18 0.730559781883359
19 0.813987905123125
20 0.901925656646122
21 0.994373036452349
22 1.09133004454181
23 1.1927966809145
24 1.29877294557042
25 1.40925883850957
26 1.52425435973195
27 1.64375950923756
28 1.7677742870264
29 1.89629869309847
30 2.02933272745377
31 2.16687639009231
32 2.30892968101407
33 2.45549260021907
34 2.60656514770729
35 2.76214732347875
36 2.92223912753343
37 3.08684055987135
38 3.2559516204925
39 3.42957230939688
40 3.60770262658449
41 3.79034257205533
42 3.9774921458094
43 4.1691513478467
44 4.36532017816723
45 4.56599863677099
46 4.77118672365799
47 4.98088443882821
48 5.19509178228166
49 5.41380875401835
50 5.63703535403826
51 5.86477158234141
52 6.09701743892778
53 6.33377292379739
54 6.57503803695023
55 6.8208127783863
56 7.0710971481056
57 7.32589114610813
58 7.58519477239389
59 7.84900802696288
60 8.1173309098151
61 8.39016342095055
62 8.66750556036923
63 8.94935732807114
64 9.23571872405629
65 9.52658974832466
66 9.82197040087627
67 10.1218606817111
68 10.4262605908292
69 10.7351701282305
70 11.048589293915
71 11.3665180878827
72 11.6889565101337
73 12.015904560668
74 12.3473622394854
75 12.6833295465861
76 13.02380648197
77 13.3687930456371
78 13.7182892375875
79 14.0722950578211
80 14.430810506338
81 14.793835583138
82 15.1613702882213
83 15.5334146215878
84 15.9099685832376
85 16.2910321731706
86 16.6766053913868
87 17.0666882378862
88 17.4612807126689
89 17.8603828157348
90 18.263994547084
91 18.6721159067163
92 19.0847468946319
93 19.5018875108308
94 19.9235377553128
95 20.3496976280781
96 20.7803671291267
97 21.2155462584584
98 21.6552350160734
99 22.0994334019716
100 22.548141416153
101 23.0013590586177
102 23.4590863293656
103 23.9213232283968
104 24.3880697557111
105 24.8593259113087
106 25.3350916951896
107 25.8153671073536
108 26.3001521478009
109 26.7894468165314
110 27.2832511135452
111 27.7815650388422
112 28.2843885924224
113 28.7917217742858
114 29.3035645844325
115 29.8199170228624
116 30.3407790895755
117 30.8661507845719
118 31.3960321078515
119 31.9304230594143
120 32.4693236392604
121 33.0127338473897
122 33.5606536838022
123 34.1130831484979
124 34.6700222414769
125 35.2314709627391
126 35.7974293122846
127 36.3678972901133
128 36.9428748962252
129 37.5223621306203
130 38.1063589932987
131 38.6948654842602
132 39.2878816035051
133 39.8854073510331
134 40.4874427268444
135 41.0939877309389
136 41.7050423633167
137 42.3206066239777
138 42.9406805129219
139 43.5652640301493
140 44.19435717566
141 44.8279599494539
142 45.466072351531
143 46.1086943818914
144 46.755826040535
145 47.4074673274618
146 48.0636182426718
147 48.7242787861651
148 49.3894489579416
149 50.0591287580014
150 50.7333181863444
151 51.4120172429706
152 52.09522592788
153 52.7829442410727
154 53.4751721825486
155 54.1719097523077
156 54.8731569503501
157 55.5789137766756
158 56.2891802312845
159 57.0039563141765
160 57.7232420253518
161 58.4470373648103
162 59.1753423325521
163 59.908156928577
164 60.6454811528852
165 61.3873150054767
166 62.1336584863513
167 62.8845115955092
168 63.6398743329504
169 64.3997466986747
170 65.1641286926823
171 65.9330203149731
172 66.7064215655472
173 67.4843324444045
174 68.266752951545
175 69.0536830869687
176 69.8451228506757
177 70.6410722426659
178 71.4415312629393
179 72.246499911496
180 73.0559781883359
181 73.869966093459
182 74.6884636268654
183 75.5114707885549
184 76.3389875785278
185 77.1710139967838
186 78.0075500433231
187 78.8485957181456
188 79.6941510212514
189 80.5442159526403
190 81.3987905123125
191 82.2578747002679
192 83.1214685165066
193 83.9895719610285
194 84.8621850338336
195 85.739307734922
196 86.6209400642935
197 87.5070820219483
198 88.3977336078864
199 89.2928948221077
200 90.1925656646122
201 91.0967461353999
202 92.0054362344709
203 92.9186359618251
204 93.8363453174625
205 94.7585643013832
206 95.6852929135871
207 96.6165311540742
208 97.5522790228445
209 98.4925365198981
210 99.437303645235
211 100.386580398855
212 101.340366780758
213 102.298662790945
214 103.261468429415
215 104.228783696167
216 105.200608591204
217 106.176943114523
218 107.157787266126
219 108.143141046012
220 109.133004454181
221 110.127377490633
222 111.126260155369
223 112.129652448387
224 113.13755436969
225 114.149965919275
226 115.166887097143
227 116.188317903295
228 117.21425833773
229 118.244708400448
230 119.27966809145
231 120.319137410734
232 121.363116358302
233 122.411604934153
234 123.464603138288
235 124.522110970705
236 125.584128431406
237 126.65065552039
238 127.721692237657
239 128.797238583208
240 129.877294557042
241 130.961860159159
242 132.050935389559
243 133.144520248242
244 134.242614735209
245 135.345218850459
246 136.452332593992
247 137.563955965808
248 138.680088965908
249 139.800731594291
250 140.925883850957
251 142.055545735906
252 143.189717249138
253 144.328398390654
254 145.471589160453
255 146.619289558535
256 147.771499584901
257 148.928219239549
258 150.089448522481
259 151.255187433696
260 152.425435973195
261 153.600194140976
262 154.779461937041
263 155.963239361389
264 157.15152641402
265 158.344323094935
266 159.541629404132
267 160.743445341613
268 161.949770907378
269 163.160606101425
270 164.375950923756
271 165.59580537437
272 166.820169453267
273 168.049043160447
274 169.282426495911
275 170.520319459657
276 171.762722051687
277 173.009634272001
278 174.261056120597
279 175.516987597477
280 176.77742870264
281 178.042379436086
282 179.311839797816
283 180.585809787828
284 181.864289406124
285 183.147278652703
286 184.434777527565
287 185.726786030711
288 187.02330416214
289 188.324331921852
290 189.629869309847
291 190.939916326126
292 192.254472970687
293 193.573539243532
294 194.89711514466
295 196.225200674072
296 197.557795831767
297 198.894900617744
298 200.236515032006
299 201.58263907455
300 202.933272745377
301 204.288416044488
302 205.648068971882
303 207.01223152756
304 208.38090371152
305 209.754085523764
306 211.131776964291
307 212.513978033101
308 213.900688730194
309 215.291909055571
310 216.687639009231
311 218.087878591174
312 219.4926278014
313 220.90188663991
314 222.315655106703
315 223.733933201779
316 225.156720925138
317 226.58401827678
318 228.015825256706
319 229.452141864915
320 230.892968101407
321 232.338303966183
322 233.788149459241
323 235.242504580583
324 236.701369330208
325 238.164743708117
326 239.632627714308
327 241.105021348783
328 242.581924611541
329 244.063337502582
330 245.549260021907
331 247.039692169514
332 248.534633945405
333 250.034085349579
334 251.538046382037
335 253.046517042778
336 254.559497331801
337 256.076987249109
338 257.598986794699
339 259.125495968572
340 260.656514770729
341 262.192043201169
342 263.732081259893
343 265.276628946899
344 266.825686262189
345 268.379253205762
346 269.937329777618
347 271.499915977757
348 273.06701180618
349 274.638617262886
350 276.214732347875
351 277.795357061147
352 279.380491402703
353 280.970135372542
354 282.564288970663
355 284.162952197069
356 285.766125051757
357 287.373807534729
358 288.985999645984
359 290.602701385522
360 292.223912753343
361 293.849633749448
362 295.479864373836
363 297.114604626507
364 298.753854507461
365 300.397614016699
366 302.04588315422
367 303.698661920024
368 305.355950314111
369 307.017748336481
370 308.684055987135
371 310.354873266072
372 312.030200173292
373 313.710036708796
374 315.394382872582
375 317.083238664652
376 318.776604085005
377 320.474479133642
378 322.176863810561
379 323.883758115764
380 325.59516204925
381 327.311075611019
382 329.031498801072
383 330.756431619407
384 332.485874066026
385 334.219826140929
386 335.958287844114
387 337.701259175583
388 339.448740135334
389 341.20073072337
390 342.957230939688
391 344.718240784289
392 346.483760257174
393 348.253789358342
394 350.028328087793
395 351.807376445528
396 353.590934431546
397 355.379002045847
398 357.171579288431
399 358.968666159298
400 360.770262658449
401 362.576368785883
402 364.3869845416
403 366.2021099256
404 368.021744937884
405 369.84588957845
406 371.6745438473
407 373.507707744434
408 375.34538126985
409 377.18756442355
410 379.034257205533
411 380.885459615799
412 382.741171654348
413 384.601393321181
414 386.466124616297
415 388.335365539696
416 390.209116091378
417 392.087376271344
418 393.970146079593
419 395.857425516125
420 397.74921458094
421 399.645513274038
422 401.54632159542
423 403.451639545085
424 405.361467123033
425 407.275804329264
426 409.194651163779
427 411.118007626577
428 413.045873717658
429 414.978249437022
430 416.91513478467
431 418.856529760601
432 420.802434364815
433 422.752848597312
434 424.707772458092
435 426.667205947156
436 428.631149064503
437 430.599601810133
438 432.572564184046
439 434.550036186243
440 436.532017816723
441 438.518509075486
442 440.509509962532
443 442.505020477862
444 444.505040621475
445 446.509570393371
446 448.51860979355
447 450.532158822012
448 452.550217478758
449 454.572785763787
450 456.599863677099
451 458.631451218695
452 460.667548388573
453 462.708155186735
454 464.75327161318
455 466.802897667908
456 468.85703335092
457 470.915678662215
458 472.978833601793
459 475.046498169654
460 477.118672365799
461 479.195356190226
462 481.276549642937
463 483.362252723931
464 485.452465433209
465 487.547187770769
466 489.646419736613
467 491.75016133074
468 493.858412553151
469 495.971173403844
470 498.088443882821
471 500.210223990081
472 502.336513725624
473 504.467313089451
474 506.60262208156
475 508.742440701953
476 510.886768950629
477 513.035606827589
478 515.188954332831
479 517.346811466357
480 519.509178228166
481 521.676054618259
482 523.847440636634
483 526.023336283293
484 528.203741558235
485 530.38865646146
486 532.578080992968
487 534.77201515276
488 536.970458940835
489 539.173412357193
490 541.380875401835
491 543.592848074759
492 545.809330375967
493 548.030322305458
494 550.255823863233
495 552.48583504929
496 554.720355863631
497 556.959386306255
498 559.202926377162
499 561.450976076353
};
\addlegendentry{$\frac{I_c}{I_z} = 1.1$}
\end{axis}

\end{tikzpicture}

%% file: failure_PILC.tex
\begin{tikzpicture}

\definecolor{cornflowerblue86180233}{RGB}{86,180,233}
\definecolor{darkcyan0158115}{RGB}{0,158,115}
\definecolor{darkslategray38}{RGB}{38,38,38}
\definecolor{lightgray204}{RGB}{204,204,204}
\definecolor{orange2301590}{RGB}{230,159,0}

\begin{axis}[height=0.1\textheight,width=0.8\columnwidth, scale only axis,
axis line style={darkslategray38},
legend cell align={left},
legend pos=south east, 
  legend style={
  legend columns=3,
    fill opacity=0.8, draw opacity=1, text opacity=1,
    font=\footnotesize, draw=lightgray204
  },
log basis y={10},
tick align=outside,
x grid style={lightgray204},
xlabel=\textcolor{darkslategray38}{Age (years)},
xmajorticks=true,
xmin=0, xmax=70,
xtick style={color=darkslategray38},
y grid style={lightgray204},
ylabel=\textcolor{darkslategray38}{Failure rate (per year)},
ymajorticks=true,
ymin=0.000001, ymax=10000,
ymode=log,
ytick style={color=darkslategray38},
ytick={1e-06,0.0001,0.01,1,100,10000},
yticklabels={
  ${10^{-6}}$,
  ${10^{-4}}$,
  ${10^{-2}}$,
  ${10^{0}}$,
  ${10^{2}}$,
 ${10^{4}}$
}
]
\addplot [very thick, orange2301590, dashed]
table {%
0 0
1 0.000235514948841737
2 0.000942059795366948
3 0.00211963453957563
4 0.00376823918146779
5 0.00588787372104343
6 0.00847853815830253
7 0.0115402324932451
8 0.0150729567258712
9 0.0190767108561807
10 0.0235514948841737
11 0.0284973088098502
12 0.0339141526332101
13 0.0398020263542536
14 0.0461609299729805
15 0.0529908634893908
16 0.0602918269034847
17 0.068063820215262
18 0.0763068434247228
19 0.0850208965318671
20 0.0942059795366948
21 0.103862092439206
22 0.113989235239401
23 0.124587407937279
24 0.135656610532841
25 0.147196843026086
26 0.159208105417014
27 0.171690397705626
28 0.184643719891922
29 0.198068071975901
30 0.211963453957563
31 0.226329865836909
32 0.241167307613939
33 0.256475779288652
34 0.272255280861048
35 0.288505812331128
36 0.305227373698891
37 0.322419964964338
38 0.340083586127468
39 0.358218237188282
40 0.376823918146779
41 0.39590062900296
42 0.415448369756824
43 0.435467140408372
44 0.455956940957603
45 0.476917771404517
46 0.498349631749116
47 0.520252521991397
48 0.542626442131362
49 0.565471392169011
50 0.588787372104343
51 0.612574381937358
52 0.636832421668057
53 0.661561491296439
54 0.686761590822505
55 0.712432720246254
56 0.738574879567687
57 0.765188068786804
58 0.792272287903603
59 0.819827536918087
60 0.847853815830253
61 0.876351124640104
62 0.905319463347637
63 0.934758831952854
64 0.964669230455755
65 0.995050658856339
66 1.02590311715461
67 1.05722660535056
68 1.08902112344419
69 1.12128667143551
70 1.15402324932451
71 1.1872308571112
72 1.22090949479556
73 1.25505916237762
74 1.28967985985735
75 1.32477158723477
76 1.36033434450987
77 1.39636813168266
78 1.43287294875313
79 1.46984879572128
80 1.50729567258712
81 1.54521357935064
82 1.58360251601184
83 1.62246248257073
84 1.6617934790273
85 1.70159550538155
86 1.74186856163349
87 1.78261264778311
88 1.82382776383041
89 1.8655139097754
90 1.90767108561807
91 1.95029929135842
92 1.99339852699646
93 2.03696879253218
94 2.08101008796559
95 2.12552241329668
96 2.17050576852545
97 2.2159601536519
98 2.26188556867604
99 2.30828201359786
100 2.35514948841737
101 2.40248799313456
102 2.45029752774943
103 2.49857809226199
104 2.54732968667223
105 2.59655231098015
106 2.64624596518576
107 2.69641064928905
108 2.74704636329002
109 2.79815310718868
110 2.84973088098502
111 2.90177968467904
112 2.95429951827075
113 3.00729038176014
114 3.06075227514721
115 3.11468519843197
116 3.16908915161441
117 3.22396413469454
118 3.27931014767235
119 3.33512719054784
120 3.39141526332101
121 3.44817436599187
122 3.50540449856041
123 3.56310566102664
124 3.62127785339055
125 3.67992107565214
126 3.73903532781142
127 3.79862060986838
128 3.85867692182302
129 3.91920426367535
130 3.98020263542536
131 4.04167203707305
132 4.10361246861843
133 4.16602393006149
134 4.22890642140223
135 4.29225994264066
136 4.35608449377677
137 4.42038007481056
138 4.48514668574204
139 4.5503843265712
140 4.61609299729805
141 4.68227269792257
142 4.74892342844479
143 4.81604518886468
144 4.88363797918226
145 4.95170179939752
146 5.02023664951047
147 5.0892425295211
148 5.15871943942941
149 5.2286673792354
150 5.29908634893908
151 5.36997634854045
152 5.44133737803949
153 5.51316943743622
154 5.58547252673063
155 5.65824664592273
156 5.73149179501251
157 5.80520797399998
158 5.87939518288512
159 5.95405342166795
160 6.02918269034847
161 6.10478298892667
162 6.18085431740255
163 6.25739667577611
164 6.33441006404736
165 6.41189448221629
166 6.4898499302829
167 6.5682764082472
168 6.64717391610919
169 6.72654245386885
170 6.8063820215262
171 6.88669261908123
172 6.96747424653395
173 7.04872690388435
174 7.13045059113243
175 7.2126453082782
176 7.29531105532165
177 7.37844783226278
178 7.4620556391016
179 7.5461344758381
180 7.63068434247228
181 7.71570523900415
182 7.8011971654337
183 7.88716012176093
184 7.97359410798585
185 8.06049912410845
186 8.14787517012873
187 8.2357222460467
188 8.32404035186235
189 8.41282948757569
190 8.50208965318671
191 8.59182084869541
192 8.68202307410179
193 8.77269632940586
194 8.86384061460761
195 8.95545592970705
196 9.04754227470417
197 9.14009964959897
198 9.23312805439146
199 9.32662748908163
200 9.42059795366948
201 9.51503944815502
202 9.60995197253824
203 9.70533552681914
204 9.80119011099773
205 9.897515725074
206 9.99431236904795
207 10.0915800429196
208 10.1893187466889
209 10.2875284803559
210 10.3862092439206
211 10.485361037383
212 10.584983860743
213 10.6850777140008
214 10.7856425971562
215 10.8866785102093
216 10.9881854531601
217 11.0901634260086
218 11.1926124287547
219 11.2955324613985
220 11.3989235239401
221 11.5027856163793
222 11.6071187387162
223 11.7119228909507
224 11.817198073083
225 11.9229442851129
226 12.0291615270406
227 12.1358497988659
228 12.2430091005889
229 12.3506394322095
230 12.4587407937279
231 12.5673131851439
232 12.6763566064577
233 12.7858710576691
234 12.8958565387782
235 13.0063130497849
236 13.1172405906894
237 13.2286391614915
238 13.3405087621914
239 13.4528493927889
240 13.5656610532841
241 13.6789437436769
242 13.7926974639675
243 13.9069222141557
244 14.0216179942417
245 14.1367848042253
246 14.2524226441066
247 14.3685315138855
248 14.4851114135622
249 14.6021623431365
250 14.7196843026086
251 14.8376772919783
252 14.9561413112457
253 15.0750763604107
254 15.1944824394735
255 15.3143595484339
256 15.4347076872921
257 15.5555268560479
258 15.6768170547014
259 15.7985782832526
260 15.9208105417014
261 16.043513830048
262 16.1666881482922
263 16.2903334964341
264 16.4144498744737
265 16.539037282411
266 16.6640957202459
267 16.7896251879786
268 16.9156256856089
269 17.0420972131369
270 17.1690397705626
271 17.296453357886
272 17.4243379751071
273 17.5526936222258
274 17.6815202992422
275 17.8108180061564
276 17.9405867429682
277 18.0708265096776
278 18.2015373062848
279 18.3327191327897
280 18.4643719891922
281 18.5964958754924
282 18.7290907916903
283 18.8621567377859
284 18.9956937137791
285 19.1297017196701
286 19.2641807554587
287 19.399130821145
288 19.534551916729
289 19.6704440422107
290 19.8068071975901
291 19.9436413828671
292 20.0809465980419
293 20.2187228431143
294 20.3569701180844
295 20.4956884229522
296 20.6348777577176
297 20.7745381223808
298 20.9146695169416
299 21.0552719414001
300 21.1963453957563
301 21.3378898800102
302 21.4799053941618
303 21.622391938211
304 21.765349512158
305 21.9087781160026
306 22.0526777497449
307 22.1970484133849
308 22.3418901069225
309 22.4872028303579
310 22.6329865836909
311 22.7792413669216
312 22.92596718005
313 23.0731640230761
314 23.2208318959999
315 23.3689707988214
316 23.5175807315405
317 23.6666616941573
318 23.8162136866718
319 23.966236709084
320 24.1167307613939
321 24.2676958436014
322 24.4191319557067
323 24.5710390977096
324 24.7234172696102
325 24.8762664714085
326 25.0295867031044
327 25.1833779646981
328 25.3376402561894
329 25.4923735775785
330 25.6475779288652
331 25.8032533100495
332 25.9593997211316
333 26.1160171621114
334 26.2731056329888
335 26.4306651337639
336 26.5886956644367
337 26.7471972250072
338 26.9061698154754
339 27.0656134358413
340 27.2255280861048
341 27.385913766266
342 27.5467704763249
343 27.7080982162815
344 27.8698969861358
345 28.0321667858878
346 28.1949076155374
347 28.3581194750847
348 28.5218023645297
349 28.6859562838724
350 28.8505812331128
351 29.0156772122508
352 29.1812442212866
353 29.34728226022
354 29.5137913290511
355 29.6807714277799
356 29.8482225564064
357 30.0161447149305
358 30.1845379033524
359 30.3534021216719
360 30.5227373698891
361 30.692543648004
362 30.8628209560166
363 31.0335692939268
364 31.2047886617348
365 31.3764790594404
366 31.5486404870437
367 31.7212729445447
368 31.8943764319434
369 32.0679509492398
370 32.2419964964338
371 32.4165130735255
372 32.5915006805149
373 32.766959317402
374 32.9428889841868
375 33.1192896808693
376 33.2961614074494
377 33.4735041639272
378 33.6513179503027
379 33.8296027665759
380 34.0083586127468
381 34.1875854888154
382 34.3672833947816
383 34.5474523306456
384 34.7280922964072
385 34.9092032920665
386 35.0907853176234
387 35.2728383730781
388 35.4553624584305
389 35.6383575736805
390 35.8218237188282
391 36.0057608938736
392 36.1901690988167
393 36.3750483336574
394 36.5603985983959
395 36.746219893032
396 36.9325122175658
397 37.1192755719973
398 37.3065099563265
399 37.4942153705534
400 37.6823918146779
401 37.8710392887002
402 38.0601577926201
403 38.2497473264377
404 38.4398078901529
405 38.6303394837659
406 38.8213421072766
407 39.0128157606849
408 39.2047604439909
409 39.3971761571946
410 39.590062900296
411 39.7834206732951
412 39.9772494761918
413 40.1715493089862
414 40.3663201716784
415 40.5615620642682
416 40.7572749867557
417 40.9534589391408
418 41.1501139214236
419 41.3472399336042
420 41.5448369756824
421 41.7429050476583
422 41.9414441495319
423 42.1404542813032
424 42.3399354429721
425 42.5398876345387
426 42.7403108560031
427 42.9412051073651
428 43.1425703886248
429 43.3444066997821
430 43.5467140408372
431 43.7494924117899
432 43.9527418126403
433 44.1564622433884
434 44.3606537040342
435 44.5653161945777
436 44.7704497150188
437 44.9760542653577
438 45.1821298455942
439 45.3886764557284
440 45.5956940957603
441 45.8031827656899
442 46.0111424655171
443 46.2195731952421
444 46.4284749548647
445 46.637847744385
446 46.847691563803
447 47.0580064131186
448 47.268792292332
449 47.480049201443
450 47.6917771404518
451 47.9039761093582
452 48.1166461081622
453 48.329787136864
454 48.5433991954635
455 48.7574822839606
456 48.9720364023554
457 49.1870615506479
458 49.4025577288381
459 49.618524936926
460 49.8349631749115
461 50.0518724427948
462 50.2692527405757
463 50.4871040682543
464 50.7054264258306
465 50.9242198133046
466 51.1434842306762
467 51.3632196779456
468 51.5834261551126
469 51.8041036621773
470 52.0252521991397
471 52.2468717659998
472 52.4689623627575
473 52.691523989413
474 52.9145566459661
475 53.1380603324169
476 53.3620350487654
477 53.5864807950116
478 53.8113975711554
479 54.036785377197
480 54.2626442131362
481 54.4889740789731
482 54.7157749747077
483 54.94304690034
484 55.1707898558699
485 55.3990038412976
486 55.6276888566229
487 55.8568449018459
488 56.0864719769666
489 56.316570081985
490 56.5471392169011
491 56.7781793817148
492 57.0096905764262
493 57.2416728010354
494 57.4741260555422
495 57.7070503399466
496 57.9404456542488
497 58.1743119984486
498 58.4086493725461
499 58.6434577765414
};
\addlegendentry{$\frac{I_c}{I_z} = 0.6$}
\addplot [very thick, cornflowerblue86180233]
table {%
0 0
1 0.013262447687235
2 0.05304979074894
3 0.119362029185115
4 0.21219916299576
5 0.331561192180875
6 0.47744811674046
7 0.649859936674515
8 0.848796651983041
9 1.07425826266604
10 1.3262447687235
11 1.60475617015544
12 1.90979246696184
13 2.24135365914272
14 2.59943974669806
15 2.98405072962788
16 3.39518660793216
17 3.83284738161092
18 4.29703305066414
19 4.78774361509184
20 5.304979074894
21 5.84873943007064
22 6.41902468062175
23 7.01583482654732
24 7.63916986784737
25 8.28902980452188
26 8.96541463657087
27 9.66832436399432
28 10.3977589867922
29 11.1537185049646
30 11.9362029185115
31 12.7452122274328
32 13.5807464317287
33 14.4428055313989
34 15.3313895264437
35 16.2464984168629
36 17.1881322026566
37 18.1562908838247
38 19.1509744603674
39 20.1721829322844
40 21.219916299576
41 22.2941745622421
42 23.3949577202826
43 24.5222657736975
44 25.676098722487
45 26.8564565666509
46 28.0633393061893
47 29.2967469411021
48 30.5566794713895
49 31.8431368970513
50 33.1561192180875
51 34.4956264344983
52 35.8616585462835
53 37.2542155534431
54 38.6732974559773
55 40.1189042538859
56 41.591035947169
57 43.0896925358266
58 44.6148740198586
59 46.1665803992651
60 47.744811674046
61 49.3495678442015
62 50.9808489097314
63 52.6386548706358
64 54.3229857269146
65 56.0338414785679
66 57.7712221255957
67 59.535127667998
68 61.3255581057747
69 63.1425134389259
70 64.9859936674516
71 66.8559987913517
72 68.7525288106263
73 70.6755837252754
74 72.6251635352989
75 74.6012682406969
76 76.6038978414694
77 78.6330523376164
78 80.6887317291378
79 82.7709360160337
80 84.8796651983041
81 87.0149192759489
82 89.1766982489682
83 91.365002117362
84 93.5798308811302
85 95.821184540273
86 98.0890630947901
87 100.383466544682
88 102.704394889948
89 105.051848130588
90 107.425826266604
91 109.826329297993
92 112.253357224757
93 114.706910046896
94 117.186987764409
95 119.693590377296
96 122.226717885558
97 124.786370289194
98 127.372547588205
99 129.98524978259
100 132.62447687235
101 135.290228857484
102 137.982505737993
103 140.701307513876
104 143.446634185134
105 146.218485751766
106 149.016862213773
107 151.841763571154
108 154.693189823909
109 157.571140972039
110 160.475617015544
111 163.406617954423
112 166.364143788676
113 169.348194518304
114 172.358770143306
115 175.395870663683
116 178.459496079434
117 181.54964639056
118 184.66632159706
119 187.809521698935
120 190.979246696184
121 194.175496588808
122 197.398271376806
123 200.647571060178
124 203.923395638926
125 207.225745113047
126 210.554619482543
127 213.910018747413
128 217.291942907658
129 220.700391963278
130 224.135365914272
131 227.59686476064
132 231.084888502383
133 234.5994371395
134 238.140510671992
135 241.708109099858
136 245.302232423099
137 248.922880641714
138 252.570053755703
139 256.243751765068
140 259.943974669806
141 263.670722469919
142 267.423995165407
143 271.203792756269
144 275.010115242505
145 278.842962624116
146 282.702334901101
147 286.588232073461
148 290.500654141196
149 294.439601104305
150 298.405072962788
151 302.397069716645
152 306.415591365878
153 310.460637910484
154 314.532209350466
155 318.630305685821
156 322.754926916551
157 326.906073042656
158 331.083744064135
159 335.287939980988
160 339.518660793216
161 343.775906500819
162 348.059677103796
163 352.369972602147
164 356.706792995873
165 361.070138284973
166 365.460008469448
167 369.876403549297
168 374.319323524521
169 378.788768395119
170 383.284738161092
171 387.807232822439
172 392.356252379161
173 396.931796831257
174 401.533866178727
175 406.162460421572
176 410.817579559792
177 415.499223593386
178 420.207392522354
179 424.942086346697
180 429.703305066414
181 434.491048681506
182 439.305317191972
183 444.146110597813
184 449.013428899029
185 453.907272095618
186 458.827640187582
187 463.774533174921
188 468.747951057634
189 473.747893835722
190 478.774361509184
191 483.82735407802
192 488.906871542231
193 494.012913901817
194 499.145481156777
195 504.304573307111
196 509.49019035282
197 514.702332293903
198 519.940999130361
199 525.206190862194
200 530.497907489401
201 535.816149011982
202 541.160915429937
203 546.532206743267
204 551.930022951972
205 557.354364056051
206 562.805230055505
207 568.282620950333
208 573.786536740536
209 579.316977426113
210 584.873943007064
211 590.45743348339
212 596.06744885509
213 601.703989122165
214 607.367054284614
215 613.056644342438
216 618.772759295636
217 624.515399144209
218 630.284563888157
219 636.080253527478
220 641.902468062175
221 647.751207492245
222 653.62647181769
223 659.52826103851
224 665.456575154704
225 671.411414166273
226 677.392778073215
227 683.400666875533
228 689.435080573225
229 695.496019166291
230 701.583482654732
231 707.697471038547
232 713.837984317737
233 720.005022492301
234 726.19858556224
235 732.418673527554
236 738.665286388241
237 744.938424144303
238 751.23808679574
239 757.564274342551
240 763.916986784737
241 770.296224122297
242 776.701986355231
243 783.13427348354
244 789.593085507224
245 796.078422426282
246 802.590284240714
247 809.128670950521
248 815.693582555702
249 822.285019056258
250 828.902980452188
251 835.547466743493
252 842.218477930172
253 848.916014012226
254 855.640074989654
255 862.390660862457
256 869.167771630634
257 875.971407294185
258 882.801567853111
259 889.658253307412
260 896.541463657087
261 903.451198902136
262 910.38745904256
263 917.350244078359
264 924.339554009531
265 931.355388836079
266 938.397748558
267 945.466633175297
268 952.562042687967
269 959.683977096013
270 966.832436399432
271 974.007420598227
272 981.208929692395
273 988.436963681938
274 995.691522566856
275 1002.97260634715
276 1010.28021502281
277 1017.61434859385
278 1024.97500706027
279 1032.36219042206
280 1039.77589867922
281 1047.21613183176
282 1054.68288987968
283 1062.17617282296
284 1069.69598066163
285 1077.24231339566
286 1084.81517102507
287 1092.41455354986
288 1100.04046097002
289 1107.69289328556
290 1115.37185049646
291 1123.07733260275
292 1130.80933960441
293 1138.56787150144
294 1146.35292829385
295 1154.16450998163
296 1162.00261656478
297 1169.86724804331
298 1177.75840441722
299 1185.6760856865
300 1193.62029185115
301 1201.59102291118
302 1209.58827886658
303 1217.61205971736
304 1225.66236546351
305 1233.73919610504
306 1241.84255164194
307 1249.97243207421
308 1258.12883740186
309 1266.31176762489
310 1274.52122274328
311 1282.75720275706
312 1291.0197076662
313 1299.30873747073
314 1307.62429217062
315 1315.96637176589
316 1324.33497625654
317 1332.73010564256
318 1341.15175992395
319 1349.59993910072
320 1358.07464317286
321 1366.57587214038
322 1375.10362600327
323 1383.65790476154
324 1392.23870841518
325 1400.8460369642
326 1409.47989040859
327 1418.14026874835
328 1426.82717198349
329 1435.540600114
330 1444.28055313989
331 1453.04703106116
332 1461.84003387779
333 1470.6595615898
334 1479.50561419719
335 1488.37819169995
336 1497.27729409808
337 1506.20292139159
338 1515.15507358048
339 1524.13375066473
340 1533.13895264437
341 1542.17067951937
342 1551.22893128976
343 1560.31370795551
344 1569.42500951664
345 1578.56283597315
346 1587.72718732503
347 1596.91806357228
348 1606.13546471491
349 1615.37939075291
350 1624.64984168629
351 1633.94681751504
352 1643.27031823917
353 1652.62034385867
354 1661.99689437354
355 1671.39996978379
356 1680.82957008942
357 1690.28569529041
358 1699.76834538679
359 1709.27752037854
360 1718.81322026566
361 1728.37544504815
362 1737.96419472602
363 1747.57946929927
364 1757.22126876789
365 1766.88959313188
366 1776.58444239125
367 1786.305816546
368 1796.05371559611
369 1805.82813954161
370 1815.62908838247
371 1825.45656211871
372 1835.31056075033
373 1845.19108427732
374 1855.09813269968
375 1865.03170601742
376 1874.99180423054
377 1884.97842733902
378 1894.99157534289
379 1905.03124824212
380 1915.09744603674
381 1925.19016872672
382 1935.30941631208
383 1945.45518879282
384 1955.62748616893
385 1965.82630844041
386 1976.05165560727
387 1986.3035276695
388 1996.58192462711
389 2006.88684648009
390 2017.21829322844
391 2027.57626487218
392 2037.96076141128
393 2048.37178284576
394 2058.80932917561
395 2069.27340040084
396 2079.76399652144
397 2090.28111753742
398 2100.82476344877
399 2111.3949342555
400 2121.9916299576
401 2132.61485055508
402 2143.26459604793
403 2153.94086643615
404 2164.64366171975
405 2175.37298189872
406 2186.12882697307
407 2196.91119694279
408 2207.72009180789
409 2218.55551156836
410 2229.4174562242
411 2240.30592577543
412 2251.22092022202
413 2262.16243956399
414 2273.13048380133
415 2284.12505293405
416 2295.14614696214
417 2306.19376588561
418 2317.26790970445
419 2328.36857841867
420 2339.49577202826
421 2350.64949053322
422 2361.82973393356
423 2373.03650222927
424 2384.26979542036
425 2395.52961350682
426 2406.81595648866
427 2418.12882436587
428 2429.46821713846
429 2440.83413480642
430 2452.22657736975
431 2463.64554482846
432 2475.09103718255
433 2486.563054432
434 2498.06159657684
435 2509.58666361704
436 2521.13825555263
437 2532.71637238358
438 2544.32101410991
439 2555.95218073162
440 2567.6098722487
441 2579.29408866115
442 2591.00482996898
443 2602.74209617218
444 2614.50588727076
445 2626.29620326471
446 2638.11304415404
447 2649.95640993874
448 2661.82630061881
449 2673.72271619427
450 2685.64565666509
451 2697.59512203129
452 2709.57111229286
453 2721.57362744981
454 2733.60266750213
455 2745.65823244983
456 2757.7403222929
457 2769.84893703134
458 2781.98407666516
459 2794.14574119436
460 2806.33393061893
461 2818.54864493887
462 2830.78988415419
463 2843.05764826488
464 2855.35193727095
465 2867.67275117239
466 2880.02008996921
467 2892.3939536614
468 2904.79434224896
469 2917.2212557319
470 2929.67469411021
471 2942.1546573839
472 2954.66114555296
473 2967.1941586174
474 2979.75369657721
475 2992.3397594324
476 3004.95234718296
477 3017.5914598289
478 3030.2570973702
479 3042.94925980689
480 3055.66794713895
481 3068.41315936638
482 3081.18489648919
483 3093.98315850737
484 3106.80794542092
485 3119.65925722986
486 3132.53709393416
487 3145.44145553384
488 3158.37234202889
489 3171.32975341932
490 3184.31368970513
491 3197.3241508863
492 3210.36113696286
493 3223.42464793478
494 3236.51468380208
495 3249.63124456476
496 3262.77433022281
497 3275.94394077623
498 3289.14007622503
499 3302.36273656921
};
\addlegendentry{$\frac{I_c}{I_z} = 0.8$}
\addplot [very thick, darkcyan0158115]
table {%
0 0
1 2.36266313669474
2 9.45065254677894
3 21.2639682302526
4 37.8026101871158
5 59.0665784173684
6 85.0558729210105
7 115.770493698042
8 151.210440748463
9 191.375714072274
10 236.266313669474
11 285.882239540063
12 340.223491684042
13 399.29007010141
14 463.081974792168
15 531.599205756315
16 604.841762993852
17 682.809646504778
18 765.502856289094
19 852.921392346799
20 945.065254677894
21 1041.93444328238
22 1143.52895816025
23 1249.84879931151
24 1360.89396673617
25 1476.66446043421
26 1597.16028040564
27 1722.38142665046
28 1852.32789916867
29 1986.99969796027
30 2126.39682302526
31 2270.51927436364
32 2419.36705197541
33 2572.94015586057
34 2731.23858601911
35 2894.26234245105
36 3062.01142515638
37 3234.48583413509
38 3411.6855693872
39 3593.61063091269
40 3780.26101871158
41 3971.63673278385
42 4167.73777312951
43 4368.56413974857
44 4574.11583264101
45 4784.39285180684
46 4999.39519724606
47 5219.12286895867
48 5443.57586694467
49 5672.75419120406
50 5906.65784173684
51 6145.28681854301
52 6388.64112162256
53 6636.72075097551
54 6889.52570660185
55 7147.05598850157
56 7409.31159667469
57 7676.2925311212
58 7947.99879184109
59 8224.43037883437
60 8505.58729210105
61 8791.46953164111
62 9082.07709745456
63 9377.40998954141
64 9677.46820790163
65 9982.25175253526
66 10291.7606234423
67 10605.9948206227
68 10924.9543440765
69 11248.6391938036
70 11577.0493698042
71 11910.1848720782
72 12248.0457006255
73 12590.6318554462
74 12937.9433365404
75 13289.9801439079
76 13646.7422775488
77 14008.2297374631
78 14374.4425236508
79 14745.3806361118
80 15121.0440748463
81 15501.4328398542
82 15886.5469311354
83 16276.38634869
84 16670.951092518
85 17070.2411626195
86 17474.2565589943
87 17882.9972816424
88 18296.463330564
89 18714.654705759
90 19137.5714072274
91 19565.2134349691
92 19997.5807889842
93 20434.6734692728
94 20876.4914758347
95 21323.03480867
96 21774.3034677787
97 22230.2974531608
98 22691.0167648162
99 23156.4614027451
100 23626.6313669474
101 24101.526657423
102 24581.147274172
103 25065.4932171944
104 25554.5644864903
105 26048.3610820595
106 26546.8830039021
107 27050.130252018
108 27558.1028264074
109 28070.8007270702
110 28588.2239540063
111 29110.3725072158
112 29637.2463866988
113 30168.8455924551
114 30705.1701244848
115 31246.2199827879
116 31791.9951673644
117 32342.4956782142
118 32897.7215153375
119 33457.6726787341
120 34022.3491684042
121 34591.7509843476
122 35165.8781265644
123 35744.7305950547
124 36328.3083898182
125 36916.6115108552
126 37509.6399581656
127 38107.3937317494
128 38709.8728316065
129 39317.0772577371
130 39929.007010141
131 40545.6620888184
132 41167.0424937691
133 41793.1482249932
134 42423.9792824907
135 43059.5356662616
136 43699.8173763058
137 44344.8244126235
138 44994.5567752145
139 45649.014464079
140 46308.1974792168
141 46972.105820628
142 47640.7394883126
143 48314.0984822706
144 48992.182802502
145 49674.9924490068
146 50362.527421785
147 51054.7877208365
148 51751.7733461615
149 52453.4842977598
150 53159.9205756315
151 53871.0821797767
152 54586.9691101952
153 55307.5813668871
154 56032.9189498523
155 56762.981859091
156 57497.7700946031
157 58237.2836563885
158 58981.5225444474
159 59730.4867587796
160 60484.1762993852
161 61242.5911662642
162 62005.7313594166
163 62773.5968788424
164 63546.1877245416
165 64323.5038965142
166 65105.5453947601
167 65892.3122192795
168 66683.8043700722
169 67480.0218471383
170 68280.9646504779
171 69086.6327800907
172 69897.0262359771
173 70712.1450181367
174 71531.9891265698
175 72356.5585612763
176 73185.8533222561
177 74019.8734095094
178 74858.618823036
179 75702.089562836
180 76550.2856289094
181 77403.2070212562
182 78260.8537398764
183 79123.22578477
184 79990.323155937
185 80862.1458533773
186 81738.693877091
187 82619.9672270782
188 83505.9659033387
189 84396.6899058726
190 85292.13923468
191 86192.3138897606
192 87097.2138711147
193 88006.8391787422
194 88921.1898126431
195 89840.2657728173
196 90764.067059265
197 91692.593671986
198 92625.8456109804
199 93563.8228762482
200 94506.5254677894
201 95453.953385604
202 96406.106629692
203 97362.9852000533
204 98324.5890966881
205 99290.9183195963
206 100261.972868778
207 101237.752744233
208 102218.257945961
209 103203.488473963
210 104193.444328238
211 105188.125508786
212 106187.532015608
213 107191.663848703
214 108200.521008072
215 109214.103493714
216 110232.41130563
217 111255.444443818
218 112283.202908281
219 113315.686699016
220 114352.895816025
221 115394.830259308
222 116441.490028863
223 117492.875124692
224 118548.985546795
225 119609.821295171
226 120675.38236982
227 121745.668770743
228 122820.680497939
229 123900.417551409
230 124984.879931152
231 126074.067637168
232 127167.980669457
233 128266.619028021
234 129369.982712857
235 130478.071723967
236 131590.88606135
237 132708.425725007
238 133830.690714937
239 134957.68103114
240 136089.396673617
241 137225.837642367
242 138367.00393739
243 139512.895558687
244 140663.512506258
245 141818.854780101
246 142978.922380219
247 144143.715306609
248 145313.233559273
249 146487.47713821
250 147666.446043421
251 148850.140274905
252 150038.559832662
253 151231.704716693
254 152429.574926998
255 153632.170463575
256 154839.491326426
257 156051.537515551
258 157268.309030948
259 158489.80587262
260 159716.028040564
261 160946.975534782
262 162182.648355273
263 163423.046502038
264 164668.169975076
265 165918.018774388
266 167172.592899973
267 168431.892351831
268 169695.917129963
269 170964.667234368
270 172238.142665046
271 173516.343421998
272 174799.269505223
273 176086.920914722
274 177379.297650494
275 178676.399712539
276 179978.227100858
277 181284.77981545
278 182596.057856316
279 183912.061223455
280 185232.789916867
281 186558.243936553
282 187888.423282512
283 189223.327954745
284 190562.957953251
285 191907.31327803
286 193256.393929083
287 194610.199906409
288 195968.731210008
289 197331.987839881
290 198699.969796027
291 200072.677078447
292 201450.10968714
293 202832.267622106
294 204219.150883346
295 205610.759470859
296 207007.093384646
297 208408.152624706
298 209813.937191039
299 211224.447083646
300 212639.682302526
301 214059.64284768
302 215484.328719107
303 216913.739916807
304 218347.876440781
305 219786.738291028
306 221230.325467548
307 222678.637970342
308 224131.675799409
309 225589.43895475
310 227051.927436364
311 228519.141244251
312 229991.080378412
313 231467.744838846
314 232949.134625554
315 234435.249738535
316 235926.090177789
317 237421.655943317
318 238921.947035118
319 240426.963453193
320 241936.705197541
321 243451.172268162
322 244970.364665057
323 246494.282388225
324 248022.925437667
325 249556.293813381
326 251094.38751537
327 252637.206543631
328 254184.750898166
329 255737.020578975
330 257294.015586057
331 258855.735919412
332 260422.181579041
333 261993.352564943
334 263569.248877118
335 265149.870515567
336 266735.217480289
337 268325.289771284
338 269920.087388553
339 271519.610332096
340 273123.858601911
341 274732.832198001
342 276346.531120363
343 277964.955368999
344 279588.104943908
345 281215.979845091
346 282848.580072547
347 284485.905626276
348 286127.956506279
349 287774.732712555
350 289426.234245105
351 291082.461103928
352 292743.413289024
353 294409.090800394
354 296079.493638037
355 297754.621801954
356 299434.475292144
357 301119.054108607
358 302808.358251344
359 304502.387720354
360 306201.142515638
361 307904.622637195
362 309612.828085025
363 311325.758859129
364 313043.414959506
365 314765.796386156
366 316492.90313908
367 318224.735218277
368 319961.292623748
369 321702.575355492
370 323448.583413509
371 325199.3167978
372 326954.775508364
373 328714.959545202
374 330479.868908313
375 332249.503597697
376 334023.863613355
377 335802.948955286
378 337586.759623491
379 339375.295617968
380 341168.55693872
381 342966.543585744
382 344769.255559043
383 346576.692858614
384 348388.855484459
385 350205.743436577
386 352027.356714969
387 353853.695319634
388 355684.759250572
389 357520.548507784
390 359361.063091269
391 361206.303001028
392 363056.26823706
393 364910.958799365
394 366770.374687944
395 368634.515902796
396 370503.382443922
397 372376.974311321
398 374255.291504993
399 376138.334024938
400 378026.101871158
401 379918.59504365
402 381815.813542416
403 383717.757367455
404 385624.426518768
405 387535.820996354
406 389451.940800213
407 391372.785930346
408 393298.356386752
409 395228.652169432
410 397163.673278385
411 399103.419713611
412 401047.891475111
413 402997.088562884
414 404951.010976931
415 406909.658717251
416 408873.031783844
417 410841.130176711
418 412813.953895851
419 414791.502941264
420 416773.777312951
421 418760.777010912
422 420752.502035145
423 422748.952385652
424 424750.128062433
425 426756.029065487
426 428766.655394814
427 430782.007050414
428 432802.084032288
429 434826.886340436
430 436856.413974857
431 438890.666935551
432 440929.645222518
433 442973.348835759
434 445021.777775273
435 447074.932041061
436 449132.811633122
437 451195.416551457
438 453262.746796065
439 455334.802366946
440 457411.583264101
441 459493.089487529
442 461579.32103723
443 463670.277913205
444 465765.960115453
445 467866.367643975
446 469971.50049877
447 472081.358679838
448 474195.94218718
449 476315.251020795
450 478439.285180684
451 480568.044666846
452 482701.529479281
453 484839.73961799
454 486982.675082972
455 489130.335874228
456 491282.721991757
457 493439.833435559
458 495601.670205634
459 497768.232301984
460 499939.519724606
461 502115.532473502
462 504296.270548671
463 506481.733950114
464 508671.92267783
465 510866.836731819
466 513066.476112082
467 515270.840818618
468 517479.930851428
469 519693.746210511
470 521912.286895867
471 524135.552907497
472 526363.5442454
473 528596.260909576
474 530833.702900026
475 533075.87021675
476 535322.762859746
477 537574.380829016
478 539830.72412456
479 542091.792746377
480 544357.586694467
481 546628.105968831
482 548903.350569468
483 551183.320496378
484 553468.015749562
485 555757.436329019
486 558051.58223475
487 560350.453466754
488 562654.050025031
489 564962.371909582
490 567275.419120406
491 569593.191657503
492 571915.689520874
493 574242.912710519
494 576574.861226436
495 578911.535068628
496 581252.934237092
497 583599.05873183
498 585949.908552841
499 588305.483700126
};
\addlegendentry{$\frac{I_c}{I_z} = 1.0$}
\end{axis}

\end{tikzpicture}

%% file: Failures_both.tex
\begin{tikzpicture}

\definecolor{cornflowerblue86180233}{RGB}{86,180,233}
\definecolor{darkslategray38}{RGB}{38,38,38}
\definecolor{lightgray204}{RGB}{204,204,204}
\definecolor{orange2301590}{RGB}{230,159,0}

\begin{axis}[height=0.1\textheight,width=0.75\columnwidth,
scale only axis,
axis line style={darkslategray38},
  legend cell align={left},
  legend pos=south east, 
  legend style={
    fill opacity=0.8, draw opacity=1, text opacity=1,
    font=\footnotesize, draw=lightgray204
  },
log basis y={10},
tick align=outside,
x grid style={lightgray204},
xlabel=\textcolor{darkslategray38}{$I_c/I_z$},
xmajorticks=true,
xmin=0.75, xmax=1.25,
xtick style={color=darkslategray38},
y grid style={lightgray204},
ylabel=\textcolor{darkslategray38}{Failure rate $\frac{F}{|\Omega^C|}$},
ymajorticks=true,
ymin=1e-07, ymax=1e07,
ymode=log,
ytick style={color=darkslategray38},
ytick={1e-6,1e-03,1e0,1e03,1e6},
]
\addplot [very thick, orange2301590, dashed] 
table {%
0.498402555910543 9.56578548160577e-13
0.501597444089457 1.07089810405687e-12
0.504792332268371 1.1995483138746e-12
0.507987220447284 1.34439911590589e-12
0.511182108626198 1.50757304357265e-12
0.514376996805112 1.69148040077486e-12
0.517571884984026 1.89885947009387e-12
0.520766773162939 2.13282249168366e-12
0.523961661341853 2.39690826125844e-12
0.527156549520767 2.69514232306908e-12
0.530351437699681 3.03210588078326e-12
0.533546325878594 3.41301471879687e-12
0.536741214057508 3.8438096222408e-12
0.539936102236422 4.33126000990013e-12
0.543130990415335 4.88308275518398e-12
0.546325878594249 5.50807847168257e-12
0.549520766773163 6.21628788810132e-12
0.552715654952077 7.01917133989151e-12
0.55591054313099 7.92981487030703e-12
0.559105431309904 8.96316697188475e-12
0.562300319488818 1.01363106220775e-11
0.565495207667732 1.14687759874492e-11
0.568690095846645 1.29829000050937e-11
0.571884984025559 1.47042400160111e-11
0.575079872204473 1.66620497441608e-11
0.578274760383387 1.88898272114704e-11
0.5814696485623 2.14259456817904e-11
0.584664536741214 2.43143804689553e-11
0.587859424920128 2.7605546464481e-11
0.591054313099042 3.13572635841629e-11
0.594249201277955 3.56358700523375e-11
0.597444089456869 4.05175065991732e-11
0.600638977635783 4.608959831122e-11
0.603833865814696 5.24525651316641e-11
0.60702875399361 5.97217969511424e-11
0.610223642172524 6.80299349754058e-11
0.613418530351438 7.75295077342142e-11
0.616613418530351 8.83959778602582e-11
0.619808306709265 1.00831264796481e-10
0.623003194888179 1.15067819094485e-10
0.626198083067093 1.31373336189216e-10
0.629392971246006 1.500562117614e-10
0.63258785942492 1.71471857361463e-10
0.635782747603834 1.96030014256263e-10
0.638977635782748 2.24203225926513e-10
0.642172523961661 2.56536655819805e-10
0.645367412140575 2.93659467471583e-10
0.648562300319489 3.36298019670258e-10
0.651757188498403 3.85291170811369e-10
0.654952076677316 4.41608034951695e-10
0.65814696485623 5.06368588499048e-10
0.661341853035144 5.8086759231596e-10
0.664536741214057 6.66602370866895e-10
0.667731629392971 7.65305079760389e-10
0.670926517571885 8.78980197812249e-10
0.674121405750799 1.009948102136e-09
0.677316293929712 1.16089572774726e-09
0.680511182108626 1.33493548025271e-09
0.68370607028754 1.53567376549694e-09
0.686900958466454 1.76729072837082e-09
0.690095846645367 2.03463306000191e-09
0.693290734824281 2.34332204486406e-09
0.696485623003195 2.69987938473644e-09
0.699680511182109 3.11187376402234e-09
0.702875399361022 3.58809162140006e-09
0.706070287539936 4.13873617867579e-09
0.70926517571885 4.77565946376941e-09
0.712460063897764 5.51263286829063e-09
0.715654952076677 6.36566272146465e-09
0.718849840255591 7.35335846509505e-09
0.722044728434505 8.49736230683209e-09
0.725239616613419 9.82285074417659e-09
0.728434504792332 1.13591201280589e-08
0.731629392971246 1.31402705179867e-08
0.73482428115016 1.52060045240984e-08
0.738019169329074 1.76025606977963e-08
0.741214057507987 2.03838043959356e-08
0.744408945686901 2.36125029906418e-08
0.747603833865815 2.73618169300111e-08
0.750798722044728 3.17170435944608e-08
0.753993610223642 3.67776572808527e-08
0.757188498402556 4.26596961483221e-08
0.76038338658147 4.94985557719478e-08
0.763578274760383 5.74522593038909e-08
0.766773162939297 6.67052864080272e-08
0.769968051118211 7.74730574334473e-08
0.773162939297125 9.00071861013255e-08
0.776357827476038 1.04601633742422e-07
0.779552715654952 1.21599921361887e-07
0.782747603833866 1.41403583140214e-07
0.78594249201278 1.64482077130268e-07
0.789137380191693 1.91384406735937e-07
0.792332268370607 2.22752751067145e-07
0.795527156549521 2.59338454647432e-07
0.798722044728435 3.02020788606732e-07
0.801916932907348 3.51828968075913e-07
0.805111821086262 4.09967995962268e-07
0.808306709265176 4.7784900392423e-07
0.811501597444089 5.57124879894942e-07
0.814696485623003 6.49732111125646e-07
0.817891373801917 7.57939936210798e-07
0.821086261980831 8.8440809337089e-07
0.824281150159744 1.03225468066908e-06
0.827476038338658 1.2051359130298e-06
0.830670926517572 1.40733987823971e-06
0.833865814696486 1.64389676819815e-06
0.837060702875399 1.92070850275726e-06
0.840255591054313 2.24470118361611e-06
0.843450479233227 2.62400442915316e-06
0.846645367412141 3.06816236461826e-06
0.849840255591054 3.58838189563842e-06
0.853035143769968 4.19782489998822e-06
0.856230031948882 4.91195216077289e-06
0.859424920127795 5.74892826634142e-06
0.862619808306709 6.73009835709637e-06
0.865814696485623 7.88054955266489e-06
0.869009584664537 9.22977219869853e-06
0.87220447284345 1.08124387948062e-05
0.875399361022364 1.26693216793208e-05
0.878594249201278 1.48483743420626e-05
0.881789137380192 1.74060057185662e-05
0.884984025559105 2.04085821131814e-05
0.888178913738019 2.39341976516273e-05
0.891373801916933 2.80747615504685e-05
0.894568690095847 3.29384592179599e-05
0.89776357827476 3.86526545117705e-05
0.900958466453674 4.53673126629159e-05
0.904153354632588 5.32590377731589e-05
0.907348242811502 6.25358358088878e-05
0.910543130990415 7.34427341258729e-05
0.913738019169329 8.62684123311136e-05
0.916932907348243 0.000101353027388958
0.920127795527157 0.000119097449099389
0.92332268370607 0.000139974161352796
0.926517571884984 0.000164540131004575
0.929712460063898 0.000193452001124837
0.932907348242812 0.000227484030313395
0.936102236421725 0.000267549276562617
0.939297124600639 0.000314724614971799
0.942492012779553 0.000370280286035537
0.945686900958466 0.000435714798286253
0.94888178913738 0.000512796159366301
0.952076677316294 0.000603610587398309
0.955271565495208 0.000710620064854675
0.958466453674121 0.000836730345967861
0.961661341853035 0.000985371323135264
0.964856230031949 0.00116059200612044
0.968051118210863 0.00136717278003294
0.971246006389776 0.00161075809580281
0.97444089456869 0.00189801332402503
0.977635782747604 0.00223681018605272
0.980830670926518 0.00263644598451623
0.984025559105431 0.00310790281204336
0.987220447284345 0.00366415404911392
0.990415335463259 0.00432052680195272
0.993610223642173 0.00509513051734134
0.996805111821086 0.00600936388843366
1 0.00708851438761693
1.00319488817891 0.0083624673925527
1.00638977635783 0.00986654498479523
1.00958466453674 0.0116424981856537
1.01277955271565 0.0137396807564034
1.01597444089457 0.0162164378541028
1.01916932907348 0.0191417489475083
1.0223642172524 0.0225971716344726
1.02555910543131 0.0266791415693431
1.02875399361022 0.0315016938508545
1.03194888178914 0.0371996832274561
1.03514376996805 0.0439325946906503
1.03833865814696 0.0518890528536925
1.04153354632588 0.0612921584331955
1.04472843450479 0.0724058037336476
1.04792332268371 0.0855421469525609
1.05111821086262 0.10107045817409
1.05431309904153 0.119427589044443
1.05750798722045 0.141130364439832
1.06070287539936 0.166790249268643
1.06389776357827 0.197130708457783
1.06709265175719 0.23300775501015
1.0702875399361 0.275434271976253
1.07348242811502 0.325608801851715
1.07667731629393 0.384949624360743
1.07987220447284 0.455135094445698
1.08306709265176 0.538151390850532
1.08626198083067 0.636349037044428
1.08945686900958 0.752509806398092
1.0926517571885 0.889925919613954
1.09584664536741 1.05249379284504
1.09904153354633 1.24482500967518
1.10223642172524 1.47237768096027
1.10543130990415 1.74161193738737
1.10862619808307 2.06017398697985
1.11182108626198 2.43711398317193
1.11501597444089 2.88314391155245
1.11821086261981 3.41094284222195
1.12140575079872 4.03551824217186
1.12460063897764 4.77463363633467
1.12779552715655 5.64931479206374
1.13099041533546 6.68444883312323
1.13418530351438 7.90949332885211
1.13738019169329 9.35931552653648
1.1405750798722 11.0751855883198
1.14376996805112 13.1059520623263
1.14696485623003 15.5094329841929
1.15015974440895 18.3540621153594
1.15335463258786 21.7208370502133
1.15654952076677 25.7056244686656
1.15974440894569 30.4218879139906
1.1629392971246 36.0039154214567
1.16613418530351 42.6106384463872
1.16932907348243 50.4301502366746
1.17252396166134 59.685051531918
1.17571884984026 70.6387748013747
1.17891373801917 83.60306580798
1.18210862619808 98.946833875647
1.185303514377 117.106620751435
1.18849840255591 138.598983465905
1.19169329073482 164.035140371944
1.19488817891374 194.138293079832
1.19808306709265 229.764112070474
1.20127795527157 271.924962442781
1.20447284345048 321.81855099477
1.20766773162939 380.861799552014
1.21086261980831 450.730895565368
1.21405750798722 533.40864354063
1.21725239616613 631.240444600504
1.22044728434505 747.000472031519
1.22364217252396 883.969894663215
1.22683706070288 1046.02933517718
1.23003194888179 1237.76814616285
1.2332268370607 1464.61355378736
1.23642172523962 1732.9832701332
1.23961661341853 2050.4658256707
1.24281150159744 2426.03364076178
1.24600638977636 2870.2947604933
1.24920127795527 3395.79024521539
1.25239616613419 4017.34546899925
1.2555910543131 4752.48506411614
1.25878594249201 5621.92300193197
1.26198083067093 6650.14136686435
1.26517571884984 7866.07381623609
1.26837060702875 9303.9125909101
1.27156549520767 11004.0613270858
1.27476038338658 13014.2599099424
1.2779552715655 15390.9123124274
1.28115015974441 18200.653903897
1.28434504792332 21522.2012424451
1.28753993610224 25448.5350566528
1.29073482428115 30089.4761832427
1.29392971246006 35574.7248989887
1.29712460063898 42057.4466531783
1.30031948881789 49718.5020060302
1.30351437699681 58771.4360025991
1.30670926517572 69468.3627240449
1.30990415335463 82106.9049027599
1.31309904153355 97038.3769046342
1.31629392971246 114677.432821923
1.31948881789137 135513.440767082
1.32268370607029 160123.890748407
1.3258785942492 189190.197960863
1.32907348242811 223516.327370432
1.33226837060703 264050.74078764
1.33546325878594 311912.256187202
1.33865814696486 368420.51315126
1.34185303514377 435131.86070927
1.34504792332268 513881.627706986
1.3482428115016 606833.904899363
1.35143769968051 716540.166615783
1.35463258785942 846008.29323665
1.35782747603834 998783.829883024
1.36102236421725 1179045.63873196
1.36421725239617 1391718.48052557
1.36741214057508 1642605.5048646
1.37060702875399 1938544.15016616
1.37380191693291 2287589.56607057
1.37699680511182 2699230.3892547
1.38019169329073 3184642.54636992
1.38338658146965 3756987.74664323
1.38658146964856 4431764.48668975
1.38977635782748 5227220.75072312
1.39297124600639 6164839.18505307
1.3961661341853 7269907.39686832
1.39936102236422 8572188.22099075
1.40255591054313 10106707.3697693
1.40575079872204 11914678.8951458
1.40894568690096 14044592.4237205
1.41214057507987 16553490.2636893
1.41533546325879 19508467.3301584
1.4185303514377 22988432.5132548
1.42172523961661 27086176.7628324
1.42492012779553 31910800.9493449
1.42811501597444 37590565.6752825
1.43130990415335 44276235.8806691
1.43450479233227 52145005.5723319
1.43769968051118 61405102.6171022
1.4408945686901 72301190.6323361
1.44408945686901 85120705.001526
1.44728434504792 100201283.427022
1.45047923322684 117939478.775758
1.45367412140575 138800973.942552
1.45686900958466 163332555.82493
1.46006389776358 192176149.177982
1.46325878594249 226085262.152863
1.46645367412141 265944254.948177
1.46964856230032 312790912.65315
1.47284345047923 367842884.706679
1.47603833865815 432528648.383633
1.47923322683706 508523764.617573
1.48242811501597 597793323.919326
1.48562300319489 702641631.230447
1.4888178913738 825770354.841226
1.49201277955272 970346570.175183
1.49520766773163 1140082369.14928
};
\addlegendentry{XLPE}
\addplot [very thick, cornflowerblue86180233 ]
table {%
0 0.00259205379331919
0.004 0.00259265091093556
0.008 0.00259444308924027
0.012 0.00259743280650217
0.016 0.00260162419951597
0.02 0.00260702307314369
0.024 0.00261363691371567
0.028 0.00262147490633476
0.032 0.00263054795613936
0.036 0.00264086871359442
0.04 0.00265245160389251
0.044 0.00266531286056063
0.048 0.00267947056338231
0.052 0.0026949446807591
0.056 0.00271175711665048
0.06 0.00272993176224649
0.064 0.0027494945525439
0.068 0.00277047352801336
0.072 0.00279289890156283
0.076 0.00281680313102102
0.08 0.00284222099738429
0.084 0.00286918968909081
0.088 0.00289774889260775
0.092 0.00292794088964002
0.096 0.00295981066129359
0.1 0.00299340599955203
0.104 0.00302877762645244
0.108 0.00306597932137595
0.112 0.00310506805689887
0.116 0.00314610414368364
0.12 0.00318915138492378
0.124 0.00323427724089443
0.128 0.00328155300420016
0.132 0.00333105398635449
0.136 0.00338285971637087
0.14 0.00343705415209429
0.144 0.0034937259050542
0.148 0.00355296847967597
0.152 0.00361488052774769
0.156 0.00367956611910344
0.16 0.00374713502955286
0.164 0.00381770304716098
0.168 0.00389139229806109
0.172 0.00396833159306884
0.176 0.00404865679645687
0.18 0.00413251121834748
0.184 0.00422004603228623
0.188 0.00431142071967272
0.192 0.00440680354284664
0.196 0.00450637204875843
0.2 0.00461031360529461
0.204 0.00471882597248014
0.208 0.00483211791094332
0.212 0.00495040983020482
0.216 0.0050739344795424
0.22 0.00520293768438706
0.224 0.00533767913142665
0.228 0.00547843320583102
0.232 0.00562548988426851
0.236 0.00577915568766059
0.24 0.0059397546979196
0.244 0.00610762964323676
0.248 0.00628314305683561
0.252 0.00646667851448163
0.256 0.00665864195644482
0.26 0.00685946310005091
0.264 0.00706959694943105
0.268 0.00728952540959276
0.272 0.00751975901248973
0.276 0.0077608387633685
0.28 0.00801333811631947
0.284 0.00827786508866306
0.288 0.00855506452456376
0.292 0.00884562051908922
0.296 0.00915025901482595
0.3 0.00946975058413161
0.304 0.00980491341115421
0.308 0.0101566164888879
0.312 0.0105257830477692
0.316 0.0109133942336603
0.32 0.0113204930545186
0.324 0.011748188616632
0.328 0.0121976606730172
0.332 0.0126701645084372
0.336 0.0131670361875232
0.34 0.0136896981946851
0.344 0.0142396654968877
0.348 0.0148185520629734
0.352 0.0154280778760405
0.356 0.0160700764784718
0.36 0.0167465030925571
0.364 0.0174594433633073
0.368 0.0182111227740353
0.372 0.0190039167896095
0.376 0.0198403617870121
0.38 0.0207231668379773
0.384 0.0216552264141027
0.388 0.0226396340909478
0.392 0.0236796973343151
0.396 0.0247789534592024
0.4 0.0259411868598758
0.404 0.0271704476182068
0.408 0.0284710716069176
0.412 0.0298477022147548
0.416 0.0313053138319611
0.42 0.0328492372468187
0.424 0.0344851871176066
0.428 0.0362192916991628
0.432 0.0380581250194885
0.436 0.040008741719619
0.44 0.0420787147894694
0.444 0.0442761764536996
0.448 0.0466098624850275
0.452 0.0490891602480523
0.456 0.0517241608047553
0.46 0.0545257154436656
0.464 0.0575054970285085
0.468 0.060676066599263
0.472 0.0640509456993139
0.476 0.06764469494713
0.48 0.0714729994200675
0.484 0.0755527614719175
0.488 0.0799022016652069
0.492 0.0845409685645587
0.496 0.0894902582092436
0.5 0.0947729441620911
0.504 0.100413719118904
0.508 0.106439249158292
0.512 0.112878341817326
0.516 0.119762129294596
0.52 0.127124268210345
0.524 0.135001157494509
0.528 0.143432176129201
0.532 0.152459942643906
0.536 0.162130598451186
0.54 0.172494117319908
0.544 0.183604643513976
0.548 0.195520861379763
0.552 0.208306399447344
0.556 0.222030272422261
0.56 0.236767364789124
0.564 0.252598960129442
0.568 0.269613320677693
0.572 0.28790632210632
0.576 0.307582149046978
0.58 0.328754057427535
0.584 0.351545210338235
0.588 0.376089594842967
0.592 0.40253302793037
0.596 0.431034260663206
0.6 0.461766190542599
0.604 0.494917193166939
0.608 0.53069258544568
0.612 0.569316233938887
0.616 0.611032323349488
0.62 0.656107301813127
0.624 0.704832021429051
0.628 0.757524094475413
0.632 0.81453048797703
0.636 0.876230381769095
0.64 0.943038317955799
0.644 1.0154076727311
0.648 1.09383448494645
0.652 1.17886167961857
0.656 1.27108372881489
0.66 1.37115179708771
0.664 1.47977942390701
0.668 1.59774880143284
0.672 1.72591771254325
0.676 1.86522720137526
0.68 2.01671005683688
0.684 2.18150019871073
0.688 2.36084306621138
0.692 2.55610712030996
0.696 2.76879658394823
0.7 3.00056555859612
0.704 3.25323367164856
0.708 3.52880342711932
0.712 3.82947945220888
0.716 4.15768985486646
0.72 4.51610993273527
0.724 4.90768850220284
0.728 5.33567714805935
0.732 5.80366272992717
0.736 6.31560352165339
0.74 6.87586940480304
0.744 7.48928658787874
0.748 8.16118737962358
0.752 8.89746560853526
0.756 9.70463835242737
0.76 10.5899147225389
0.764 11.5612725374624
0.768 12.6275438243402
0.772 13.798510199842
0.776 15.0850093130562
0.78 16.499053678496
0.784 18.0539633920854
0.788 19.7645144086904
0.792 21.6471042692474
0.796 23.7199374019624
0.8 26.003232388957
0.804 28.5194538911706
0.808 31.2935722648762
0.812 34.3533542880502
0.816 37.7296888499865
0.82 41.4569519496961
0.824 45.5734159054785
0.828 50.1217083082937
0.832 55.1493269651713
0.836 60.7092178871706
0.84 66.8604242922506
0.844 73.6688156315261
0.848 81.2079068245389
0.852 89.5597792244413
0.856 98.8161163492334
0.86 109.079369135239
0.864 120.464067422297
0.868 133.098296599167
0.872 147.125360859343
0.876 162.705657384453
0.88 180.018789032857
0.884 199.265946820605
0.888 220.67259770387
0.892 244.491517978929
0.896 271.006218090835
0.9 300.534810880337
0.904 333.434382409406
0.908 370.105932613968
0.912 410.999962281503
0.916 456.622793405638
0.92 507.543722018189
0.924 564.403116358492
0.928 627.921588958506
0.932 698.910389186589
0.936 778.28318333095
0.94 867.069412793903
0.944 966.429447843971
0.948 1077.6717851348
0.952 1202.2725724229
0.956 1341.89778426234
0.96 1498.42841868722
0.964 1673.98913788985
0.968 1870.98083667858
0.972 2092.11769222064
0.976 2340.46932859122
0.98 2619.50882151427
0.984 2933.16737418769
0.988 3285.89661631037
0.992 3682.73961776264
0.996 4129.41186860462
1 4632.39366134713
1.004 5199.0355235163
1.008 5837.67859265817
1.012 6557.7921070662
1.016 7370.13050940026
1.02 8286.91303364294
1.024 9322.02907620995
1.028 10491.2731484149
1.032 11812.6137802372
1.036 13306.5014064573
1.04 14996.221029633
1.044 16908.2963362728
1.048 19072.952961684
1.052 21524.6497771698
1.056 24302.6884358595
1.06 27451.9129899759
1.064 31023.5132170526
1.068 35075.9474054242
1.072 39676.0027966342
1.076 44900.0147183622
1.08 50835.2687290074
1.084 57581.613907613
1.088 65253.3198459528
1.092 73981.2150330974
1.096 83915.1502831003
1.1 95226.8377794277
1.104 108113.124354113
1.108 122799.766970601
1.112 139545.789253385
1.116 158648.510557974
1.12 180449.353797317
1.124 205340.555381823
1.128 233772.920594709
1.132 266264.790986774
1.136 303412.417490426
1.14 345901.964572692
1.144 394523.407633593
1.148 450186.628904389
1.152 513940.067358395
1.156 586992.336846907
1.16 670737.295258581
1.164 766783.127670102
1.168 876986.100202105
1.172 1003489.75096023
1.176 1148770.41278186
1.18 1315690.11276337
1.184 1507558.06952429
1.188 1728202.21535878
1.192 1982052.41211703
1.196 2274237.31307265
1.2 2610697.15550183
1.204 2998315.15887519
1.208 3445070.66164003
1.212 3960217.66758462
1.216 4554493.10492814
1.22 5240359.84434137
1.224 6032290.39588197
1.228 6947098.23273595
1.232 8004324.8993634
1.236 9226692.48588295
1.24 10640632.7280118
1.244 12276905.9684356
1.248 14171325.5453773
1.252 16365605.9216336
1.256 18908356.1086172
1.26 21856243.7652614
1.264 25275359.8680816
1.268 29242819.1832636
1.272 33848638.075115
1.276 39197938.6365749
1.28 45413536.9393237
1.284 52638983.626
1.288 61042137.4054352
1.292 70819366.6207676
1.296 82200491.3646791
1.3 95454599.1214462
1.304 110896891.224057
1.308 128896746.243434
1.312 149887220.630198
1.316 174376247.52609
1.32 202959842.866695
1.324 236337685.158547
1.328 275331503.361622
1.332 320906788.20735
1.336 374198438.49898
1.34 436541068.42266
1.344 509504838.166464
1.348 594937832.414148
1.352 695016204.590329
1.356 812303535.116468
1.36 949821126.614745
1.364 1111131286.62016
1.368 1300436039.29669
1.372 1522694174.32626
1.376 1783760098.45309
1.38 2090548620.99763
1.384 2451230600.46192
1.388 2875465330.89631
1.392 3374676684.93091
1.396 3962381392.4836
1.4 4654579464.84264
1.404 5470218726.8484
1.408 6431747759.9515
1.412 7565774363.85112
1.416 8903850007.96127
1.42 10483404779.0344
1.424 12348862173.9589
1.428 14552968901.0903
1.432 17158381837.3114
1.436 20239562680.0613
1.44 23885040921.9338
1.444 28200117908.0428
1.448 33310099333.4682
1.452 39364161107.615
1.456 46539974668.3141
1.46 55049243313.6165
1.464 65144331832.1385
1.468 77126208741.1947
1.472 91353965103.1148
1.476 108256227779.721
1.48 128344850036.988
1.484 152231340971.222
1.488 180646590138.731
1.492 214464558484.2
1.496 254730745368.111
1.5 302696409278.555
1.504 359859722858.732
};
\addlegendentry{PILC}
\end{axis}

\end{tikzpicture}

%% file: Distribution_change.tex
\begin{tikzpicture}

\definecolor{cornflowerblue86180233}{RGB}{86,180,233}
\definecolor{darkslategray38}{RGB}{38,38,38}
\definecolor{lightgray204}{RGB}{204,204,204}
\definecolor{orange2301590}{RGB}{230,159,0}

\begin{axis}[height=0.13\textheight,width=0.75\columnwidth,
scale only axis,
axis line style={darkslategray38},
legend cell align={left},
legend columns=2,
legend style={fill opacity=0.8, draw opacity=1, text opacity=1, draw=lightgray204},
tick align=outside,
x grid style={lightgray204},
xlabel=\textcolor{darkslategray38}{Time $t$ (years)},
xmajorticks=true,
xmin=0, xmax=100,
xtick style={color=darkslategray38},
y grid style={lightgray204},
ylabel=\textcolor{darkslategray38}{$|\Omega^C|$},
ytick={0.5,1},
ymajorticks=true,
ymin=0, ymax=1.3,
ytick style={color=darkslategray38}
]
\path[
  draw=white,
  pattern=north east lines,            
  pattern color=orange2301590          
]
(axis cs:0,1)
--(axis cs:0,0.25)
--(axis cs:1,0.226179642869932)
--(axis cs:2,0.204362438382824)
--(axis cs:3,0.184422752344599)
--(axis cs:4,0.16623617679425)
--(axis cs:5,0.149680618012522)
--(axis cs:6,0.134637231962974)
--(axis cs:7,0.120991208122156)
--(axis cs:8,0.108632405763821)
--(axis cs:9,0.0974558494204998)
--(axis cs:10,0.0873620924264698)
--(axis cs:11,0.0782574591398911)
--(axis cs:12,0.0700541776572006)
--(axis cs:13,0.062670415592891)
--(axis cs:14,0.0560302318385415)
--(axis cs:15,0.0500634571817465)
--(axis cs:16,0.044705516310541)
--(axis cs:17,0.0398972031079133)
--(axis cs:18,0.0355844203111265)
--(axis cs:19,0.0317178936277035)
--(axis cs:20,0.0282528693168181)
--(axis cs:21,0.0251488031094926)
--(axis cs:22,0.022369047195628)
--(axis cs:23,0.0198805408861233)
--(axis cs:24,0.0176535094929728)
--(axis cs:25,0.0156611749812344)
--(axis cs:26,0.0138794810495989)
--(axis cs:27,0.0122868345004799)
--(axis cs:28,0.0108638640703577)
--(axis cs:29,0.00959319730637987)
--(axis cs:30,0.00845925559220915)
--(axis cs:31,0.00744806703832728)
--(axis cs:32,0.0065470966510482)
--(axis cs:33,0.0057450929707348)
--(axis cs:34,0.0050319502129629)
--(axis cs:35,0.00439858484638425)
--(axis cs:36,0.00383682548792899)
--(axis cs:37,0.00333931498055122)
--(axis cs:38,0.00289942353264069)
--(axis cs:39,0.00251117183418803)
--(axis cs:40,0.00216916311654876)
--(axis cs:41,0.00186852318499978)
--(axis cs:42,0.00160484752201206)
--(axis cs:43,0.00137415463097129)
--(axis cs:44,0.00117284486246151)
--(axis cs:45,0.000997664036361475)
--(axis cs:46,0.000845671241628747)
--(axis cs:47,0.00071421026094777)
--(axis cs:48,0.000600884128920796)
--(axis cs:49,0.000503532389957708)
--(axis cs:50,0.000420210675413982)
--(axis cs:51,0.00034917226888008)
--(axis cs:52,0.000288851373940696)
--(axis cs:53,0.000237847840317514)
--(axis cs:54,0.000194913142200425)
--(axis cs:55,0.000158937436859731)
--(axis cs:56,0.000128937562397547)
--(axis cs:57,0.000104045860807219)
--(axis cs:58,8.34997364252657e-05)
--(axis cs:59,6.66318804456767e-05)
--(axis cs:60,5.28611095019098e-05)
--(axis cs:61,4.16837805154215e-05)
--(axis cs:62,3.26657552047653e-05)
--(axis cs:63,2.54348960318335e-05)
--(axis cs:64,1.96740811613797e-05)
--(axis cs:65,1.51147294985333e-05)
--(axis cs:66,1.1530828355496e-05)
--(axis cs:67,8.73345611994628e-06)
--(axis cs:68,6.56579080766716e-06)
--(axis cs:69,4.8985929381508e-06)
--(axis cs:70,3.62614812230287e-06)
--(axis cs:71,2.66265142090094e-06)
--(axis cs:72,1.93901221158378e-06)
--(axis cs:73,1.40005523720834e-06)
--(axis cs:74,1.00209089494943e-06)
--(axis cs:75,7.10825804748324e-07)
--(axis cs:76,4.99583354219867e-07)
--(axis cs:77,3.47803289597231e-07)
--(axis cs:78,2.39789496924052e-07)
--(axis cs:79,1.63675843633632e-07)
--(axis cs:80,1.10581246767729e-07)
--(axis cs:81,7.3926898440961e-08)
--(axis cs:82,4.88906983296508e-08)
--(axis cs:83,3.19763006973406e-08)
--(axis cs:84,2.06766681099801e-08)
--(axis cs:85,1.32145344009327e-08)
--(axis cs:86,8.34462919023971e-09)
--(axis cs:87,5.20483610570242e-09)
--(axis cs:88,3.20559563130405e-09)
--(axis cs:89,1.94878764779402e-09)
--(axis cs:90,1.16902069438689e-09)
--(axis cs:91,6.91710974011825e-10)
--(axis cs:92,4.03561378000889e-10)
--(axis cs:93,2.32064733664005e-10)
--(axis cs:94,1.31476947121296e-10)
--(axis cs:95,7.33586035886846e-11)
--(axis cs:96,4.02927512363719e-11)
--(axis cs:97,2.17762755059187e-11)
--(axis cs:98,1.15749725951559e-11)
--(axis cs:99,6.04821035968738e-12)
--(axis cs:99,0.999999999999998)
--(axis cs:99,0.999999999999998)
--(axis cs:98,0.999999999999999)
--(axis cs:97,1)
--(axis cs:96,1)
--(axis cs:95,1)
--(axis cs:94,1)
--(axis cs:93,1)
--(axis cs:92,1)
--(axis cs:91,1)
--(axis cs:90,1)
--(axis cs:89,1)
--(axis cs:88,1)
--(axis cs:87,1)
--(axis cs:86,1)
--(axis cs:85,1)
--(axis cs:84,1)
--(axis cs:83,1)
--(axis cs:82,1)
--(axis cs:81,1)
--(axis cs:80,1)
--(axis cs:79,0.999999999999999)
--(axis cs:78,0.999999999999999)
--(axis cs:77,1)
--(axis cs:76,0.999999999999999)
--(axis cs:75,1)
--(axis cs:74,0.999999999999999)
--(axis cs:73,0.999999999999999)
--(axis cs:72,0.999999999999999)
--(axis cs:71,1)
--(axis cs:70,1)
--(axis cs:69,0.999999999999999)
--(axis cs:68,1)
--(axis cs:67,1)
--(axis cs:66,1)
--(axis cs:65,1)
--(axis cs:64,1)
--(axis cs:63,1)
--(axis cs:62,1)
--(axis cs:61,1)
--(axis cs:60,1)
--(axis cs:59,1)
--(axis cs:58,1)
--(axis cs:57,1)
--(axis cs:56,1)
--(axis cs:55,1)
--(axis cs:54,1)
--(axis cs:53,1)
--(axis cs:52,1)
--(axis cs:51,1)
--(axis cs:50,1)
--(axis cs:49,1)
--(axis cs:48,1)
--(axis cs:47,1)
--(axis cs:46,1)
--(axis cs:45,1)
--(axis cs:44,1)
--(axis cs:43,1)
--(axis cs:42,1)
--(axis cs:41,1)
--(axis cs:40,1)
--(axis cs:39,0.999999999999999)
--(axis cs:38,1)
--(axis cs:37,1)
--(axis cs:36,1)
--(axis cs:35,1)
--(axis cs:34,1)
--(axis cs:33,0.999999999999999)
--(axis cs:32,1)
--(axis cs:31,1)
--(axis cs:30,1)
--(axis cs:29,1)
--(axis cs:28,1)
--(axis cs:27,1)
--(axis cs:26,1)
--(axis cs:25,1)
--(axis cs:24,1)
--(axis cs:23,1)
--(axis cs:22,1)
--(axis cs:21,1)
--(axis cs:20,1)
--(axis cs:19,1)
--(axis cs:18,1)
--(axis cs:17,1)
--(axis cs:16,1)
--(axis cs:15,1)
--(axis cs:14,1)
--(axis cs:13,1)
--(axis cs:12,1)
--(axis cs:11,1)
--(axis cs:10,1)
--(axis cs:9,1)
--(axis cs:8,1)
--(axis cs:7,1)
--(axis cs:6,1)
--(axis cs:5,0.999999999999999)
--(axis cs:4,1)
--(axis cs:3,1)
--(axis cs:2,1)
--(axis cs:1,1)
--(axis cs:0,1)
--cycle;
\addlegendimage{area legend, draw=white, pattern=north east lines, pattern color=orange2301590}
\addlegendentry{XLPE}
\path [draw=white, fill=cornflowerblue86180233]
(axis cs:0,0.25)
--(axis cs:0,0)
--(axis cs:1,0)
--(axis cs:2,0)
--(axis cs:3,0)
--(axis cs:4,0)
--(axis cs:5,0)
--(axis cs:6,0)
--(axis cs:7,0)
--(axis cs:8,0)
--(axis cs:9,0)
--(axis cs:10,0)
--(axis cs:11,0)
--(axis cs:12,0)
--(axis cs:13,0)
--(axis cs:14,0)
--(axis cs:15,0)
--(axis cs:16,0)
--(axis cs:17,0)
--(axis cs:18,0)
--(axis cs:19,0)
--(axis cs:20,0)
--(axis cs:21,0)
--(axis cs:22,0)
--(axis cs:23,0)
--(axis cs:24,0)
--(axis cs:25,0)
--(axis cs:26,0)
--(axis cs:27,0)
--(axis cs:28,0)
--(axis cs:29,0)
--(axis cs:30,0)
--(axis cs:31,0)
--(axis cs:32,0)
--(axis cs:33,0)
--(axis cs:34,0)
--(axis cs:35,0)
--(axis cs:36,0)
--(axis cs:37,0)
--(axis cs:38,0)
--(axis cs:39,0)
--(axis cs:40,0)
--(axis cs:41,0)
--(axis cs:42,0)
--(axis cs:43,0)
--(axis cs:44,0)
--(axis cs:45,0)
--(axis cs:46,0)
--(axis cs:47,0)
--(axis cs:48,0)
--(axis cs:49,0)
--(axis cs:50,0)
--(axis cs:51,0)
--(axis cs:52,0)
--(axis cs:53,0)
--(axis cs:54,0)
--(axis cs:55,0)
--(axis cs:56,0)
--(axis cs:57,0)
--(axis cs:58,0)
--(axis cs:59,0)
--(axis cs:60,0)
--(axis cs:61,0)
--(axis cs:62,0)
--(axis cs:63,0)
--(axis cs:64,0)
--(axis cs:65,0)
--(axis cs:66,0)
--(axis cs:67,0)
--(axis cs:68,0)
--(axis cs:69,0)
--(axis cs:70,0)
--(axis cs:71,0)
--(axis cs:72,0)
--(axis cs:73,0)
--(axis cs:74,0)
--(axis cs:75,0)
--(axis cs:76,0)
--(axis cs:77,0)
--(axis cs:78,0)
--(axis cs:79,0)
--(axis cs:80,0)
--(axis cs:81,0)
--(axis cs:82,0)
--(axis cs:83,0)
--(axis cs:84,0)
--(axis cs:85,0)
--(axis cs:86,0)
--(axis cs:87,0)
--(axis cs:88,0)
--(axis cs:89,0)
--(axis cs:90,0)
--(axis cs:91,0)
--(axis cs:92,0)
--(axis cs:93,0)
--(axis cs:94,0)
--(axis cs:95,0)
--(axis cs:96,0)
--(axis cs:97,0)
--(axis cs:98,0)
--(axis cs:99,0)
--(axis cs:99,6.04821035968738e-12)
--(axis cs:99,6.04821035968738e-12)
--(axis cs:98,1.15749725951559e-11)
--(axis cs:97,2.17762755059187e-11)
--(axis cs:96,4.02927512363719e-11)
--(axis cs:95,7.33586035886846e-11)
--(axis cs:94,1.31476947121296e-10)
--(axis cs:93,2.32064733664005e-10)
--(axis cs:92,4.03561378000889e-10)
--(axis cs:91,6.91710974011825e-10)
--(axis cs:90,1.16902069438689e-09)
--(axis cs:89,1.94878764779402e-09)
--(axis cs:88,3.20559563130405e-09)
--(axis cs:87,5.20483610570242e-09)
--(axis cs:86,8.34462919023971e-09)
--(axis cs:85,1.32145344009327e-08)
--(axis cs:84,2.06766681099801e-08)
--(axis cs:83,3.19763006973406e-08)
--(axis cs:82,4.88906983296508e-08)
--(axis cs:81,7.3926898440961e-08)
--(axis cs:80,1.10581246767729e-07)
--(axis cs:79,1.63675843633632e-07)
--(axis cs:78,2.39789496924052e-07)
--(axis cs:77,3.47803289597231e-07)
--(axis cs:76,4.99583354219867e-07)
--(axis cs:75,7.10825804748324e-07)
--(axis cs:74,1.00209089494943e-06)
--(axis cs:73,1.40005523720834e-06)
--(axis cs:72,1.93901221158378e-06)
--(axis cs:71,2.66265142090094e-06)
--(axis cs:70,3.62614812230287e-06)
--(axis cs:69,4.8985929381508e-06)
--(axis cs:68,6.56579080766716e-06)
--(axis cs:67,8.73345611994628e-06)
--(axis cs:66,1.1530828355496e-05)
--(axis cs:65,1.51147294985333e-05)
--(axis cs:64,1.96740811613797e-05)
--(axis cs:63,2.54348960318335e-05)
--(axis cs:62,3.26657552047653e-05)
--(axis cs:61,4.16837805154215e-05)
--(axis cs:60,5.28611095019098e-05)
--(axis cs:59,6.66318804456767e-05)
--(axis cs:58,8.34997364252657e-05)
--(axis cs:57,0.000104045860807219)
--(axis cs:56,0.000128937562397547)
--(axis cs:55,0.000158937436859731)
--(axis cs:54,0.000194913142200425)
--(axis cs:53,0.000237847840317514)
--(axis cs:52,0.000288851373940696)
--(axis cs:51,0.00034917226888008)
--(axis cs:50,0.000420210675413982)
--(axis cs:49,0.000503532389957708)
--(axis cs:48,0.000600884128920796)
--(axis cs:47,0.00071421026094777)
--(axis cs:46,0.000845671241628747)
--(axis cs:45,0.000997664036361475)
--(axis cs:44,0.00117284486246151)
--(axis cs:43,0.00137415463097129)
--(axis cs:42,0.00160484752201206)
--(axis cs:41,0.00186852318499978)
--(axis cs:40,0.00216916311654876)
--(axis cs:39,0.00251117183418803)
--(axis cs:38,0.00289942353264069)
--(axis cs:37,0.00333931498055122)
--(axis cs:36,0.00383682548792899)
--(axis cs:35,0.00439858484638425)
--(axis cs:34,0.0050319502129629)
--(axis cs:33,0.0057450929707348)
--(axis cs:32,0.0065470966510482)
--(axis cs:31,0.00744806703832728)
--(axis cs:30,0.00845925559220915)
--(axis cs:29,0.00959319730637987)
--(axis cs:28,0.0108638640703577)
--(axis cs:27,0.0122868345004799)
--(axis cs:26,0.0138794810495989)
--(axis cs:25,0.0156611749812344)
--(axis cs:24,0.0176535094929728)
--(axis cs:23,0.0198805408861233)
--(axis cs:22,0.022369047195628)
--(axis cs:21,0.0251488031094926)
--(axis cs:20,0.0282528693168181)
--(axis cs:19,0.0317178936277035)
--(axis cs:18,0.0355844203111265)
--(axis cs:17,0.0398972031079133)
--(axis cs:16,0.044705516310541)
--(axis cs:15,0.0500634571817465)
--(axis cs:14,0.0560302318385415)
--(axis cs:13,0.062670415592891)
--(axis cs:12,0.0700541776572006)
--(axis cs:11,0.0782574591398911)
--(axis cs:10,0.0873620924264698)
--(axis cs:9,0.0974558494204998)
--(axis cs:8,0.108632405763821)
--(axis cs:7,0.120991208122156)
--(axis cs:6,0.134637231962974)
--(axis cs:5,0.149680618012522)
--(axis cs:4,0.16623617679425)
--(axis cs:3,0.184422752344599)
--(axis cs:2,0.204362438382824)
--(axis cs:1,0.226179642869932)
--(axis cs:0,0.25)
--cycle;
\addlegendimage{area legend, draw=white, fill=cornflowerblue86180233}
\addlegendentry{PILC}
\end{axis}

\end{tikzpicture}

%% file: Failures_change.tex
\begin{tikzpicture}

\definecolor{cornflowerblue86180233}{RGB}{86,180,233}
\definecolor{darkslategray38}{RGB}{38,38,38}
\definecolor{lightgray204}{RGB}{204,204,204}
\definecolor{orange2301590}{RGB}{230,159,0}

\begin{axis}[height=0.13\textheight,width=0.75\columnwidth,
scale only axis,
axis line style={darkslategray38},
legend cell align={left},
legend columns=2,
legend style={fill opacity=0.8, draw opacity=1, text opacity=1, draw=lightgray204},
tick align=outside,
x grid style={lightgray204},
xlabel=\textcolor{darkslategray38}{Time $t$ (years)},
xmajorticks=true,
xmin=0, xmax=100,
xtick style={color=darkslategray38},
y grid style={lightgray204},
ylabel=\textcolor{darkslategray38}{Failure rate $\frac{F}{|\Omega^C|}$},
ymajorticks=true,
ymin=0, ymax=0.03,
ytick style={color=darkslategray38}
]
\path[
  draw=white,
  pattern=north east lines,            
  pattern color=orange2301590          
]
(axis cs:0,0)
--(axis cs:0,0)
--(axis cs:1,0.0238203571300677)
--(axis cs:2,0.0218172044871087)
--(axis cs:3,0.0199396860382244)
--(axis cs:4,0.0181865755503488)
--(axis cs:5,0.0165555587817282)
--(axis cs:6,0.0150433860495485)
--(axis cs:7,0.0136460238408174)
--(axis cs:8,0.0123588023583358)
--(axis cs:9,0.0111765563433207)
--(axis cs:10,0.01009375699403)
--(axis cs:11,0.0091046332865787)
--(axis cs:12,0.00820328148269062)
--(axis cs:13,0.00738376206430955)
--(axis cs:14,0.00664018375434956)
--(axis cs:15,0.00596677465679492)
--(axis cs:16,0.00535794087120548)
--(axis cs:17,0.00480831320262772)
--(axis cs:18,0.00431278279678684)
--(axis cs:19,0.00386652668342294)
--(axis cs:20,0.0034650243108854)
--(axis cs:21,0.00310406620732552)
--(axis cs:22,0.00277975591386454)
--(axis cs:23,0.00248850630950469)
--(axis cs:24,0.00222703139315052)
--(axis cs:25,0.00199233451173837)
--(axis cs:26,0.00178169393163552)
--(axis cs:27,0.00159264654911902)
--(axis cs:28,0.00142297043012225)
--(axis cs:29,0.00127066676397779)
--(axis cs:30,0.00113394171417072)
--(axis cs:31,0.00101118855388188)
--(axis cs:32,0.000900970387279085)
--(axis cs:33,0.000802003680313395)
--(axis cs:34,0.000713142757771901)
--(axis cs:35,0.000633365366578647)
--(axis cs:36,0.000561759358455258)
--(axis cs:37,0.000497510507377774)
--(axis cs:38,0.000439891447910533)
--(axis cs:39,0.000388251698452656)
--(axis cs:40,0.000342008717639273)
--(axis cs:41,0.000300639931548976)
--(axis cs:42,0.00026367566298772)
--(axis cs:43,0.000230692891040776)
--(axis cs:44,0.000201309768509779)
--(axis cs:45,0.000175180826100032)
--(axis cs:46,0.000151992794732729)
--(axis cs:47,0.000131460980680977)
--(axis cs:48,0.000113326132026974)
--(axis cs:49,9.73517389630876e-05)
--(axis cs:50,8.33217145437264e-05)
--(axis cs:51,7.10384065339018e-05)
--(axis cs:52,6.03208949393837e-05)
--(axis cs:53,5.1003533623182e-05)
--(axis cs:54,4.29346981170892e-05)
--(axis cs:55,3.59757053406941e-05)
--(axis cs:56,2.99998744621834e-05)
--(axis cs:57,2.4891701590328e-05)
--(axis cs:58,2.05461243819535e-05)
--(axis cs:59,1.68678559795891e-05)
--(axis cs:60,1.37707709437669e-05)
--(axis cs:61,1.11773289864883e-05)
--(axis cs:62,9.01802531065626e-06)
--(axis cs:63,7.23085917293167e-06)
--(axis cs:64,5.76081487045383e-06)
--(axis cs:65,4.55935166284641e-06)
--(axis cs:66,3.58390114303728e-06)
--(axis cs:67,2.79737223554976e-06)
--(axis cs:68,2.16766531227911e-06)
--(axis cs:69,1.66719786951637e-06)
--(axis cs:70,1.27244481584793e-06)
--(axis cs:71,9.63496701401931e-07)
--(axis cs:72,7.23639209317164e-07)
--(axis cs:73,5.38956974375442e-07)
--(axis cs:74,3.97964342258907e-07)
--(axis cs:75,2.91265090201105e-07)
--(axis cs:76,2.11242450528457e-07)
--(axis cs:77,1.51780064622636e-07)
--(axis cs:78,1.08013792673179e-07)
--(axis cs:79,7.61136532904206e-08)
--(axis cs:80,5.30945968659032e-08)
--(axis cs:81,3.66543483267676e-08)
--(axis cs:82,2.50362001113103e-08)
--(axis cs:83,1.69143976323101e-08)
--(axis cs:84,1.12996325873605e-08)
--(axis cs:85,7.46213370904739e-09)
--(axis cs:86,4.86990521069296e-09)
--(axis cs:87,3.13979308453729e-09)
--(axis cs:88,1.99924047439837e-09)
--(axis cs:89,1.25680798351003e-09)
--(axis cs:90,7.79766953407135e-10)
--(axis cs:91,4.77309720375061e-10)
--(axis cs:92,2.88149596010936e-10)
--(axis cs:93,1.71496644336884e-10)
--(axis cs:94,1.0058778654271e-10)
--(axis cs:95,5.81183435326112e-11)
--(axis cs:96,3.30658523523127e-11)
--(axis cs:97,1.85164757304533e-11)
--(axis cs:98,1.02013029107628e-11)
--(axis cs:99,5.52676223546851e-12)
--(axis cs:99,0.0179049387732839)
--(axis cs:99,0.0179049387732839)
--(axis cs:98,0.0178912441751571)
--(axis cs:97,0.0178772442389864)
--(axis cs:96,0.0178630430660335)
--(axis cs:95,0.0178487539018749)
--(axis cs:94,0.0178344989928307)
--(axis cs:93,0.0178204093491151)
--(axis cs:92,0.0178066244004341)
--(axis cs:91,0.0177932915293222)
--(axis cs:90,0.0177805654675007)
--(axis cs:89,0.0177686075410165)
--(axis cs:88,0.017757584750923)
--(axis cs:87,0.0177476686778324)
--(axis cs:86,0.0177390342008113)
--(axis cs:85,0.0177318580238119)
--(axis cs:84,0.0177263170061072)
--(axis cs:83,0.0177225862969832)
--(axis cs:82,0.0177208372791855)
--(axis cs:81,0.0177212353302214)
--(axis cs:80,0.0177239374155015)
--(axis cs:79,0.0177290895323378)
--(axis cs:78,0.0177368240288738)
--(axis cs:77,0.017747256826978)
--(axis cs:76,0.0177604845828361)
--(axis cs:75,0.0177765818232933)
--(axis cs:74,0.0177955980998063)
--(axis cs:73,0.0178175552050256)
--(axis cs:72,0.0178424444994602)
--(axis cs:71,0.0178702243972839)
--(axis cs:70,0.0179008180610673)
--(axis cs:69,0.0179341113550416)
--(axis cs:68,0.0179699511053935)
--(axis cs:67,0.0180081437140985)
--(axis cs:66,0.0180484541699484)
--(axis cs:65,0.0180906054967997)
--(axis cs:64,0.018134278674747)
--(axis cs:63,0.0181791130650047)
--(axis cs:62,0.0182247073638929)
--(axis cs:61,0.0182706211055709)
--(axis cs:60,0.0183163767271861)
--(axis cs:59,0.0183614622040262)
--(axis cs:58,0.0184053342561995)
--(axis cs:57,0.0184474221224502)
--(axis cs:56,0.0184871318910529)
--(axis cs:55,0.0185238513724329)
--(axis cs:54,0.0185569554933404)
--(axis cs:53,0.0185858121881568)
--(axis cs:52,0.0186097887593449)
--(axis cs:51,0.0186282586762513)
--(axis cs:50,0.0186406087795331)
--(axis cs:49,0.0186462468574881)
--(axis cs:48,0.0186446095606113)
--(axis cs:47,0.0186351706218263)
--(axis cs:46,0.0186174493521308)
--(axis cs:45,0.0185910193848393)
--(axis cs:44,0.0185555176462401)
--(axis cs:43,0.018510653536223)
--(axis cs:42,0.0184562183092058)
--(axis cs:41,0.0183920946533087)
--(axis cs:40,0.0183182664739516)
--(axis cs:39,0.0182348288965535)
--(axis cs:38,0.0181419985113278)
--(axis cs:37,0.0180401238907644)
--(axis cs:36,0.0179296964165656)
--(axis cs:35,0.0178113614567715)
--(axis cs:34,0.0176859299346425)
--(axis cs:33,0.0175543903275412)
--(axis cs:32,0.017417921125457)
--(axis cs:31,0.0172779037637878)
--(axis cs:30,0.0171359360223696)
--(axis cs:29,0.0169938458514257)
--(axis cs:28,0.0168537055441198)
--(axis cs:27,0.0167178461239974)
--(axis cs:26,0.0165888717533645)
--(axis cs:25,0.0164696738955824)
--(axis cs:24,0.0163634448809006)
--(axis cs:23,0.016273690432997)
--(axis cs:22,0.0162042406137741)
--(axis cs:21,0.0161592585399232)
--(axis cs:20,0.0161432461199161)
--(axis cs:19,0.016161045958898)
--(axis cs:18,0.0162178384867141)
--(axis cs:17,0.0163191332870314)
--(axis cs:16,0.0164707535497805)
--(axis cs:15,0.0166788125418291)
--(axis cs:14,0.0169496809988588)
--(axis cs:13,0.0172899443914605)
--(axis cs:12,0.0177063491164163)
--(axis cs:11,0.018205736814788)
--(axis cs:10,0.0187949662250086)
--(axis cs:9,0.0194808222429273)
--(axis cs:8,0.0202699121805813)
--(axis cs:7,0.0211685495875919)
--(axis cs:6,0.0221826264168825)
--(axis cs:5,0.0233174747703188)
--(axis cs:4,0.0245777199374273)
--(axis cs:3,0.0259671269264548)
--(axis cs:2,0.0274884431643392)
--(axis cs:1,0.0291432404915963)
--(axis cs:0,0)
--cycle;
\addlegendimage{area legend, draw=white, pattern=north east lines, pattern color=orange2301590}
\addlegendentry{XLPE}
\path [draw=white, fill=cornflowerblue86180233]
(axis cs:0,0)
--(axis cs:0,0)
--(axis cs:1,0)
--(axis cs:2,0)
--(axis cs:3,0)
--(axis cs:4,0)
--(axis cs:5,0)
--(axis cs:6,0)
--(axis cs:7,0)
--(axis cs:8,0)
--(axis cs:9,0)
--(axis cs:10,0)
--(axis cs:11,0)
--(axis cs:12,0)
--(axis cs:13,0)
--(axis cs:14,0)
--(axis cs:15,0)
--(axis cs:16,0)
--(axis cs:17,0)
--(axis cs:18,0)
--(axis cs:19,0)
--(axis cs:20,0)
--(axis cs:21,0)
--(axis cs:22,0)
--(axis cs:23,0)
--(axis cs:24,0)
--(axis cs:25,0)
--(axis cs:26,0)
--(axis cs:27,0)
--(axis cs:28,0)
--(axis cs:29,0)
--(axis cs:30,0)
--(axis cs:31,0)
--(axis cs:32,0)
--(axis cs:33,0)
--(axis cs:34,0)
--(axis cs:35,0)
--(axis cs:36,0)
--(axis cs:37,0)
--(axis cs:38,0)
--(axis cs:39,0)
--(axis cs:40,0)
--(axis cs:41,0)
--(axis cs:42,0)
--(axis cs:43,0)
--(axis cs:44,0)
--(axis cs:45,0)
--(axis cs:46,0)
--(axis cs:47,0)
--(axis cs:48,0)
--(axis cs:49,0)
--(axis cs:50,0)
--(axis cs:51,0)
--(axis cs:52,0)
--(axis cs:53,0)
--(axis cs:54,0)
--(axis cs:55,0)
--(axis cs:56,0)
--(axis cs:57,0)
--(axis cs:58,0)
--(axis cs:59,0)
--(axis cs:60,0)
--(axis cs:61,0)
--(axis cs:62,0)
--(axis cs:63,0)
--(axis cs:64,0)
--(axis cs:65,0)
--(axis cs:66,0)
--(axis cs:67,0)
--(axis cs:68,0)
--(axis cs:69,0)
--(axis cs:70,0)
--(axis cs:71,0)
--(axis cs:72,0)
--(axis cs:73,0)
--(axis cs:74,0)
--(axis cs:75,0)
--(axis cs:76,0)
--(axis cs:77,0)
--(axis cs:78,0)
--(axis cs:79,0)
--(axis cs:80,0)
--(axis cs:81,0)
--(axis cs:82,0)
--(axis cs:83,0)
--(axis cs:84,0)
--(axis cs:85,0)
--(axis cs:86,0)
--(axis cs:87,0)
--(axis cs:88,0)
--(axis cs:89,0)
--(axis cs:90,0)
--(axis cs:91,0)
--(axis cs:92,0)
--(axis cs:93,0)
--(axis cs:94,0)
--(axis cs:95,0)
--(axis cs:96,0)
--(axis cs:97,0)
--(axis cs:98,0)
--(axis cs:99,0)
--(axis cs:99,5.52676223546851e-12)
--(axis cs:99,5.52676223546851e-12)
--(axis cs:98,1.02013029107628e-11)
--(axis cs:97,1.85164757304533e-11)
--(axis cs:96,3.30658523523127e-11)
--(axis cs:95,5.81183435326112e-11)
--(axis cs:94,1.0058778654271e-10)
--(axis cs:93,1.71496644336884e-10)
--(axis cs:92,2.88149596010936e-10)
--(axis cs:91,4.77309720375061e-10)
--(axis cs:90,7.79766953407135e-10)
--(axis cs:89,1.25680798351003e-09)
--(axis cs:88,1.99924047439837e-09)
--(axis cs:87,3.13979308453729e-09)
--(axis cs:86,4.86990521069296e-09)
--(axis cs:85,7.46213370904739e-09)
--(axis cs:84,1.12996325873605e-08)
--(axis cs:83,1.69143976323101e-08)
--(axis cs:82,2.50362001113103e-08)
--(axis cs:81,3.66543483267676e-08)
--(axis cs:80,5.30945968659032e-08)
--(axis cs:79,7.61136532904206e-08)
--(axis cs:78,1.08013792673179e-07)
--(axis cs:77,1.51780064622636e-07)
--(axis cs:76,2.11242450528457e-07)
--(axis cs:75,2.91265090201105e-07)
--(axis cs:74,3.97964342258907e-07)
--(axis cs:73,5.38956974375442e-07)
--(axis cs:72,7.23639209317164e-07)
--(axis cs:71,9.63496701401931e-07)
--(axis cs:70,1.27244481584793e-06)
--(axis cs:69,1.66719786951637e-06)
--(axis cs:68,2.16766531227911e-06)
--(axis cs:67,2.79737223554976e-06)
--(axis cs:66,3.58390114303728e-06)
--(axis cs:65,4.55935166284641e-06)
--(axis cs:64,5.76081487045383e-06)
--(axis cs:63,7.23085917293167e-06)
--(axis cs:62,9.01802531065626e-06)
--(axis cs:61,1.11773289864883e-05)
--(axis cs:60,1.37707709437669e-05)
--(axis cs:59,1.68678559795891e-05)
--(axis cs:58,2.05461243819535e-05)
--(axis cs:57,2.4891701590328e-05)
--(axis cs:56,2.99998744621834e-05)
--(axis cs:55,3.59757053406941e-05)
--(axis cs:54,4.29346981170892e-05)
--(axis cs:53,5.1003533623182e-05)
--(axis cs:52,6.03208949393837e-05)
--(axis cs:51,7.10384065339018e-05)
--(axis cs:50,8.33217145437264e-05)
--(axis cs:49,9.73517389630876e-05)
--(axis cs:48,0.000113326132026974)
--(axis cs:47,0.000131460980680977)
--(axis cs:46,0.000151992794732729)
--(axis cs:45,0.000175180826100032)
--(axis cs:44,0.000201309768509779)
--(axis cs:43,0.000230692891040776)
--(axis cs:42,0.00026367566298772)
--(axis cs:41,0.000300639931548976)
--(axis cs:40,0.000342008717639273)
--(axis cs:39,0.000388251698452656)
--(axis cs:38,0.000439891447910533)
--(axis cs:37,0.000497510507377774)
--(axis cs:36,0.000561759358455258)
--(axis cs:35,0.000633365366578647)
--(axis cs:34,0.000713142757771901)
--(axis cs:33,0.000802003680313395)
--(axis cs:32,0.000900970387279085)
--(axis cs:31,0.00101118855388188)
--(axis cs:30,0.00113394171417072)
--(axis cs:29,0.00127066676397779)
--(axis cs:28,0.00142297043012225)
--(axis cs:27,0.00159264654911902)
--(axis cs:26,0.00178169393163552)
--(axis cs:25,0.00199233451173837)
--(axis cs:24,0.00222703139315052)
--(axis cs:23,0.00248850630950469)
--(axis cs:22,0.00277975591386454)
--(axis cs:21,0.00310406620732552)
--(axis cs:20,0.0034650243108854)
--(axis cs:19,0.00386652668342294)
--(axis cs:18,0.00431278279678684)
--(axis cs:17,0.00480831320262772)
--(axis cs:16,0.00535794087120548)
--(axis cs:15,0.00596677465679492)
--(axis cs:14,0.00664018375434956)
--(axis cs:13,0.00738376206430955)
--(axis cs:12,0.00820328148269062)
--(axis cs:11,0.0091046332865787)
--(axis cs:10,0.01009375699403)
--(axis cs:9,0.0111765563433207)
--(axis cs:8,0.0123588023583358)
--(axis cs:7,0.0136460238408174)
--(axis cs:6,0.0150433860495485)
--(axis cs:5,0.0165555587817282)
--(axis cs:4,0.0181865755503488)
--(axis cs:3,0.0199396860382244)
--(axis cs:2,0.0218172044871087)
--(axis cs:1,0.0238203571300677)
--(axis cs:0,0)
--cycle;
\addlegendimage{area legend, draw=white, fill=cornflowerblue86180233}
\addlegendentry{PILC}
\end{axis}

\end{tikzpicture}

%% file: Oberrheincableages.tex
\begin{tikzpicture}

\definecolor{cornflowerblue86180233}{RGB}{86,180,233}
\definecolor{darkcyan0158115}{RGB}{0,158,115}
\definecolor{darkgray176}{RGB}{176,176,176}
\definecolor{lightgray204}{RGB}{204,204,204}
\definecolor{orange2301590}{RGB}{230,159,0}

\begin{axis}[height=0.13\textheight,width=0.8\columnwidth,
scale only axis,
legend cell align={left},
legend columns=3,
legend style={fill opacity=0.8, draw opacity=1, text opacity=1, draw=lightgray204},
tick align=outside,
tick pos=left,
x grid style={darkgray176},
xlabel={Age (years)},
xmin=-3.2, xmax=67.2,
xtick style={color=black},
xtick={-10,0,10,20,30,40,50,60,70},
xticklabels={
  ${−10}$,
  ${0}$,
  ${10}$,
  ${20}$,
  ${30}$,
  ${40}$,
  ${50}$,
  ${60}$,
  ${70}$
},
y grid style={darkgray176},
ylabel={Frequency},
ymin=0, ymax=40,
ytick style={color=black},
ytick={0,10,20,30,40},
]
\path [draw=orange2301590, semithick]
(axis cs:0,0)
--(axis cs:0,0)
--(axis cs:3.2,0)
--(axis cs:3.2,2)
--(axis cs:6.4,2)
--(axis cs:6.4,9)
--(axis cs:9.6,9)
--(axis cs:9.6,14)
--(axis cs:12.8,14)
--(axis cs:12.8,13)
--(axis cs:16,13)
--(axis cs:16,31)
--(axis cs:19.2,31)
--(axis cs:19.2,15)
--(axis cs:22.4,15)
--(axis cs:22.4,11)
--(axis cs:25.6,11)
--(axis cs:25.6,13)
--(axis cs:28.8,13)
--(axis cs:28.8,7)
--(axis cs:32,7)
--(axis cs:32,13)
--(axis cs:35.2,13)
--(axis cs:35.2,7)
--(axis cs:38.4,7)
--(axis cs:38.4,15)
--(axis cs:41.6,15)
--(axis cs:41.6,5)
--(axis cs:44.8,5)
--(axis cs:44.8,6)
--(axis cs:48,6)
--(axis cs:48,2)
--(axis cs:51.2,2)
--(axis cs:51.2,5)
--(axis cs:54.4,5)
--(axis cs:54.4,4)
--(axis cs:57.6,4)
--(axis cs:57.6,5)
--(axis cs:60.8,5)
--(axis cs:60.8,4)
--(axis cs:64,4)
--(axis cs:64,0);
\addlegendimage{line legend, draw=orange2301590, semithick}
\addlegendentry{$t={0}$}

\path [draw=cornflowerblue86180233, semithick, dashed]
(axis cs:0,0)
--(axis cs:0,15)
--(axis cs:3.2,15)
--(axis cs:3.2,9)
--(axis cs:6.4,9)
--(axis cs:6.4,9)
--(axis cs:9.6,9)
--(axis cs:9.6,11)
--(axis cs:12.8,11)
--(axis cs:12.8,10)
--(axis cs:16,10)
--(axis cs:16,5)
--(axis cs:19.2,5)
--(axis cs:19.2,4)
--(axis cs:22.4,4)
--(axis cs:22.4,19)
--(axis cs:25.6,19)
--(axis cs:25.6,0)
--(axis cs:28.8,0)
--(axis cs:28.8,2)
--(axis cs:32,2)
--(axis cs:32,13)
--(axis cs:35.2,13)
--(axis cs:35.2,14)
--(axis cs:38.4,14)
--(axis cs:38.4,18)
--(axis cs:41.6,18)
--(axis cs:41.6,22)
--(axis cs:44.8,22)
--(axis cs:44.8,15)
--(axis cs:48,15)
--(axis cs:48,15)
--(axis cs:51.2,15)
--(axis cs:51.2,0)
--(axis cs:54.4,0)
--(axis cs:54.4,0)
--(axis cs:57.6,0)
--(axis cs:57.6,0)
--(axis cs:60.8,0)
--(axis cs:60.8,0)
--(axis cs:64,0)
--(axis cs:64,0);
\addlegendimage{line legend, draw=cornflowerblue86180233, semithick, dashed}
\addlegendentry{$t={25}$}

\path [draw=darkcyan0158115, semithick, dash pattern=on 1pt off 3pt on 3pt off 3pt]
(axis cs:0,0)
--(axis cs:0,2)
--(axis cs:3.2,2)
--(axis cs:3.2,7)
--(axis cs:6.4,7)
--(axis cs:6.4,11)
--(axis cs:9.6,11)
--(axis cs:9.6,12)
--(axis cs:12.8,12)
--(axis cs:12.8,21)
--(axis cs:16,21)
--(axis cs:16,30)
--(axis cs:19.2,30)
--(axis cs:19.2,9)
--(axis cs:22.4,9)
--(axis cs:22.4,13)
--(axis cs:25.6,13)
--(axis cs:25.6,9)
--(axis cs:28.8,9)
--(axis cs:28.8,9)
--(axis cs:32,9)
--(axis cs:32,11)
--(axis cs:35.2,11)
--(axis cs:35.2,11)
--(axis cs:38.4,11)
--(axis cs:38.4,10)
--(axis cs:41.6,10)
--(axis cs:41.6,3)
--(axis cs:44.8,3)
--(axis cs:44.8,4)
--(axis cs:48,4)
--(axis cs:48,19)
--(axis cs:51.2,19)
--(axis cs:51.2,0)
--(axis cs:54.4,0)
--(axis cs:54.4,0)
--(axis cs:57.6,0)
--(axis cs:57.6,0)
--(axis cs:60.8,0)
--(axis cs:60.8,0)
--(axis cs:64,0)
--(axis cs:64,0);
\addlegendimage{line legend, draw=darkcyan0158115, semithick, dash pattern=on 1pt off 3pt on 3pt off 3pt}
\addlegendentry{$t={50}$}
\end{axis}

\end{tikzpicture}

%% file: Oberrhein_maintenance.tex
\begin{tikzpicture}

\definecolor{darkgray176}{RGB}{176,176,176}
\definecolor{orange2301590}{RGB}{230,159,0}

\begin{axis}[height=0.13\textheight,width=0.8\columnwidth,
scale only axis,
tick align=outside,
tick pos=left,
x grid style={darkgray176},
xlabel={Time $t$ (years)},
xmin=-4.95, xmax=103.95,
xtick style={color=black},
xtick={0,20,40,60,80,100,120},
xticklabels={
  ${0}$,
  ${20}$,
  ${40}$,
  ${60}$,
  ${80}$,
  ${100}$,
  ${120}$
},
y grid style={darkgray176},
ylabel={$N(t)$},
ymin=-0.9, ymax=20.1,
ytick style={color=black},
ytick={0,5,10,15,20},
]
\addplot[
  only marks,
  orange2301590,
  mark=o,
  mark size=2.2,
  mark options={solid, fill=orange2301590},
]
table {%
0 18
1 1
2 1
3 0
4 4
5 2
6 0
7 1
8 2
9 2
10 6
11 2
12 7
13 2
14 2
15 3
16 3
17 3
18 2
19 5
20 2
21 1
22 4
23 4
24 6
25 4
26 3
27 5
28 3
29 1
30 7
31 7
32 9
33 7
34 6
35 9
36 6
37 3
38 4
39 5
40 5
41 4
42 2
43 3
44 4
45 0
46 2
47 0
48 0
49 0
50 0
51 0
52 18
53 1
54 1
55 0
56 3
57 2
58 0
59 1
60 2
61 2
62 6
63 2
64 8
65 2
66 2
67 3
68 3
69 3
70 2
71 5
72 2
73 1
74 4
75 4
76 5
77 5
78 3
79 5
80 3
81 1
82 7
83 7
84 9
85 7
86 6
87 9
88 6
89 4
90 4
91 5
92 5
93 4
94 2
95 3
96 4
97 0
98 2
99 0
};
\end{axis}

\end{tikzpicture}

%% file: betasenpilc.tex
\begin{tikzpicture}

\definecolor{cornflowerblue86180233}{RGB}{86,180,233}
\definecolor{darkslategray38}{RGB}{38,38,38}
\definecolor{lightgray204}{RGB}{204,204,204}
\definecolor{orange2301590}{RGB}{230,159,0}

\begin{axis}[height=0.13\textheight,width=0.75\columnwidth,
scale only axis,
axis line style={darkslategray38},
legend cell align={left},
legend style={
  fill opacity=0.8,
  draw opacity=1,
  text opacity=1,
  at={(0.03,0.97)},
  anchor=north west,
  draw=lightgray204
},
tick align=outside,
x grid style={lightgray204},
xlabel=\textcolor{darkslategray38}{$\beta$},
xmajorticks=true,
xmin=2.7255, xmax=3.26449999999999,
xtick style={color=darkslategray38},
y grid style={lightgray204},
ylabel=\textcolor{darkslategray38}{$\frac{N(t_1)}{N(t_0)}$},
ymajorticks=true,
ymin=16.0967586655301, ymax=508.001593321145,
ytick style={color=darkslategray38}
]
\addplot [semithick, orange2301590, mark=Mercedes star, mark size=3.5, mark options={solid,rotate=270}, only marks]
table {%
2.75 190.569141658527
2.76 194.242240202475
2.77 197.98613537591
2.78 201.802191738667
2.79 205.691800151603
2.8 209.656378283527
2.81 213.697371127912
2.82 217.816251529556
2.83 222.014520721402
2.84 226.293708871698
2.85 230.655375641707
2.86 235.101110754171
2.87 239.632534572719
2.88 244.251298692458
2.89 248.959086541932
2.9 253.757613996699
2.91 258.648630004722
2.92 263.633917223819
2.93 268.7152926714
2.94 273.894608386725
2.95 279.173752105928
2.96 284.554647950053
2.97 290.039257126348
2.98 295.629578643079
2.98999999999999 301.32765003812
2.99999999999999 307.135548121586
3.00999999999999 313.055389732783
3.01999999999999 319.089332511741
3.02999999999999 325.239575685625
3.03999999999999 331.5083608703
3.04999999999999 337.897972887345
3.05999999999999 344.410740596818
3.06999999999999 351.049037746065
3.07999999999999 357.8152838349
3.08999999999999 364.711944997447
3.09999999999999 371.741534900996
3.10999999999999 378.906615662164
3.11999999999999 386.209798780735
3.12999999999999 393.653746091481
3.13999999999999 401.24117073434
3.14999999999999 408.974838143291
3.15999999999999 416.857567054288
3.16999999999999 424.892230532621
3.17999999999999 433.081757020077
3.18999999999999 441.429131402292
3.19999999999999 449.937396096664
3.20999999999999 458.609652161247
3.21999999999999 467.449060425007
3.22999999999999 476.458842639872
3.23999999999999 485.642282654981
};
\addlegendentry{PV-dominated}
\addplot [semithick, cornflowerblue86180233, mark=Mercedes star, mark size=3.5, mark options={solid}, only marks]
table {%
2.75 38.4560693316944
2.76 38.9698200890082
2.77 39.4904342581379
2.78 40.01800353028
2.79 40.5526208215725
2.8 41.0943802894589
2.81 41.6433773492716
2.82 42.1997086910362
2.83 42.763472296501
2.84 43.3347674563937
2.85 43.9136947879083
2.86 44.5003562524263
2.87 45.0948551734737
2.88 45.6972962549192
2.89 46.3077855994142
2.9 46.9264307270798
2.91 47.5533405944437
2.92 48.1886256136296
2.93 48.8323976718027
2.94 49.4847701508761
2.95 50.1458579474794
2.96 50.8157774931943
2.97 51.4946467750608
2.98 52.1825853563576
2.98999999999999 52.8797143976589
2.99999999999999 53.5861566781741
3.00999999999999 54.3020366173716
3.01999999999999 55.0274802968915
3.02999999999999 55.7626154827515
3.03999999999999 56.507571647849
3.04999999999999 57.2624799947642
3.05999999999999 58.0274734788677
3.06999999999999 58.8026868317364
3.07999999999999 59.5882565848831
3.08999999999999 60.3843210938022
3.09999999999999 61.1910205623371
3.10999999999999 62.0084970673734
3.11999999999999 62.8368945838611
3.12999999999999 63.676359010172
3.13999999999999 64.5270381937957
3.14999999999999 65.3890819573781
3.15999999999999 66.2626421251088
3.16999999999999 67.1478725494604
3.17999999999999 68.0449291382851
3.18999999999999 68.9539698822736
3.19999999999999 69.8751548827805
3.20999999999999 70.8086463800216
3.21999999999999 71.7546087816475
3.22999999999999 72.7132086916998
3.23999999999999 73.6846149399533
};
\addlegendentry{Wind-dominated}
\end{axis}

\end{tikzpicture}